\documentclass[a4paper,11pt]{article}
\pdfoutput=1 % if your are submitting a pdflatex (i.e. if you have
\usepackage{pub} % for details on the use of the package, please
\usepackage[T1]{fontenc} % if needed
\usepackage{float} % for option to force figure positions
\usepackage{feynmp-auto}

\usepackage{booktabs}

\usepackage{multirow}

\usepackage{makecell}

\usepackage[utf8]{inputenc}

\usepackage{subcaption}

\usepackage{xspace}

\usepackage{placeins}

\usepackage{siunitx}
\usepackage{physics}
\usepackage{nicefrac}
\newcommand{\mhh}{\ensuremath{m_{hh}}\xspace}

\newcommand{\lamhhh}{\ensuremath{\lambda_{hhh}}\xspace}
\newcommand{\topyuk}{\ensuremath{y_{t}}\xspace}
\newcommand{\pt}{\ensuremath{p_\text{T}}\xspace}
\newcommand{\klam}{\ensuremath{\kappa_{\lambda}}\xspace}
\newcommand{\kapt}{\ensuremath{\kappa_{t}}\xspace}

\title{\boldmath Higgs self-coupling measurements using deep learning in the $b\bar{b}b\bar{b}$ final state}

\author[a]{Jacob Amacker,}
\author[a]{William Balunas,}
\author[a]{Lydia Beresford,}
\author[a]{Daniela Bortoletto,}
\author[a]{James Frost,}
\author[a,b,c]{Cigdem Issever,}
\author[d]{Jesse Liu,}
\author[a]{James McKee,}
\author[a]{Alessandro Micheli,}
\author[a]{Santiago Paredes Saenz,}
\author[e]{Michael Spannowsky,}
\author[a]{and Beojan Stanislaus}

\affiliation[a]{Department of Physics, University of Oxford, 1 Keble Road, Oxford OX1 3RH, UK}
\affiliation[b]{Humboldt-Universit\"{a}t zu Berlin,
Institut f\"{u}r Physik, Newtonstra\ss e 15, 12489 Berlin, Germany}
\affiliation[c]{DESY, Platanenallee 6, D-15738 Zeuthen, Germany}
\affiliation[d]{Department of Physics, University of Chicago, 933 E 56th St, Chicago IL 60637, USA}
\affiliation[e]{Institute of Particle Physics Phenomenology, Durham University, Durham DH1 3LE, UK}

\emailAdd{william.balunas@physics.ox.ac.uk}
\emailAdd{lydia.beresford@physics.ox.ac.uk}
\emailAdd{daniela.bortoletto@physics.ox.ac.uk}
\emailAdd{james.frost@physics.ox.ac.uk}
\emailAdd{isseverc@physik.hu-berlin.de}
\emailAdd{jesseliu@uchicago.edu}
\emailAdd{santiago.paredes@physics.ox.ac.uk}
\emailAdd{michael.spannowsky@durham.ac.uk}
\emailAdd{beojan.stanislaus@physics.ox.ac.uk}

\abstract{
Measuring the Higgs trilinear self-coupling $\lamhhh$ is experimentally demanding but fundamental for understanding the shape of the Higgs potential. We present a comprehensive analysis strategy for the HL-LHC using di-Higgs events in the four $b$-quark channel ($hh\to 4b$), extending current methods in several directions. We perform deep learning to suppress the formidable multijet background with dedicated optimisation for BSM \lamhhh scenarios. We compare the \lamhhh constraining power of events using different multiplicities of large radius jets with a two-prong structure that reconstruct boosted $h\to bb$ decays. We show that current uncertainties in the SM top Yukawa coupling $y_t$ can modify \lamhhh constraints by $\sim 20\%$. For SM $y_t$, we find prospects of $-0.8 < \lambda_{hhh} / \lambda_{hhh}^\text{SM} < 6.6$ at 68\% CL under simplified assumptions for 3000~fb$^{-1}$ of HL-LHC data. Our results provide a careful assessment of di-Higgs identification and machine learning techniques for all-hadronic measurements of the Higgs self-coupling and sharpens the requirements for future improvement.}

\begin{document} 
\maketitle
\flushbottom

\section{Introduction}
\label{sec:intro}

Discovering Higgs boson pair production $pp\to hh$ opens the only direct laboratory probe of the Higgs trilinear self-coupling \lamhhh, which is a principal goal of the LHC and its upgrades~\cite{Baglio:2012np,deFlorian:2016spz,ATLAS:2013hta,CMS:2013xfa,Cepeda:2019klc,Atlas:2019qfx,DiMicco:2019ngk}. Measuring \lamhhh is critical for characterising the dynamics of electroweak symmetry breaking that could be modified by beyond the Standard Model (BSM) physics~\cite{Grojean:2004xa,Cao:2013si,Gouzevitch:2013qca,Gupta:2013zza,Han:2013sga,Nishiwaki:2013cma,Goertz:2014qta,Hespel:2014sla,Cao:2014kya,Azatov:2015oxa,Carena:2015moc,Grober:2015cwa,Wu:2015nba,He:2015spf,Carvalho:2015ttv,Zhang:2015mnh,Huang:2015tdv,Nakamura:2017irk,DiLuzio:2017tfn,Huang:2017nnw,Buchalla:2018yce,Borowka:2018pxx,Chang:2019vez,Blanke:2019hpe,Li:2019tfd,Capozi:2019xsi,Alves:2019igs,Kozaczuk:2019pet,Barducci:2019xkq,Huang:2019bcs,Cheung:2020xij}. Recent experimental~\cite{Aad:2015xja,Aaboud:2016xco,Aaboud:2018knk,Aad:2020kub,Sirunyan:2017isc,Sirunyan:2018qca,Sirunyan:2018tki} and phenomenological~\cite{Behr:2015oqq,Wardrope:2014kya,deLima:2014dta} advances in the four bottom quark channel $hh \to 4b$ suggest that this final state is competitive for discovery~\cite{Aaboud:2018sfw,Sirunyan:2017tqo,Aaboud:2018ewm,Aaboud:2018ftw,Khachatryan:2016sey}. Current 95\% CL limits from ATLAS~\cite{Aad:2019uzh} (CMS~\cite{Sirunyan:2018two}) on the dominant gluon fusion cross-section reach 6.9 (22.2) times the SM, with the $4b$ channel being second most constraining in ATLAS. 

For direct self-coupling $\lamhhh$ constraints, the present best combined 95\% CL limit with respect to the SM value $\lamhhh^\text{SM}$ is $-5.0 < \lamhhh/\lamhhh^\text{SM} < 12.0$, where $4b$ is among the most competitive channels for values near $\lamhhh^\text{SM}$~\cite{Aad:2019uzh}.
ATLAS also reports tighter limits of $-2.3 < \lamhhh/\lamhhh^\text{SM} < 10.3$ when combining with indirect constraints from single Higgs channels~\cite{ATLAS-CONF-2019-049}.
In the $4b$ channel, these constraints and High Luminosity LHC (HL-LHC) projections~\cite{ATL-PHYS-PUB-2018-053,CMS-PAS-FTR-18-019} only optimise for the SM coupling and without using events with boosted Higgs bosons. However, non-SM \lamhhh values can modify Higgs boson kinematics such that these analyses are no longer optimal for constraining \lamhhh. Boosted Higgs decays~\cite{Butterworth:2008iy,ATL-PHYS-PUB-2014-013,Asquith:2018igt,Aad:2019uoz} are targeted in resonant di-Higgs searches~\cite{Aaboud:2018sfw,Sirunyan:2018qca}, but not for self-coupling constraints. Interestingly, Ref.~\cite{Behr:2015oqq} suggests that these boosted reconstruction techniques can improve the discovery significance of $hh\to 4b$. 
Recent progress in $h\to bb$ analyses~\cite{Aaboud:2018zhk,Sirunyan:2018kst,Sirunyan:2017dgc,ATLAS-CONF-2018-052,Aaboud:2018urx,Aaboud:2017rss,Sirunyan:2018hoz,Sirunyan:2018mvw} and the large signal statistics from the high branching ratio $\mathcal{B}(h \to bb) \simeq 58\%$ motivate use of advanced techniques for signal characterisation.

This paper synthesises these separate advances to present a comprehensive assessment of HL-LHC analysis strategies across all kinematic regimes of the $hh\to 4b$ channel to evaluate and improve \lamhhh constraints. We now discuss the specific novelties of this paper. 

\begin{figure}
    \centering
    \begin{subfigure}[b]{0.5\textwidth}
        \centering
        \includegraphics[height=2in]{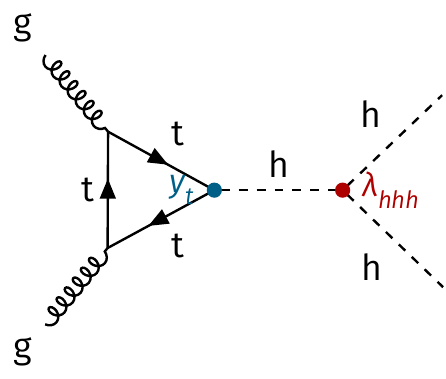}
        \caption{`Triangle' amplitude $\propto y_t \lambda_{hhh}$}
    \end{subfigure}%
    \begin{subfigure}[b]{0.5\textwidth}
        \centering
        \includegraphics[height=2in]{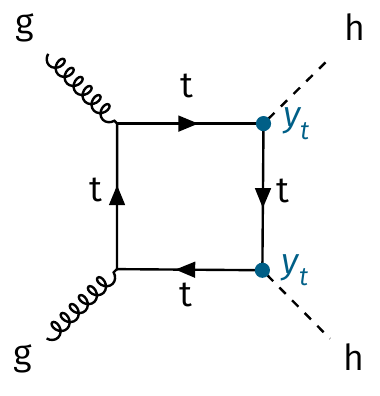}
        \caption{`Box' amplitude $\propto y_t^2$}
    \end{subfigure}
    \caption{Leading order Feynman diagrams of Higgs boson pair production in gluon fusion. The (a) `triangle' amplitude features the trilinear self-coupling \lamhhh and one power of the top Yukawa \topyuk, and (b) `box' amplitude does not feature \lamhhh and has two powers of \topyuk. These two diagrams interfere destructively in the SM. }
    \label{fig:feyngraphs}
\end{figure}

We show that analyses optimised for discovery sensitivity of $hh$ with SM couplings are suboptimal for constraining \lamhhh at the boundaries of projected limits. This is because SM $hh$ production has more boosted Higgs bosons than scenarios where $\lamhhh / \lamhhh^\text{SM} \sim 5$ due to the relative contributions of the interfering production amplitudes (Fig.~\ref{fig:feyngraphs}). Reconstructing Higgs bosons with lower boosts is more limited by trigger thresholds, but we nonetheless demonstrate that optimising for signals with non-SM \lamhhh rather than SM values can improve \lamhhh constraints. Furthermore, due to the small signal-to-background ratios, our sensitivity is limited by systematics, whose size we vary to quantify its impact on \lamhhh constraints. Reducing the formidable multijet background can mitigate the impact of systematics, whose suppression relies on modern $b$-tagging algorithms~\cite{ATL-PHYS-PUB-2015-022,Sirunyan:2017ezt,Aaboud:2018xwy}. Our study adopts expected improvements of the impact parameter resolution from inner tracker upgrades crucial for $b$-tagging. 

To further enhance sensitivity, we use neural networks~\cite{Lippmann1987AnIT,Hornik:1989,Hornik1991ApproximationCO} for deep learning, which are witnessing widespread applications in particle physics~\cite{Baldi:2014kfa,Baldi:2014pta,deOliveira:2015xxd,Baldi:2016fzo,Caron:2016hib,Chang:2017kvc,Lin:2018cin,Albertsson:2018maf,Guest:2018yhq,Abdughani:2019wuv,Windischhofer:2019ltt}. We apply these state-of-the-art techniques to reject backgrounds and improve constraining power of \lamhhh. We utilise a recently developed framework called SHapley Additive exPlanations (SHAP)~\cite{NIPS2017_7062} to show what physics information the neural network learns.

Additionally, we investigate the impact of experimental effects such as the finite resolution of jet reconstruction and trigger limitations. This presents a more complete picture of the limitations on the sensitivity of the $hh \to 4b$ channel beyond previous studies. We also study the composition of the background in detail, including small backgrounds (such as single Higgs production) and the impact of a neural network selection on this composition and shape of background distributions. This is important as uncertainties in background composition affect the experimental systematics of current analyses.

Finally, Fig.~\ref{fig:feyngraphs} shows that only the triangle diagram contains \lamhhh while the box diagram contains two factors of the top Yukawa \topyuk. Current \topyuk uncertainties are $\sim 20\%$ from $pp \to t\bar{t}h$ measurements~\cite{Aaboud:2018urx,Sirunyan:2018hoz}, which we account for in our results due to it being sufficiently large to impact \lamhhh constraints as demonstrated in Ref.~\cite{ATLAS-CONF-2019-049}. Our \topyuk-dependent HL-LHC projections of \lamhhh constraints allowing equal-footing comparison of sensitivity for different boosted Higgs topologies in the $4b$ channel are new.

While we focus on $4b$ final states, techniques from these studies are readily applied to other channels, such as $bb\tau\tau$ and $bb\gamma\gamma$ that involve $h\to bb$ decays. Complementary production and decay modes for $hh$ production are explored in Refs.~\cite{Dolan:2012rv,Papaefstathiou:2012qe,Barr:2013tda,Chen:2014xra,Dawson:2015oha,Lu:2015jza,Kling:2016lay,Bizon:2016wgr,Bishara:2016kjn,Adhikary:2017jtu,Alves:2017ued,Huang:2017jws,Kim:2018uty,Chang:2018uwu,Kim:2018cxf,Basler:2018dac,Chang:2019ncg,Arganda:2018ftn,Cao:2015oxx}. 

This paper is structured as follows. Section~\ref{sec:samples} outlines the Monte Carlo simulation of the signal and background processes together with discussion of detector emulation. Section~\ref{sec:analyses} describes the baseline and neural network analysis strategies. Section~\ref{sec:constraints} presents the results and statistical analysis before section~\ref{sec:conclusion} summarises our conclusions.
\section{Signal and background modelling}
\label{sec:samples}

In the SM after electroweak symmetry breaking, the Higgs self-coupling is determined by the Higgs potential 
\begin{equation}
    V(h) = m_h^2 h^2 + \lambda_{hhh}vh^3 + \lambda_{hhhh}h^4.
\end{equation}
This paper focuses on measuring the trilinear coupling \lamhhh. This will directly test the SM prediction $m_h^2 = \lamhhh v^2$, and any deviation from this value is indicative of BSM physics. The Higgs boson mass $m_h=125$~GeV and electroweak vacuum expectation value $v=246$~GeV are fixed to their independently measured values. Directly probing the quartic self-coupling $\lambda_{hhhh}$ requires triple Higgs boson production~\cite{Chen:2015gva,Liu:2018peg,Bizon:2018syu,Papaefstathiou:2019ofh}, which is beyond the scope of this work due to its highly suppressed cross-section. 
This section details the Monte Carlo simulation of the signal and background used for our analyses. We outline the phenomenology of Higgs pair production along with its simulation  (subsection~\ref{sec:hhpheno}), background processes (subsection~\ref{sec:bkg_proc}), and detector emulation (subsection~\ref{sec:detector}).

%------------------------------------------
\subsection{\label{sec:hhpheno}Higgs pair production }
%------------------------------------------
At leading order, the Higgs pair production cross-section has contributions from amplitudes we refer to as `triangle' and `box', shown in Fig.~\ref{fig:feyngraphs}, and their interference. These three contributions scale with the top Yukawa and trilinear self-coupling as
\begin{align}
    \sigma_\text{triangle} \sim \lamhhh^2 \topyuk^2, \quad 
    \sigma_\text{box} \sim \topyuk^4, \quad 
    \sigma_\text{interference} \sim -\lamhhh \topyuk^3.
    \label{eq:diHiggs_xsec}
\end{align} 
Sensitivity to \lamhhh therefore depends on probing the triangle and interference amplitudes. The comparatively stronger dependence of the total cross-section on \topyuk highlights the importance of top Yukawa constraints for \lamhhh determination. The interference term probes the sign of \lamhhh, as it has a relative negative sign due to an additional fermion propagator in the box loop with respect to the triangle. This implies destructive interference for $\lamhhh > 0$. In the $|\lamhhh|/ \topyuk \to \infty$ limit, the total cross-section asymptotes to the same value for either sign of \lamhhh as the triangle contribution dominates over the interference.

\begin{figure}
    \centering
    \includegraphics[width=0.67\textwidth]{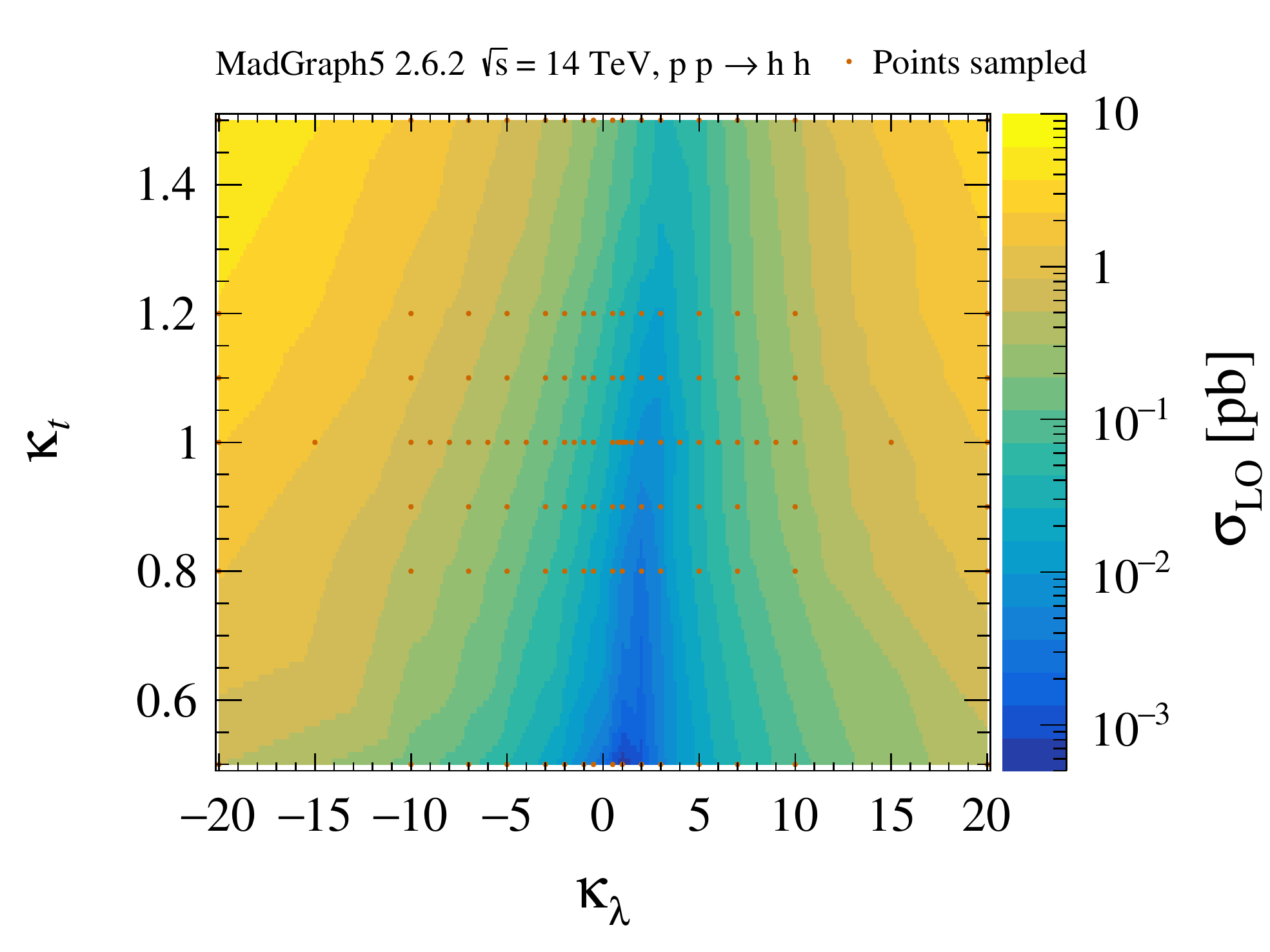}
    \caption{Leading order cross-sections $\sigma_\text{LO}$ from \textsc{MadGraph}~\cite{Alwall:2011uj,Alwall:2014hca} as a function of the variation of Higgs trilinear self-coupling \klam and top Yukawa coupling \kapt from their SM values $\lamhhh^\text{SM}, \topyuk^\text{SM}$ using the \textsc{HeavyHiggsTHDM} model~\cite{ATL-HeavyHiggsTHDM}. The orange markers indicate the points sampled from this parameter space.}
    \label{fig:xsec_14TeV_hh_contour}
\end{figure}

To modify couplings ($\lamhhh, \topyuk$) away from SM values $\left(\lamhhh^\text{SM},\topyuk^\text{SM} \right)$, keeping all other SM parameters fixed, we use the \textsc{HeavyHiggsTHDM} model~\cite{ATL-HeavyHiggsTHDM}. This adopts the `improved effective field theory' prescription, which uses gluon--Higgs vertices in the large quark mass limit and is improved by accounting for finite top mass effects as discussed in Ref.~\cite{Frederix:2014hta}. Such corrections are important given loop momenta are comparable to the top mass pole. We set couplings to all BSM particles to zero. We vary the top Yukawa $y_t$ while fixing the top mass to $m_t=172$~GeV. We define $\klam = \lamhhh/ \lamhhh^\text{SM}$ and $\kapt = y_t/ y_t^\text{SM}$ to parameterise variations from the SM couplings, as is conventional in the so-called `kappa framework'~\cite{deFlorian:2016spz}. This is a standard prescription, but to consistently study global constraints with other measurements, a full effective field theory treatment is recommended~\cite{Grzadkowski:2010es,Brivio:2017vri,Ellis:2018gqa}; this is deferred to future work.

To sample points in the two-dimensional signal parameter space $\left(\lamhhh, \topyuk \right)$, we employ \textsc{MadGraph}~2.6.2~\cite{Alwall:2011uj,Alwall:2014hca}. We consider only the dominant production mode via gluon fusion $gg\to hh$, inclusive of all Higgs boson decays. The NNPDF3.0 parton distribution functions~\cite{Ball:2014uwa} at next-to-leading order (NLO) are used from the LHAPDF package~\cite{Buckley:2014ana}. We generate 100k Monte Carlo (MC) events per point and calculate cross-sections at leading order (LO) in the strong coupling constant $\alpha_s$. Figure~\ref{fig:xsec_14TeV_hh_contour} illustrates the sampled points and LO cross-sections calculated by \textsc{MadGraph}, where the SM value is around $\sigma_\text{LO}^\text{SM} = 16$~fb at $\sqrt{s}=14$~TeV, and Table~\ref{tab:bkgs} shows LO cross-sections for example points.  For $\kapt = 1$, the LO cross-section falls to a minimum of 6.2~fb at around $\klam = 2.5$ due to destructive interference. The cross-section gradient is also shallow around this minimum, rising by only $\sim 15\%$ to 7.0~fb and 7.3~fb for $\klam = 2$ and $\klam = 3$ respectively. This makes \lamhhh constraints around such values challenging using inclusive cross-section measurements alone. For the training of the neural network (subsection~\ref{sec:NNsel}), we use an identical generator configuration to produce a dedicated set of high statistics samples with 250k events per \klam variation for fixed $\kapt = 1$ and use \textsc{MadGraph} to decay both Higgs bosons via $h\to bb$. 

\begin{figure}
    \centering
    \begin{subfigure}[b]{0.5\textwidth}
        \includegraphics[width=\textwidth]{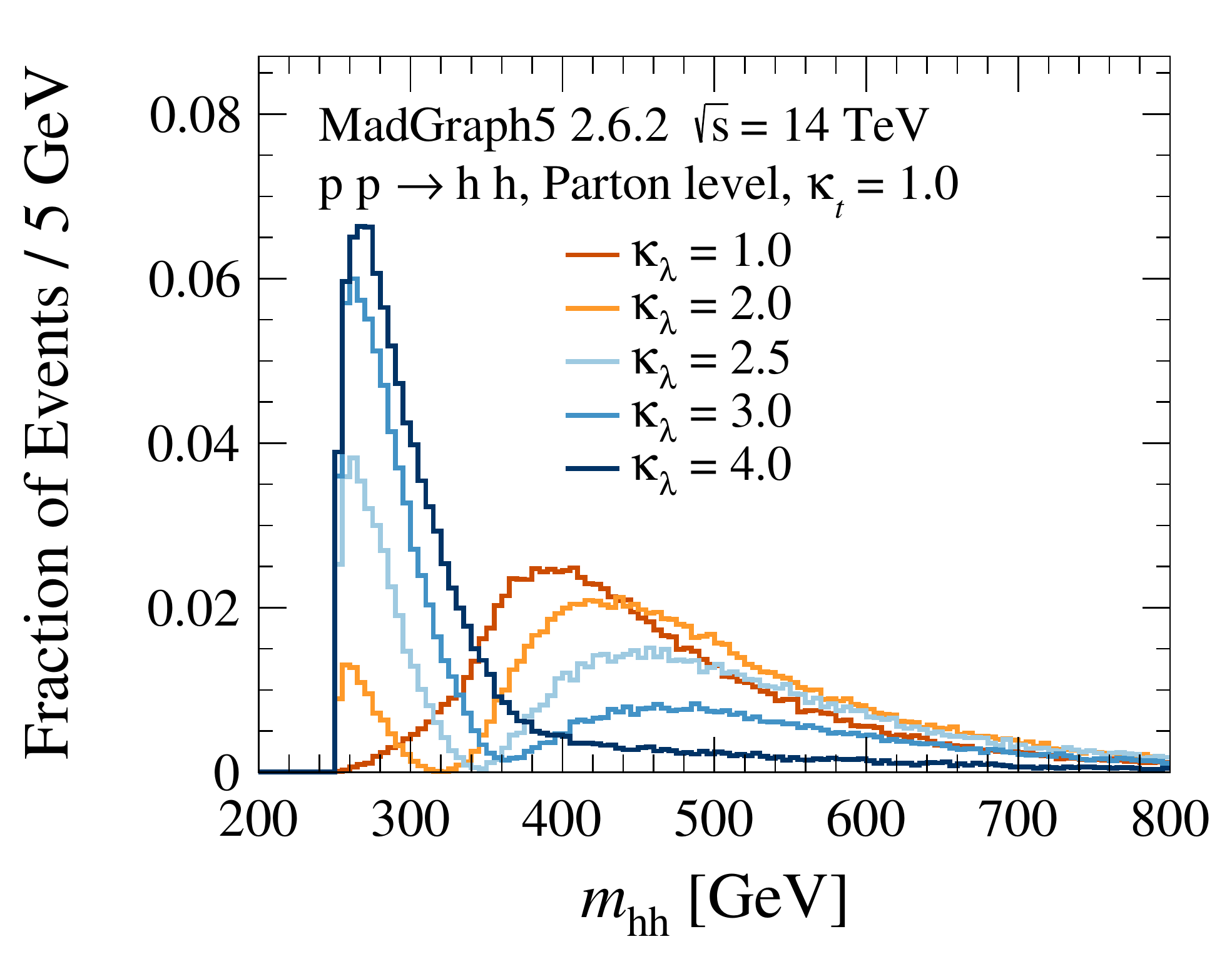}\\
        \includegraphics[width=\textwidth]{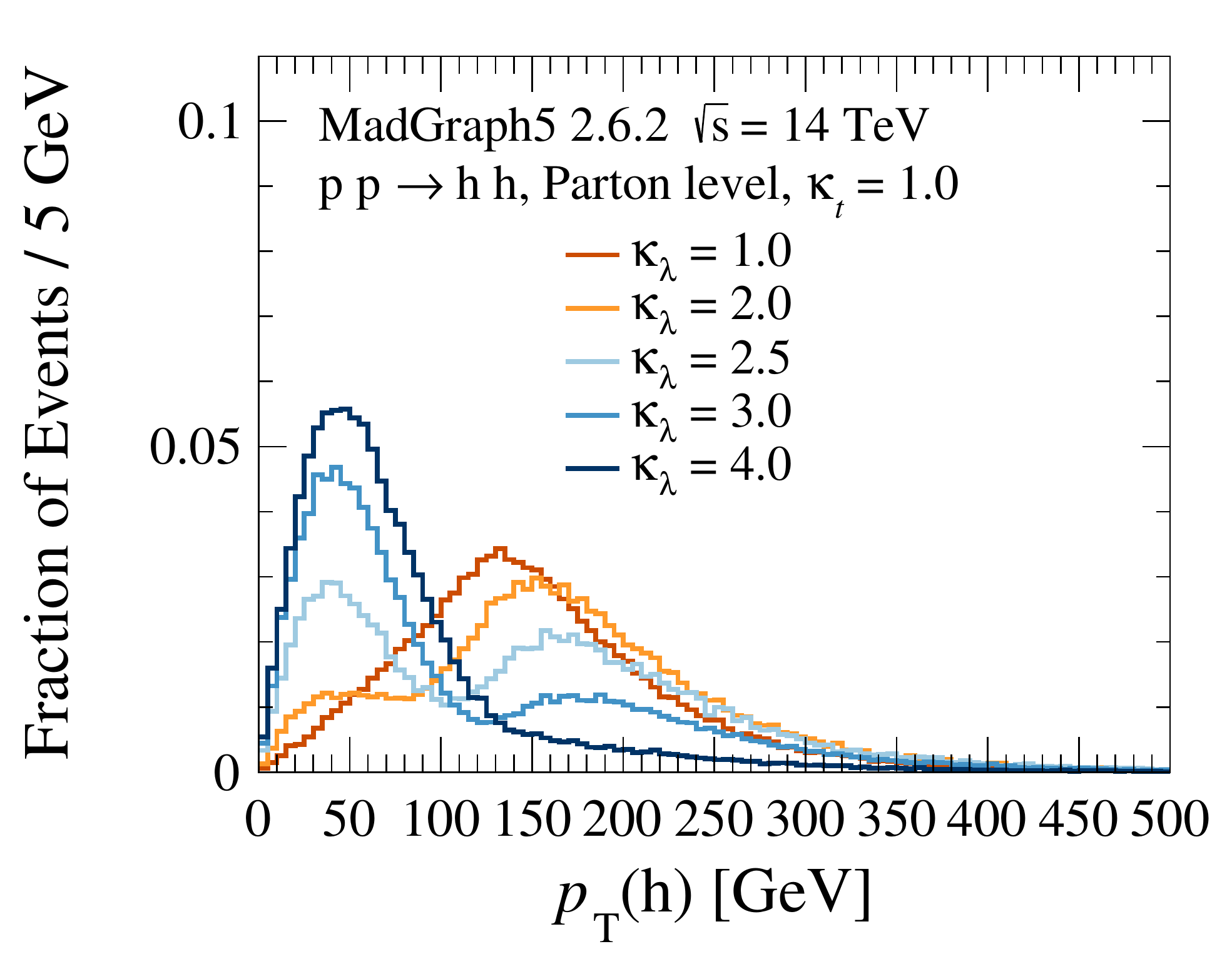}
        \caption{\label{fig:Norm_DiHiggsM_a}$\klam$ near maximal destructive interference}
    \end{subfigure}%
    \begin{subfigure}[b]{0.5\textwidth}
        \includegraphics[width=\textwidth]{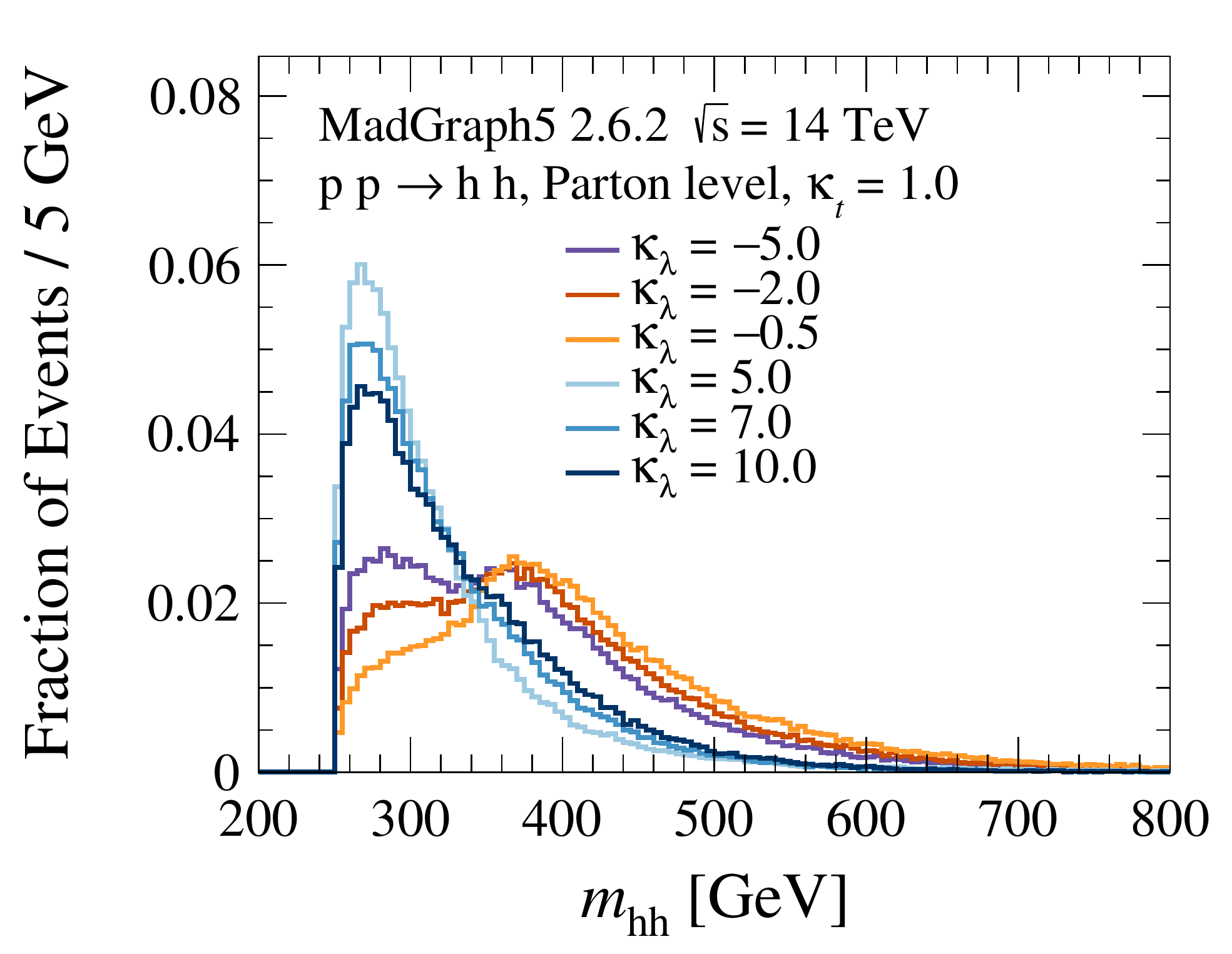}\\
        \includegraphics[width=\textwidth]{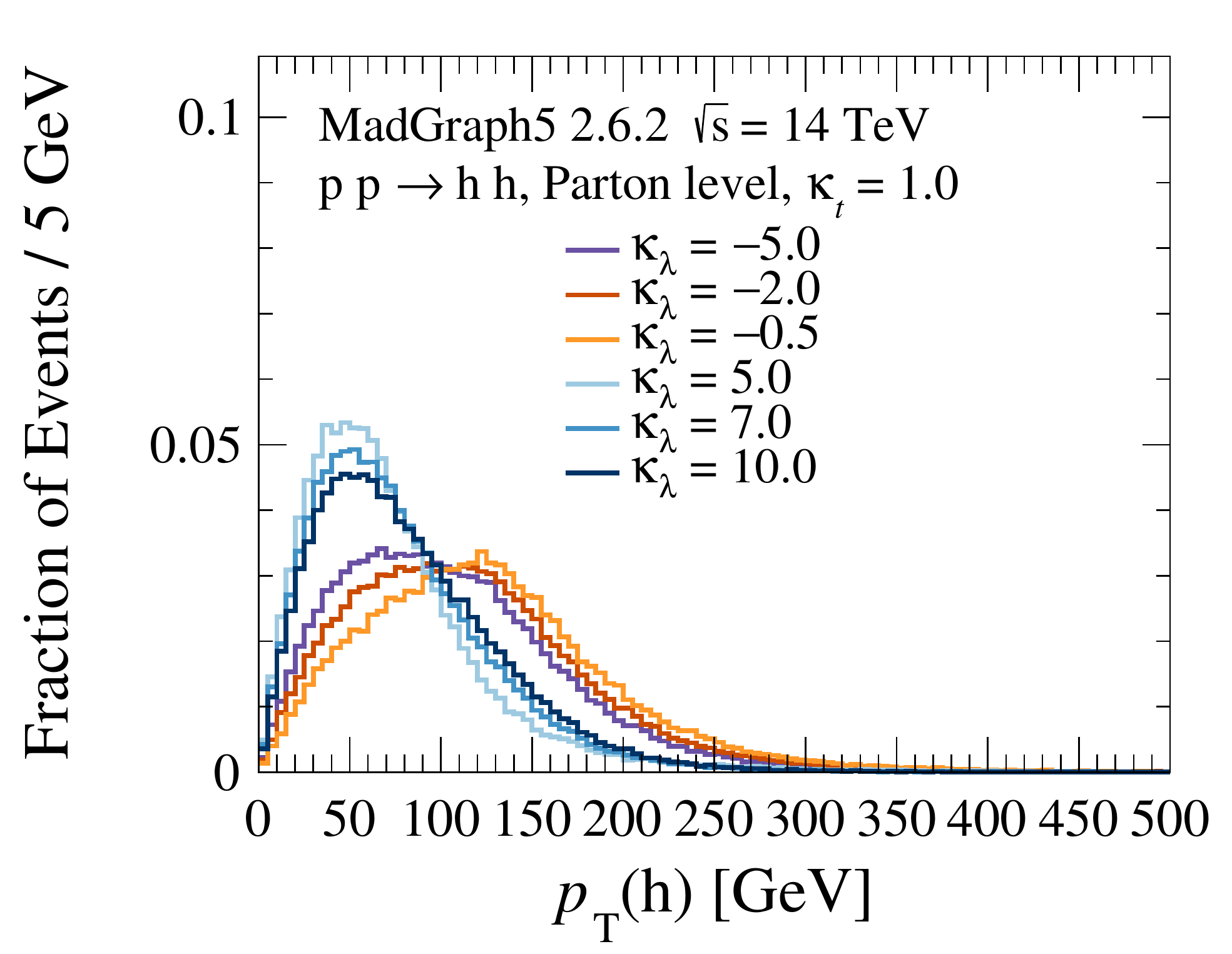}
    \caption{\label{fig:Norm_DiHiggsM_b}$\klam$ constructive vs destructive interference}
    \end{subfigure}
    \caption{Unit normalised distributions of the (upper) di-Higgs invariant mass \mhh and (lower) Higgs transverse momentum $\pt(h)$ at parton level. The \klam values highlight differential effects (a) around the point of maximal destructive interference $\klam \simeq 2.5$ and (b) of constructive vs destructive interference arising from the sign of \klam. 
    }
    \label{fig:Norm_DiHiggsM}
\end{figure}

We normalise the signal rate to NNLO\footnote{The cross-section for $\klam =1$ has recently been calculated to N3LO accuracy~\cite{Chen:2019lzz,Chen:2019fhs}.}. We start from the LO \textsc{MadGraph} cross-sections and improve their accuracy by applying $\lamhhh$-dependent LO-to-NLO k-factors $K_\lambda' = \sigma_\text{NLO} / \sigma_\text{LO}$ and a $\lamhhh$-independent NLO-to-NNLO k-factor $K'' = \sigma_\text{NNLO} / \sigma_\text{NLO}$ such that $\sigma_\text{NNLO} = K_\lambda' K'' \sigma_\text{LO}^\text{MG}$. For $K_\lambda'$, we use the NLO cross-sections for non-SM variations of \lamhhh from \textsc{Powheg-Box-v2}~\cite{Heinrich:2019bkc}. The $\lamhhh$-dependent LO-to-NLO k-factor $K_\lambda' = \sigma_\text{NLO} / \sigma_\text{LO}$ is 1.66 for $\klam = 1$ and varies by 35\% between $-1< \klam < 5$, ranging from $K_\lambda' = 1.56$ at $\klam = 2$ to 2.14 for $\klam = 5$~\cite{Heinrich:2019bkc}. For $\klam < -1$ and $>5$, we smoothly extrapolate the k-factors in Ref.~\cite{Heinrich:2019bkc} to asymptote at 1.95 and 2.1 respectively. In principle, there is a \topyuk-dependence on these NLO k-factors, but for simplicity we take these to be $\kapt$-independent. For the NLO-to-NNLO k-factor $K''$, the cross-section for $\klam =1$ at NNLO and next-to-next-to-leading logarithm (NNLL) accuracy~\cite{deFlorian:2013jea,deFlorian:2015moa}, with finite top mass corrections to NLO accuracy~\cite{Borowka:2016ypz,Borowka:2016ehy,Davies:2019dfy,Baglio:2018lrj,Baglio:2020ini}, gives $\sigma_\text{NNLO} = 39.6$~fb~\cite{deFlorian:2016spz}. Based on this, $\sigma_\text{LO}^\text{MG} = 16$~fb and $K_\lambda' = 1.66$, we derive $K'' = (\sigma_\text{NNLO}/\sigma_\text{LO}^\text{MG}) / K_\lambda' = 1.45$. Table~\ref{tab:bkgs} shows the LO cross-sections and overall k-factors  $K_\lambda' K''$ applied for some example signals. 

\begin{table}[t]
\centering
\begin{tabular}{llllll}
\toprule
Process    & $\pt(j_1^\text{gen})$ [GeV] & $\sigma_\text{LO}^\text{MG}$ [pb] & $N_\text{events}^\text{gen} (\times 10^6)$ & $\mathcal{L}_\text{eff}$ [fb$^{-1}$] & k-factor\\
\midrule
$2b2j$           & [20, 200]                & $2.3\times 10^4$                  & 2                                                & 0.087                                       & \multirow{4}{*}{1.3}\\
$2b2j$           & [200, 500]               & $1.5 \times 10^3$                 & 2                                                & 1.3                                        \\
$2b2j$           & [500, 1000]              & 35.3                              & 2                                                & 56.7                                       \\
$2b2j$           & $>1000$                  & 0.706                             & 1.1                                              & 1560                                       \\
\midrule
$4b$             & [20, 200]                & 63.2                              & 2                                                & 31.7                                       & \multirow{4}{*}{1.6}\\
$4b$             & [200, 500]               & 2.82                              & 2                                                & 710                                        \\
$4b$             & [500, 1000]              & 0.041                             & 2                                                & $4.9 \times 10^4$                          \\
$4b$             & $>1000$                  & $5.5\times 10^{-4}$               & 2                                                & $3.6 \times 10^6$                          \\
\midrule
$t\bar{t}$       & ---                      & 532                               & 2.1                                              & 3.95                                       & 1.4\\
$t\bar{t}+b\bar{b}$ & ---                      & 2.7                               & 1.0    
    & 370                                   & ---\\
$t\bar{t}h$      & ---                      & 0.44                              & 1.0                                              & 2300                                       & ---\\
$b\bar{b}h$      & ---                      & 0.076                             & 1.0                                              & $1.3\times 10^4$                           & ---\\
\midrule
$ZZ$             & ---                      & 11.5                              & 1.0                                              & 87                                       & ---\\
$Zh$             & ---                      & 0.72                              & 1.0                                              & 1400                                       & ---\\
$Wh$             & ---                      & 1.4                               & 1.0                                              & 710                                       & ---\\
\midrule\midrule
\multicolumn{4}{l}{$hh$ signal $(\klam, \kapt)$}\\
\midrule
$(1, 1)_\text{SM}$ & --- & 0.016   & 0.1 & 6200  & $2.4$\\
$(2, 1)$           & --- & 0.0076  & 0.1 & 13000 & $2.3$\\
$(3, 1)$           & --- & 0.0082  & 0.1 & 12000 & $2.7$\\
$(5, 1)$           & --- & 0.037   & 0.1 & 2700  & $3.1$\\
$(10, 1)$          & --- & 0.26    & 0.1 & 380  & $3.1$\\
\bottomrule
\end{tabular}
\caption{\label{tab:bkgs}Summary of Monte Carlo samples. The columns indicate any explicit parton level generator preselection on the leading parton $\pt(j_1^\text{gen})$, the leading order \textsc{MadGraph} cross-section $\sigma_\text{LO}^\text{MG}$, the number of events generated $N_\text{events}^\text{gen}$, and an estimate of the effective luminosity $\mathcal{L}_\text{eff} = N_\text{events}^\text{gen} / \sigma_\text{LO}^\text{MG}$. The k-factor is a scaling factor applied to account for NLO corrections for the dominant backgrounds, and overall higher order corrections for the signal described in the main text. Information for a representative set of signals are displayed below the double rule. For the signals, higher statistics samples of 250k events per point are generated for neural network training as discussed in the main text. }
\end{table}

Turning to the differential distributions, variations in ($\lamhhh, \topyuk$) away from SM values induce striking features illustrated in Fig.~\ref{fig:Norm_DiHiggsM}. These are shown at parton level (`parton level' refers to the $b$-quarks produced by the Higgs decays, before hadronisation or showering), with other variables found in appendix~\ref{sec:distributions}. Ideally, analyses should have a high signal-to-background ratio $S/B$ across different kinematic regimes of the Higgs boson as their boosts can vary rapidly with variations in coupling.

Figure~\ref{fig:Norm_DiHiggsM_a} shows \klam values between 1 and 4 inclusive, where destructive interference is near maximal and small \klam variations cause dramatic changes to the \mhh and $\pt(h)$ distributions. For SM couplings, events are suppressed at low values $250 < \mhh \lesssim 300$~GeV and peaks at $\mhh \simeq 400$~GeV, resulting in higher \pt Higgs bosons. For $\klam = 2$, destructive interference causes the $\mhh$ shape to vanish at $\mhh \simeq 320$~GeV. The signal instead occupies localised regions on either side of this minimum: near the kinematic threshold and at higher values where the Higgs bosons are more boosted. As \klam increases to $\klam = 3$ and 4, many events shift to low $\mhh \lesssim 350$~GeV, where the Higgs bosons have comparatively lower $\pt$. 

Figure~\ref{fig:Norm_DiHiggsM_b} shows how the sign of \klam impacts the \mhh and $\pt(h)$ distributions. For scenarios with destructive interference $\klam > 0$, the signals $\klam \geq 5$ all occupy low \mhh values peaking near 250~GeV. Interestingly, as \klam increases from 5 to 10, the \mhh and $\pt(h)$ spectra become slightly harder. Meanwhile, scenarios with constructive interference $\klam < 0$ have a greater proportion of signal events occupying comparatively higher \mhh and $\pt(h)$ values. These qualitatively different kinematic features can help lift degeneracies in the sign and value of \lamhhh that give the same total cross-sections. 

Physically, these features arises due to kinematic thresholds of the interfering amplitudes. The triangle amplitude tends to dominate at lower \mhh, while the box diagram has a kinematic threshold at approximately twice the top mass $2m_\text{top}$ and impacts larger \mhh values. Reconstructing these differential features, especially where the cross-section gradient is small, can improve \lamhhh sensitivity.

%------------------------------------------
\subsection{\label{sec:bkg_proc}Background processes}
%------------------------------------------

We calculate cross-sections and produce MC events at LO for background processes using the same generator configuration as the signals, which are summarised in Table~\ref{tab:bkgs}. We refer to processes with four real $b$-quarks at parton level as `irreducible' backgrounds. Processes with fewer than four real $b$-quarks can enter the analysis selection if a jet is misidentified as a $b$-jet, which we refer to as `reducible' backgrounds. We apply k-factors from Ref.~\cite{Behr:2015oqq} to account for NLO corrections for the dominant backgrounds shown in Table~\ref{tab:bkgs}.

We generate two sets of multijet processes: irreducible $b\bar{b}b\bar{b}$ $(4b)$ and reducible $b\bar{b}j\bar{j}$ $(2b2j)$. We define light-flavour partons $j$ as all quarks and gluons except bottom and top $j\in\{u,d,s,c,g\}$. The $2b2j$ cross-section is two orders of magnitude larger than $4b$, so is substantial even with powerful light jet rejection from $b$-tagging algorithms. We do not simulate $4j$ to preserve computational resources and Ref.~\cite{Behr:2015oqq} showed that this process is subdominant compared to $4b$ and $2b2j$. To improve statistics in high $\pt$ tails, we generate events for ranges of leading parton $\pt(j_1^\text{gen})$ in \textsc{MadGraph}, sliced with lower bin edges at $20, 200, 500, 1000$~GeV, where leading parton refers to $b$-quarks or light partons $j$. 

We generate top quark pairs $t\bar{t}$ and the irreducible $t\bar{t}+b\bar{b}$ process at the matrix element level, which together are expected to comprise around 10\% of the background rates. For single Higgs processes in associated production, we consider $Wh, Zh, t\bar{t}h, b\bar{b}h$. These processes typically have cross-sections that are an order of magnitude greater than di-Higgs, and the presence of a Higgs boson increases the probability of events passing analysis selections than other electroweak processes. Finally, we consider the diboson $ZZ$ process, which has a modest cross-section and its $4b$ final state constitutes an irreducible background. We generate at least one million events for each of these processes.

%------------------------------------------
\subsection{\label{sec:detector}Detector emulation }
%------------------------------------------

For all signal and background samples, the decay, parton shower, hadronisation, and underlying event are modelled by \textsc{Pythia}~8.230~\cite{Sjostrand:2007gs}. To save computational resources for our simplified study, we do not emulate pileup. We expect that detector upgrades~\cite{Collaboration:2285584,Collaboration:2623663,CMS:2667167} and improvements in pileup mitigation techniques~\cite{Tseng:2013dva,Bertolini:2014bba,Cacciari:2014gra,Komiske:2017ubm,Berta:2019hnj} will reduce the impact of pileup. To emulate reconstruction effects, we use \textsc{Delphes}~3.4.1~\cite{deFavereau:2013fsa} and assume the default ATLAS configuration card unless stated otherwise. We define three sets of reconstructed jets using the anti-$k_t$ clustering algorithm~\cite{Cacciari:2008gp,Cacciari:2011ma} with radius parameter $R$: 
\begin{itemize}
    \item \emph{Small jets} ($j_S$) are defined with $R=0.4$ and cluster only calorimeter towers. We impose $\pt > 40$~GeV  on all these jets to emulate the detector trigger requirement. We require these to be within a tracking acceptance of $|\eta| < 2.5$ to allow $b$-tagging, where $\eta$ is the pseudorapidity.
    \item \emph{Large jets} ($j_L$) are defined with $R=1.0$, also clustering only calorimeter towers. Based on the expected kinematics of boosted Higgs bosons $\pt \gtrsim 2m_h$, we require these to satisfy $\pt > 250$~GeV and be central $|\eta| < 2.0$.
    \item \emph{Track jets} ($j_T$) are defined by clustering only tracking information using $R=0.2$. We impose kinematic requirements of $\pt > 20$~GeV and $|\eta|< 2.5$. We associate these track jets to large jets if their distance satisfies $\Delta R(j_T, j_L) < 1.0$, where $\Delta R = \sqrt{(\Delta \phi)^2 + (\Delta \eta)^2}$.
\end{itemize}
For simplicity, we tag the flavour of both small jets and track jets using the same default implementation in \textsc{Delphes}, which parameterises tagging efficiencies based on the truth quark flavour of the jet. The $\pt$-dependent efficiencies of bottom, charm, and light jet ($u, d, s, g$) mistag rates are based on the `70\% working point' of the ATLAS multivariate MV2c20 algorithm~\cite{ATL-PHYS-PUB-2015-022}. The $b$-tagging efficiency peaks at 74\% for $\pt \simeq 150$~GeV, falling to 50\% at $\pt \simeq 500$~GeV. The charm mistag rate is around 10\%, peaking at 14\% for $\pt \simeq 100$~GeV before falling to 7\% at $\pt \simeq 500$~GeV. The light mistag rate remains on the order of 1\% throughout the $\pt$ regimes of interest. We do not increase the $b$-tagging range to $|\eta|< 4.0$ expected from the detector upgrades~\cite{ATLAS:ITk:Pixel:TDR} because we expect the region $|\eta|>2.5$ to be dominated by the gluon splitting background $g\to bb$ whereas $b$-quarks from Higgs are more central. However, we follow Ref.~\cite{ATL-PHYS-PUB-2018-053} to emulate expected $b$-tagging improvements of 8\% per jet in $|\eta|<2.5$ by scaling the $4b$ background and $hh$ signals by a factor of 1.36, and the $2b2j$ and $t\bar{t}$ backgrounds by 1.17. To improve MC statistics after $b$-tagging, we multiply events by these efficiencies as a weight rather than discarding non-$b$-tagged events. For processes with a low number of real $b$-quarks such as $2b2j$, this can improve the number of raw MC events by up to three orders of magnitude when requiring four $b$-tagged jets.

Electrons and muons must be within the tracking acceptance $|\eta| < 2.5$, satisfy $\pt>10$~GeV, together with default \textsc{Delphes} efficiencies, smearing and isolation. We modify the default calculation of missing transverse momentum ($\mathbf{p}_\text{T}^\text{miss}$) to ensure muons are included. 

\begin{figure}
    \centering
    \begin{subfigure}[b]{0.5\textwidth}
    \includegraphics[width=\textwidth]{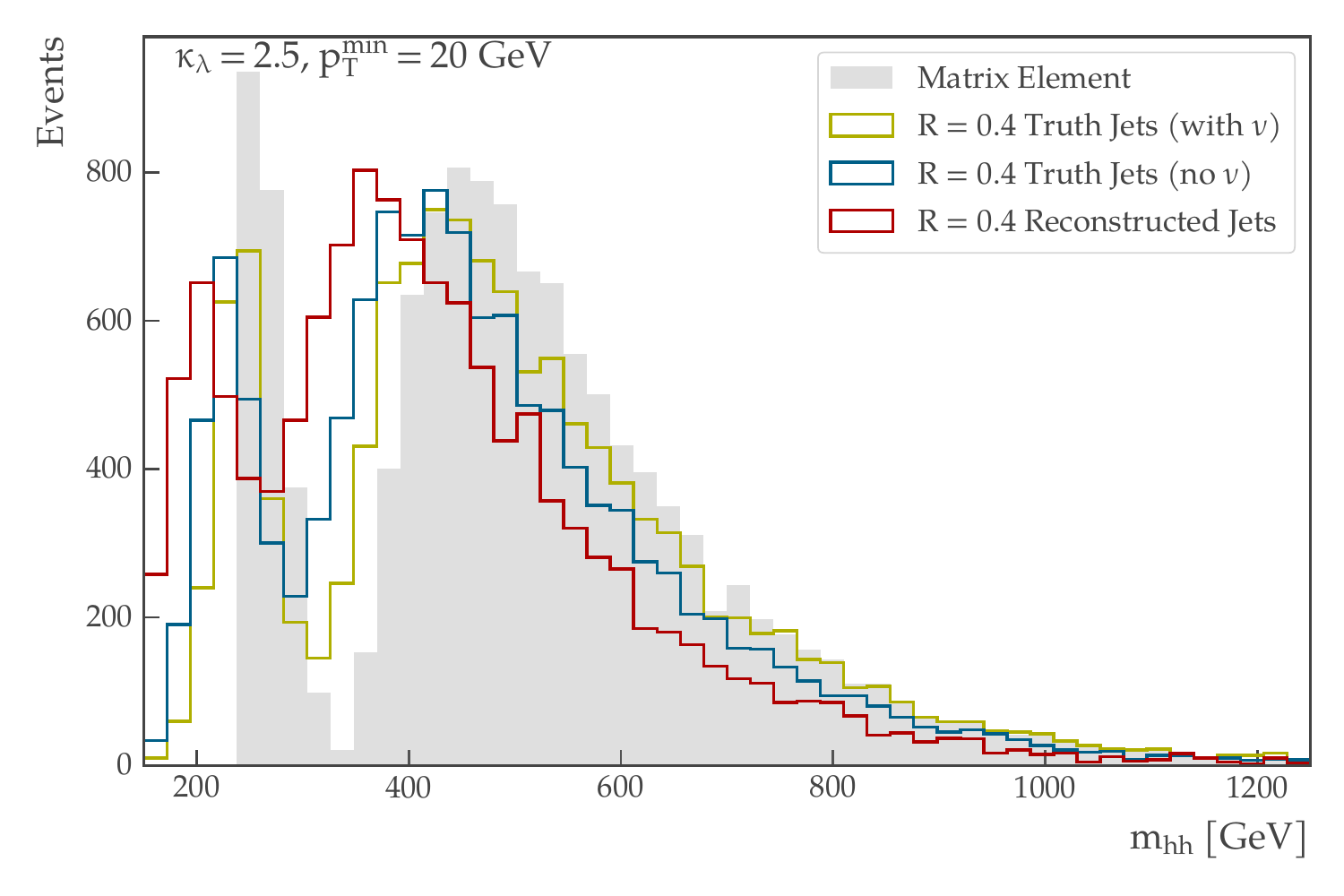}
    \caption{
    \label{fig:mhh-headliner-20}
    $\pt > 20$~GeV reconstructed jets
    }
    \end{subfigure}%
    \begin{subfigure}[b]{0.5\textwidth}
    \includegraphics[width=\textwidth]{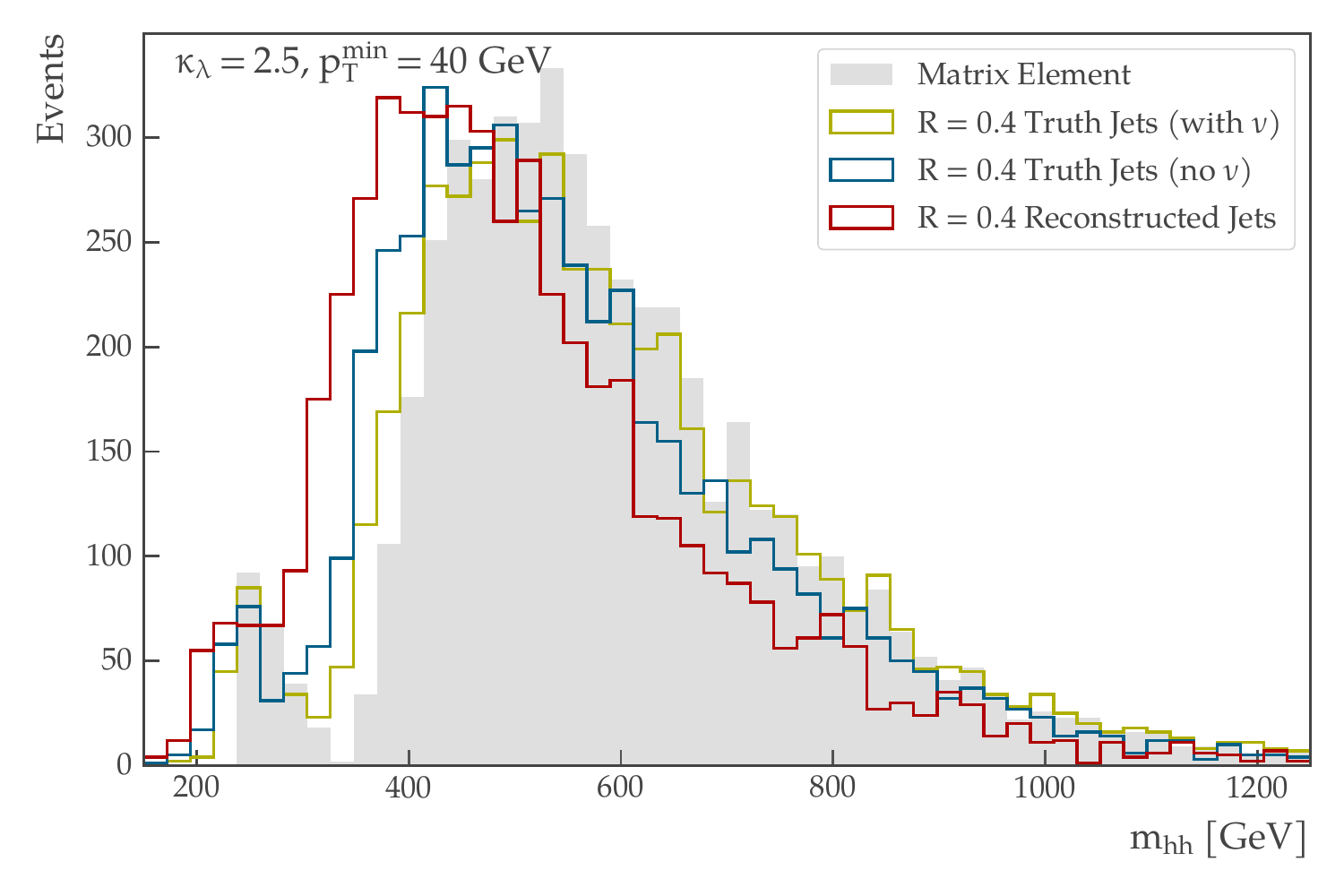}
    \caption{\label{fig:mhh-headliner-40}
    $\pt > 40$~GeV reconstructed jets}
    \end{subfigure}
    \caption{Distributions of the di-Higgs invariant mass \mhh for $\klam = 2.5$ where destructive interference is near-maximal. These are shown at different levels of reconstruction: the parton level predictions (grey shaded), truth jets including neutrinos (yellow line), truth jets without including neutrinos (blue line), and reconstructed jets after the detector emulation described in subsection~\ref{sec:detector} (red line). Small $R=0.4$ radius jets are used that satisfy $|\eta|<2.5$. The $\pt$ thresholds in the subfigure caption are applied to the reconstructed jets.
    }
    \label{fig:mhh_headliner}
\end{figure}

It is relevant to study how experimental effects such as jet clustering, missing energy due to neutrinos and finite detector resolution impact the discriminating variable \mhh. As \mhh is sensitive to \klam, this can motivate improvements to its reconstruction for future work.  Figure~\ref{fig:mhh_headliner} shows the impact of these effects for the $\klam = 2.5$ signal with near-maximal destructive interference to accentuate the shape differences. First, four small jets (with the $\pt$ and $\eta$ requirements defined above applied) are identified by association to the $b$-quarks from the Higgs bosons found in the truth record. For these jets, we then compare their parton level \mhh distribution (shaded grey) to three classes of jets at different levels of reconstruction: 1) truth-level jets clustered from \textit{all} final state truth particles (yellow line), 2) truth-level jets clustered from all final state truth particles \textit{except neutrinos} (blue line), and 3) reconstructed-level jets (red line). The shape of \mhh for all three jet definitions is mildly distorted to lower values than the parton level, with the reconstructed jets being generally lower than their truth counterparts. While this shows the degradation of \mhh resolution, the qualitative features of the non-trivial \mhh shape are preserved.

Figure~\ref{fig:mhh-headliner-20} considers a $\pt > 20$~GeV threshold on the reconstructed jets, while Fig.~\ref{fig:mhh-headliner-40} raises this threshold to the nominal  $\pt > 40$~GeV considered in this work. This illustrates that the signal rate at low \mhh is substantially driven by trigger thresholds. Future work may also consider jets with variable radius, which may improve signal-to-background ratios~\cite{Krohn:2009zg}. 
This comparison highlights the following important experimental considerations:
\begin{itemize}
  \item Maintaining sufficiently low trigger thresholds for the HL-LHC upgrades~\cite{Collaboration:2285584} is of key importance for both discovery of the di-Higgs process as well as constraining \lamhhh. In particular, Fig.~\ref{fig:Norm_DiHiggsM_a} shows the $1 \leq \klam \leq 4$ scenarios exhibit large qualitative changes in shape at low $\mhh \lesssim 400$~GeV. A rise in trigger thresholds can reduce sensitivity to this region. Figure~\ref{fig:Norm_DiHiggsM_b} shows the majority of $\klam \geq 5$ signals reside at $\mhh \lesssim 350$~GeV, but this is also compensated by the faster changes in total cross-section.
  
  \item Corrections should be applied to compensate for energy loss in $b$-jets, which is not accounted for by standard jet calibration. These techniques have been recently deployed by the LHC collaborations~\cite{Sirunyan:2019wwa} but are beyond the scope of this work. Applying these corrections at the trigger level remains an open problem.
\end{itemize}

%\newpage 
\section{Analysis strategies}
\label{sec:analyses}

This section presents our analysis strategies to constrain the trilinear self-coupling \lamhhh in the $hh \to 4b$ channel. We first outline the figures of merit that motivate our analysis design. The goal of measurement is to maximise discrimination between different $(\lamhhh, \topyuk)$ values, which requires solving two conceptually distinct classification problems:
\begin{itemize}
\item \emph{signal characterisation} i.e.\ the measurement power to discriminate $\lamhhh^i$ vs $\lamhhh^{j\neq i}$. This is most simply quantified by the difference (squared) of the signal rates $(S_i - S_{j\neq i})^2$.
\item \emph{background suppression} i.e.\ signal $S$ vs background $B$ discrimination. Intuitively, if the changes $(S_i - S_{j\neq i})^2$ between two couplings are comparable or smaller than the background uncertainties, this reduces the ability to discriminate $\lamhhh^i$ vs $\lamhhh^{j\neq i}$.
\end{itemize} 
We quantify how well our analyses simultaneously achieves these goals in a statistically meaningful way using the chi-square
\begin{equation}
    \chi^2_{ij} =\frac{(S_i-S_j)^2}{\varsigma_{S_i}^2 + \varsigma_B^2 }.
    \label{eq:generic_chiSq}
\end{equation}
Here, $S$ denotes the signal yields after analysis selections for the nominal $(\lamhhh^i, \topyuk^i)$ and alternative $(\lamhhh^j, \topyuk^j)$ coupling hypotheses. The $\varsigma_S$ ($\varsigma_B$) are the combined absolute uncertainties on the signal (background), where $\varsigma_B \gg \varsigma_{S_i}$ is typical in the $4b$ channel. 

To benchmark SM measurement precision, we can fix $S_j = S_\text{SM}$, which assumes the observed data will correspond to that of the SM. Nonetheless, it is important to consider dedicated optimisation assuming $S_i = S_\text{BSM}$ should nature prefer BSM couplings. Higher $\chi^2$ values indicate better discrimination power between two coupling hypotheses, which is achieved by maximising the numerator $(S_i-S_j)^2$ while minimising the uncertainties $\varsigma_{S, B}$. As background rates are large, the systematic component of the uncertainty dominates the denominator. One way to reduce the impact of background systematics is to suppress $B$. We design two classes of analysis strategies to fulfil these objectives:

\begin{itemize}
\item The \emph{baseline analysis} (subsection \ref{sec:cutbasedsel}) uses conventional rectangular cuts on variables reconstructing Higgs bosons, inspired by recent ATLAS and CMS strategies. This serves as a baseline to benchmark the performance of neural network optimisation.

\item The \emph{neural network analysis} (subsection \ref{sec:NNsel}) demonstrates the use of a multivariate strategy to optimise sensitivity beyond the baseline analysis. This consists of an additional selection requirement based on the output of an artificial neural network. 
\end{itemize}

%--------------------------------
\subsection{Baseline analysis}
\label{sec:cutbasedsel}
%--------------------------------

The \emph{baseline analysis} is loosely inspired by a recent ATLAS analysis~\cite{Aaboud:2018knk}. Our event selection is summarised in Table~\ref{tab:preselNN}. In all categories, we require at least four $b$-tagged small or track jets to suppress multijet backgrounds. We also require the pseudorapidity difference of the two reconstructed Higgs candidates to be small $|\Delta \eta(h_1, h_2)| < 1.5$ because high mass objects occupy more central regions than multijet backgrounds dominated by gluon--gluon scattering. To suppress $W\to \ell\nu$ decays from top quarks, we veto electrons or muons and require $E_\text{T}^\text{miss}< 150$~GeV in all categories. 

\begin{figure}
    \centering
    \includegraphics[width=0.5\textwidth]{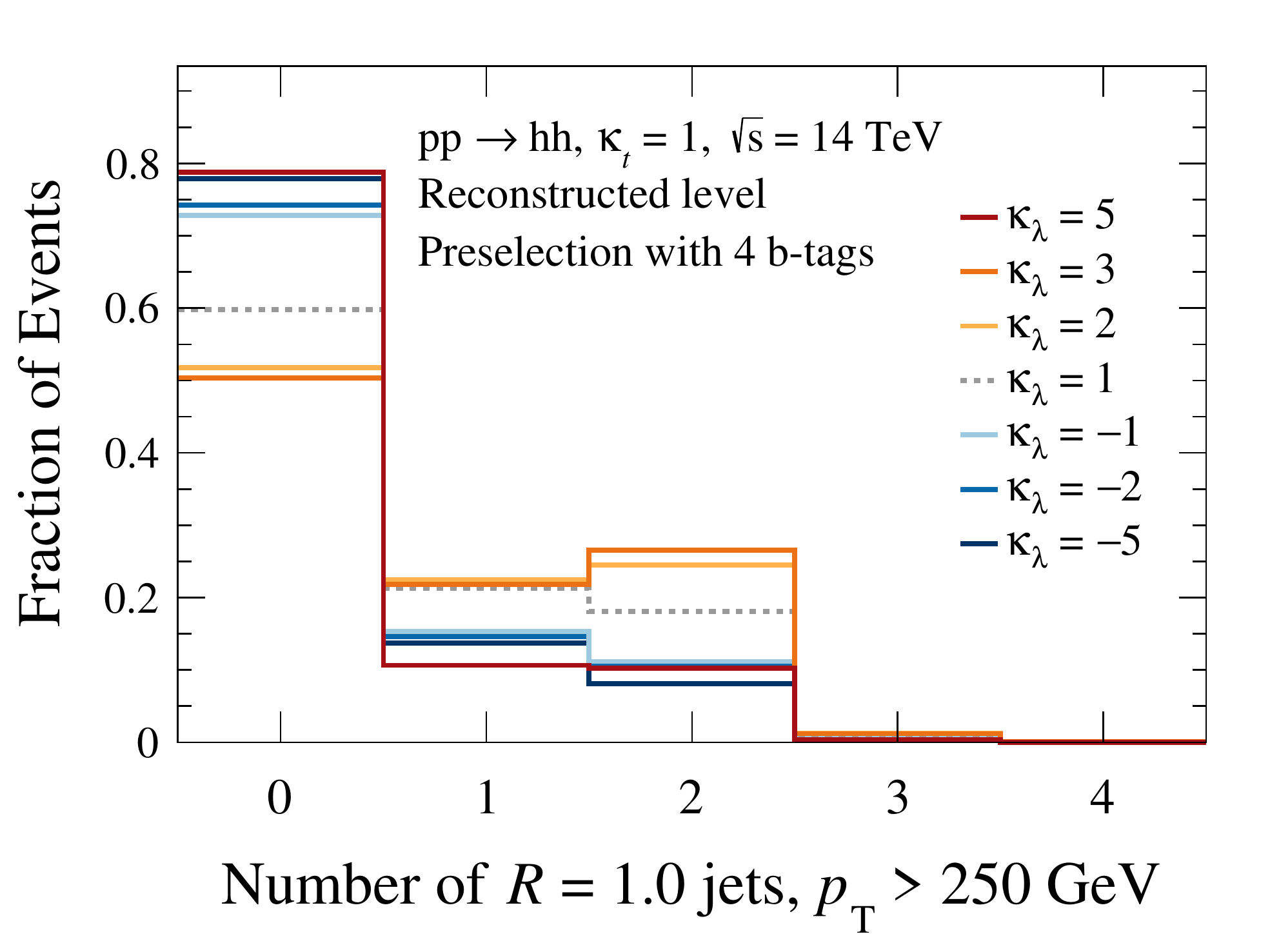}%
    \includegraphics[width=0.5\textwidth]{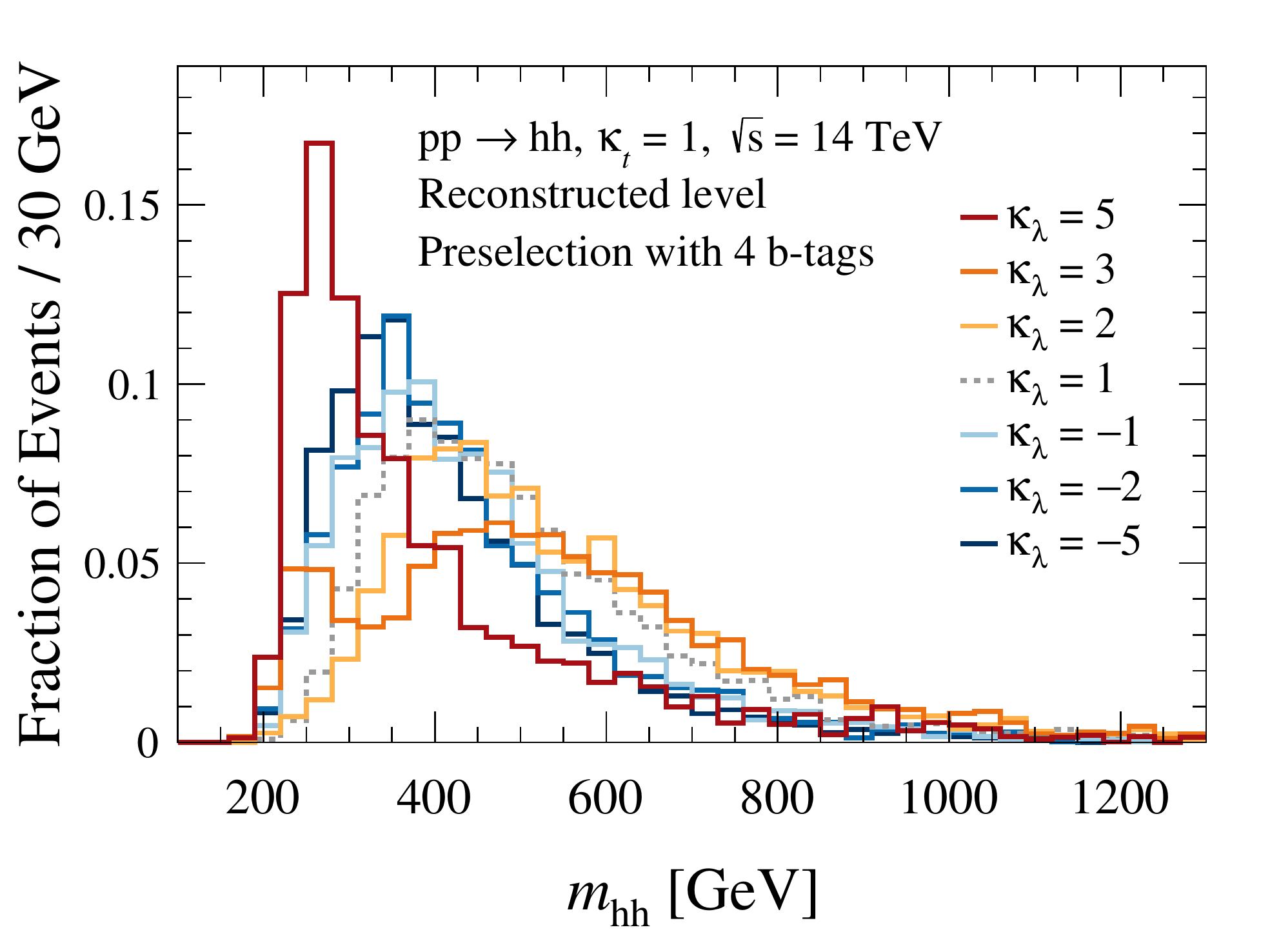}
    \caption{Unit normalised distributions of $pp \to hh$ signals with different \klam (solid lines) compared with the SM (grey dashed) at reconstructed level. The preselection with four $b$-tagged jets is applied. Displayed are (left) the number of reconstructed large jets and (right) the \mhh variable. Negative (positive) values of \klam are coloured by shades of blue (orange). Large jets have a radius parameter of $R=1.0$ and $\pt > 250$~GeV.} 
    \label{fig:n_large_jets_signals}
\end{figure}

To probe different regimes of Higgs boson kinematics, we define three categories based on the exclusive number of large jets in each event, which we refer to as \emph{resolved}, \emph{intermediate}, and \emph{boosted}. Figure~\ref{fig:n_large_jets_signals} displays the multiplicity of resolved, intermediate and boosted events for different \klam, as well as reconstructed \mhh distributions inclusive of number of large jets. The multiplicity distribution shows that this categorisation has \lamhhh discrimination. Interestingly, $\klam = 2$ and $\klam = 3$ have more events falling in the intermediate and boosted categories $N(j_L) \geq 1$ than the SM $\klam = 1$, consistent with the higher \mhh and $\pt(h)$ tails (Fig.~\ref{fig:Norm_DiHiggsM_a}). This orthogonal categorisation in $N(j_L)$ enables straightforward statistical combination to enhance sensitivity. For the three categories, the final signal region (SR) is defined by mass window requirements on the reconstructed $h\to bb$ systems of $m(h_{1,2}) \in [90, 140]$~GeV, based on the mass resolution and background rejection.

\paragraph{Resolved} The resolved category requires exactly zero reconstructed large jets. This targets Higgs bosons with low transverse momenta, $\pt \lesssim 2 m_h$, which we reconstruct as four distinct small jets. To identify pairs of jets consistent with a $h\to bb$ decay, we consider the leading four small jets and construct Higgs boson candidates from small jet pairs that minimise the mass difference between the candidates $\Delta m(h_1, h_2)$. The (sub)leading Higgs candidate is defined as the system of a pair of small jets with the (lower) higher $\pt$. As the triangle amplitude dominates at low \mhh and therefore low Higgs $\pt$, this category is particularly important for \lamhhh sensitivity. We implement a selection defined in Ref.~\cite{Aaboud:2018knk}, where the angular distance $\Delta R_{jj}^{h_{1,2}}$ between the jet pair of the Higgs candidates satisfy
\begin{align}
m_{4j} < 1250~\text{GeV}&:
    \begin{cases}
- 0.5 + \tfrac{ 360~\text{GeV} }{ m_{4j} } < \Delta R^{h_1}_{jj} < \tfrac{ 653~\text{GeV} }{ m_{4j} } + 0.475,\\ 
\tfrac{ 235~\text{GeV} }{ m_{4j} } < \Delta R^{h_2}_{jj} < \tfrac{ 875~\text{GeV} }{ m_{4j} } + 0.35,
 \label{eq:dRjj_m4j_less_1250}
\end{cases}\\
 m_{4j} \geq 1250~\text{GeV}&:
 \Delta R^{h_{1,2}}_{jj} < 1.
 \label{eq:dRjj_m4j_greater_1250}
\end{align}
This adjusts the angular distance between the jets of each Higgs candidate according to the boost of the system characterised by the invariant mass of the 4-jet system $m_{4j}$. 

\paragraph{Intermediate} The intermediate category requires exactly one large jet in the event. This targets regimes where exactly one Higgs boson is sufficiently boosted $\pt \gtrsim 2 m_h$ that the $b$-jets in the $h \to bb$ system become merged so are more efficiently reconstructed as one large jet. The two $b$-quarks are reconstructed as two track jets $j_T$, which we require to be associated to this large jet by $\Delta R(j_T, j_L) < 1.0$. The remaining small jets $j_S$, separated from the large jet by $\Delta R(j_S, j_L)>1.2$, are paired to form the subleading Higgs candidate, where the $j_S$ pair is chosen to minimise the mass difference of the two Higgs candidates. 

\paragraph{Boosted} The boosted category requires exactly two large jets targeting $hh \to 4b$ events where both Higgs bosons have high Lorentz boosts. These events typically reside in the tails of the $m_{hh}$ distribution ($m_{hh} \gtrsim 500$~GeV), which can be important for probing BSM $\lambda_{hhh}$ couplings that enhance $m_{hh}$ and $\pt(h)$ at high values. Importantly, the high expected signal rate due to the large branching ratio $\mathcal{B}(h\to bb) \simeq 58\%$ implies greater statistical power in these tails compared with lower rate channels such as $bb\tau\tau$ or $bb\gamma\gamma$. We require each large jet to have two $b$-tagged track jets associated to this large jet. 

\begin{figure}
    \centering 
    \begin{subfigure}[b]{0.49\textwidth}
        \includegraphics[width=\textwidth]{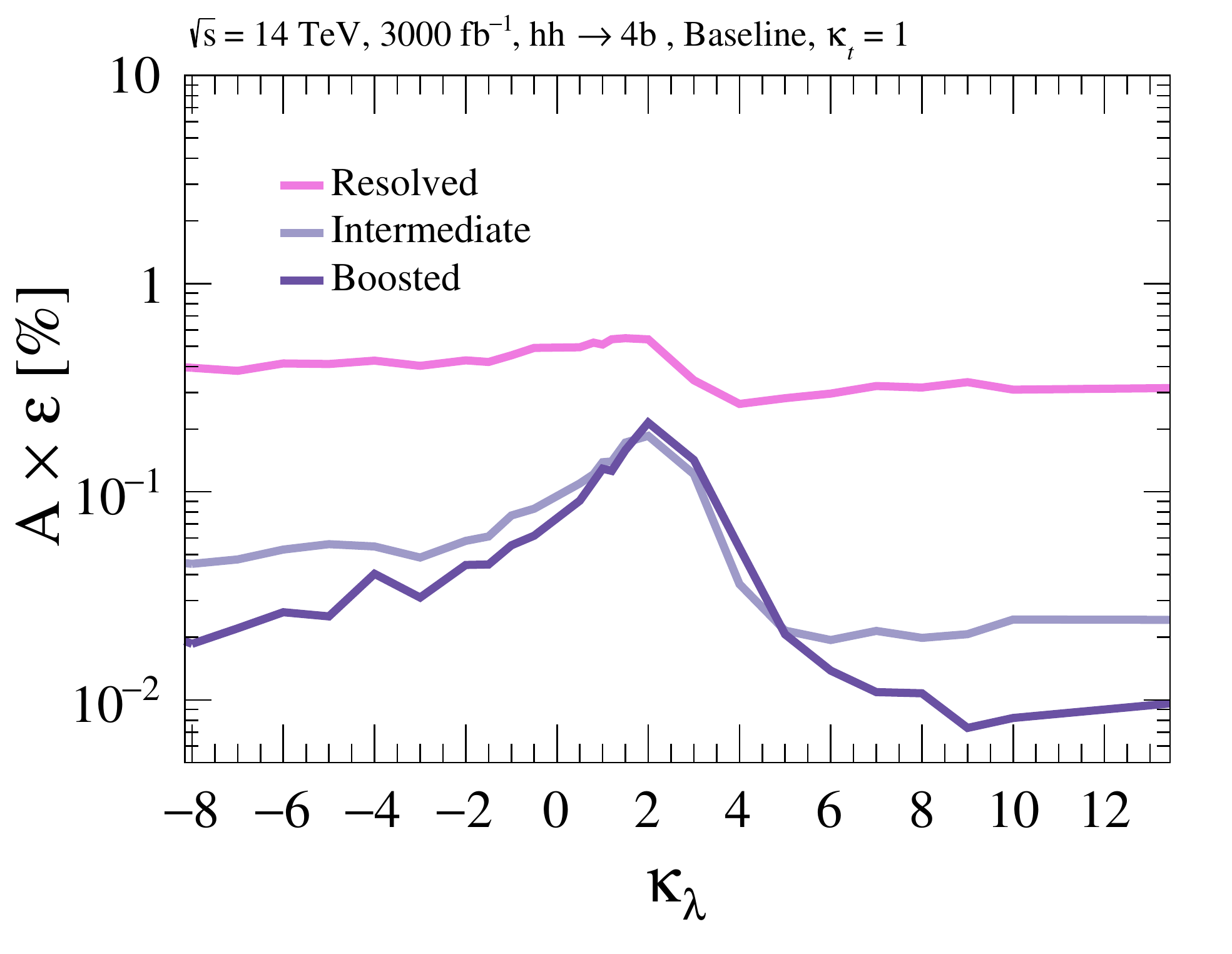}
    \caption{\label{fig:baseline_acceptance_a}$\klam $ variations (fixed $\kapt = 1$)}
    \end{subfigure}%
    \begin{subfigure}[b]{0.49\textwidth}
        \includegraphics[width=\textwidth]{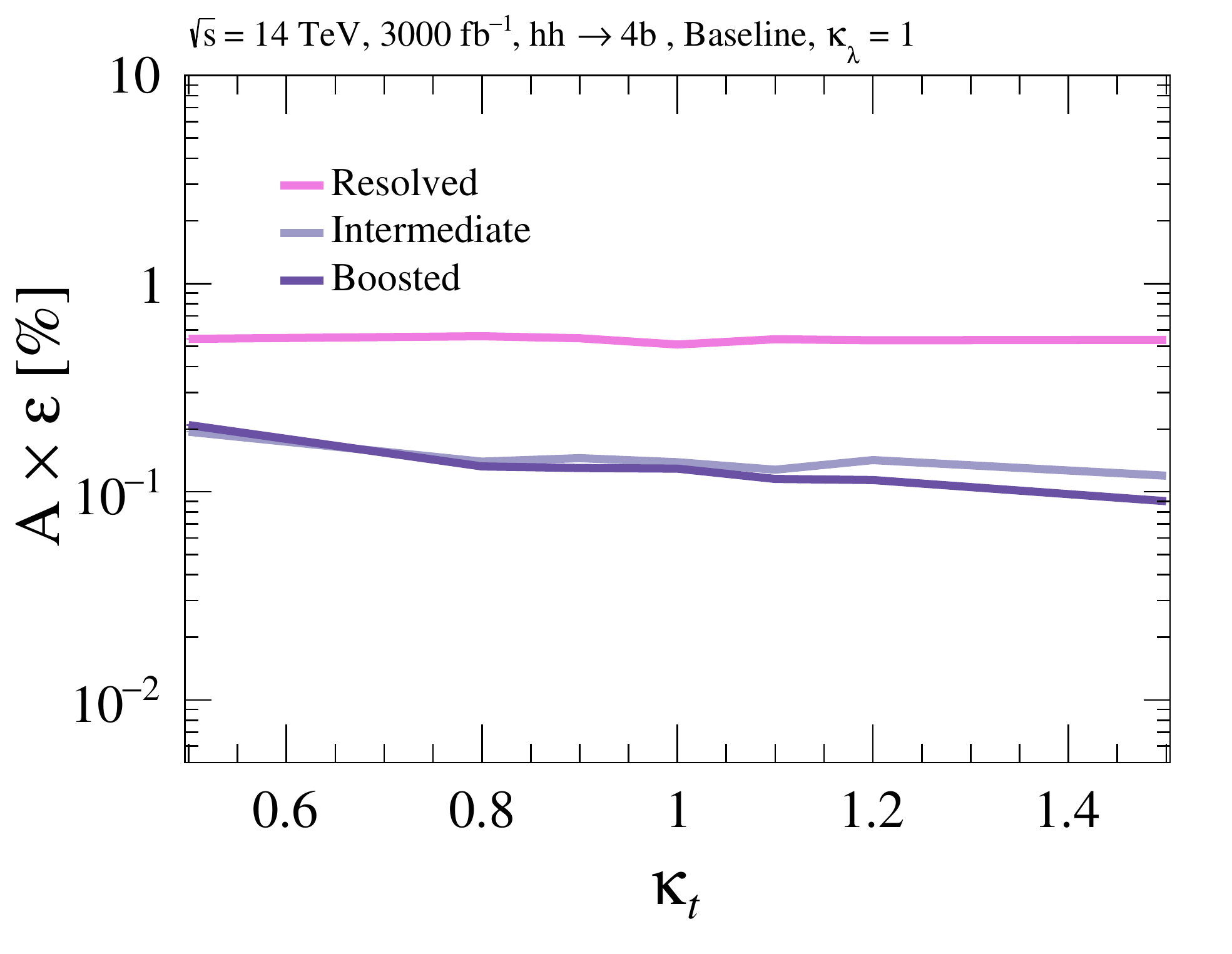}
    \caption{\label{fig:baseline_acceptance_b}$\kapt$ variations (fixed $\klam = 1$)}
    \end{subfigure}
    \caption{Signal acceptance times efficiency $A \times \varepsilon$ in percent for the \emph{baseline analysis} for variations in (a) $\klam$ and (b) $\kapt$. This is shown for the resolved (pink), intermediate (lilac) and boosted (purple) categories. The $A \times \varepsilon$ is equivalent to the number of signal events $S$ divided by initial number of events $\sigma \times \mathcal{L}$. }
    \label{fig:baseline_acceptance}
\end{figure}

Figure~\ref{fig:baseline_acceptance} shows the signal acceptance times efficiency $A \times \varepsilon$ for the \emph{baseline analysis} signal region in the three categories. Due to the different Higgs boson kinematics when varying \klam, the $A \times \varepsilon$ strongly depends on \klam. We find the highest $A \times \varepsilon = 0.55\%$ at $\klam = 1.5$ in the resolved category while the largest $A \times \varepsilon = 0.19\%~(0.21\%)$ for the intermediate (boosted) category peaks at $\klam = 2$, where destructive interference is near maximal and the Higgs bosons have the greatest boost (Fig.~\ref{fig:Norm_DiHiggsM_a}). The $A \times \varepsilon$ values in the intermediate and boosted categories decrease precipitously outside $1\lesssim \klam \lesssim 3$ and become an order of magnitude lower than those of the resolved category. In particular, we find the lowest $A \times \varepsilon = 0.26\%, 0.019\%, 0.007\%$ at $\klam = 4, 6, 9$ for the resolved, intermediate, and boosted categories respectively. This suppression is dominantly due to the lower $h\to bb$ boosts causing events to pass the jet (trigger) $\pt$ requirements with lower efficiency. Figure~\ref{fig:baseline_acceptance_b} shows that the acceptance decreases slowly for increasing \kapt due to the $m_{hh}$ distribution being shifted toward lower values, as displayed in appendix~\ref{sec:distributions}. 

Figure~\ref{fig:mHH_cutana} shows the \mhh and leading Higgs \pt distributions for signal and background in the signal region of the three categories. Notably, the resolved category has the greatest shape discrimination between \klam hypotheses, where $\klam = 5$ has a visibly higher proportion of events at lower \mhh values than $\klam = 1$. To exploit this, our final selection divides events into non-overlapping \mhh bins whose lower edges are defined in Table~\ref{tab:preselNN}. The resolved category uses six bins chosen for simplicity, while fewer bins are used for the intermediate and boosted categories due to lower expected event rates. When performing the statistical analysis in section~\ref{sec:constraints}, these orthogonal \mhh bins are combined to enhance sensitivity. Future work could consider dedicated optimisation of this binning scheme. Appendix~\ref{sec:distributions} shows further distributions involving the subleading Higgs $\pt$ and di-Higgs system $\pt(hh)$. 

\begin{table}
\centering
\begin{tabular}{llll}
\toprule
Observable               & \multicolumn{3}{l}{Preselection}              \\
\midrule
Large jet $j_L$          & \multicolumn{3}{l}{$R=1.0, \pt>250$~GeV, $|\eta| < 2.0$ }  \\
Small jet $j_S$          & \multicolumn{3}{l}{$R=0.4, \pt>40$~GeV,  $|\eta| < 2.5$  } \\
Track jet $j_T$          & \multicolumn{3}{l}{$R=0.2, \pt>20$~GeV,  $|\eta| < 2.5$  } \\
$j_T \in j_L$            & \multicolumn{3}{l}{$\Delta R(j_T, j_L) < 1.0$  } \\
\midrule
                         & Resolved                        & Intermediate                                 & Boosted                    \\
\midrule
$N(j_L)$                 & $=0$                            & $=1$                                         & $= 2$   \\
$N(j_S)$                 & $\geq 4$                        & $\geq 2$                                     & $\geq 0$   \\
%$N(j_T)$                 & $\geq 0$                        & $\geq 1 \in j_L$                             & $\geq 1\in j_L$ \\
$h_1^\text{cand}$        & $j_S^{(i)}$ pair                & $j_L$                                        & $j_L^{(1)}$       \\
$h_2^\text{cand}$        & $j_S^{(i)}$ pair                & $j_S^{(i)}$ pair, $\Delta R(j_S^{(i)}, j_L) > 1.2$  & $j_L^{(2)}$     \\

$\Delta R_{jj}$          & See Eqs.~\ref{eq:dRjj_m4j_less_1250}, \ref{eq:dRjj_m4j_greater_1250} & --- & --- \\
\midrule\midrule 
& \multicolumn{3}{l}{Signal region}\\
\midrule
$j_T \in h_1^\text{cand}$ & --- & $\geq 2$ & $\geq 2$  \\
$j_T \in h_2^\text{cand}$ & --- & --- & $\geq 2$  \\
$b$-tagging                    & \multicolumn{3}{l}{Two $b$-tags for each $h_i^\text{cand}$}   \\
$|\Delta \eta(h_1, h_2)|$      & < 1.5               \\
$E_\text{T}^\text{miss}$       & $< 150$~GeV         \\
$\pt^\ell, |\eta_\ell|$ & $> 10$~GeV, $< 2.5$ \\
$N_\ell$                       & $=0$                \\
$p_\text{signal}^\text{DNN}$   & \multicolumn{3}{l}{$>0.75$ (\emph{neural network analysis} only)} \\
\midrule
 & Resolved & Intermediate & Boosted \\
\midrule 
$m(h_1)$ [GeV]           & [90, 140] & [90, 140] & [90, 140] \\
$m(h_2)$ [GeV]           & [90, 140] & [90, 140] & [90, 140] \\
\midrule\midrule
& \multicolumn{3}{l}{Lower bin edges for $m_{hh}$ binning [GeV]}\\
\midrule
Resolved     & \multicolumn{3}{l}{$[200, 250, 300, 350, 400, 500]$}\\
Intermediate & \multicolumn{3}{l}{$[200, 500, 600]$}\\
Boosted      & \multicolumn{3}{l}{$[500, 800]$}\\
\bottomrule
\end{tabular}
\caption{Overview of event selection for the \emph{baseline analysis} in the resolved, intermediate and boosted categories. The requirements above the upper double rule are the same as the preselection used for the \emph{neural network analysis} training. The requirements below the upper double rule are the signal region requirements. The lower bin edges for the \mhh binning scheme is shown in the bottom three rows. See the main text for details.}
\label{tab:preselNN}
\end{table}
\clearpage

\begin{figure}
    \centering    
    \includegraphics[width=0.5\textwidth]{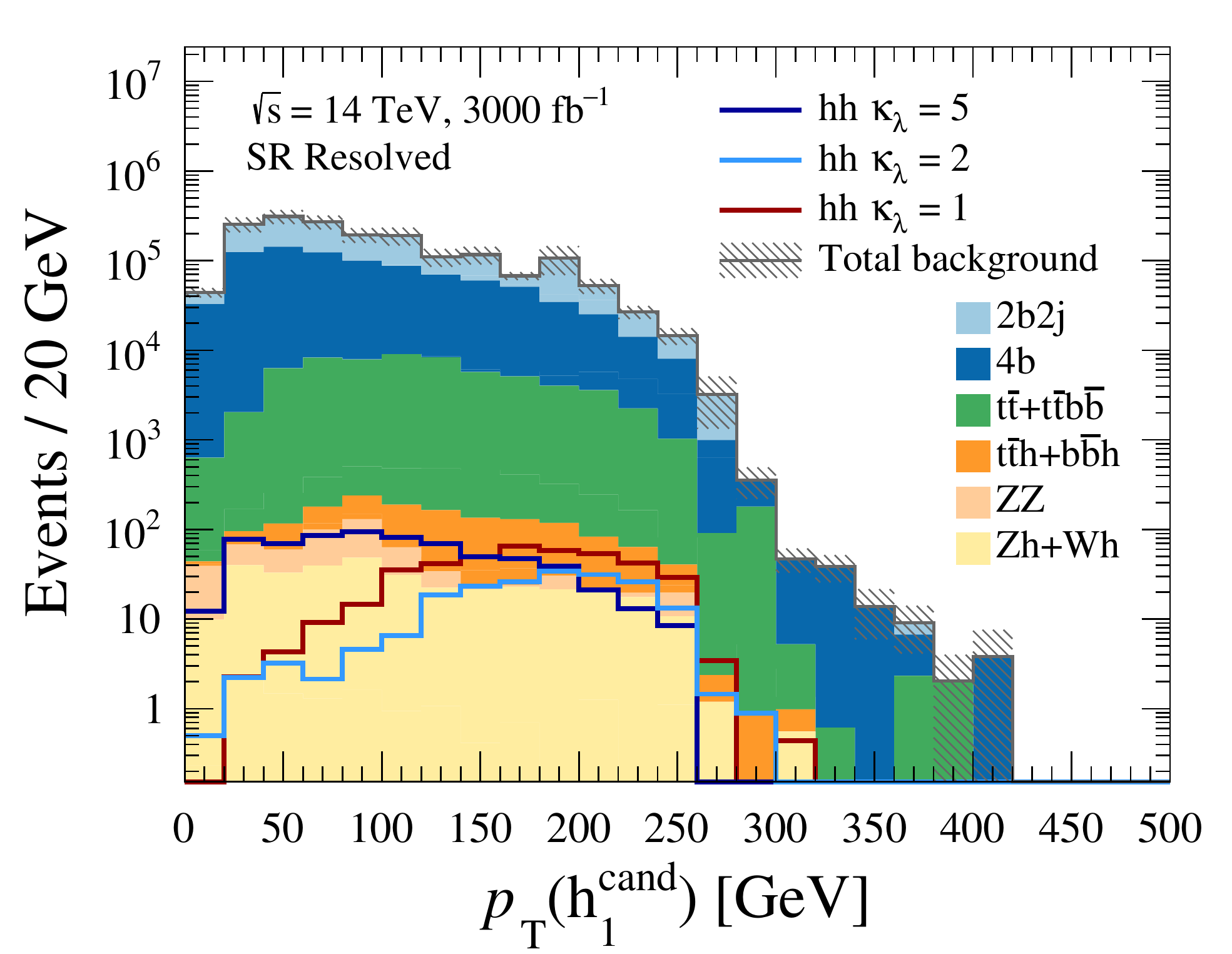}%
    \includegraphics[width=0.5\textwidth]{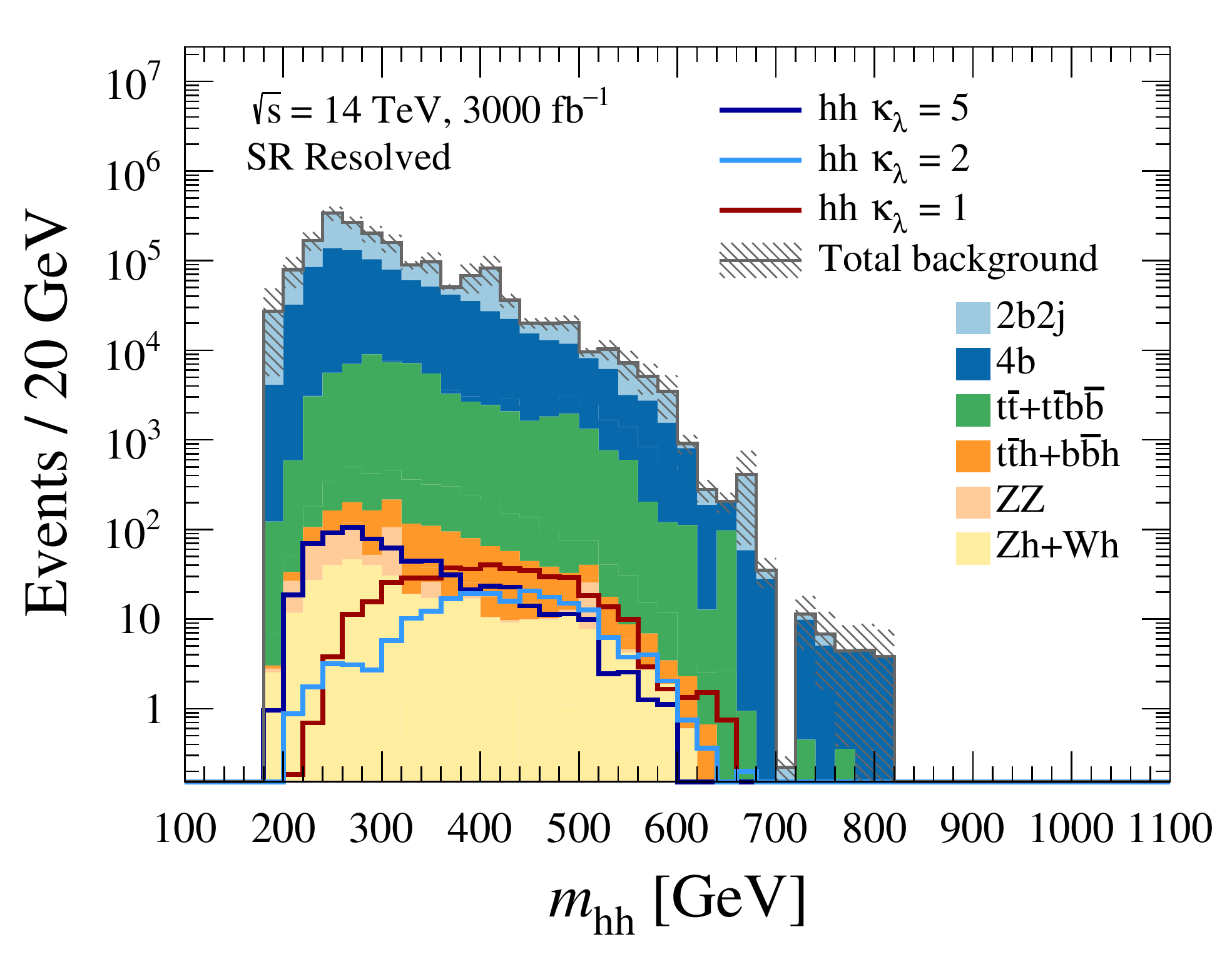}\\    \includegraphics[width=0.5\textwidth]{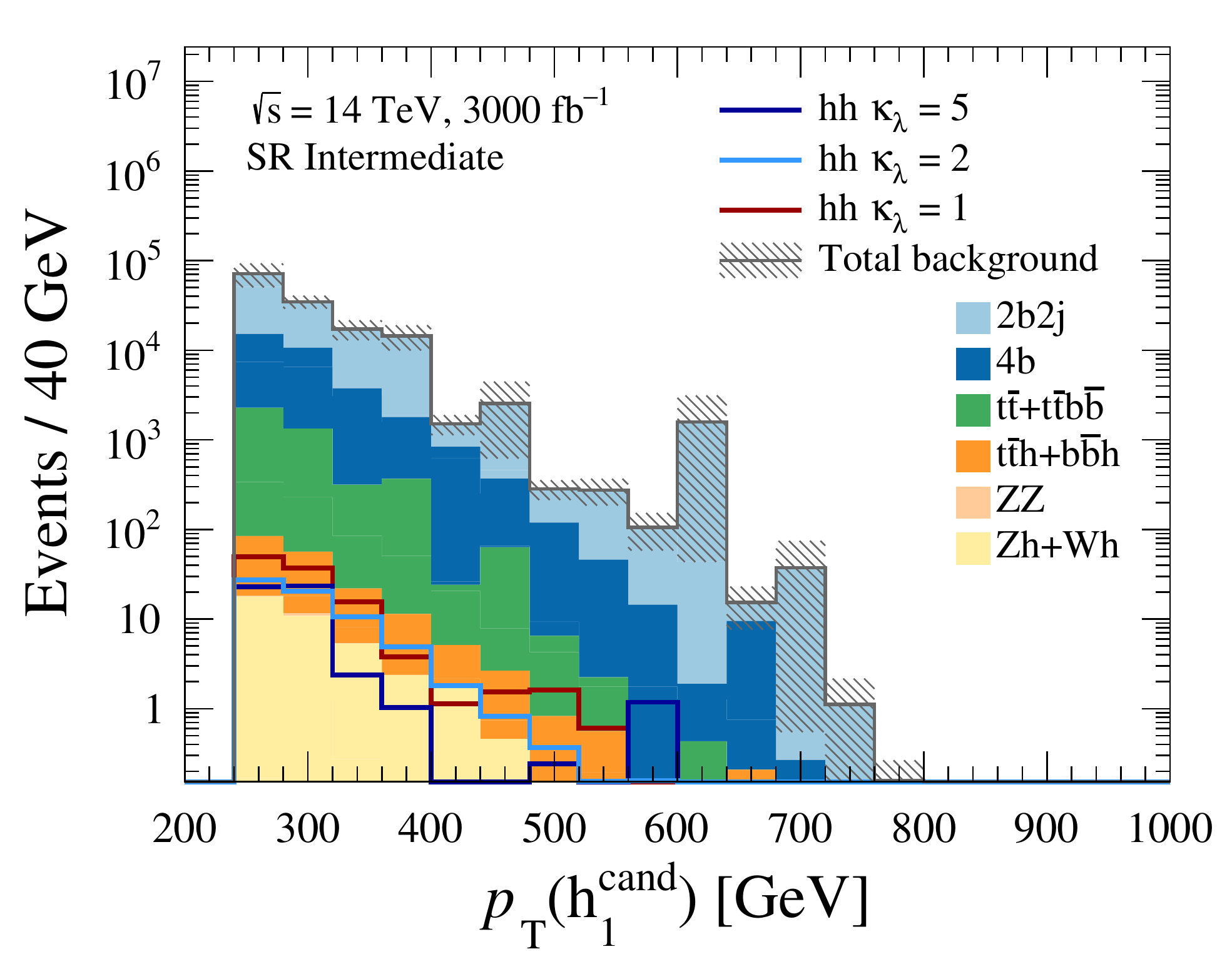}%
    \includegraphics[width=0.5\textwidth]{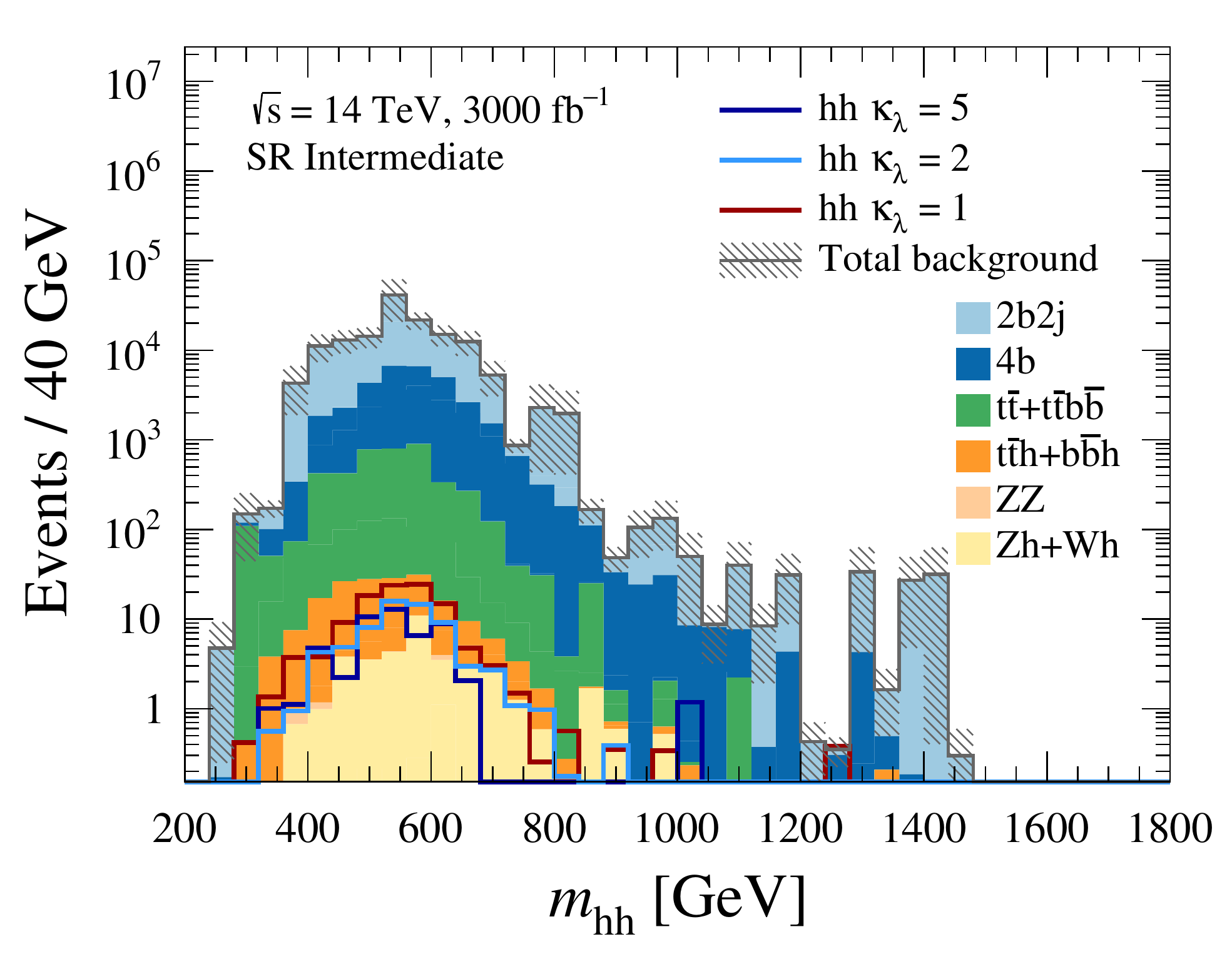}\\
    \includegraphics[width=0.5\textwidth]{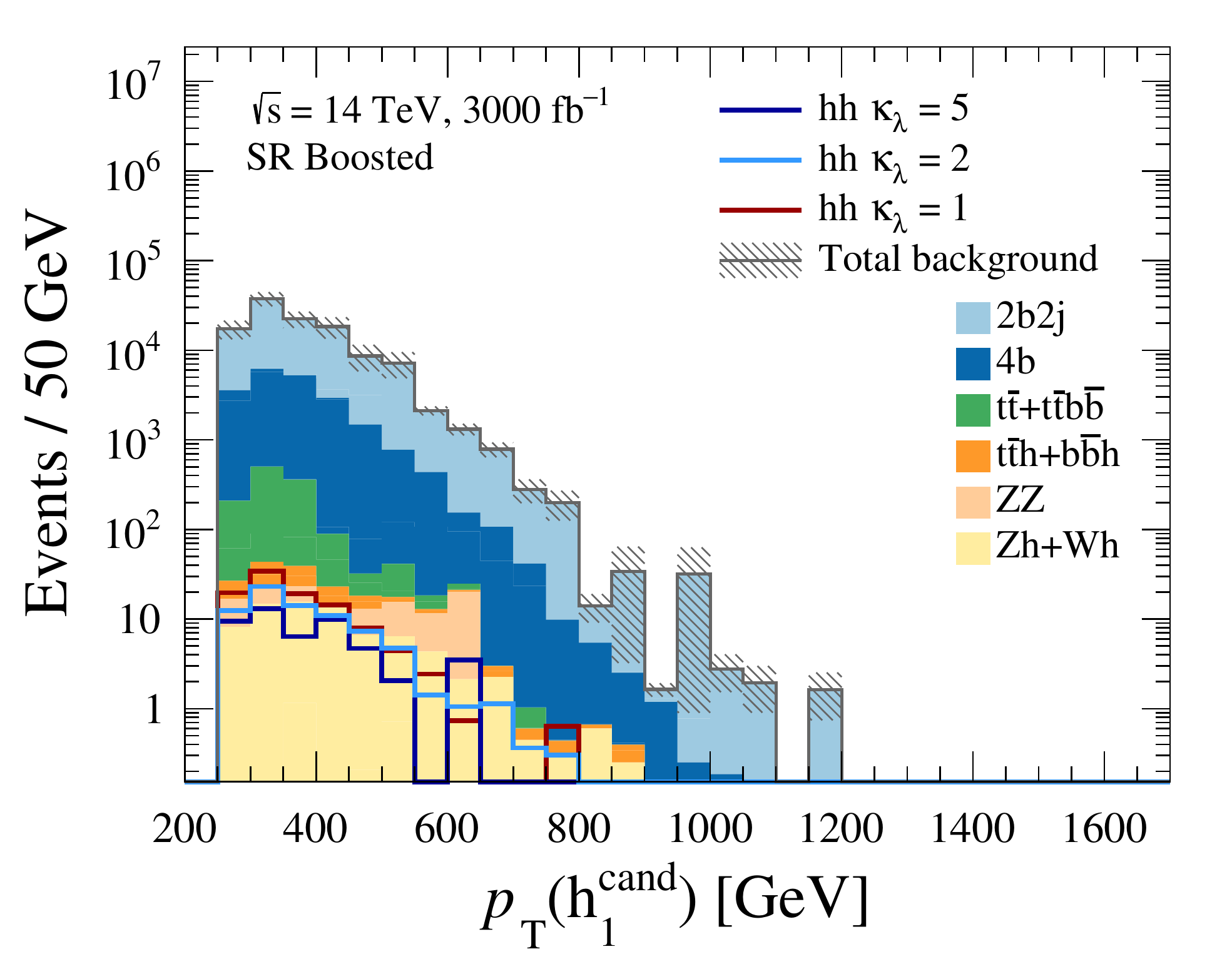}%
    \includegraphics[width=0.5\textwidth]{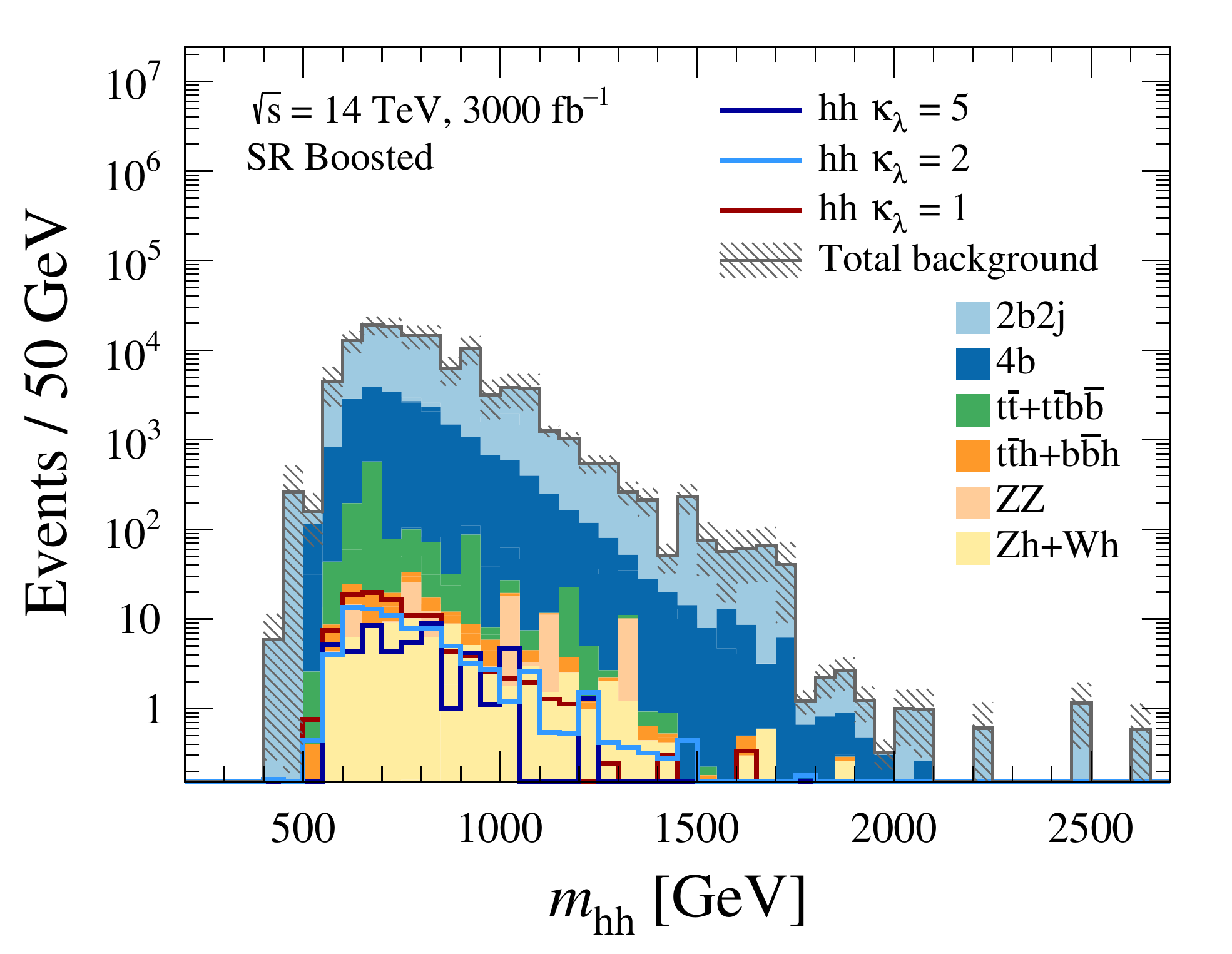}\\
    \caption{The \emph{baseline analysis} leading Higgs candidate transverse momentum $\pt(h_1)$ (left) and invariant mass of the pair of Higgs boson candidates \mhh (right), for benchmark signals (lines) and background (filled). These are displayed for resolved (upper), intermediate (middle) and boosted (lower) categories after all kinematic selections, including $m(h_i)$ mass window cuts, are applied.}
    \label{fig:mHH_cutana}
\end{figure}

%--------------------------------
\subsection{Neural network analysis}
\label{sec:NNsel}
%--------------------------------

Signal characterisation ($\lamhhh^i$ vs $\lamhhh^{j\neq i}$) and background suppression ($S$ vs $B$) are demanding classification problems seeing promising adoption of deep neural network (DNN) solutions in particle physics~\cite{Baldi:2014kfa,Baldi:2014pta,deOliveira:2015xxd,Baldi:2016fzo,Caron:2016hib,Chang:2017kvc,Lin:2018cin,Albertsson:2018maf,Guest:2018yhq,Abdughani:2019wuv,Windischhofer:2019ltt}. We implement our \emph{neural network analysis} using a single network architecture with the \textsc{Keras} library~\cite{chollet2015keras}. This comprises a feed-forward network~\cite{Lippmann1987AnIT,Hornik:1989,Hornik1991ApproximationCO} with a depth of two internal (`hidden') layers each with 200 nodes, similar to Ref.~\cite{Aad:2019yxi}, densely connected to each other and to the input and output nodes. The internal nodes use the rectified linear unit (ReLU) defined as $\text{max}(0, x)$ for the activation function, whose advantages over the traditionally used sigmoid function are discussed in Ref.~\cite{pmlr-v15-glorot11a}.

As input, the network uses a comprehensive set of 20 variables summarised in Table~\ref{tab:nnInputs}. This comprises the four-momenta of the two Higgs candidates, the $\Delta R$ distance between the two subjets associated to each Higgs candidate, the $b$-tagging state of these subjets, the missing transverse momentum with magnitude $E_\text{T}^\text{miss}$ and azimuthal angle $\phi$, the number of reconstructed electrons and muons, and the mass and transverse momentum of the di-Higgs system. These variables are chosen for their signal vs background discrimination power. A separate neural network is trained for each of the resolved, intermediate and boosted categories using the preselection defined in Table~\ref{tab:preselNN}. For the training, the input samples are normalised such that equal weight is given to signal, multijet ($2b2j$ and $4b$) background, and $t\bar{t}$ background.

For outputs, we construct three nodes corresponding to signal, multijet ($2b2j$ and $4b$) background, and $t\bar{t}$ background. This is referred to as a `multi-class classification network'. The model assigns a score $p_i$ corresponding to how likely an event corresponds to one of the three processes. The output nodes use a normalised exponential activation function known as `softmax', which is a standard choice for multi-class configurations. This constrains each output score\footnote{Despite the notation and its properties, $p_i$ is not a true posterior probability.} to
$p_i \in [0, 1]$ and their sum to unity $\sum_i p_i = 1$. We constructed a three-output classifier to explore its utility in classifying background processes for designing control regions, but section~\ref{sec:constraints} will only use the binary signal classifier for simplicity. In typical experimental implementations, multijet processes are estimated using data-driven methods while $t\bar{t}$ employs MC, which have different systematics.

Half of the MC events for the multijet and $t\bar{t}$ background samples, and all of the high statistics $hh \rightarrow 4b$ signal samples are set aside for training (${20\%}$ of the training events are used for training validation and are not used for actual training). The categorical cross-entropy is used as the loss function, which quantifies the accuracy of the model predictions at each training step, and is minimised using the \textsc{Adamax} algorithm~\cite{kingma2014adam}. The step size used in this minimisation is controlled by a \emph{learning rate} hyperparameter. The cross-entropy $H$ between the network prediction and the true classes in $N$ events from the training set is defined by $H(p^{\text{label}},p^{\text{model}}) = -\frac{1}{N}\sum_{i=1}^N p^{\text{label}}_i \log p^{\text{model}}_i.$
%\begin{equation}
%    H(p^{\text{label}},p^{\text{model}}) = -\frac{1}{N}\sum_{i=1}^N p^{\text{label}}_i \log p^{\text{model}}_i.
%\end{equation}
Here, $p^{\text{label}}$ is the vector containing the class of each event (1 for the true class and 0 for the other two) and $p^{\text{model}}$ is a vector containing the score assigned to each class for each event.
The initial learning rate in the \textsc{Adamax} optimiser is set to $5\times 10^{-3}$ for the resolved and intermediate neural networks, and $5\times 10^{-5}$ for the boosted ones. The training set is divided into batches during training. Once all batches are finished processing, one training epoch is complete and the next one starts by processing the first batch again. The number of events in each batch is a tunable hyperparameter, which we set to 100 in this study. To mitigate overfitting unphysical features such as statistical fluctuations, we apply dropout~\cite{JMLR:v15:srivastava14a} at a rate of 30\% to both internal layers. This means 30\% of internal nodes are randomly masked during each training iteration. We find 20 training epochs gives close to optimal performance.  
The learning rate, batch size and dropout rate are optimised using a random search method. 

\begin{table}
    \centering
    \begin{tabular}{l|l}
        \toprule
        Reconstructed objects & Variables used for training \\
        \midrule
        Higgs candidates $h^\text{cand}_{1,2}$ & $(\pt, \eta, \phi, m)$ \\
        Subjets $\in  h^\text{cand}_{1,2}$ & $\Delta R(j_1, j_2) $  \\
        Missing transverse momentum & $E_\text{T}^\text{miss}, \phi(\textbf{p}_\text{T}^\text{miss})$ \\
        Leptons & $N_e, N_\mu$ \\ 
        $b$-tagging & Boolean for $j_i \in h_{1, 2}^\text{cand}$ \\
        Di-Higgs system & $\pt^{hh}, m_{hh}$\\
        \bottomrule
    \end{tabular}
    \caption{Input variables used to train the neural network. }
    \label{tab:nnInputs}
\end{table}

\begin{figure}[tb]
    \centering    
    \begin{subfigure}[b]{0.5\textwidth}
        \includegraphics[width=\textwidth]{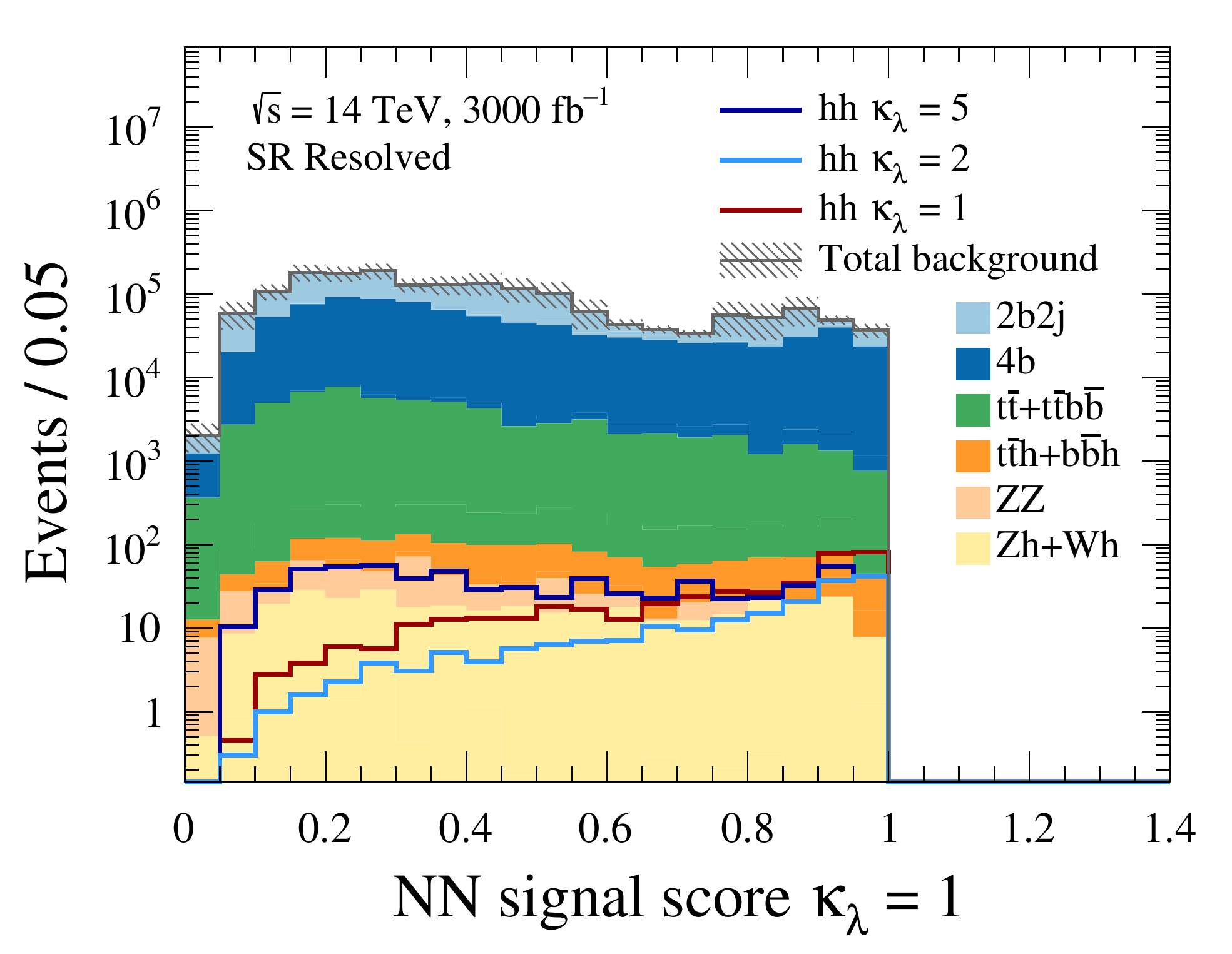}\\
        \includegraphics[width=\textwidth]{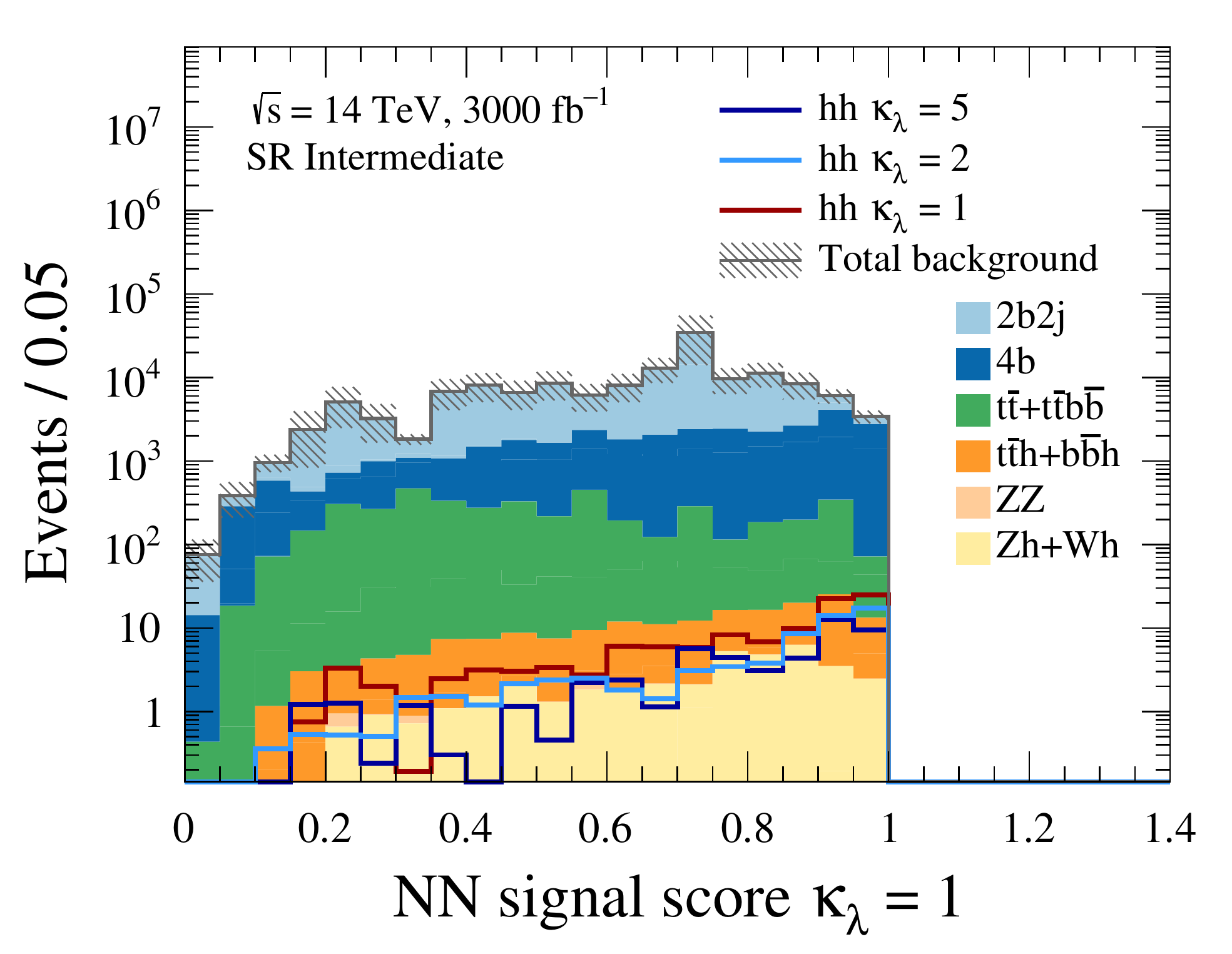}\\
        \includegraphics[width=\textwidth]{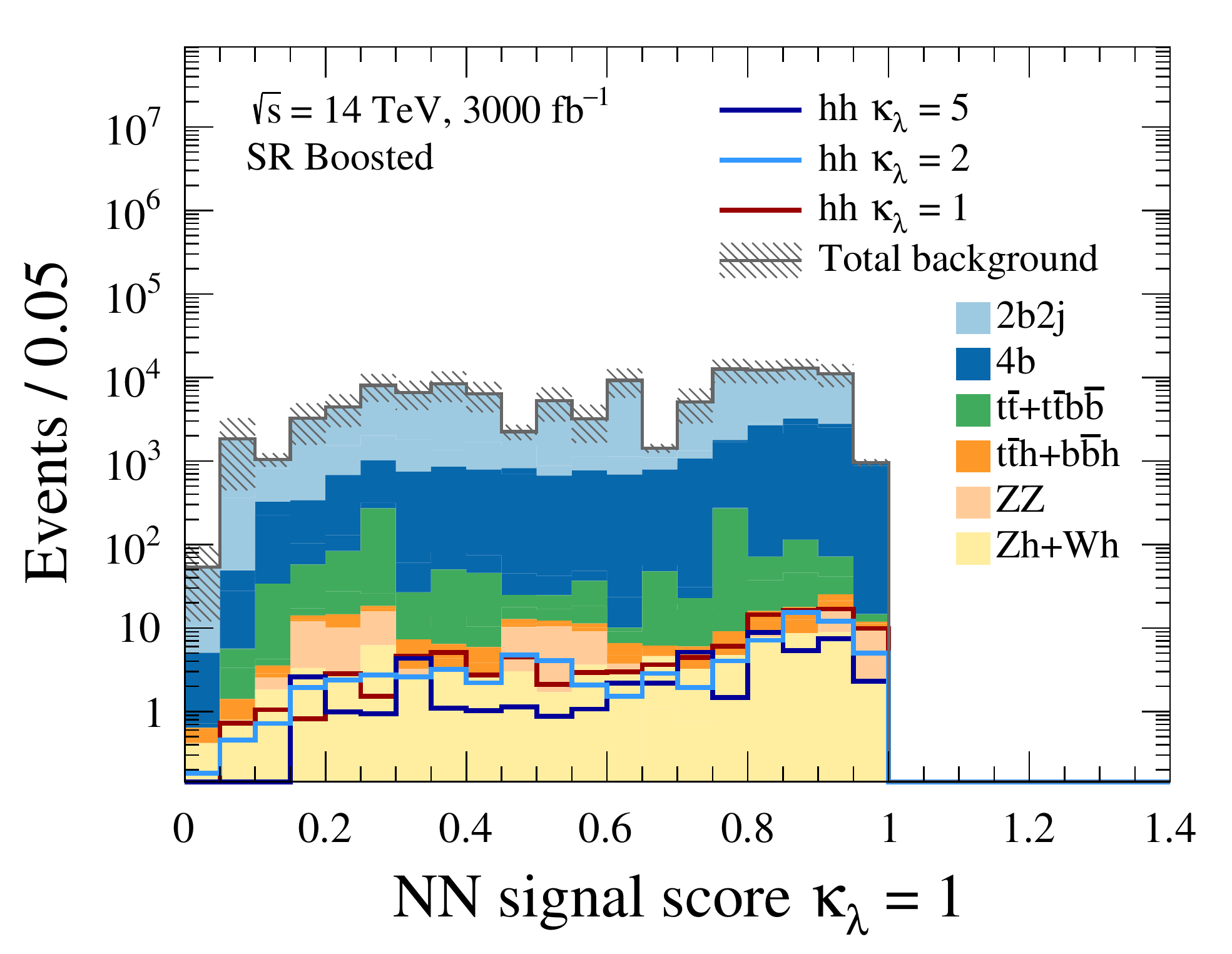} 
        \caption{DNN trained on $\klam = 1$}
    \end{subfigure}%
    \begin{subfigure}[b]{0.5\textwidth}
        \includegraphics[width=\textwidth]{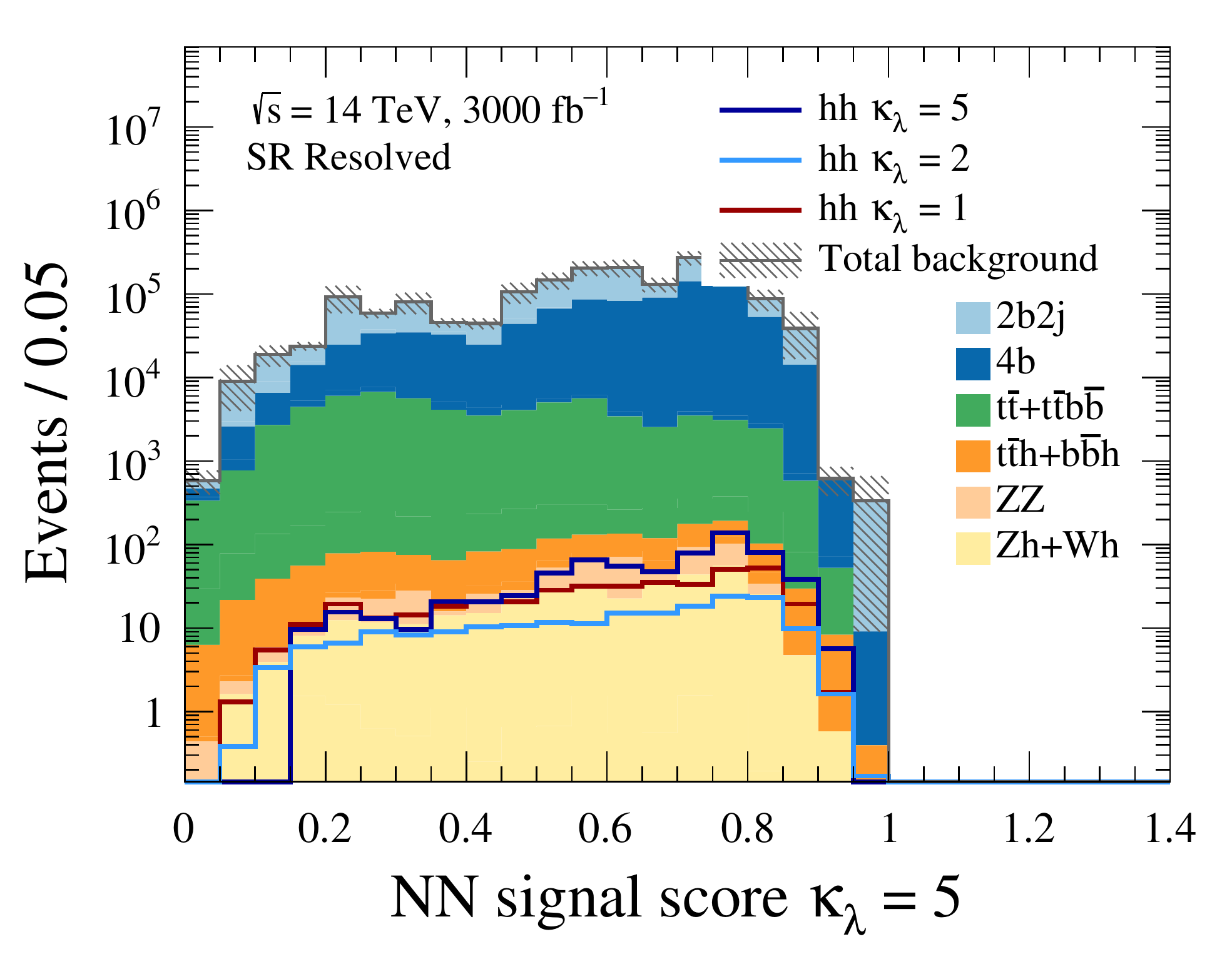}\\
        \includegraphics[width=\textwidth]{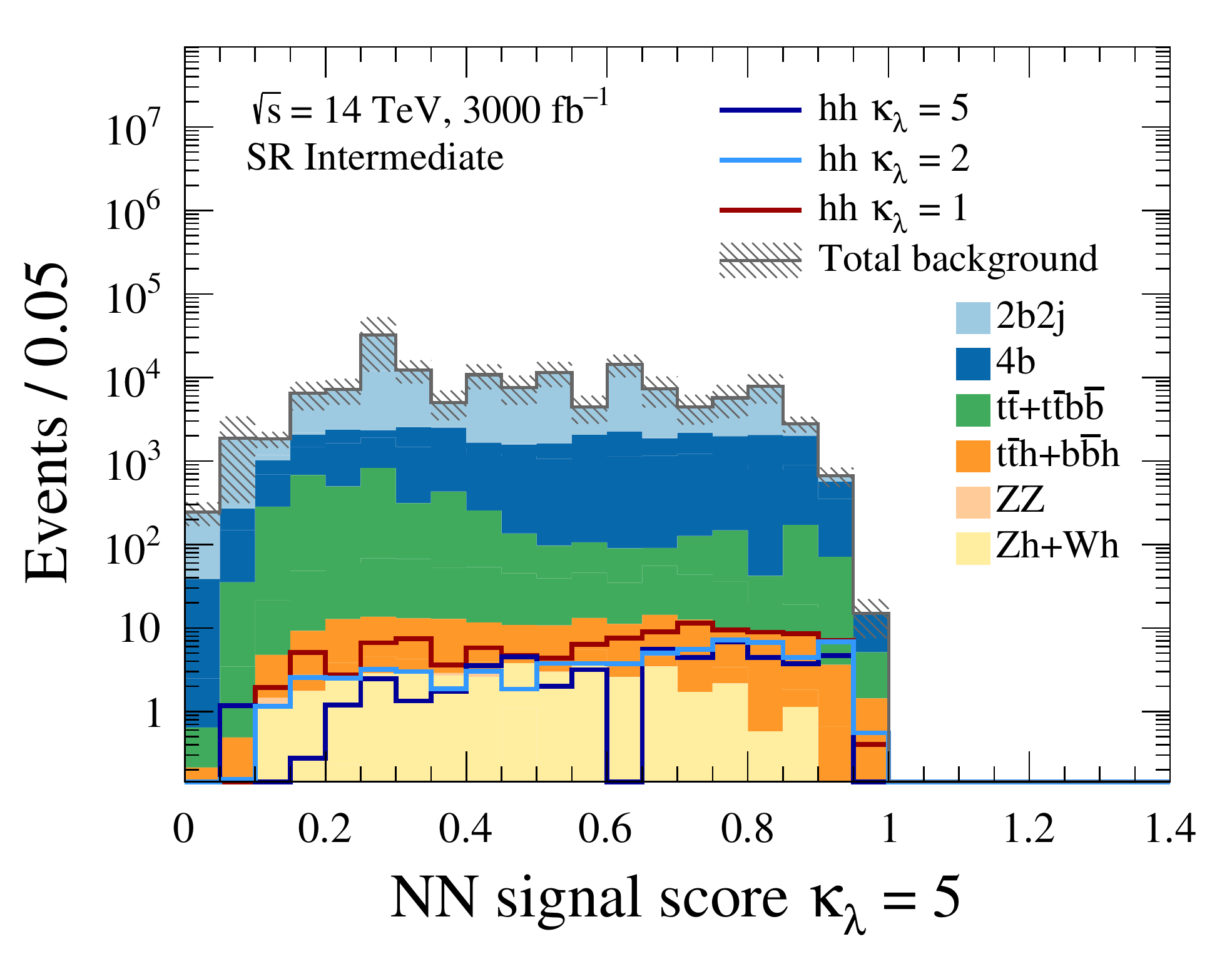}\\
        \includegraphics[width=\textwidth]{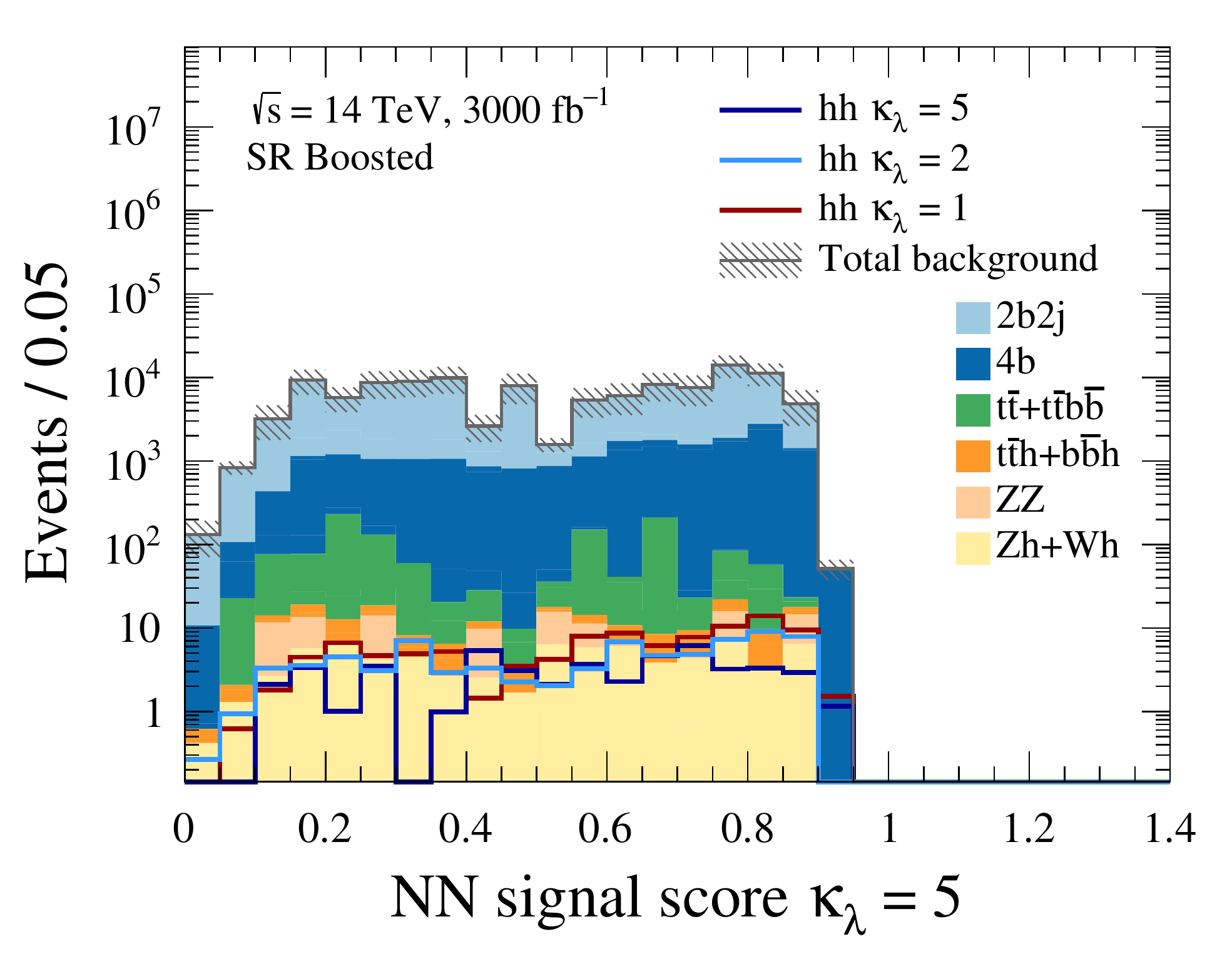} 
        \caption{DNN trained on $\klam = 5$}
    \end{subfigure}
    \caption{Neural network score distributions $p_\text{signal}^\text{DNN}$ of benchmark signals (solid lines) and background processes (filled stacked) displayed in the legend. All event selection criteria of the \emph{neural network analysis} except the $p_\text{signal}^\text{DNN} > 0.75$ requirement are imposed.
    The DNN  is trained on (a) $\klam = 1$ and (b) $\klam = 5$ signals. These are displayed for (upper) resolved, (middle) intermediate and (lower) boosted categories. The plots are normalised to $\mathcal{L} = 3000$~fb$^{-1}$.
    }
    \label{fig:DNN_score_distro}
\end{figure}

\begin{figure}[tb]
    \centering    
    \begin{subfigure}[b]{0.5\textwidth}
        \includegraphics[width=\textwidth]{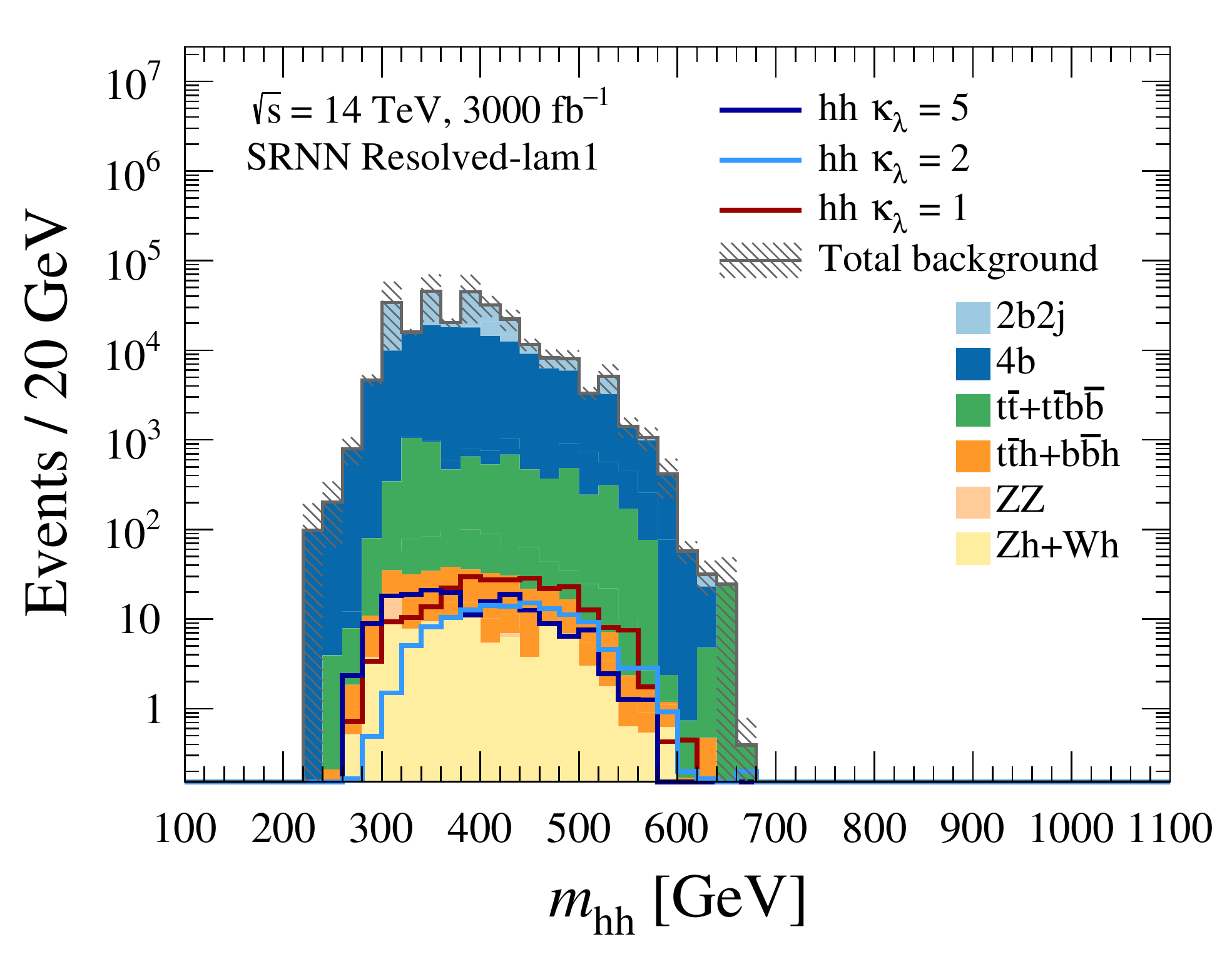}\\
        \includegraphics[width=\textwidth]{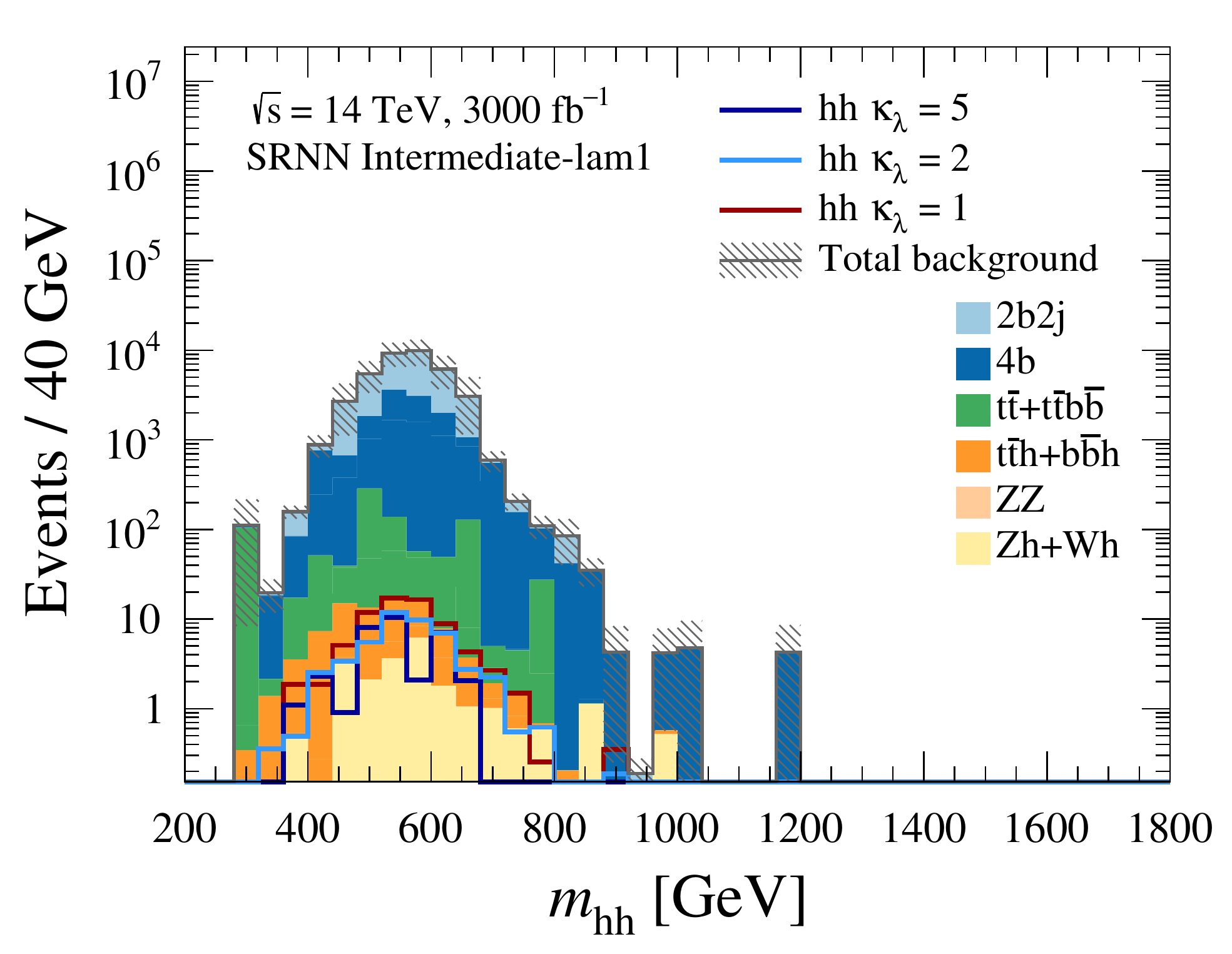}\\
        \includegraphics[width=\textwidth]{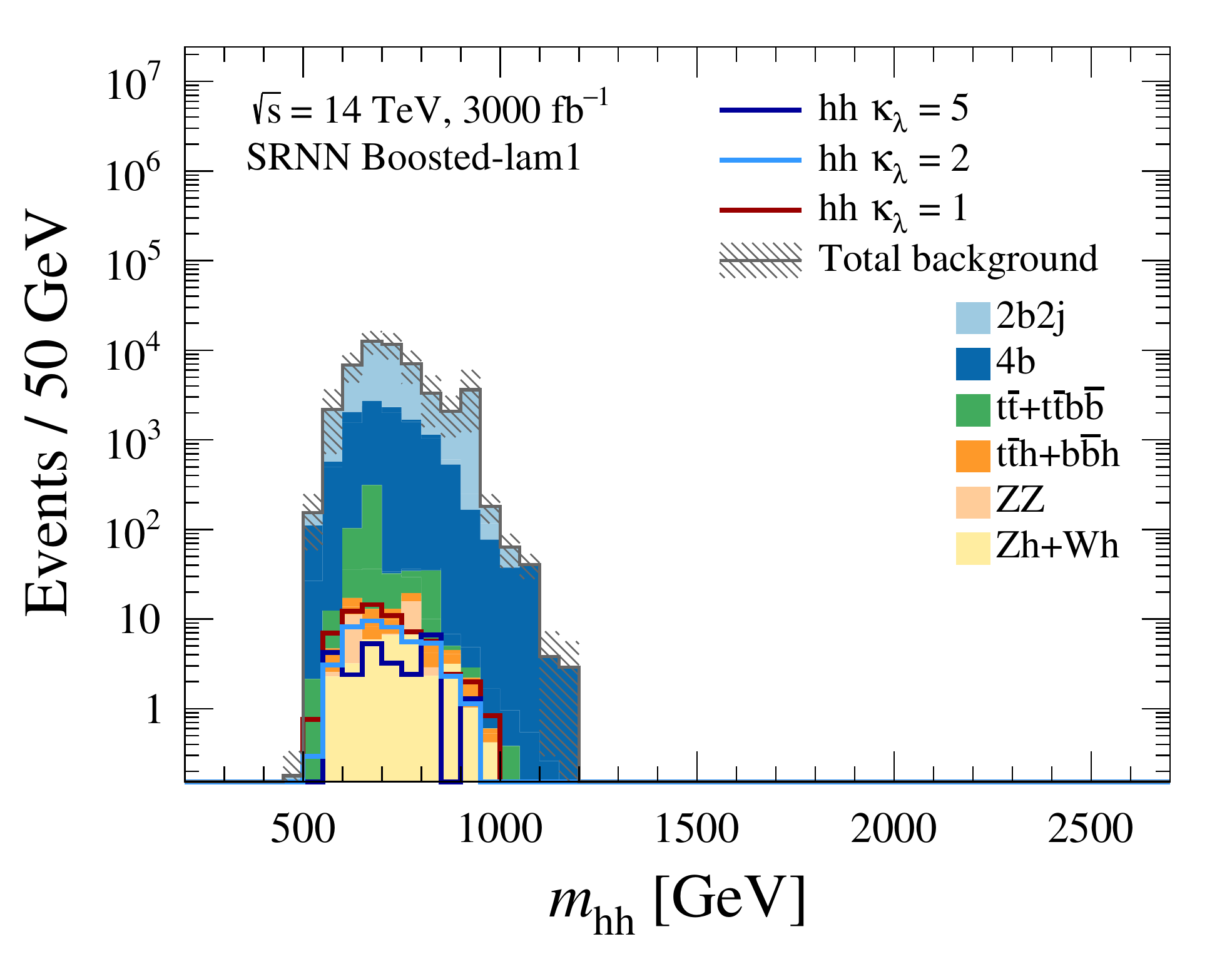}
        \caption{$p_\text{signal}^\text{DNN} > 0.75$ trained on $\klam = 1$}
    \end{subfigure}%
    \begin{subfigure}[b]{0.5\textwidth}
        \includegraphics[width=\textwidth]{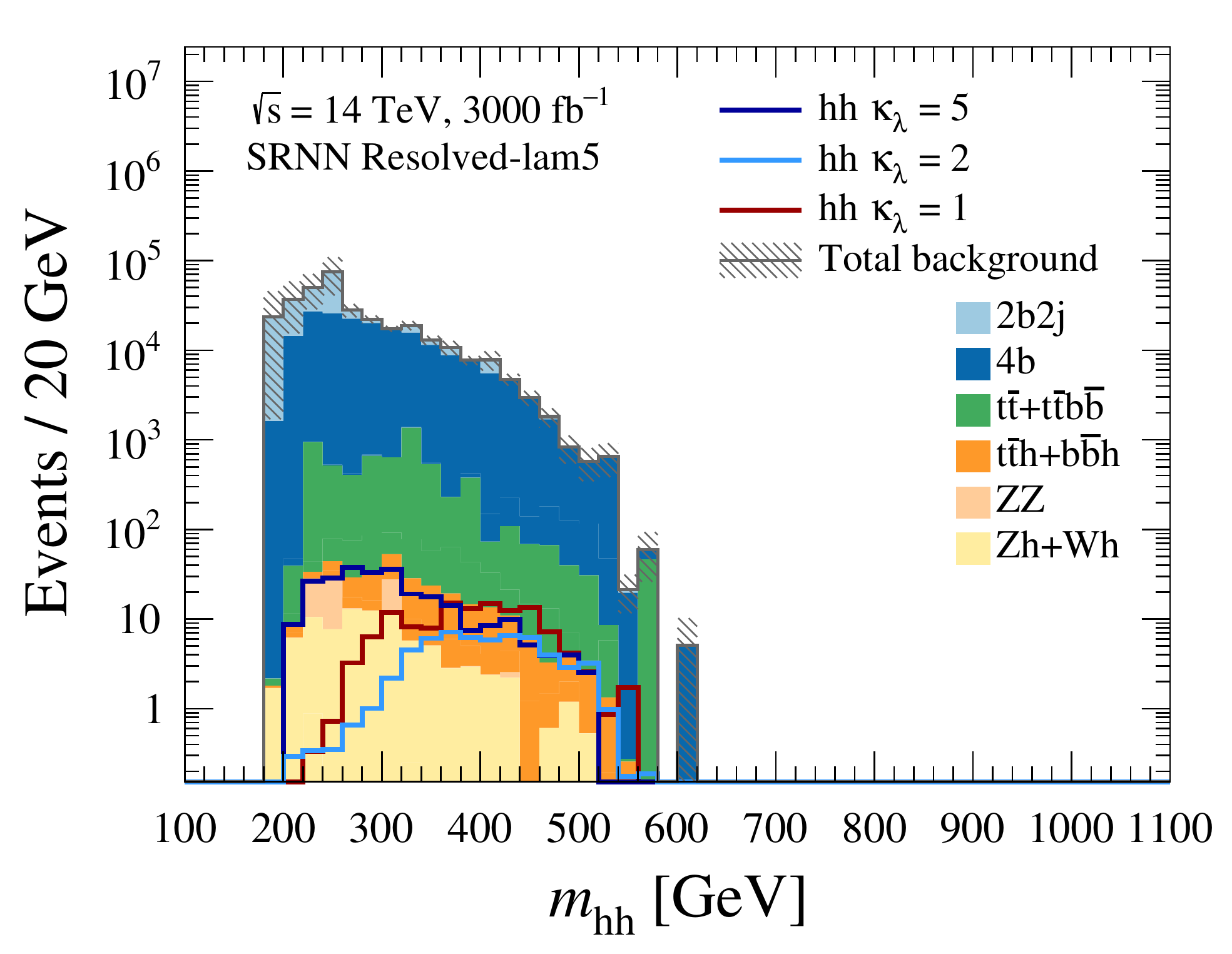}\\        \includegraphics[width=\textwidth]{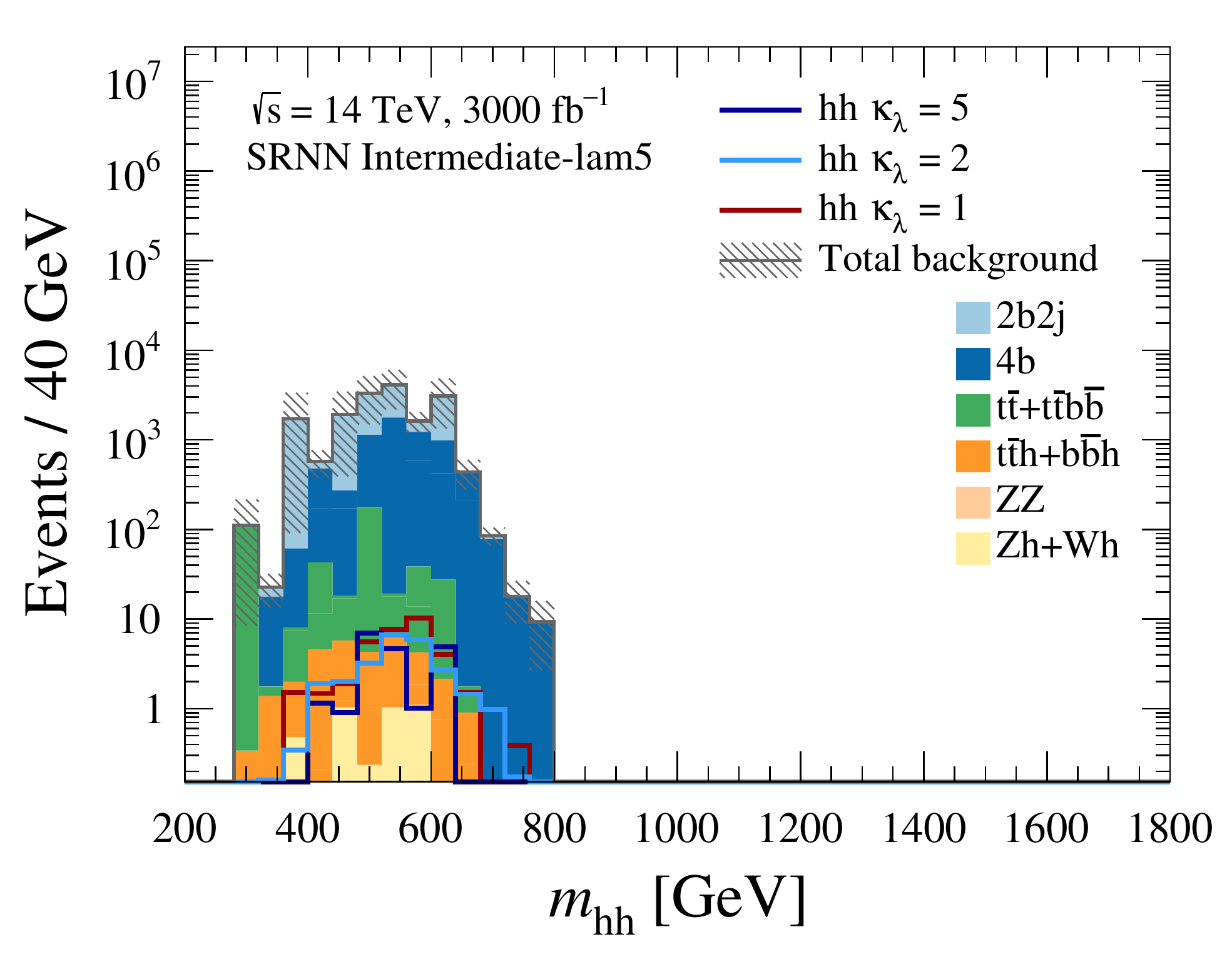}\\
        \includegraphics[width=\textwidth]{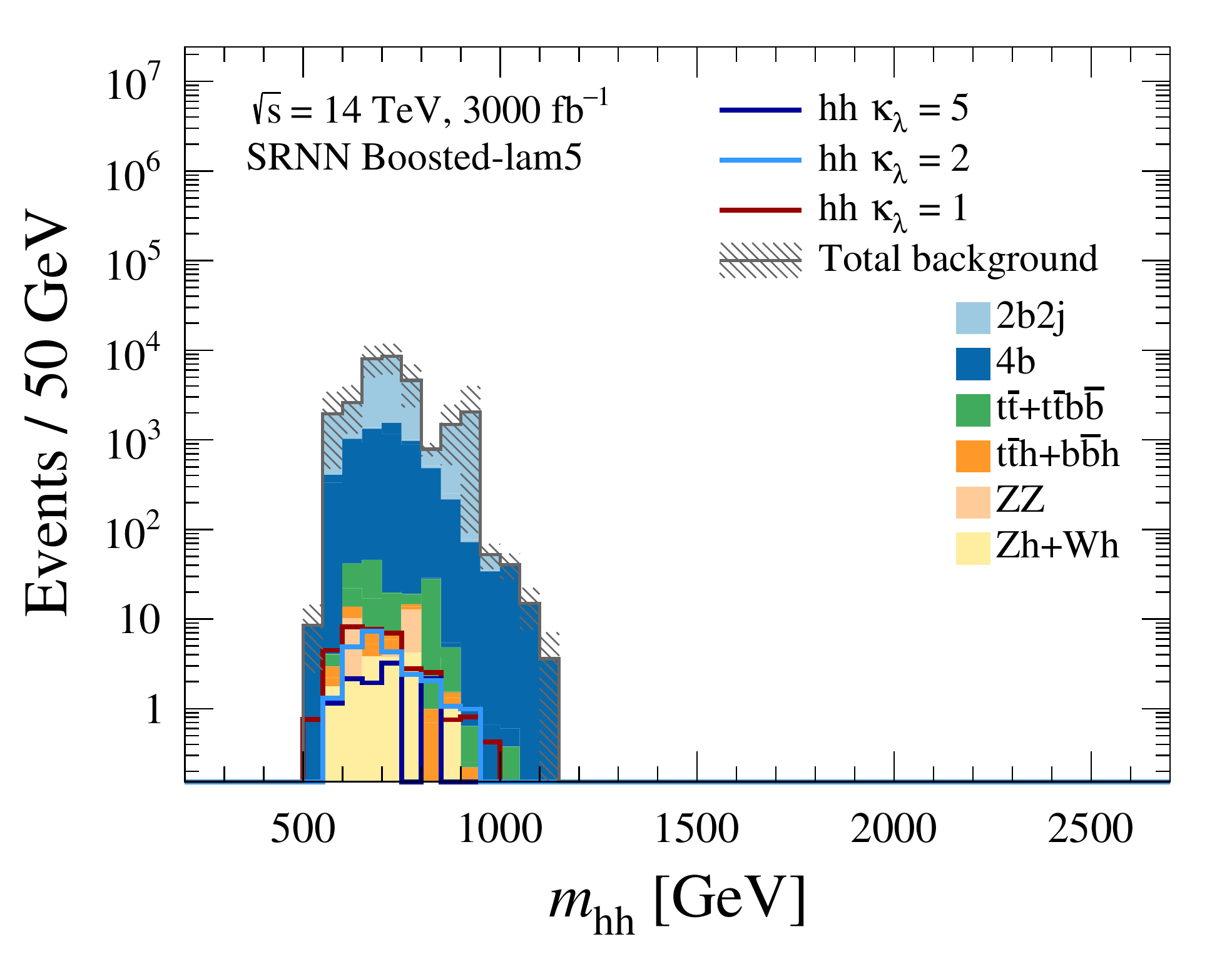}
        \caption{$p_\text{signal}^\text{DNN} > 0.75$  trained on $\klam = 5$}
    \end{subfigure}
    \caption{The \mhh distributions of benchmark signals (solid lines) and background processes (filled stacked) displayed in the legend.
    All event selection criteria of the \emph{neural network analysis} are imposed. The DNN is trained on (a) $\klam = 1$ and (b) $\klam = 5$ signals. 
    These are displayed for (upper) resolved, (middle) intermediate and (lower) boosted categories. The plots are normalised to $\mathcal{L} = 3000$~fb$^{-1}$.
    }
    \label{fig:mhh_post_DNN_score}
\end{figure}

Figure~\ref{fig:DNN_score_distro} shows the signal score $p_\text{signal}^\text{DNN}$ distributions for the DNN trained on $\klam = 1$ and $\klam = 5$ for background and benchmark signals in the three categories. The signal vs background discrimination is improved across the categories, suggesting that our neural networks capture kinematic information beyond the cuts of the \emph{baseline analysis}. However, this depends on the value of $\klam$. For example, the upper-left plot shows that the DNN trained on $\klam = 1$ adds substantial discrimination power for a $\klam = 1$ signal, but not for a $\klam = 5$ signal. This will be further discussed in section~\ref{sec:constraints}. Our \emph{neural network analysis} imposes a universal requirement of $p_{\rm{signal}}^\text{DNN} > 0.75$ for simplicity. Future work could consider optimising requirements on $p_{\rm{signal}}^\text{DNN}$ that are different for each category. Additionally, one could extend the analysis by fitting the $p_\text{signal}^\text{DNN}$ variable instead of using only one bin.

Figure~\ref{fig:mhh_post_DNN_score} shows the \mhh distributions after the $p_{\rm{signal}}^\text{DNN} > 0.75$ requirement is imposed for the DNN trained on $\klam = 1$ and $\klam = 5$. We note that the shape of the $\klam = 5$ signal (as well as that of the background) in the resolved category is particularly different between the two trainings. Since the low-\mhh regime offers the most discrimination between different \klam values, we use $\klam = 5$ as the nominal signal training sample to ensure the DNN gives more weight to these events with respect to a SM optimisation. As we will see in section~\ref{sec:constraints}, this indeed performs better than training on $\klam = 1$ when setting $\klam$ coupling limits. Note that the same is not necessarily true when setting cross-section limits assuming SM kinematics.

%--------------------------------
\subsubsection{What is the DNN learning?}
\label{sec:feature_importance_DNN}
%--------------------------------

As machine learning techniques become increasingly widespread, it is important to understand how our neural network exploits the given physics information~\cite{Chang:2017kvc}. Specifically, we evaluate the ranked \emph{feature importance} to quantify how much each input variable (model \emph{feature}) changes the signal score in both signal and background events. Interpretability of deep neural networks has seen rapid development in the computational sciences. We adopt a recently developed framework for interpreting predictions called SHapley Additive exPlanations (SHAP)~\cite{NIPS2017_7062}. This framework combines several feature importance tests available for machine learning models in the literature into a single value as detailed in Ref.~\cite{NIPS2017_7062}. These values are consistent for all types of inputs to the neural network, and can be compared with one another. A heuristic description of SHAP values can be found in Appendix~\ref{sec:shapley_heuristics} that provides intuition for this approach.

The SHAP framework requires that the trained model is applied to a specific set of events and only evaluates the feature importance for these events. We construct a new subset of events with half signal and half background to reflect the importance of distinguishing them from each other rather than discrimination between the background components had the $S/B$ mirrored the preselection rates. For this evaluation, we construct the background sample to be composed of 80\% $2b2j$, 15\% $4b$ and 5\% $t\bar{t}$ events to approximate the background composition of the three categories.

\begin{figure}[htb]
    \centering
    \begin{subfigure}[b]{0.48\textwidth}
       \includegraphics[width=\textwidth]{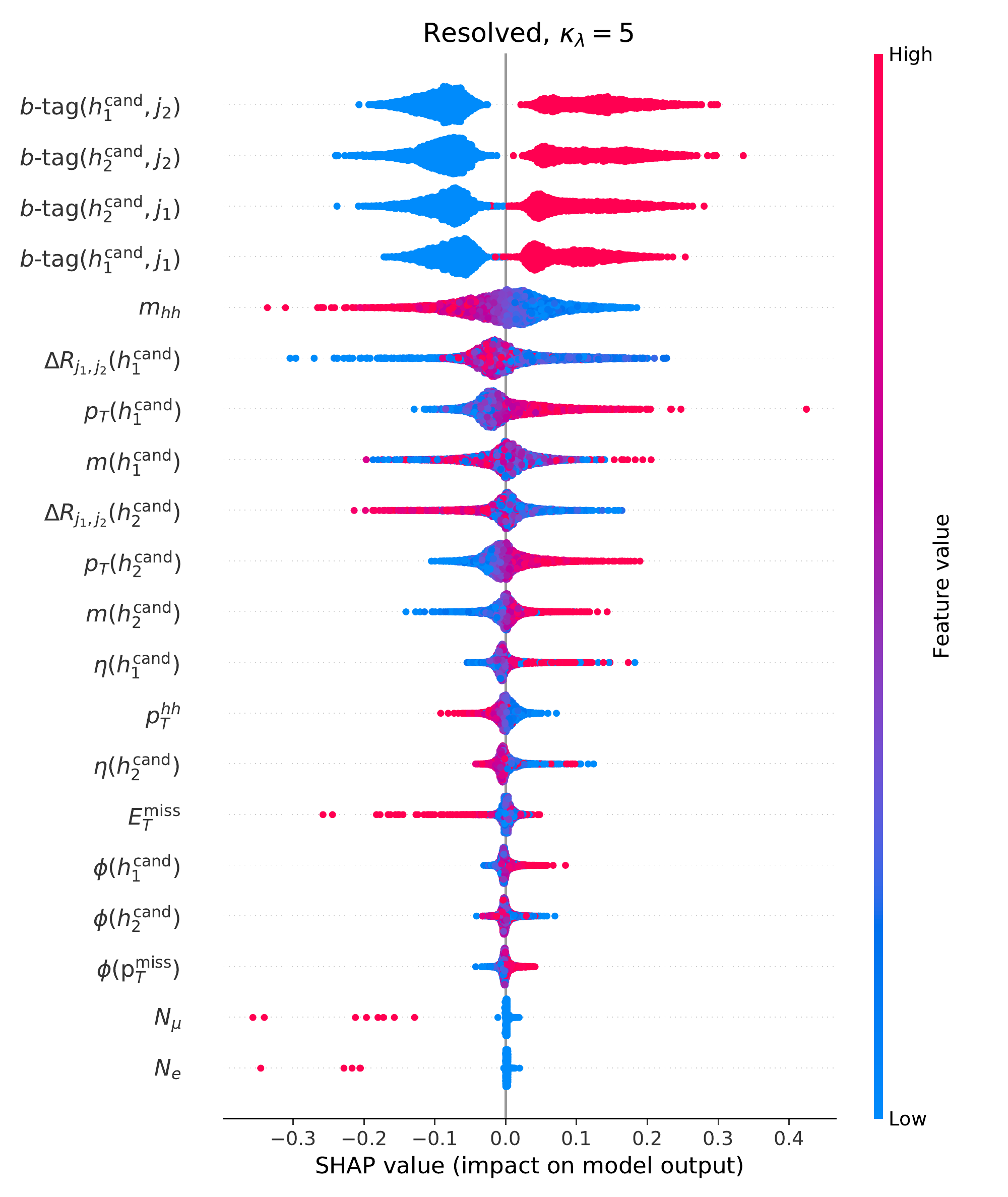}
      \caption{Resolved}
    \end{subfigure}% 
    \begin{subfigure}[b]{0.48\textwidth}
        \includegraphics[width=\textwidth]{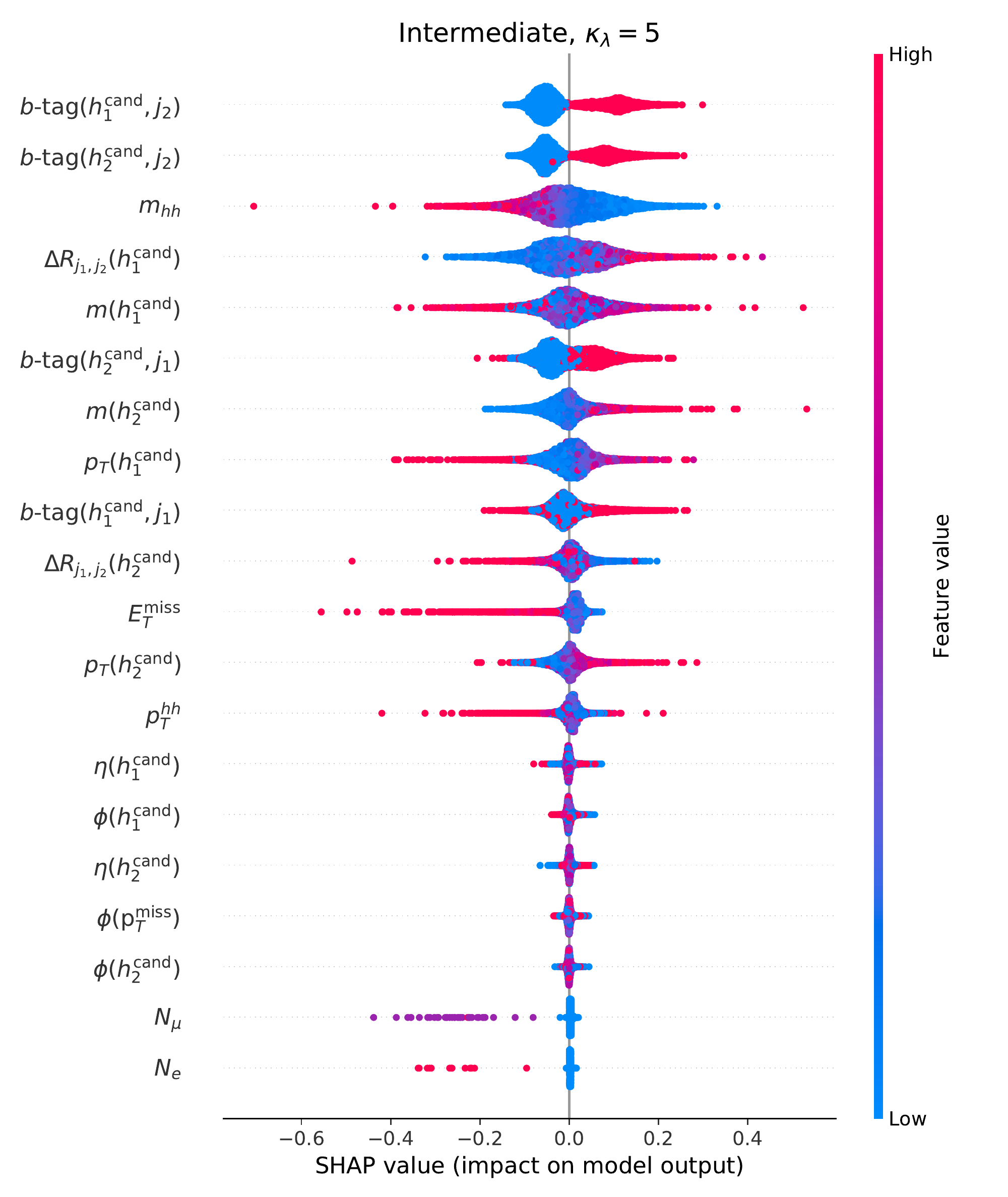}
    \caption{Intermediate}
    \end{subfigure} \\
    \begin{subfigure}[b]{0.48\textwidth}
        \includegraphics[width=\textwidth]{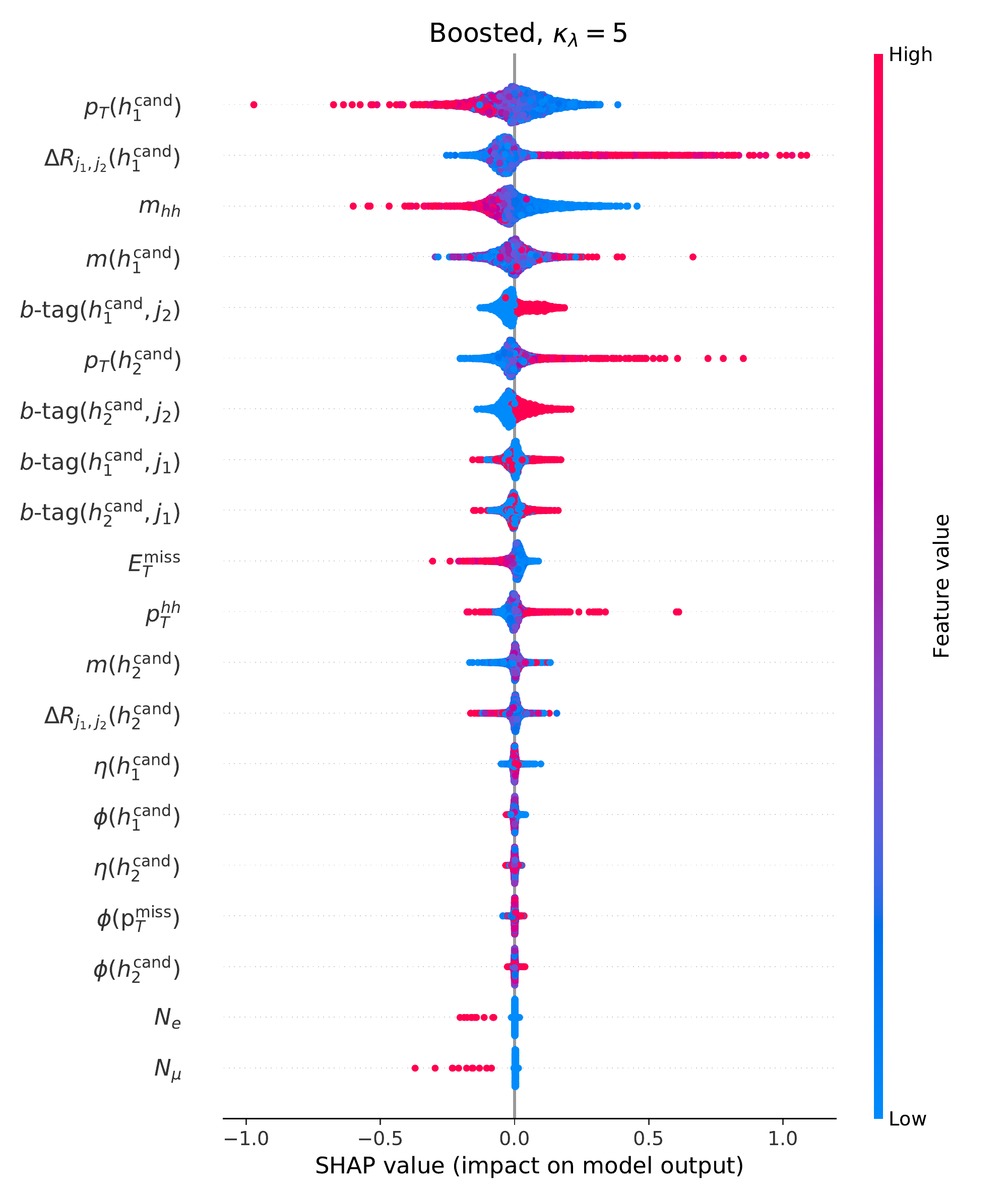}
    \caption{Boosted}
    \end{subfigure}  
    \caption{SHAP value plots representing ranked variable (feature) importance for the DNN trained on $\klam=5$ signals in the (a) resolved, (b) intermediate and (c) boosted categories. The input variables are ranked in descending order by their average absolute SHAP value plotted on the $x$-axis. The colour scale indicates the value of the variable on the specific event for which the SHAP value is plotted. Analogous plots for the DNN trained on $\klam = 1$ are found in appendix~\ref{sec:NNperform}. }
    \label{fig:shap_values_5}
\end{figure}

Figure \ref{fig:shap_values_5} shows the 20 input variables of the neural networks, ranked by their impact on the final signal score for the \klam = 5 training. This impact is measured by the magnitude of their SHAP values averaged over the whole dataset given to the framework, which is plotted on the $x$-axis. Each point plotted per row corresponds to one event fed to the framework, and its location along the $x$-axis represents what impact that variable has on the signal score of the event. The relative magnitude represents how much the value of that variable changes the signal score compared to all other variables in all events. Points with a larger positive SHAP value increase the signal score more while the inverse is true for negative SHAP values. The absolute scale of the SHAP value is arbitrary in these plots. The colour scale indicates the value of the feature on the specific event e.g.\ a blue dot on the $b$-tag($h_1^\text{cand}, j_1)$ indicates the leading subjet associated to the leading Higgs candidate is not $b$-tagged. Meanwhile, a pink dot in the $m_{hh}$ row indicates that event has high di-Higgs invariant mass for the plotted SHAP value.

The $b$-tagging state of (sub-)jets are among the most important features in the resolved and intermediate categories. This is less important in the boosted category possibly due to lower $b$-tagging efficiencies at high $\pt$, which could be improved by future work in novel $b$-tagging techniques~\cite{Huffman:2016wjk}. Angular and mass variables are stronger discriminants against multijet processes in this regime. The opening angles between these (sub-)jets and the invariant mass of the di-Higgs system carry a large amount of information in all three categories. Variables sensitive to semi-leptonic top decays, such as the number of leptons or missing transverse momentum, are effective at rejecting background (large negative SHAP value for high feature value) but less so at identifying signal (small positive SHAP values for any feature value).

Appendix~\ref{sec:NNperform} presents supplementary material characterising the performance of our neural networks. This includes receiver operator characteristic (ROC) curves, signal acceptance times efficiency, ranked feature importance for the DNN trained on $\klam = 1$, together with correlations between the neural network scores and reconstructed (di-)Higgs mass variables $m(h_1)$ and $\mhh$.

\FloatBarrier
%\newpage
\section{Higgs self-coupling constraints}
\label{sec:constraints}

This section presents the results and discussion for the \emph{baseline} and \emph{neural network} analyses. We compare the performance of the resolved, intermediate, boosted categories, and their combined constraints. We set the luminosity to the target HL-LHC dataset of $\mathcal{L} = 3000$~fb$^{-1}$. Subsection~\ref{sec:event_rates} presents the signal and background rates in the signal regions of our analyses. Subsection~\ref{sec:1d_limits} then performs the $\chi^2$ statistical analysis to determine \lamhhh constraints assuming $\kapt = 1$ and discusses the impact of systematic uncertainties. In subsection~\ref{sec:discrimination_power}, we evaluate $\lamhhh^i$ vs $\lamhhh^{j\neq i}$ discrimination power and discuss strategies for improvement. Finally,  subsection~\ref{sec:2d_limits} lifts the assumption of $\kapt = 1$ to examine the impact of top Yukawa \topyuk uncertainties on \klam constraints.

%-------------------------------------------
\subsection{\label{sec:event_rates}Signal and background rates}
%-------------------------------------------

\begin{table}
\centering
\resizebox{\linewidth}{!}{%
\begin{tabular}{lrrr|rrr|rrr}
\toprule
Category            &\multicolumn{3}{c}{Resolved} & \multicolumn{3}{c}{Intermediate} & \multicolumn{3}{c}{Boosted} \\
\midrule
Analysis            & Baseline & DNN  & DNN   & Baseline & DNN  & DNN   & Baseline & DNN  & DNN  \\
Trained on          &--- & $\klam = 5$ &  $\klam = 1$   & --- &  $\klam = 5$ & $\klam = 1$  & --- &  $\klam = 5$ & $\klam = 1$ \\
\midrule
$2b2j$              & 889000 & 134000 & 116000    & 112000 & 10400 & 24600    & 95300  & 24100 & 38400\\
$4b$                & 810000 & 183000 & 137000    & 28500  & 6170  & 13300    & 19700  & 6000  & 10900\\
$t\bar{t}$          & 60400  & 5430   & 6100      & 3650   & 339   & 639      & 915    & 76.8  & 385\\
$t\bar{t}+b\bar{b}$ & 2420   & 398    & 474       & 554    & 65.1  & 181      & 160    & 37.9  & 77  \\
$t\bar{t}h$         & 818    & 151    & 189       & 125    & 24.4  & 59       & 37     & 8.42  & 16\\
$Zh$                & 329    & 84.6   & 87        & 37     & 3.68  & 22       & 73     & 16.8  & 32  \\
\midrule
Total bkg $B$       & $1.8\times10^6$ & 323064 & 259000    & 144000 & 17041 & 38800   & 116000 & 30237 & 49800\\
$\varsigma_\text{stat}=\sqrt{B}$     & 1330  & 568   & 509      & 380  & 131 & 197    & 341  & 174  & 223\\
$\varsigma_\text{syst}=0.3\%\cdot B$ & 5290  & 969   & 778      & 433  & 51  & 116    & 349  & 91   & 149\\
$\varsigma_\text{syst}=1\%\cdot B$   & 17600 & 3230  & 2590     & 1440 & 170 & 388    & 1160 & 302  & 498 \\
$\varsigma_\text{syst}=5\%\cdot B$   & 88000 & 16200 & 13000    & 7200 & 852 & 1940   & 5800 & 1510 & 2490\\
\midrule\midrule
\multicolumn{2}{l}{$hh$ signal $(\klam, \kapt)$ } &   \\

\midrule
$(1, 1)_\text{SM}$  & 408  & 124   & 249      & 111 & 34.4 & 72      & 104 & 35.4 & 64 \\
$(2, 1)$            & 194  & 58.9  & 127      & 67  & 25.7 & 47      & 77  & 24.5 & 44\\
$(5, 1)$            & 669  & 263   & 175      & 51  & 19.6 & 34      & 49  & 10.6 & 26\\
$(10, 1)$           & 5230 & 2000  & 1890     & 411 & 134  & 275     & 138 & 71.9 & 97\\
$(-5, 1)$           & 6210 & 2210  & 3050     & 847 & 361  & 595     & 381 & 102  & 181\\
$(1, 1.2)$          & 1010 & 316   & 626      & 270 & 83.6 & 179     & 216 & 89.1 & 139\\
$(1, 0.8)$          & 149  & 48.2  & 97       & 37  & 13.7 & 26      & 35  & 12.0 & 21 \\

\bottomrule
\end{tabular}
}
\caption{\label{tab:SR_yields} Summary of signal region yields for the backgrounds and benchmark signals for the \emph{baseline analysis} and \emph{neural network analyses} (DNN) trained on $\klam = 1$ and $\klam = 5$. The yields shown are prior to binning in \mhh and normalised to integrated luminosity of $\mathcal{L} = 3000$~fb$^{-1}$. Dominant contributions to backgrounds are displayed together with the absolute size of the statistical $\varsigma_\text{stat}$ and benchmark systematic $\varsigma_\text{syst}$ uncertainties. Signals with benchmark couplings $(\klam, \kapt)$ are shown, where $(1, 1)$ is the SM value.}
\end{table}

\begin{table}
\centering
\resizebox{\linewidth}{!}{%
\begin{tabular}{lrrrr|rrrr|rrrr}
\toprule
Analysis           & \multicolumn{4}{c}{Baseline}    & \multicolumn{4}{|c}{DNN trained on $\klam = 5$}  & \multicolumn{4}{|c}{DNN trained on $\klam = 1$}      \\
\midrule
\mhh bin [GeV]     & $B$    & $S_{\klam =1}$ & $S_{\klam =2}$ & $S_{\klam =5}$ & $B$    & $S_{\klam =1}$ & $S_{\klam =2}$ & $S_{\klam =5}$ & $B$    & $S_{\klam =1}$ & $S_{\klam =2}$ & $S_{\klam =5}$ \\
\midrule\midrule
\multicolumn{13}{l}{Resolved}          \\
\midrule
$[200, 250]$       & 369000 & 3.1   & 3.6   & 133 & 100000    & 0.76   & 0.662   & 50.9    & 98    & 0.0   & 0.0   & 0.0  \\
$[250, 300]$       & 687000 & 28.4  & 7.9   & 231  & 112000   & 10.0   & 2.0  & 83.9   & 5660  & 4.1   & 0.7   & 11.2  \\
$[300, 350]$       & 310000 & 68.6  & 21.6  & 137  & 42500    & 23.6   & 9.6  & 69.1   & 85200 & 25.6  & 10.2  & 55.4 \\
$[350, 400]$       & 154000 & 87.8  & 42.7  & 66.5 & 25300    & 32.5   & 16.6 & 25.2   & 75200 & 59.7  & 27.5  & 33.5  \\
$[400, 500]$       & 179000 & 170   & 88.3  & 82.7 & 18200    & 52.1   & 25.5 & 31.4   & 81900 & 128.3 & 67.7  & 62.2  \\
$[>500]$           & 37600  & 50.1  & 30.0  & 17.3 & 1310     & 5.2    & 4.6  & 2.5    & 11400 & 30.9  & 21.0  & 12.5  \\
\midrule\midrule
\multicolumn{13}{l}{Intermediate}   \\
\midrule
$[200, 500]$       & 36600  & 28.7   & 14.1   & 14.9 & 5090    & 8.5   & 6.0   & 5.53 & 8250  & 16.7  & 9.0   & 8.9  \\
$[500, 600]$       & 69300  & 56.4   & 35.3   & 24.2 & 8310    & 20.0  & 14.2  & 9.2  & 20200 & 37.6  & 25.0  & 16.0   \\
$[>600]$           & 38600  & 26.0   & 17.4   & 12.1 & 3640    & 5.9   & 5.5   & 4.89 & 10300 & 18.0  & 13.3  & 9.1 \\
\midrule\midrule
\multicolumn{13}{l}{Boosted} \\
\midrule
$[500, 800]$       & 69400  & 74.3  & 49.8   & 27.9 & 25800 & 30.8  & 20.4 & 8.5   & 40500 & 52.5  & 34.7  & 17.6   \\
$[>800]$           & 46600  & 29.1  & 27.3   & 21.2 & 4430  & 4.6   & 4.1  & 2.15  & 9300  & 11.0  & 8.9   & 7.9  \\
\bottomrule
\end{tabular}
}
\caption{\label{tab:SR_yields_mhh_binned}Summary of signal region event yields binned by $\mhh$ for the total background $B$ and benchmark signals for different $\klam$ couplings $S_{\klam = i}$. These are displayed for the \emph{baseline} and \emph{neural network} (DNN) trained on $\klam = 5$, and trained on $\klam = 1$ for comparison. The yields are normalised to an integrated luminosity of $\mathcal{L} = 3000$~fb$^{-1}$ for the three categories. }
\end{table}

\begin{table}
\centering
\begin{tabular}{lc}
\toprule
Category     & Systematic $\zeta_b$ \\
\midrule
Resolved     & 0.3\%                        \\
Intermediate & 1\%                          \\
Boosted      & 5\%     \\
\bottomrule
\end{tabular}
\caption{\label{tab:assumed_syst}
The nominal relative systematic uncertainties on the background estimate $\zeta_b$ assumed when performing the $\chi^2$ analysis for the different analysis categories. }
\end{table}

\begin{table}
\small
\centering
\begin{tabular}{lrr|rr|rr}
\toprule
Analysis 
 & \multicolumn{2}{c}{Baseline} & \multicolumn{2}{|c}{DNN } & \multicolumn{2}{|c}{DNN } \\
Trained on & \multicolumn{2}{c}{---} & \multicolumn{2}{|c}{$\klam = 1$} & \multicolumn{2}{|c}{$\klam = 5$}\\
 \midrule\midrule
 Resolved      \\
\midrule
\mhh bin [GeV] & $Z_{\klam = 1}$ &  $Z_{\klam = 5}$ &  $Z_{\klam = 1}$ &  $Z_{\klam = 5}$ & $Z_{\klam = 1}$ &  $Z_{\klam = 5}$  \\
\midrule
$[200, 250]$ & 0.00   & 0.11   & 0.00   & 0.00   & 0.00 & 0.12 \\
$[250, 300]$ & 0.01   & 0.10   & 0.05   & 0.15   & 0.02 & 0.18 \\
$[300, 350]$ & 0.06   & 0.13   & 0.07   & 0.14   & 0.10 & 0.29 \\
$[350, 400]$ & 0.14   & 0.11   & 0.17   & 0.09   & 0.18 & 0.14 \\
$[400, 500]$ & 0.25   & 0.12   & 0.34   & 0.17   & 0.36 & 0.22 \\
> 500        & 0.22   & 0.08   & 0.28   & 0.11   & 0.14 & 0.07 \\
\midrule
Quadrature sum $\mhh$ bins (0\% syst) & 0.54 & 0.51 & 0.59 & 0.37 & 0.47 & 0.53 \\
Quadrature sum $\mhh$ bins & 0.37 & 0.27 & 0.48 & 0.30 & 0.44 & 0.44 \\
One bin $\mhh > 200$ GeV & 0.08 & 0.12 & 0.27 & 0.19 & 0.12 & 0.25 \\
\midrule
\midrule
Intermediate  \\
\midrule
$[200, 500]$ & 0.07 & 0.04 & 0.14 & 0.07 & 0.10 & 0.06\\
$[500, 600]$ & 0.08 & 0.03 & 0.15 & 0.06 & 0.16 & 0.07 \\
> 600        & 0.06 & 0.03 & 0.12 & 0.06 & 0.08 & 0.07\\
\midrule
Quadrature sum $\mhh$ bins (0\% syst) & 0.29 & 0.14 & 0.37 & 0.17 & 0.27 & 0.15 \\
Quadrature sum $\mhh$ bins & 0.12 & 0.06 & 0.24 & 0.12 & 0.21 & 0.12 \\
One bin $\mhh > 200$ GeV & 0.07 & 0.03 & 0.17 & 0.08 & 0.16 & 0.09\\
\midrule
\midrule
Boosted  \\
\midrule
$[500, 800]$ & 0.02 & 0.01 & 0.03 & 0.01 & 0.02 & 0.01\\
> 800        & 0.01 & 0.01 & 0.02 & 0.02 & 0.02 & 0.01 \\
\midrule
Quadrature sum $\mhh$ bins (0\% syst) & 0.31 & 0.14 & 0.28 & 0.12 & 0.20 & 0.062\\
Quadrature sum $\mhh$ bins & 0.02 & 0.01 & 0.03 & 0.02 & 0.03 & 0.01\\
One bin $\mhh > 500$ GeV & 0.02 & 0.01 & 0.03 & 0.01 & 0.02 & 0.01\\
\midrule
\midrule
Combined quadrature sum (0\% syst) & 0.69 & 0.55 & 0.75 & 0.42 & 0.58 & 0.56 \\
Combined quadrature sum & 0.39 & 0.27 & 0.53 & 0.32 & 0.49 & 0.46 \\
Combined one bin & 0.11 & 0.13 & 0.32 & 0.20 & 0.20 & 0.27 \\
\midrule
\bottomrule
\end{tabular}
\normalsize
\caption{
\label{tab:SR_significances_mhh_binned}
Statistical significances defined by $Z=S/\sqrt{B+(\zeta_b B)^2}$ for $\mathcal{L} = 3000$~fb$^{-1}$ using the signal and background counts of Table~\ref{tab:SR_yields_mhh_binned}. Here, $Z_i$ corresponds to using the $\klam = i$ signal samples and the nominal systematics $\zeta_b$ are taken from Table~\ref{tab:assumed_syst}. These are evaluated for the baseline analysis, the DNN analysis where the DNN was trained with the $\klam =1$ and $\klam =5$ samples. The results are shown for the resolved, the intermediate and boosted categories which are each separated by a double rule. The lowest two rows of each category presents the significances combined by the `quadrature sum' of the significances binned in $\mhh$, and second `one bin' sums the yields into one inclusive bin before calculating $Z$. The bottom three rows shows significances combined across all three categories.
}  
\end{table}

Table~\ref{tab:SR_yields} shows the signal and background yields in the final signal region for the \emph{baseline} (`Baseline' columns) and \emph{neural network} (`DNN' columns) analyses prior to binning in \mhh. For the \emph{baseline analysis} in the resolved category, we find the reducible $2b2j$ process comprises 49\% of the background due to its formidable cross-section and only modest suppression from $b$-tagging. The irreducible $4b$ rate is comparable, comprising 45\% of the background. These multijet processes constitute 94\% of the backgrounds, similar to recent experimental analyses~\cite{Aaboud:2018knk}. 
They also dominate in the intermediate and boosted categories but interestingly, the $4b/2b2j$ ratio reduces to 0.25 and 0.21 respectively, suggesting light flavour jet systems have higher boost than those from $b$-jets. The top quark contribution, dominated by $t\bar{t}$, is around 3\% for the resolved and intermediate categories. The single Higgs backgrounds contribute less than 1\% of the total, with the most dominant contributions arising from the irreducible associated production $t\bar{t} h$ and the electroweak $Zh$ processes.

For the \emph{baseline analysis} assuming the SM $\klam = 1$ signal rate presented in Table~\ref{tab:SR_yields}, we find a signal-to-background of $S/B = 2.3\times 10^{-4}$ and significance of $S/\sqrt{B} = 0.30$ in the resolved category. For the intermediate category, the signal-to-background is higher at $S/B = 7.7\times 10^{-4}$ but the significance $S/\sqrt{B} = 0.29$ is similar. For the boosted category, we find even higher signal-to-background $S/B = 9.0\times 10^{-4}$ but again similar significance $S/\sqrt{B} = 0.31$. This pattern is consistent with the intermediate and boosted categories having higher signal purity due to better background rejection than the resolved but lower absolute yields. With $S/B \sim 10^{-3}$ to $10^{-4}$ and $B\gtrsim 10^4$ before binning in \mhh, we expect background statistical uncertainties $\varsigma_\text{stat} = \sqrt{B}$ to be below the percent level and our analyses will be limited by systematic uncertainties. 

For the \emph{neural network analyses}, the DNN signal score $p_\text{signal}$ is required to satisfy $p_\text{signal}>0.75$. Two DNNs are considered: one trained on the SM $\klam = 1$ signal and the other on $\klam = 5$ where we expect the boundary of sensitivity to be for $\klam$ limits. In Table~\ref{tab:SR_yields} for the resolved category, we find the $p_\text{signal}>0.75$ requirement of the DNN trained on $\klam = 1$ gives a 61\% SM $\klam = 1$ signal efficiency for 86\% background rejection compared to the \emph{baseline analysis}, providing a signal-to-background of $S_{\klam = 1}/B = 9.6\times 10^{-4}$ and significance of $S_{\klam = 1}/\sqrt{B} = 0.49$. When a different signal hypothesis $\klam = 5$ is considered to that used in DNN training, we find the significance is lower $S_{\klam = 5}/\sqrt{B} = 0.34$, as one would expect. The background rejection is dominated by the reduction in the $2b2j$ process such that the irreducible $4b$ component of the background now becomes dominant for the resolved category. For the DNN trained on $\klam = 5$, we find the $p_\text{signal}>0.75$ requirement results in an 82\% background rejection for 39\% $\klam = 5$ signal efficiency, giving $S_{\klam = 5}/B = 8.1\times 10^{-4}$ and $S_{\klam = 5}/\sqrt{B} = 0.46$. As one would expect, this is higher for the $\klam = 5$ signal than using the DNN trained on $\klam = 1$, but this is only the case for the resolved category. Interestingly for the intermediate and boosted categories, the DNN trained on $\klam = 1$ results in higher signal yields for all signal hypotheses listed in Table~\ref{tab:SR_yields}. Such categories have a greater signal acceptance times efficiency for $\klam = 1$ than for $\klam = 5$ scenarios (Fig.~\ref{fig:baseline_acceptance}) as the former \klam case has a harder Higgs \pt spectrum (Fig.~\ref{fig:Norm_DiHiggsM}). 

Turning to the yields binned in \mhh,   Table~\ref{tab:SR_yields_mhh_binned} compares these across the three analyses and three categories for the $\klam = 1, 2, 5$ signals. For the DNN trained on $\klam = 1$, the yields are suppressed in the low \mhh bins such that there is no signal in $\mhh \in [200, 250]$~GeV. This is expected given the background rate is high but the signal rate is low in these bins. Therefore, only training the DNN on $\klam = 1$ signals is suboptimal for signals that occupy lower \mhh values such as $\klam = 5$. We see that the $\klam = 5$ training retains the $\klam = 5$ signal in the first $\mhh \in [200, 250]$ bin, which is important for \klam constraints that will be further discussed in the next subsections. 

We now evaluate the binned statistical significances defined by $Z=S/\sqrt{B+(\zeta_b B)^2}$ including background systematics $\zeta_b$ and display these in
Table~\ref{tab:SR_significances_mhh_binned}. For each category, the lowest two rows show the significances combined in two ways for comparison: \begin{itemize} 
\item \emph{Quadrature sum}. The significance for each \mhh bin $i$ is summed in quadrature to give $Z = \sqrt{\sum_i Z_i^2}$. This accounts for \mhh shape information, where for simplicity we assume no correlations in systematics.
\item \emph{One bin}. The yields in each bin $i$ are summed $S = \sum_i S_i, B = \sum_i B_i$ before the significance $Z$ is calculated such that no \mhh shape information is considered. 
\end{itemize}
Finally, the bottom three rows of Table~\ref{tab:SR_significances_mhh_binned} combine the significances across all three categories. The quadrature sum considers all the \mhh bins in all three categories, while the combined one bin approach is the quadrature sum of the \emph{one bin} significances from each category. For simplicity, we neglect statistical and systematic uncertainties on the signal in the significances as analyses are in background-dominated $S\ll B$ regimes. Given $B \gg 1$, we assume statistical uncertainties follow the Gaussian approximation $\varsigma_\text{stat}^{B} = \sqrt{B}$. 

For the systematic uncertainties, we adopt a simple approach that assumes the nominal benchmark values $\zeta_b$ summarised in Table~\ref{tab:assumed_syst}. We assume $\zeta_b = 0.3\%$ for the resolved category, which is very demanding but in accord with Ref.~\cite{ATL-PHYS-PUB-2018-053}; this is a HL-LHC extrapolation of the recent ATLAS analysis which reports 2\% systematics~\cite{Aaboud:2018knk}. For the boosted category, a similar ATLAS analysis~\cite{Aaboud:2018knk} currently reports up to 18\% systematics, dominated by the data-driven background estimate method and jet mass resolution~\cite{Aaboud:2018knk}, while the current CMS analysis report up to 7\% systematics~\cite{Sirunyan:2017isc}. For our work, we assume that this can be controlled to the level of $\zeta_b = 5\%$ at the HL-LHC. For the intermediate category, we assume a benchmark value between what we apply for resolved and boosted of $\zeta_b = 1\%$. In experimental implementations~\cite{Aad:2015xja,Aaboud:2016xco,Aaboud:2018knk,Aad:2020kub,Sirunyan:2017isc,Sirunyan:2018qca,Sirunyan:2018tki}, the multijet background is typically modelled via data-driven methods so is not subject to theoretical uncertainties such as renormalisation scale, parton shower and parton distribution functions. Nonetheless, data-driven procedures have systematics due to unknown composition and extrapolation over correlated variables. 

For the \emph{baseline analysis} significances in Table~\ref{tab:SR_significances_mhh_binned}, the total significance is higher for the quadrature sum approach compared to the one bin counterpart, demonstrating the importance of \mhh shape information for sensitivity. With $\zeta_b = 0\%$, the $Z_{\klam = 1}$ significance evaluated using the $\klam = 1$ signal for the resolved category is 0.54 compared to 0.37 with the nominal 0.3\% systematics and just 0.08 without binning in \mhh. The intermediate and boosted categories have comparable significances of 0.29 and 0.31 respectively with $\zeta_b = 0\%$, but dilutes to 0.12 and 0.02 with their nominal systematics of Table~\ref{tab:assumed_syst}. The significance combining all categories is 0.69 for $\zeta_b = 0\%$ and 0.39 with nominal systematics. 

Moving to the \emph{neural network analyses}, we find the significances are higher than those of the corresponding \emph{baseline analysis} across all \mhh bins. This is especially important when systematics are included, where we see the resolved category $Z_{\klam = 1}$ significance improve from 0.54 to 0.59 for the DNN trained on $\klam = 1$ with 0\% systematics but the improvement is 0.37 to 0.48 with the nominal 0.3\% systematics. When systematics are included, this suggests that the background rejection is more important to improve significance. We find that the DNN trained on $\klam = 5$ provides higher significances for $\klam = 5 $ signals ($Z_{\klam = 5} = 0.44$) than the DNN trained on $\klam = 1$ ($Z_{\klam = 5} = 0.30$) in the resolved category. The intermediate category has significances that are numerically about half those of the resolved category: $Z_{\klam = 1} = 0.24$ (vs 0.48) for DNN trained on $\klam = 1$ intermediate (vs resolved). With all categories combined (bottom rows of Table~\ref{tab:SR_significances_mhh_binned}), the DNN trained on $\klam = 1$ gives $Z_{\klam = 1} = 0.75$ with 0\% systematics and is diluted to 0.53 with systematics, which improves upon the \emph{baseline analysis} significance of 0.39. The significance without binning in \mhh is lower at 0.32, suggesting that additional discrimination power remained in the \mhh shape not captured by a cut on the DNN score.  

\FloatBarrier
%-------------------------------------------
\subsection{\label{sec:1d_limits}\texorpdfstring{$\chi^2$ analysis of \lamhhh fixing $\kapt = 1$}{χ² analysis of λₕₕₕ fixing κₜ = 1}}
%-------------------------------------------

We now perform  a $\chi^2$ analysis to evaluate projections of Higgs self-coupling \lamhhh constraints. We adapt the definition in Eq.~\eqref{eq:generic_chiSq} into one for benchmarking our analyses:
\begin{equation}
    \chi^2 = \sum_i \left[\frac{(S-S_\text{SM})^2}{S+B+\left(\zeta_b B\right)^2+\left(\zeta_sS\right)^2}\right]_i.
\end{equation}
Here, $B$ is the total background rate, $S_\text{SM}$ is the SM signal rate, and $S$ is the signal rate we wish to distinguish from the SM counterpart. These are evaluated in each \mhh bin $i$ using the yields in Table~\ref{tab:SR_yields_mhh_binned}. Like the previous significance calculations of subsection~\ref{sec:event_rates}, we neglect statistical and systematic uncertainties on the signal as we are in background-dominated $S\ll B$ regimes. For simplicity, we perform a statistical combination by summing the $\chi^2$ values assuming constant and uncorrelated uncertainties across bins $\{i\}$. We first combine the yields in the \mhh bins shown in Table \ref{tab:SR_yields_mhh_binned} separately for the resolved, intermediate and boosted categories to explore their complementary sensitivity. Then we combine the three categories to obtain the final combined constraint. 

\begin{figure}
    \centering
    \begin{subfigure}[b]{0.5\textwidth}
        \includegraphics[width=\textwidth]{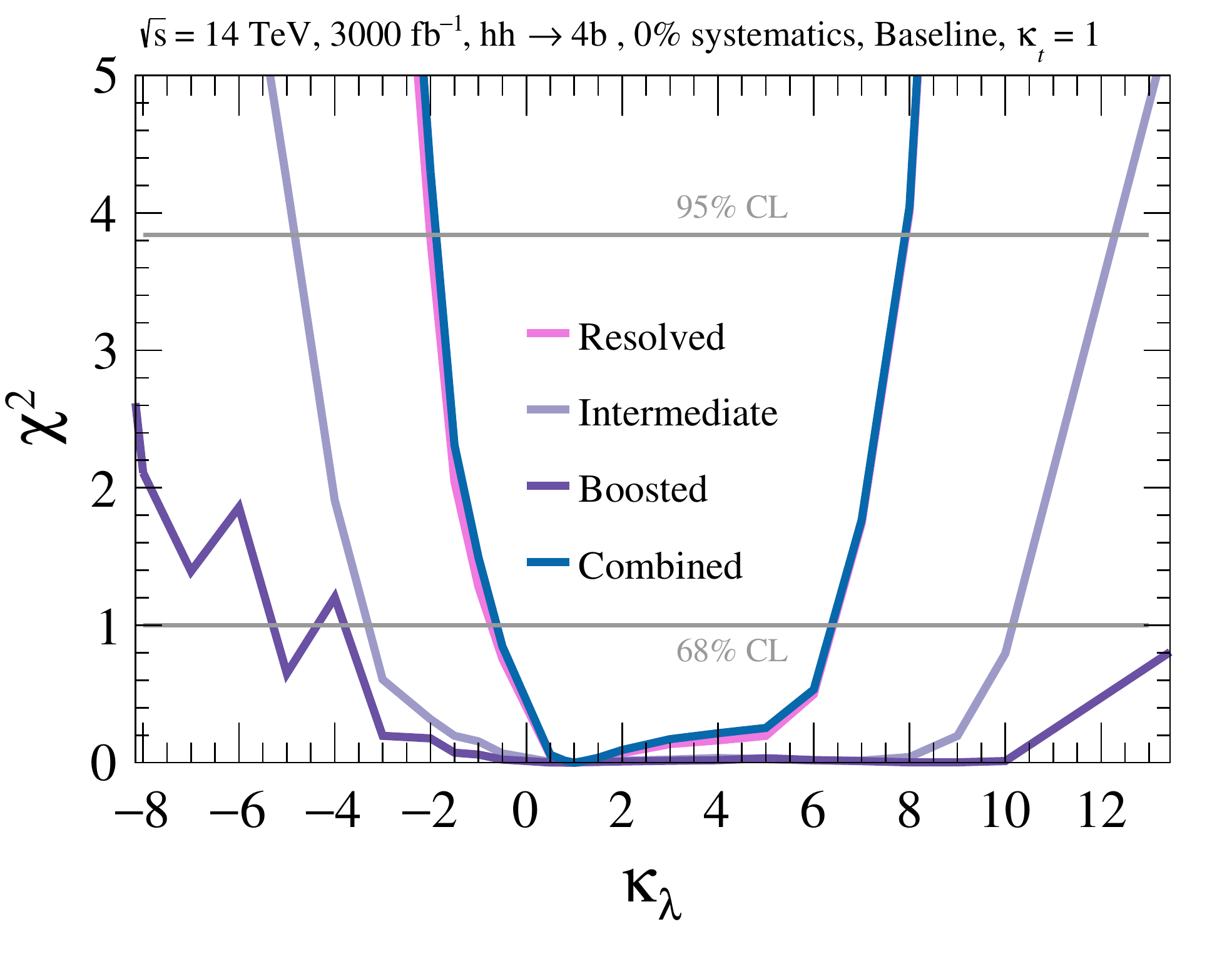}
        \caption{\label{fig:limit1d_chiSq_a}
        Baseline analysis, 0\% syst.}
    \end{subfigure}%
    \begin{subfigure}[b]{0.5\textwidth}
        \includegraphics[width=\textwidth]{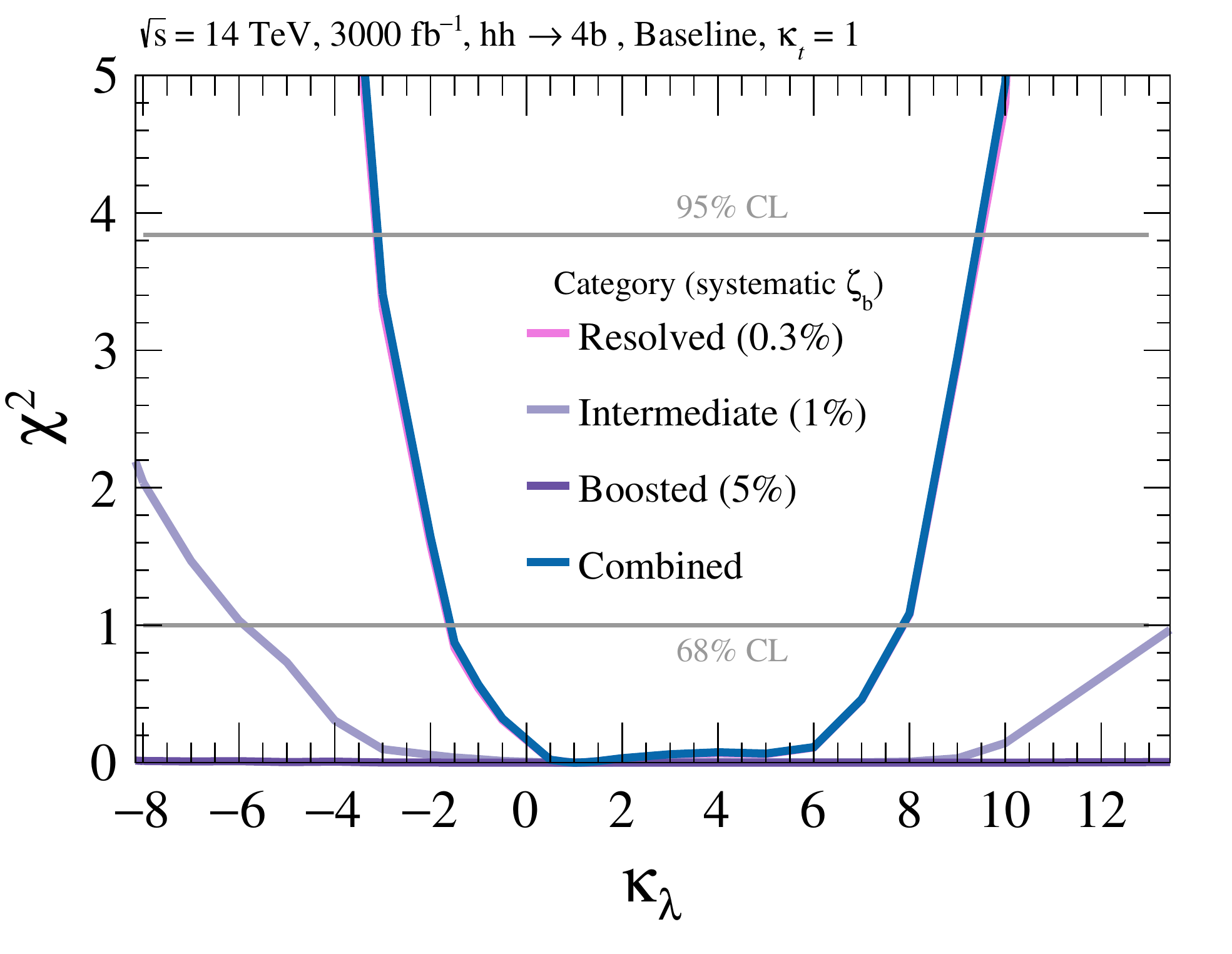}
        \caption{\label{fig:limit1d_chiSq_b}
        Baseline analysis, nominal syst.}
    \end{subfigure}\\
        \begin{subfigure}[b]{0.5\textwidth}
        \includegraphics[width=\textwidth]{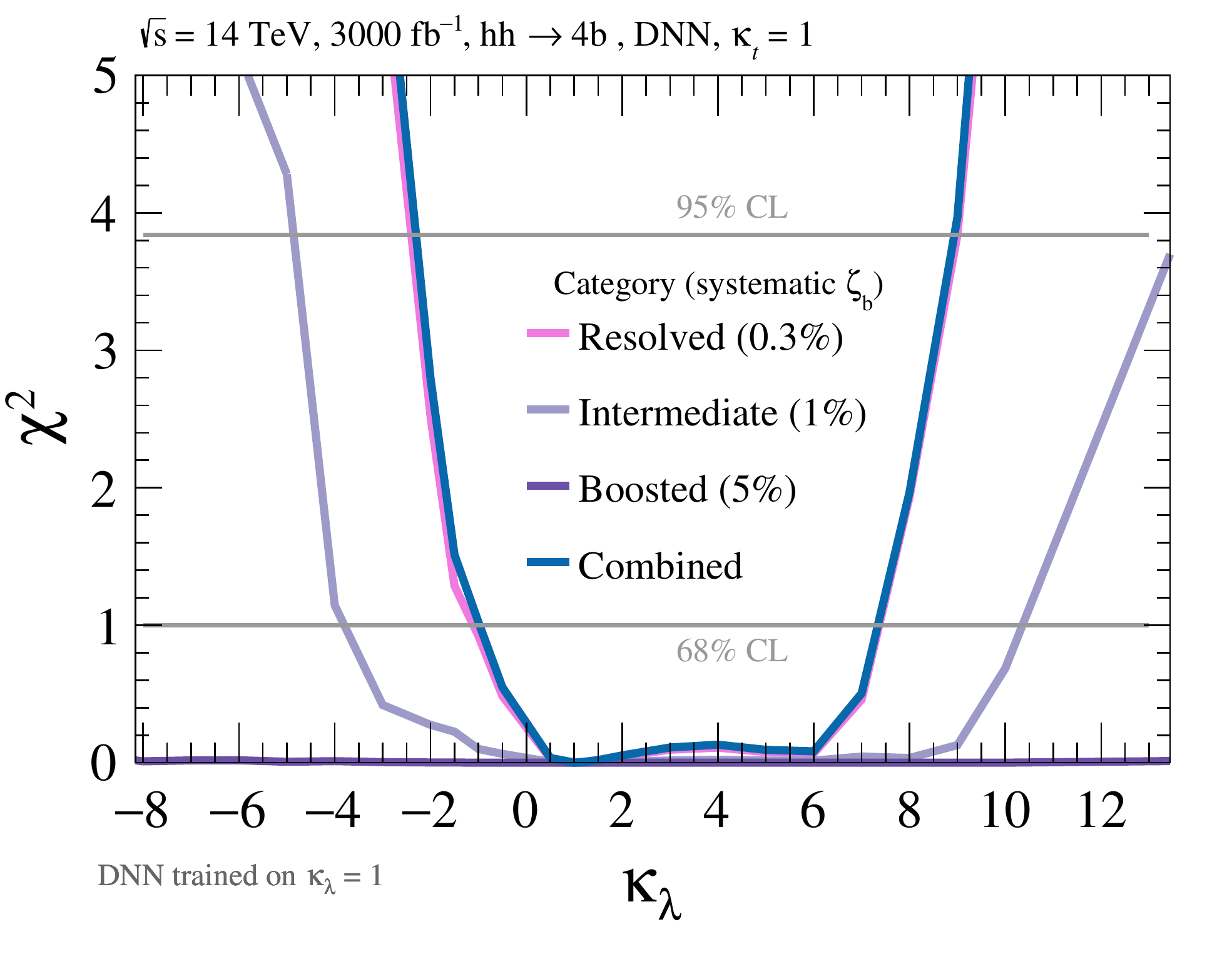}
        \caption{\label{fig:limit1d_chiSq_c}
        DNN trained on $\klam = 1$, nominal syst.}
    \end{subfigure}%
    \begin{subfigure}[b]{0.5\textwidth}
        \includegraphics[width=\textwidth]{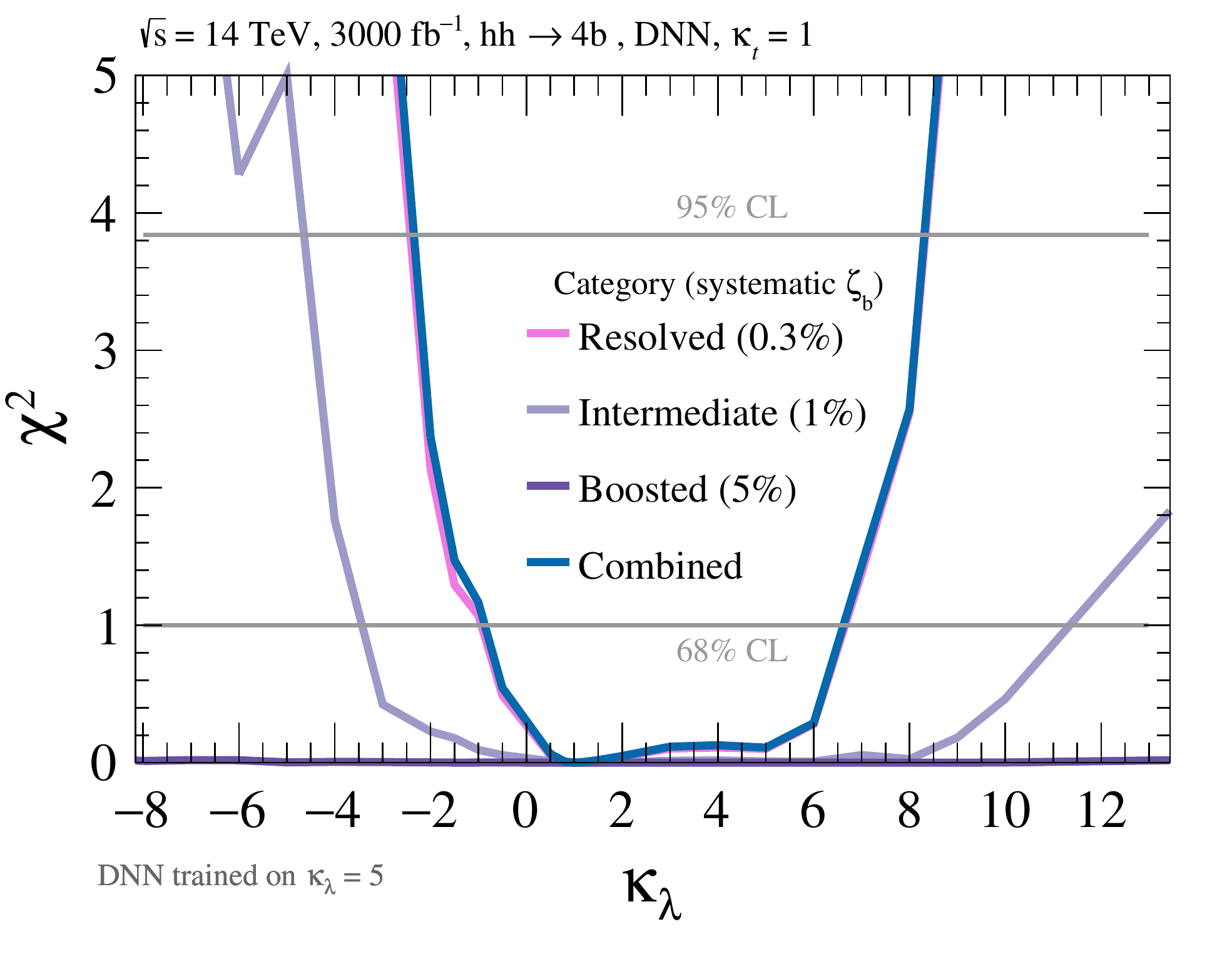}
        \caption{\label{fig:limit1d_chiSq_d}
        DNN trained on $\klam = 5$, nominal syst.}
    \end{subfigure}
    \caption{The $\chi^2$ vs $\klam$ distributions with fixed $\kapt=1$ for different analyses and assumed systematics using $\mathcal{L} = 3000$~fb$^{-1}$. This is displayed in the resolved (pink), intermediate (lilac), boosted (purple) categories, together with their combination (blue). The \emph{baseline analysis} is shown assuming (a) 0\% background systematics and (b) the nominal systematics of 0.3\%, 1\% and 5\% for the resolved, intermediate and boosted categories, respectively (Table~\ref{tab:assumed_syst}). The \emph{neural network analysis} is shown with the nominal systematics for the DNN trained on (c) $\klam = 1$ and (d) $\klam = 5$. The grey horizontal lines at $\chi^2 = 1, 3.84$ indicate the 68\% CL, 95\% CL thresholds respectively. The unphysical spikes for $\klam < 4$ for the boosted $\chi^2$ line in subfigure (a) and $\klam < -5$ for subfigure (d) are due to limited MC statistics.  }
    \label{fig:limit1d_chiSq}
\end{figure}
 
% Limits vs systematic for DNN trained on klam = 5
\begin{figure}
    \centering
    \includegraphics[width=\textwidth]{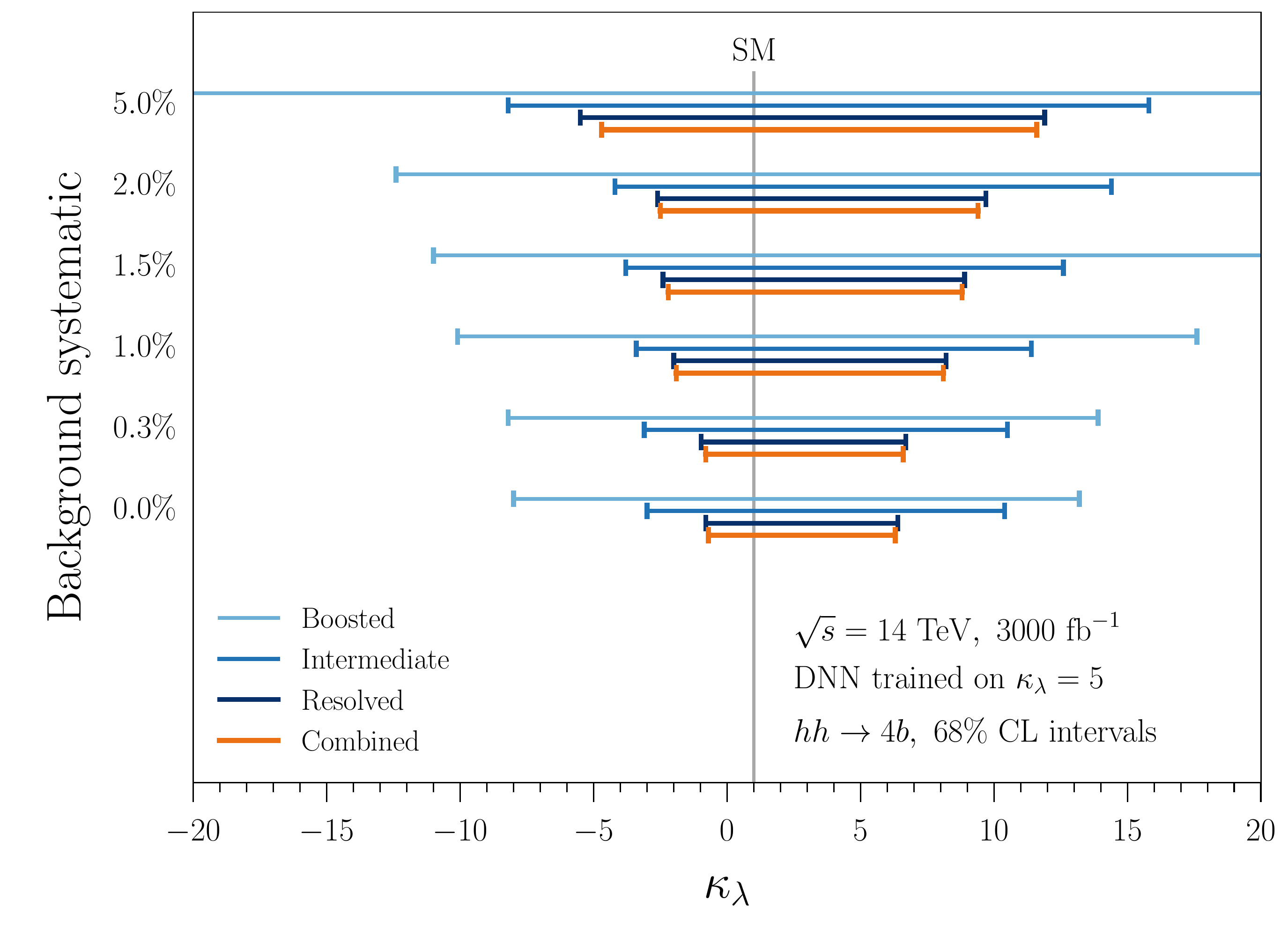}
    \caption{\label{fig:summary_1dlimits} 
    Summary of 68\% CL intervals ($\chi^2 < 1$) on \klam with fixed $\kapt = 1$ for different assumed background systematics from 0\% to 5\% for the \emph{neural network analysis} trained on the $\klam = 5$ signal. This is shown for the resolved (dark blue), intermediate (medium blue), boosted (light blue) categories along with their combination (orange). The luminosity is assumed to be 3000~fb$^{-1}$. Lines without endcaps mean the 68\% CL limit is outside the considered range of $\klam \in [-20, 20]$. The vertical grey line denotes the SM value of \klam. The nominal systematics assumed in this work are shown in Table~\ref{tab:assumed_syst}. In appendix~\ref{sec:chiSq}, Figure~\ref{fig:summary_1dlimits_DNNklam1} shows the version of this plot with the DNN trained on the $\klam = 1$ signal. }
\end{figure}

For the \emph{baseline analysis}, Fig.~\ref{fig:limit1d_chiSq_a} shows the $\chi^2$ vs \klam distribution without systematics. The resolved category is most constraining $\klam \in [-0.75, 6.5]$ at 68\% CL, followed by the intermediate $\klam \in [-3.3, 10.2]$ at 68\% CL with boosted being the weakest category. 
To gain intuition for the sensitivity between the three categories, we note that Fig.~\ref{fig:Norm_DiHiggsM} shows that signals with $\klam \gtrsim 5$ have the lowest Higgs \pt spectrum. Therefore, the intermediate and boosted scenarios have lower acceptance for these scenarios where the upper boundary of \klam sensitivity is expected. Meanwhile, for negative \klam, we see that the Higgs bosons in $\klam \lesssim -1$ scenarios have modest boost (Fig.~\ref{fig:Norm_DiHiggsM}). Therefore, the acceptance is higher for all categories and the constraints are stronger for negative \klam. Although scenarios near the SM value $\klam = 1$ have the most boosted Higgs bosons and signal acceptance is high (Fig.~\ref{fig:baseline_acceptance}), these \klam values are very challenging to constrain in a $\chi^2$ analysis because the total signal cross-sections are small and change slowly with respect to \klam due to destructive interference (subsection~\ref{sec:hhpheno}). Inspecting Table~\ref{tab:SR_yields}, we find factors of two change in signal yield between the challenging hypotheses of $\klam = 1, 2, 5$. However, the yields binned in \mhh (Table~\ref{tab:SR_yields_mhh_binned}) suggest that our shape analysis does have modest sensitivity to signal distribution changes for these $\klam$ scenarios even if it is insufficient for 68\% CL sensitivity.

Figure~\ref{fig:limit1d_chiSq_b} shows the $\chi^2$ vs $\klam$ distributions assuming the nominal systematics  of Table~\ref{tab:assumed_syst}. The 0.3\% systematics for the resolved category dilutes the constraints to $\klam \in [1.6, 7.9]$ at 68\% CL. Meanwhile, the 1\% systematics for the intermediate category weakens the constraint to $\klam \in [-5.8, 13.6]$ at 68\% CL. However, the boosted category loses all constraining power in the considered \klam range for the assumed 5\% systematics. Overall for the \emph{baseline analysis}, the statistical combination of all three categories leads to a 68\% CL constraint of $\klam \in [-0.6, 6.4]$ for no systematics. This degrades to $\klam \in [-1.6, 7.9]$ when assuming the nominal systematics, driven by the sensitivity of the resolved category.

Turning to the \emph{neural network analysis}, we evaluate the $\chi^2$ vs \klam distribution assuming the nominal systematics with the DNN trained on $\klam = 1$ (Fig.~\ref{fig:limit1d_chiSq_c}) and $\klam = 5$ (Fig.~\ref{fig:limit1d_chiSq_d}). Statistically combining all three categories achieves 68\% CL constraints of $\klam \in [-1.0,7.4]$  and $\klam \in [-0.8,6.6]$ for the DNN trained on $\klam = 1$ and $\klam = 5$ respectively. These improve upon the \emph{baseline analysis} of $\klam \in [-1.6, 7.8]$, with the $\klam = 5$ DNN training surpassing those achieved by the $\klam = 1$ DNN training. This shows that dedicated optimisation for a signal closer to the boundary of sensitivity can improve \klam constraints. We choose to train on the $\klam = 5$ signal due to its substantial triangle diagram contribution and therefore has a significant fraction of events at low \mhh. The kinematics are also reasonably representative of higher $\klam$ values (Fig.~\ref{fig:Norm_DiHiggsM_b}). Training on $\klam = 5$ instead of $\klam = 1$ encourages the neural network to retain low \mhh events, which has larger shape differences between different \klam values. This underscores the importance of optimisation away from the signals with SM couplings for \klam constraints in the $hh \to 4b$ final state. 

A future improvement may consider designing multiple independent neural networks, each optimised for one sampled value of \klam. A more sophisticated approach may extend the idea of a parameterised neural network~\cite{Baldi:2016fzo} to construct an observable that is a well-behaved function of \klam and provides good signal discrimination power, but this is beyond the scope of this work. While we find that sensitivity is driven by the resolved category, it is noteworthy that the DNN substantially improves the \klam constraints of the intermediate category from $\klam \in [-5.8, 13.6]$ of the \emph{baseline analysis} to $\klam \in [-3.8, 10.4]$ for the DNN trained on $\klam = 1$ and $\klam \in [-3.4, 11.3]$ for the DNN trained on $\klam = 5$. In a wider context, we note that intermediate category has close to comparable \lamhhh constraints with projected HL-LHC constraints for the di-$b$-quark plus diboson $hh\to bbVV$ final states~\cite{CMS-PAS-FTR-18-019}. 

While we have assumed the nominal systematics  of Table~\ref{tab:assumed_syst},  Fig.~\ref{fig:summary_1dlimits} now evaluates the 68\% CL \klam limits for different systematics ranging from 0\% to 5\% with the DNN trained on $\klam =5$. We find that the constraints for the resolved category degrade to $\klam \in [-5.5,11.9]$ for $\zeta_b = 5\%$ and the intermediate category constraints weaken to $\klam \in [-8.2,15.8]$. This shows how important the assumed level of systematics are for constraining \lamhhh due to small $S/B \simeq 10^{-3}$. Meanwhile, the boosted category has no constraining power within $\klam \in [-20, 20]$ when systematic uncertainties are assumed at the nominal 5\% level. If these could be improved to the 2\% level, the boosted category can constrain $\klam > -12.4$. We also find that if the systematics are controlled to 1\%, there is additionally sensitivity to positive $\klam < 17.6$.  The difficulty in forecasting experimental systematics at the HL-LHC leads to significant uncertainty for projections of \klam constraints. One can also interpret these systematics as benchmark targets for future analyses to improve \lamhhh constraints.

%--------------------------------
\subsection{\label{sec:discrimination_power}\texorpdfstring{$\lamhhh^i$ vs $\lamhhh^{j\neq i}$ discrimination power}{λₕₕₕ\^{}i vs λₕₕₕ\^{}j discrimination power}}
%--------------------------------

\begin{figure}
    \centering
    \includegraphics[width=\textwidth]{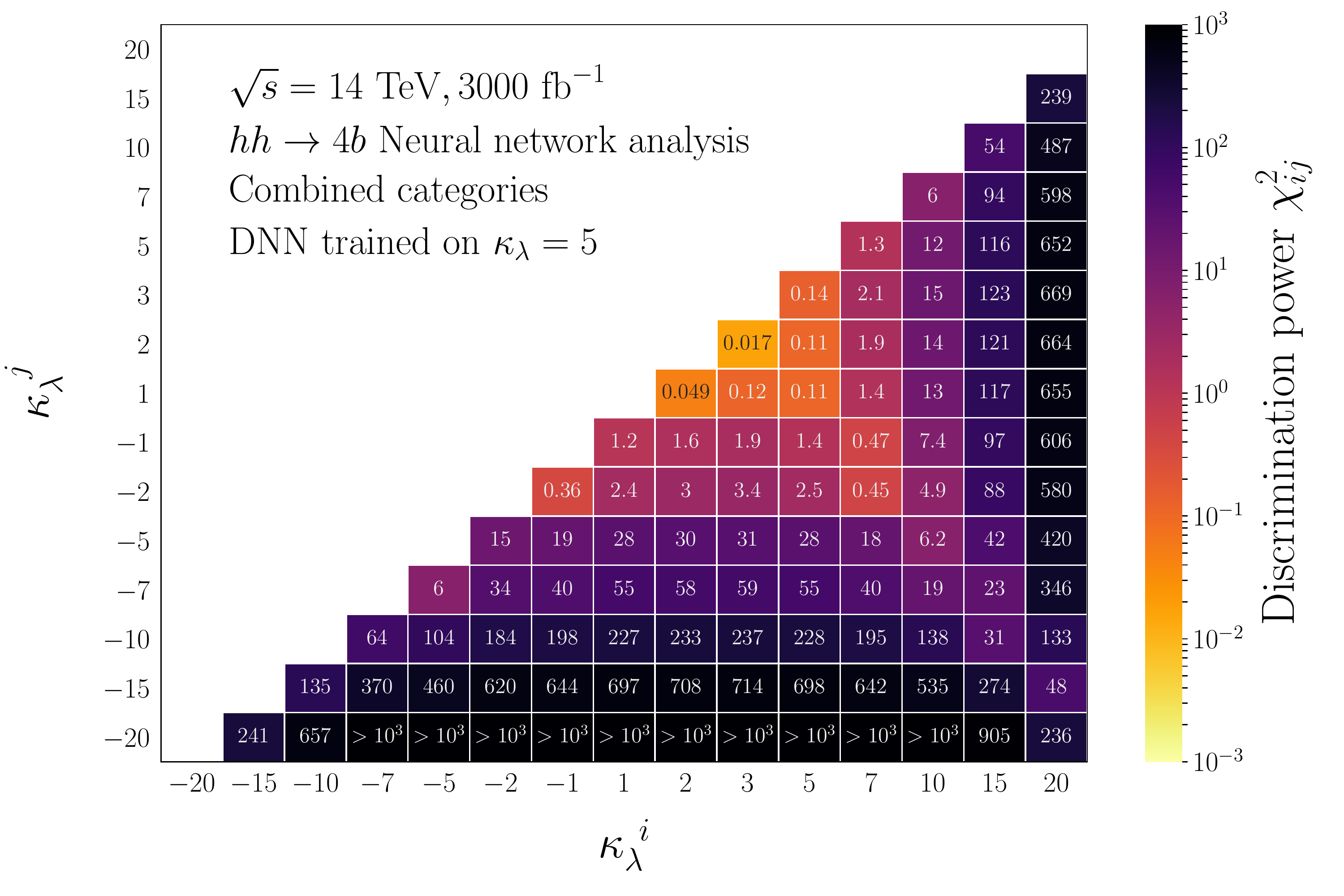}%
    \caption{Discrimination power $\chi^2_{ij}$ (higher is better) between two self-coupling hypotheses $\lamhhh^i$ vs $\lamhhh^j$ for the \emph{neural network analysis} after combining the three categories. The DNN is trained on the $\klam = 5$ sample, where the signal score is required to be $p_\text{signal} > 0.75$. The $\chi_{ij}^2$ is calculated from Eq.~\eqref{eq:generic_chiSq} assuming $\kapt = 1$ and background systematics of $\zeta_b = 0.3\%, 1\%$ and 5\% for the resolved, intermediate and boosted categories, respectively. For comparison, this figure for the \emph{baseline} and \emph{neural network} analyses trained on $\klam = 1$, are found in appendix~\ref{sec:chiSq}. }
    \label{fig:limit1d_chiSq_2Dlambda_DNN}
\end{figure}

So far, our $\chi^2$ distributions (Fig.~\ref{fig:limit1d_chiSq}) have been evaluated with respect to the SM $\klam = 1$ signal i.e.\ assuming we observe the SM signal in data. We now generalise this to evaluate the discrimination power between any two \klam hypotheses, which we present in Fig.~\ref{fig:limit1d_chiSq_2Dlambda_DNN}. While we train our DNN analysis on one class of signal (here $\klam = 5$), it is important to quantify the signal characterisation capability should \lamhhh deviate from the SM or the trained signal scenario. We calculate the $\chi_{ij}^2$ between two signal hypotheses $(i, j)$ shown on the axes of Fig.~\ref{fig:limit1d_chiSq_2Dlambda_DNN}. We present this for the \emph{neural network analysis} trained on $\klam = 5$, combining the three categories and assuming the nominal systematics in Table~\ref{tab:assumed_syst}. Larger values of $\chi^2_{ij}$ indicate that it is easier to discriminate between two hypotheses $(i, j)$, while $\chi^2_{ij}$ values around unity or below indicate difficulty in discriminating. 

Overall, we find Fig.~\ref{fig:limit1d_chiSq_2Dlambda_DNN} shows good discrimination power where $\chi^2_{ij} \gg 1$ for scenarios with $\klam \geq 10$ and $\klam \leq -5$, whose sensitivity is driven by the total cross-section. This corresponds to \klam values far away from destructive interference and the cross-section changing rapidly with respect to \klam. We also note small $\chi^2_{ij} < 1$ occurs around $\klam = -2$ and $\klam = 7$, which has $\chi^2_{ij} \simeq 0.5$. This arises physically due to the total cross-sections being similar $\sigma(\klam = -2) / \sigma(\klam = 7) \simeq 0.84$ combined with acceptance being higher for negative \klam (Fig.~\ref{fig:baseline_acceptance}) due to harder Higgs \pt spectra (Fig.~\ref{fig:Norm_DiHiggsM}). These sign degeneracies are not symmetric about zero due to maximal destructive interference being near $\klam \sim 2.5$. Generally, scenarios that are difficult to distinguish fall along a northwest-to-southeast corridor of low $\chi^2_{ij}$ values. There is also a cluster around $1 \leq \klam < 4$ with $\chi^2_{ij} < 1$ arising from both total cross-sections and changes with respect to \klam being suppressed due to near-maximal destructive interference (subsection~\ref{sec:hhpheno}).

For future work, it would be interesting to exploit \mhh shape information better via dedicated optimisations to break degeneracies in the \klam parameter space further. Moreover, one could imagine a DNN trained just on a single \klam hypothesis $\klam = 5$ is suboptimal for discriminating against two signal hypotheses unrelated to $\klam = 5$. It would be desirable to design analyses, such as generalising the current neural network strategy, to optimise the discrimination power across a wider variety of \klam values.

%-------------------------------------------
\subsection{\label{sec:2d_limits}\texorpdfstring{\klam constraints allowing $\kapt \neq 1$}{κλ constraints allowing κₜ /= 1}}
%-------------------------------------------

\begin{figure}
    \centering
    \begin{subfigure}[b]{0.5\textwidth}
        \includegraphics[width=\textwidth]{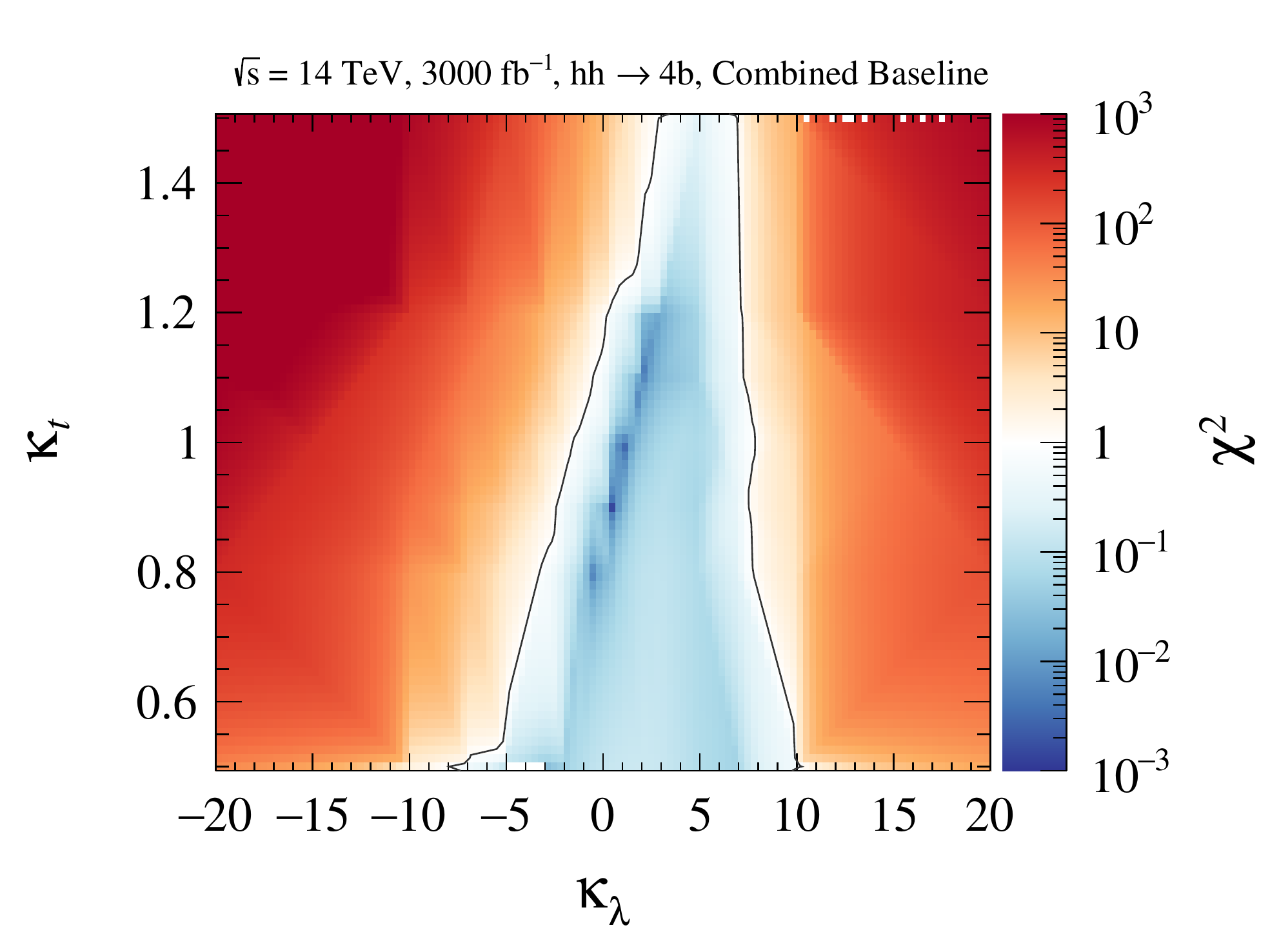}
        \caption{\label{fig:limit2d_chiSq_a}Baseline analysis}
    \end{subfigure}%
    \begin{subfigure}[b]{0.5\textwidth}
        \includegraphics[width=\textwidth]{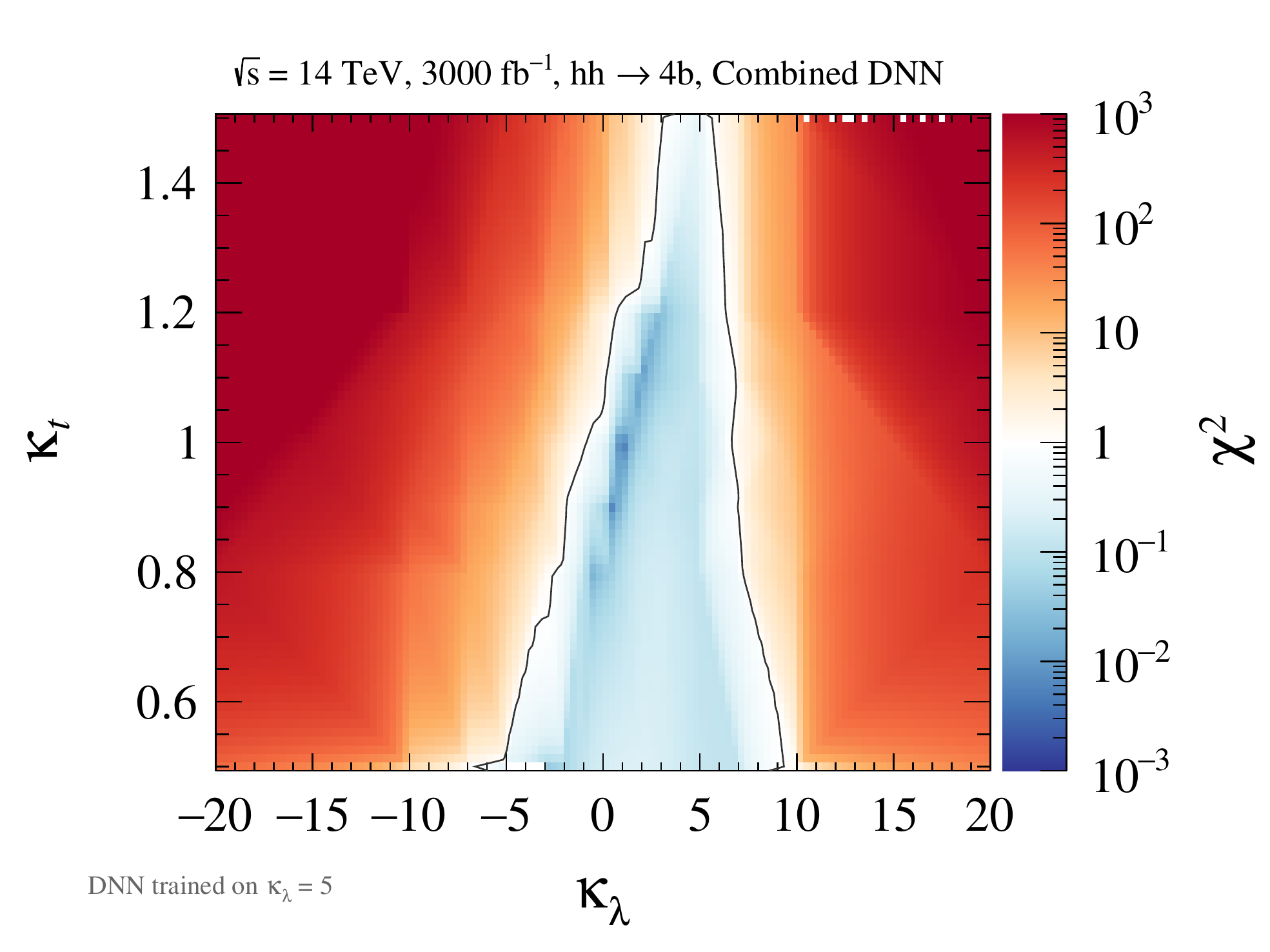}
        \caption{\label{fig:limit2d_chiSq_b}Neural network analysis}
    \end{subfigure}
    \caption{The combined $\chi^2$ distributions for the (a) \emph{baseline} and (b) \emph{neural network} analyses in the \kapt vs \klam plane, assuming $\mathcal{L} = 3000$~fb$^{-1}$. The $\chi^2$ statistically combines the resolved, intermediate and boosted categories with 0.3\%, 1\% and 5\% background systematic uncertainties, respectively. The DNN is trained on $\klam = 5$. The grey contour indicates $\chi^2 = 1$ corresponding to 68\% CL. A breakdown by analysis category is displayed in appendix~\ref{sec:chiSq}. }
    \label{fig:limit2d_chiSq}
\end{figure}

\begin{figure}
    \centering
    \includegraphics[width=\textwidth]{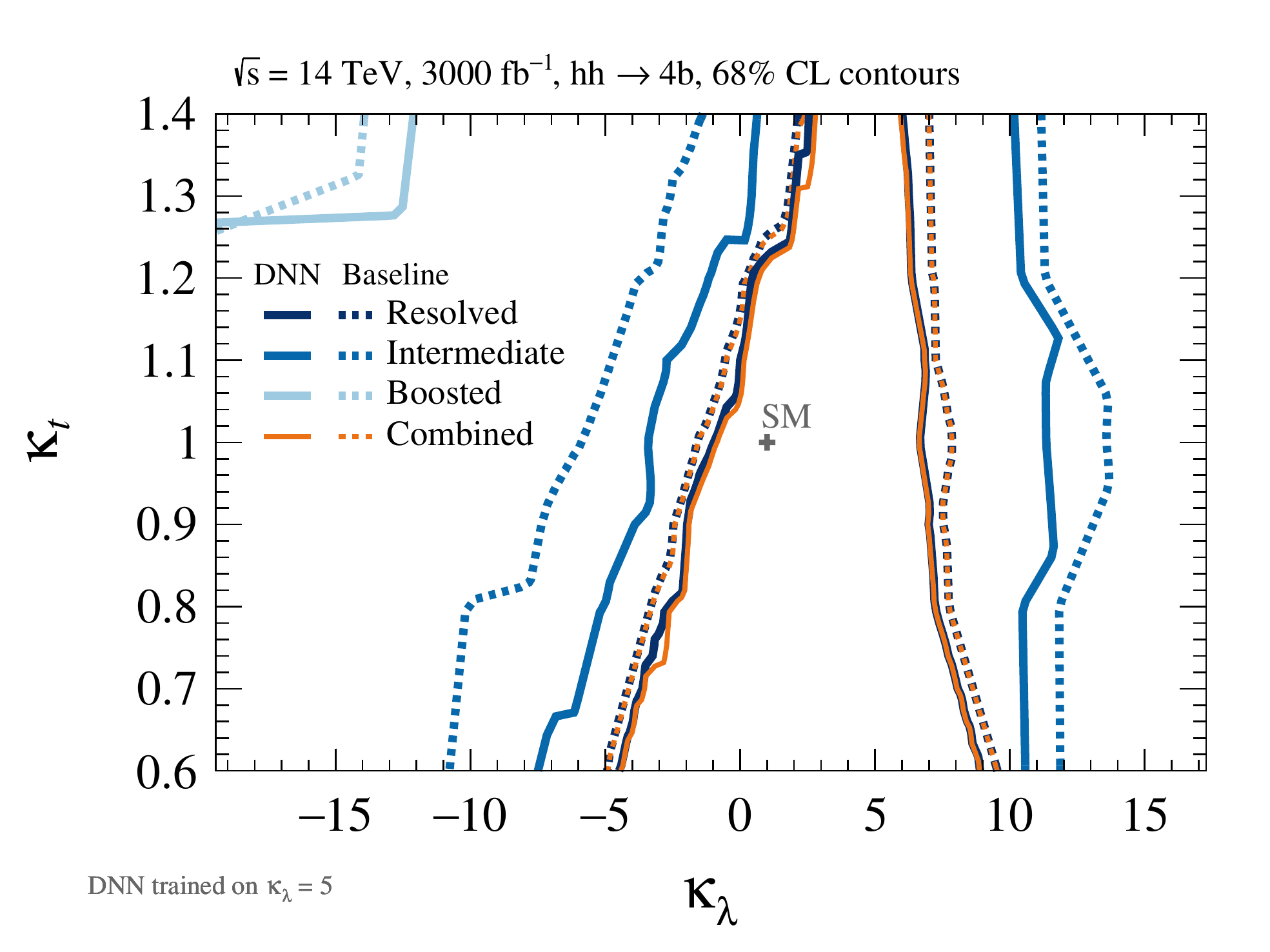}
    \caption{\label{fig:summary_contours_2d} Summary of $68\%$ CL $(\chi^2 = 1)$ contours in the $\kapt$ vs $\klam$ plane. These are displayed for the resolved (dark blue), intermediate (medium blue), boosted (light blue) categories and their combination (orange) for the \emph{baseline analysis} (dashed) and \emph{neural network analysis} (DNN) trained on $\klam = 5$ (solid). A luminosity of 3000~fb$^{-1}$ is assumed along with systematic uncertainties of 0.3\%, 1\% and 5\% for the resolved, intermediate and boosted categories, respectively. The cross indicates the SM prediction. The combined $\chi^2$ distributions are displayed in Figure~\ref{fig:limit2d_chiSq} and the $\chi^2$ distributions for each analysis category are shown in appendix~\ref{sec:chiSq}. }
\end{figure}

\begin{table}
\centering
\resizebox{\linewidth}{!}{
\begin{tabular}{lcccccc}
\toprule
\multirow{2}{*}{Analysis} & \multirow{2}{*}{DNN trained on}  & \multirow{2}{*}{\kapt} & \multicolumn{4}{c}{68\% CL on \klam} \\\cmidrule{4-7}
             &   &                & Resolved       & Intermediate   & Boosted           & Combined \\
\midrule
Baseline     & ---           & 1.2  & $[0.2, 7.2]$  & $[-3.7, 11.3]$   & ---            & $[0.2, 7.2]$  \\
DNN          & $\klam = 1$   & 1.2  & $[0.5, 7.1]$  & $[-1.2, 10.3]$   & ---            & $[0.7,7.1]$   \\
DNN          & $\klam  = 5$  & 1.2  & $[0.4, 6.3]$  & $[-3.1, 10.4]$   & ---            & $[0.6, 6.3]$  \\
DNN 0\% syst & $\klam  = 5$  & 1.2  & $[0.6, 5.8]$  & $[-0.8, 10.3]$   & $[-3.7, 11.0]$ & $[0.9, 5.8]$  \\
\midrule
Baseline     & ---           & 1.0  & $[-1.6, 7.9]$ & $[-5.8, 13.6]$  & ---            & $[-1.6, 7.9]$  \\
DNN          & $\klam = 1$   & 1.0  & $[-1.1, 7.4]$ & $[-3.8, 10.4]$   & ---            & $[-1.0, 7.4]$  \\
DNN          & $\klam = 5$   & 1.0  & $[-1.0, 6.6]$ & $[-3.4, 11.3]$   & ---            & $[-0.8, 6.6]$  \\
DNN 0\% syst & $\klam  = 5$  & 1.0  & $[-0.8, 6.4]$ & $[-3.0, 10.4]$   & $[-8.0, 13.2]$ & $[-0.7, 6.3]$  \\
\midrule
Baseline     & ---           & 0.8  & $[-3.3, 7.8]$ & $[-10.0, 11.9]$  & ---            & $[-3.3, 7.8]$  \\
DNN          & $\klam = 1$   & 0.8  & $[-2.7, 7.5]$ & $[-5.2, 10.4]$   & ---            & $[-2.6, 7.4]$  \\
DNN          & $\klam = 5$   & 0.8  & $[-2.8, 7.2]$ & $[-5.2, 10.4]$   & ---            & $[-2.6, 7.2]$  \\
DNN 0\% syst & $\klam  = 5$  & 0.8  & $[-2.3, 7.0]$ & $[-4.3, 10.3]$   & $[-8.4, 13.2]$ & $[-2.2, 6.9]$  \\

\bottomrule
\end{tabular}
}
\caption{\label{tab:summary_68CL} Summary of 68\% CL limits on \klam for \emph{baseline analysis} and \emph{neural network analysis} (DNN) by separate categories and combined. These are shown for $\kapt = 1.2, 1.0, 0.8$ to demonstrate the impact of uncertainties on the top Yukawa. Systematic uncertainties on the background of 0.3\%, 1\% and 5\% are assumed for the resolved, intermediate and boosted categories, respectively. The DNN analysis trained on $\klam = 5$ is also shown with 0\% systematics for comparison. A `---' in the 68\% CL columns indicates no constraint in the range of \klam considered, \klam $\in [-20, 20]$. 
}
\end{table}

Given the rapid scaling of di-Higgs cross-sections with the top Yukawa coupling \topyuk shown by Eq.~\eqref{eq:diHiggs_xsec}, we expect modifications in \topyuk to impact the projected \lamhhh constraints, which we now explore. Figure~\ref{fig:limit2d_chiSq_a} shows the $\chi^2$ distribution in the \kapt vs \klam plane for the \emph{baseline analysis} after combining the resolved, intermediate and boosted categories, assuming the nominal systematics of Table~\ref{tab:assumed_syst}. The contours are displayed after linear interpolation of the points sampled from this parameter space with the density shown in Fig.~\ref{fig:xsec_14TeV_hh_contour}. The grey line marks out the $\chi^2 = 1$ contour that corresponds to 68\% CL. Consistent with the one-dimensional distributions (Fig.~\ref{fig:limit1d_chiSq_b}), we see the red colours representing regions that have high $\chi^2$ values and are statistically excluded, while the blue shades indicate parameters that have low sensitivity. The dark blue strip of very low $\chi^2$ that runs from around $(\klam, \kapt) \sim (0, 0.8)$ to $(3, 1.2)$ indicates a region that is very similar to the SM (1, 1) parameters, given this is our reference value for calculating the $\chi^2$. 

Of particular interest is the shape of the $\chi^2 = 1$ contour in the two-dimensional \kapt vs \klam plane: as \kapt increases, the 68\% CL constraint on \klam improves, indicated by the region satisfying $\chi^2< 1$ decreasing. To make quantitative comparisons, we choose benchmark \kapt values of 0.8 and 1.2 based on current uncertainties from direct $pp \to t\bar{t} h$ measurements of \topyuk being on the order of 20\%~\cite{Aaboud:2018urx,Sirunyan:2018hoz}. We note that while measurements of gluon fusion production have smaller uncertainties, these indirectly probe \topyuk via the one-loop gluon--gluon--Higgs interaction. 

For the \emph{baseline analysis}, the combined 68\% CL constraint goes from $\klam(\kapt = 1) \in [-1.6, 7.9]$ to $\klam(\kapt = 1.2) \in [0.2, 7.2]$ and $\klam(\kapt = 0.8) \in [-3.3, 7.8]$. These constraints are also summarised in Table~\ref{tab:summary_68CL} to facilitate comparison. We note that the change in sensitivity is more significant for negative \lamhhh, where constraining the sign of $\klam$ to be positive is possible if $\kapt = 1.2$. To provide a clearer intuition of the impact, we can define how the absolute width $\delta \klam = |\klam^+ - \klam^-|$ of the 68\% CL intervals change, where $\klam^{+(-)}$ is the upper (lower) bound on \klam. With this metric, we find that the absolute width of the 68\% CL constraint $\delta \klam(\kapt = 1) = 9.5$ decreases by 26\% for $\kapt = 1.2$ and increases by 17\% for $\kapt = 0.8$. The results of the \emph{neural network analysis} are shown in Fig.~\ref{fig:limit2d_chiSq_b}, where the DNN trained on the $\klam = 5$ signal is used. As Table~\ref{tab:summary_68CL} also summarises, we find the 
68\% CL constraint goes from $\klam(\kapt = 1) \in [-0.8, 6.6]$ to $\klam(\kapt = 1.2) \in [0.6, 6.3]$ and $\klam(\kapt = 0.8) \in [-2.6, 7.2]$. Using the $\delta \klam$ metric, we find $\delta \klam(\kapt = 1) = 7.4$ decreases by 23\% for $\kapt = 1.2$ and increases by 32\% for $\kapt = 0.8$. These changes in $\delta \klam$ are non-trivial to interpret but they suggest that the size of the systematic on $\klam$ projections due to uncertainties on \topyuk impacting the signal modelling can be substantial. 

A further observation in Fig.~\ref{fig:limit2d_chiSq_a} is that we can extract a 68\% CL upper limit on the top Yukawa $\kapt < 1.25$ assuming fixed $\klam = 1$. This improves slightly to $\kapt < 1.22$ using the DNN trained on $\klam = 5$. It is interesting that this additional result of our work arises without considering dedicated optimisation for constraining $\kapt$. That the di-Higgs process has constraining power to put a better than 25\% upper bound on \kapt is helped by the fast quartic dependence of the box diagram contribution to the cross-section $\sigma_\text{box} \propto y_t^4$. The ability to constrain $\kapt$ depends strongly on the value of \klam: for example, we find no \kapt constraint if $\klam = 5$. Although our $hh \to 4b$ analysis is not expected to compete with other higher statistics channels in constraining \topyuk (directly via $t\bar{t} h$ or indirectly via gluon fusion), it provides an additional independent probe that could contribute in a global fit. 

Now we extend our discussion by using
Fig.~\ref{fig:summary_contours_2d}, which overlays two-dimensional 68\% CL contours to compare the \emph{baseline analysis} (dashed lines) with DNN analysis trained on $\klam = 5$ (solid lines). This figure also separates the contours by category, but as the boosted has little constraining power (in part due to the conservative 5\% systematic), only comparisons between resolved and intermediate are of interest. Similar to the one-dimensional $\chi^2$ distributions, sensitivity is driven by the resolved category across $\kapt \neq 1$ with the intermediate category contribution being non-negligible for negative \klam. 

Assuming $\kapt = 1$, the resolved category \emph{baseline analysis} is $\klam \in [-1.6, 7.9]$ and improves by 20\% (in $\delta \klam$) to $\klam \in [-1.0, 6.6]$ for the DNN trained on $\klam = 5$, which also is summarised in Table~\ref{tab:summary_68CL}. While the intermediate category has comparably weaker constraints of $\klam \in [-5.8, 13.6]$ for the \emph{baseline analysis}, the DNN trained on $\klam = 5$ improves this by 24\% to $\klam \in  [-3.4, 11.3]$. This improvement is largely uniform for different $\kapt$. For negative \klam, the slope of the intermediate contour in the \kapt vs \klam plane is similar to that of the resolved. This suggests that the two categories are sensitive to similar effects of the constructive interference occurring for negative \klam. By contrast, the fact that the intermediate contour is nearer vertical for positive \klam shows it is less sensitive to changes in \kapt even if the constraints are weaker than resolved. Additionally, we see that it is the resolved category that drives the $\kapt < 1.22$ 68\% CL constraint for $\klam = 1$.

\newpage
\section{Conclusion}
\label{sec:conclusion}

This paper presented a comprehensive analysis of Higgs pair production in $4b$ final states to evaluate and improve Higgs self-coupling ($\lamhhh$) constraints at the HL-LHC. We extended current strategies in several directions and now summarise our conclusions. 

We performed a detailed comparison of \lamhhh constraints across event categories that have zero (resolved), exactly one (intermediate) and two or more (boosted) boosted Higgs bosons in the $4b$ channel for the first time. The resolved analysis was found to be most constraining, followed by the intermediate and then the boosted. We used deep neural networks for signal--background separation and achieved an 86\% background rejection for 61\% SM signal efficiency in the resolved category compared to the \emph{baseline analysis}. This improved the signal significance from 0.69 (0.39) to 0.75 (0.53) without (with) the nominal systematics assumed, which can contribute in combination with other channels. Importantly, analyses optimised for discovery of SM $hh$ production can be suboptimal for constraining $\lamhhh$ in the $4b$ channel. We improved constraints by optimising on signals with \lamhhh values that are five times the SM and have lower boosts than those closer to SM values. We showed the impact of experimental limitations on jet reconstruction and triggering for reconstructing the principal discriminating variable \mhh. 

Assuming SM top Yukawa \topyuk, our \emph{baseline analysis} provides 68\% CL constraints of $-1.6 < \klam < 7.9$, while the corresponding constraint for the \emph{neural network analysis} is $-0.8 < \klam < 6.6$. This assumes 3000~fb$^{-1}$ of luminosity, background systematics controlled to 0.3\%, 1\% and 5\% for resolved, intermediate and boosted categories respectively, and no further advances in the analysis strategy. Interestingly, we can extract a 68\% CL upper limit on the top Yukawa of $\kapt < 1.22$ assuming $\klam = 1$ despite no dedicated optimisation, though this constraint changes rapidly for different \klam values. We quantified that current uncertainties on \topyuk constraints can modify $\lamhhh$ limits by $\sim$ 20\%. We caution that these constraints are not necessarily directly comparable with other projections in the literature due to different simplifying assumptions.

While our study suggests that the $4b$ channel is challenging for probing \lamhhh compared with for example $bb\gamma\gamma$, information from all independent channels are experimentally important in the final statistical combination. Our conclusions sharpen the experimental requirements to improve \lamhhh constraints using the $hh\to 4b$ channel, providing key performance targets for the HL-LHC programme. Given the dominant reducible $2b2j$ background, reducing the flavour mistag rate is critical, which requires both hardware (upgraded trackers improve vertex resolution) and software (e.g.\ deep learning) solutions. We quantified the impact of different scenarios of background systematic uncertainties from 5\% to statistical-only for all three categories. This showed that controlling systematics on multijet background estimates to demanding percent level or better is crucial for improving \lamhhh sensitivity. For scenarios where the boost of the Higgs bosons is low, notably around the boundary of expected sensitivity for positive $\klam \sim 5$ values, maintaining low $p_\text{T}$ jet triggers will be particularly important. Experimentally, applying and improving this rich confluence of techniques explored will accelerate progress towards constraining \lamhhh beyond gains from increased statistics alone.

\acknowledgments

% Please add people we wish to acknowledge for helpful discussions and funding sources as appropriate
We thank Luca Ambroz, Alan Barr, Mikkel Bj\o rn, Keith Ellis, James Grundy, Todd Huffman, Young-Kee Kim, Frank Krauss, David Miller, Mike Nelson, Hannah Pullen, Joe Ray, Ariel Schwartzman, Todd Seiss, Hayden Smith, Migl\.{e} Stankaityt\.{e}, Max Swiatlowski, Marc Weinberg, and Miha Zgubi\v{c} for useful discussions. We thank Katharina Behr, Nathan Hartland, Zhiyuan Li, Nurfikri Norjoharuddeen, Juan Rojo, Emma Slade, Cecilia Tosciri, Tony Weidberg, and Zachary Zajicek for earlier collaboration on this work. We are grateful to the hospitality of the Double Higgs Production at Colliders Workshop at Fermilab and the IPPP at Durham during the completion of this work. This research is supported by the STFC Particle Physics Consolidated Grant (ST/S000933/1, ST/N000447/1), IPPP Senior Experimental Fellowship, St John's College Junior Research Fellowship, Royal Society University Research Fellowship (Grant UF160190), and Grainger Fellowship. Beojan Stanislaus was supported by an STFC Studentship (ST/N504233/1) and a taberdarship from The Queen's College, Oxford. Santiago Paredes S\'{a}enz was supported by the ATLAS PhD Grant, the sub-department of particle physics of the University of Oxford, and SENESCYT. This project has received funding from the European Research Council (ERC) under the European Union's Horizon 2020 research and innovation programme (Grant agreement No. 787331).

\addcontentsline{toc}{section}{References}
\bibliographystyle{JHEP}
\bibliography{biblio/intro.bib,biblio/samples.bib,biblio/analyses.bib}

\providecommand{\href}[2]{#2}\begingroup\raggedright\begin{thebibliography}{100}

\bibitem{Baglio:2012np}
J.~Baglio, A.~Djouadi, R.~Gröber, M.~M. Mühlleitner, J.~Quevillon, and
  M.~Spira, {\it {The measurement of the Higgs self-coupling at the LHC:
  theoretical status}},  {\em JHEP} {\bf 04} (2013) 151,
  [\href{https://arxiv.org/abs/1212.5581}{{\tt arXiv:1212.5581}}].

\bibitem{deFlorian:2016spz}
{\bf LHC Higgs Cross Section Working Group} Collaboration, D.~de~Florian
  et~al., {\it {Handbook of LHC Higgs Cross Sections: 4. Deciphering the Nature
  of the Higgs Sector}},  \href{https://arxiv.org/abs/1610.07922}{{\tt
  arXiv:1610.07922}}.

\bibitem{ATLAS:2013hta}
{ATLAS Collaboration}, {\it {Physics at a High-Luminosity LHC with ATLAS}},  in
  {\em {Proceedings, Snowmass Community Planning Study 2013 (CSS2013)}}, 2013.
\newblock \href{https://arxiv.org/abs/1307.7292}{{\tt arXiv:1307.7292}}.

\bibitem{CMS:2013xfa}
{CMS collaboration}, {\it {Projected Performance of an Upgraded CMS Detector at
  the LHC and HL-LHC: Contribution to the Snowmass Process}},  in {\em
  {Proceedings, Snowmass Community Planning Study 2013 (CSS2013)}}, 2013.
\newblock \href{https://arxiv.org/abs/1307.7135}{{\tt arXiv:1307.7135}}.

\bibitem{Cepeda:2019klc}
{\bf HL/HE WG2 group} Collaboration, M.~Cepeda et~al., {\it {Higgs Physics at
  the HL-LHC and HE-LHC}},  \href{https://arxiv.org/abs/1902.00134}{{\tt
  arXiv:1902.00134}}.

\bibitem{Atlas:2019qfx}
{ATLAS and CMS Collaborations}, {\it {Report on the Physics at the HL-LHC and
  Perspectives for the HE-LHC}},  in {\em {HL/HE-LHC Physics Workshop: final
  jamboree Geneva, CERN, March 1, 2019}}, 2019.
\newblock \href{https://arxiv.org/abs/1902.10229}{{\tt arXiv:1902.10229}}.

\bibitem{DiMicco:2019ngk}
J.~Alison et~al., {\it {Higgs boson pair production at colliders: status and
  perspectives}},  in {\em {Double Higgs Production at Colliders Batavia, IL,
  USA, September 4, 2018-9, 2019}} (B.~Di~Micco, M.~Gouzevitch, J.~Mazzitelli,
  and C.~Vernieri, eds.), 2019.
\newblock \href{https://arxiv.org/abs/1910.00012}{{\tt arXiv:1910.00012}}.

\bibitem{Grojean:2004xa}
C.~Grojean, G.~Servant, and J.~D. Wells, {\it {First-order electroweak phase
  transition in the standard model with a low cutoff}},  {\em Phys. Rev.} {\bf
  D71} (2005) 036001, [\href{https://arxiv.org/abs/hep-ph/0407019}{{\tt
  hep-ph/0407019}}].

\bibitem{Cao:2013si}
J.~Cao, Z.~Heng, L.~Shang, P.~Wan, and J.~M. Yang, {\it {Pair Production of a
  125 GeV Higgs Boson in MSSM and NMSSM at the LHC}},  {\em JHEP} {\bf 04}
  (2013) 134, [\href{https://arxiv.org/abs/1301.6437}{{\tt arXiv:1301.6437}}].

\bibitem{Gouzevitch:2013qca}
M.~Gouzevitch, A.~Oliveira, J.~Rojo, R.~Rosenfeld, G.~P. Salam, and V.~Sanz,
  {\it {Scale-invariant resonance tagging in multijet events and new physics in
  Higgs pair production}},  {\em JHEP} {\bf 07} (2013) 148,
  [\href{https://arxiv.org/abs/1303.6636}{{\tt arXiv:1303.6636}}].

\bibitem{Gupta:2013zza}
R.~S. Gupta, H.~Rzehak, and J.~D. Wells, {\it {How well do we need to measure
  the Higgs boson mass and self-coupling?}},  {\em Phys. Rev.} {\bf D88} (2013)
  055024, [\href{https://arxiv.org/abs/1305.6397}{{\tt arXiv:1305.6397}}].

\bibitem{Han:2013sga}
C.~Han, X.~Ji, L.~Wu, P.~Wu, and J.~M. Yang, {\it {Higgs pair production with
  SUSY QCD correction: revisited under current experimental constraints}},
  {\em JHEP} {\bf 04} (2014) 003, [\href{https://arxiv.org/abs/1307.3790}{{\tt
  arXiv:1307.3790}}].

\bibitem{Nishiwaki:2013cma}
K.~Nishiwaki, S.~Niyogi, and A.~Shivaji, {\it {$ttH$ Anomalous Coupling in
  Double Higgs Production}},  {\em JHEP} {\bf 04} (2014) 011,
  [\href{https://arxiv.org/abs/1309.6907}{{\tt arXiv:1309.6907}}].

\bibitem{Goertz:2014qta}
F.~Goertz, A.~Papaefstathiou, L.~L. Yang, and J.~Zurita, {\it {Higgs boson pair
  production in the $D=6$ extension of the SM}},  {\em JHEP} {\bf 04} (2015)
  167, [\href{https://arxiv.org/abs/1410.3471}{{\tt arXiv:1410.3471}}].

\bibitem{Hespel:2014sla}
B.~Hespel, D.~Lopez-Val, and E.~Vryonidou, {\it {Higgs pair production via
  gluon fusion in the Two-Higgs-Doublet Model}},  {\em JHEP} {\bf 09} (2014)
  124, [\href{https://arxiv.org/abs/1407.0281}{{\tt arXiv:1407.0281}}].

\bibitem{Cao:2014kya}
J.~Cao, D.~Li, L.~Shang, P.~Wu, and Y.~Zhang, {\it {Exploring the Higgs Sector
  of a Most Natural NMSSM and its Prediction on Higgs Pair Production at the
  LHC}},  {\em JHEP} {\bf 12} (2014) 026,
  [\href{https://arxiv.org/abs/1409.8431}{{\tt arXiv:1409.8431}}].

\bibitem{Azatov:2015oxa}
A.~Azatov, R.~Contino, G.~Panico, and M.~Son, {\it {Effective field theory
  analysis of double Higgs boson production via gluon fusion}},  {\em Phys.
  Rev.} {\bf D92} (2015), no.~3 035001,
  [\href{https://arxiv.org/abs/1502.00539}{{\tt arXiv:1502.00539}}].

\bibitem{Carena:2015moc}
M.~Carena, H.~E. Haber, I.~Low, N.~R. Shah, and C.~E.~M. Wagner, {\it
  {Alignment limit of the NMSSM Higgs sector}},  {\em Phys. Rev.} {\bf D93}
  (2016), no.~3 035013, [\href{https://arxiv.org/abs/1510.09137}{{\tt
  arXiv:1510.09137}}].

\bibitem{Grober:2015cwa}
R.~Grober, M.~Muhlleitner, M.~Spira, and J.~Streicher, {\it {NLO QCD
  Corrections to Higgs Pair Production including Dimension-6 Operators}},  {\em
  JHEP} {\bf 09} (2015) 092, [\href{https://arxiv.org/abs/1504.06577}{{\tt
  arXiv:1504.06577}}].

\bibitem{Wu:2015nba}
L.~Wu, J.~M. Yang, C.-P. Yuan, and M.~Zhang, {\it {Higgs self-coupling in the
  MSSM and NMSSM after the LHC Run 1}},  {\em Phys. Lett.} {\bf B747} (2015)
  378--389, [\href{https://arxiv.org/abs/1504.06932}{{\tt arXiv:1504.06932}}].

\bibitem{He:2015spf}
H.-J. He, J.~Ren, and W.~Yao, {\it {Probing new physics of cubic Higgs boson
  interaction via Higgs pair production at hadron colliders}},  {\em Phys.
  Rev.} {\bf D93} (2016), no.~1 015003,
  [\href{https://arxiv.org/abs/1506.03302}{{\tt arXiv:1506.03302}}].

\bibitem{Carvalho:2015ttv}
A.~Carvalho, M.~Dall'Osso, T.~Dorigo, F.~Goertz, C.~A. Gottardo, and M.~Tosi,
  {\it {Higgs Pair Production: Choosing Benchmarks With Cluster Analysis}},
  {\em JHEP} {\bf 04} (2016) 126, [\href{https://arxiv.org/abs/1507.02245}{{\tt
  arXiv:1507.02245}}].

\bibitem{Zhang:2015mnh}
W.-J. Zhang, W.-G. Ma, R.-Y. Zhang, X.-Z. Li, L.~Guo, and C.~Chen, {\it {Double
  Higgs boson production and decay in Randall-Sundrum model at hadron
  colliders}},  {\em Phys. Rev.} {\bf D92} (2015) 116005,
  [\href{https://arxiv.org/abs/1512.01766}{{\tt arXiv:1512.01766}}].

\bibitem{Huang:2015tdv}
P.~Huang, A.~Joglekar, B.~Li, and C.~E.~M. Wagner, {\it {Probing the
  Electroweak Phase Transition at the LHC}},  {\em Phys. Rev.} {\bf D93}
  (2016), no.~5 055049, [\href{https://arxiv.org/abs/1512.00068}{{\tt
  arXiv:1512.00068}}].

\bibitem{Nakamura:2017irk}
K.~Nakamura, K.~Nishiwaki, K.-y. Oda, S.~C. Park, and Y.~Yamamoto, {\it
  {Di-higgs enhancement by neutral scalar as probe of new colored sector}},
  {\em Eur. Phys. J.} {\bf C77} (2017), no.~5 273,
  [\href{https://arxiv.org/abs/1701.06137}{{\tt arXiv:1701.06137}}].

\bibitem{DiLuzio:2017tfn}
L.~Di~Luzio, R.~Gröber, and M.~Spannowsky, {\it {Maxi-sizing the trilinear
  Higgs self-coupling: how large could it be?}},  {\em Eur. Phys. J.} {\bf C77}
  (2017), no.~11 788, [\href{https://arxiv.org/abs/1704.02311}{{\tt
  arXiv:1704.02311}}].

\bibitem{Huang:2017nnw}
P.~Huang, A.~Joglekar, M.~Li, and C.~E.~M. Wagner, {\it {Corrections to
  di-Higgs boson production with light stops and modified Higgs couplings}},
  {\em Phys. Rev.} {\bf D97} (2018), no.~7 075001,
  [\href{https://arxiv.org/abs/1711.05743}{{\tt arXiv:1711.05743}}].

\bibitem{Buchalla:2018yce}
G.~Buchalla, M.~Capozi, A.~Celis, G.~Heinrich, and L.~Scyboz, {\it {Higgs boson
  pair production in non-linear Effective Field Theory with full
  $m_t$-dependence at NLO QCD}},  {\em JHEP} {\bf 09} (2018) 057,
  [\href{https://arxiv.org/abs/1806.05162}{{\tt arXiv:1806.05162}}].

\bibitem{Borowka:2018pxx}
S.~Borowka, C.~Duhr, F.~Maltoni, D.~Pagani, A.~Shivaji, and X.~Zhao, {\it
  {Probing the scalar potential via double Higgs boson production at hadron
  colliders}},  {\em JHEP} {\bf 04} (2019) 016,
  [\href{https://arxiv.org/abs/1811.12366}{{\tt arXiv:1811.12366}}].

\bibitem{Chang:2019vez}
S.~Chang and M.~A. Luty, {\it {The Higgs Trilinear Coupling and the Scale of
  New Physics}},  {\em JHEP} {\bf 03} (2020) 140,
  [\href{https://arxiv.org/abs/1902.05556}{{\tt arXiv:1902.05556}}].

\bibitem{Blanke:2019hpe}
M.~Blanke, S.~Kast, J.~M. Thompson, S.~Westhoff, and J.~Zurita, {\it {Spotting
  hidden sectors with Higgs binoculars}},  {\em JHEP} {\bf 04} (2019) 160,
  [\href{https://arxiv.org/abs/1901.07558}{{\tt arXiv:1901.07558}}].

\bibitem{Li:2019tfd}
H.-L. Li, M.~Ramsey-Musolf, and S.~Willocq, {\it {Probing a Scalar
  Singlet-Catalyzed Electroweak Phase Transition with Resonant Di-Higgs
  Production in the $4b$ Channel}},  {\em Phys. Rev.} {\bf D100} (2019), no.~7
  075035, [\href{https://arxiv.org/abs/1906.05289}{{\tt arXiv:1906.05289}}].

\bibitem{Capozi:2019xsi}
M.~Capozi and G.~Heinrich, {\it {Exploring anomalous couplings in Higgs boson
  pair production through shape analysis}},  {\em JHEP} {\bf 03} (2020) 091,
  [\href{https://arxiv.org/abs/1908.08923}{{\tt arXiv:1908.08923}}].

\bibitem{Alves:2019igs}
A.~Alves, D.~Gonçalves, T.~Ghosh, H.-K. Guo, and K.~Sinha, {\it {Di-Higgs
  Production in the $4b$ Channel and Gravitational Wave Complementarity}},
  {\em JHEP} {\bf 03} (2020) 053, [\href{https://arxiv.org/abs/1909.05268}{{\tt
  arXiv:1909.05268}}].

\bibitem{Kozaczuk:2019pet}
J.~Kozaczuk, M.~J. Ramsey-Musolf, and J.~Shelton, {\it {Exotic Higgs Decays and
  the Electroweak Phase Transition}},
  \href{https://arxiv.org/abs/1911.10210}{{\tt arXiv:1911.10210}}.

\bibitem{Barducci:2019xkq}
D.~Barducci, K.~Mimasu, J.~M. No, C.~Vernieri, and J.~Zurita, {\it {Enlarging
  the scope of resonant di-Higgs searches: Hunting for Higgs-to-Higgs cascades
  in $4b$ final states at the LHC and future colliders}},  {\em JHEP} {\bf 02}
  (2020) 002, [\href{https://arxiv.org/abs/1910.08574}{{\tt
  arXiv:1910.08574}}].

\bibitem{Huang:2019bcs}
P.~Huang and Y.~H. Ng, {\it {Di-Higgs Production in SUSY models at the LHC}},
  \href{https://arxiv.org/abs/1910.13968}{{\tt arXiv:1910.13968}}.

\bibitem{Cheung:2020xij}
K.~Cheung, A.~Jueid, C.-T. Lu, J.~Song, and Y.~W. Yoon, {\it {Disentangling new
  physics effects on non-resonant Higgs boson pair production from gluon
  fusion}},  \href{https://arxiv.org/abs/2003.11043}{{\tt arXiv:2003.11043}}.

\bibitem{Aad:2015xja}
{ATLAS Collaboration}, {\it {Searches for Higgs boson pair production in the
  $hh\to bb\tau\tau, \gamma\gamma WW^*, \gamma\gamma bb, bbbb$ channels with
  the ATLAS detector}},  {\em Phys. Rev.} {\bf D92} (2015) 092004,
  [\href{https://arxiv.org/abs/1509.04670}{{\tt arXiv:1509.04670}}].

\bibitem{Aaboud:2016xco}
{ATLAS Collaboration}, {\it {Search for pair production of Higgs bosons in the
  $b\bar{b}b\bar{b}$ final state using proton--proton collisions at $\sqrt{s} =
  13$ TeV with the ATLAS detector}},  {\em Phys. Rev.} {\bf D94} (2016), no.~5
  052002, [\href{https://arxiv.org/abs/1606.04782}{{\tt arXiv:1606.04782}}].

\bibitem{Aaboud:2018knk}
{ATLAS Collaboration}, {\it {Search for pair production of Higgs bosons in the
  $b\bar{b}b\bar{b}$ final state using proton-proton collisions at $\sqrt{s} =
  13$ TeV with the ATLAS detector}},  {\em JHEP} {\bf 01} (2019) 030,
  [\href{https://arxiv.org/abs/1804.06174}{{\tt arXiv:1804.06174}}].

\bibitem{Aad:2020kub}
{ATLAS Collaboration}, {\it {Search for the $HH \rightarrow b \bar{b} b
  \bar{b}$ process via vector-boson fusion production using proton-proton
  collisions at $\sqrt{s} = 13$ TeV with the ATLAS detector}},
  \href{https://arxiv.org/abs/2001.05178}{{\tt arXiv:2001.05178}}.

\bibitem{Sirunyan:2017isc}
{CMS Collaboration}, {\it {Search for a massive resonance decaying to a pair of
  Higgs bosons in the four b quark final state in proton-proton collisions at
  $\sqrt{s}=$ 13 TeV}},  {\em Phys. Lett.} {\bf B781} (2018) 244--269,
  [\href{https://arxiv.org/abs/1710.04960}{{\tt arXiv:1710.04960}}].

\bibitem{Sirunyan:2018qca}
{CMS Collaboration}, {\it {Search for production of Higgs boson pairs in the
  four b quark final state using large-area jets in proton-proton collisions at
  $\sqrt{s}=$ 13 TeV}},  {\em JHEP} {\bf 01} (2019) 040,
  [\href{https://arxiv.org/abs/1808.01473}{{\tt arXiv:1808.01473}}].

\bibitem{Sirunyan:2018tki}
{CMS Collaboration}, {\it {Search for nonresonant Higgs boson pair production
  in the $\mathrm{b\overline{b}b\overline{b}}$ final state at $\sqrt{s} =$ 13
  TeV}},  {\em JHEP} {\bf 04} (2019) 112,
  [\href{https://arxiv.org/abs/1810.11854}{{\tt arXiv:1810.11854}}].

\bibitem{Behr:2015oqq}
J.~K. Behr, D.~Bortoletto, J.~A. Frost, N.~P. Hartland, C.~Issever, and
  J.~Rojo, {\it {Boosting Higgs pair production in the $b\bar{b}b\bar{b}$ final
  state with multivariate techniques}},  {\em Eur. Phys. J.} {\bf C76} (2016),
  no.~7 386, [\href{https://arxiv.org/abs/1512.08928}{{\tt arXiv:1512.08928}}].

\bibitem{Wardrope:2014kya}
D.~Wardrope, E.~Jansen, N.~Konstantinidis, B.~Cooper, R.~Falla, and
  N.~Norjoharuddeen, {\it {Non-resonant Higgs-pair production in the
  $b\overline{b}b\overline{b}$ final state at the LHC}},  {\em Eur. Phys. J.}
  {\bf C75} (2015), no.~5 219, [\href{https://arxiv.org/abs/1410.2794}{{\tt
  arXiv:1410.2794}}].

\bibitem{deLima:2014dta}
D.~E. Ferreira~de Lima, A.~Papaefstathiou, and M.~Spannowsky, {\it {Standard
  model Higgs boson pair production in the ($b\overline{b}$)($b\overline{b}$)
  final state}},  {\em JHEP} {\bf 08} (2014) 030,
  [\href{https://arxiv.org/abs/1404.7139}{{\tt arXiv:1404.7139}}].

\bibitem{Aaboud:2018sfw}
{ATLAS Collaboration}, {\it {Search for resonant and non-resonant Higgs boson
  pair production in the ${b\bar{b}\tau^+\tau^-}$ decay channel in $pp$
  collisions at $\sqrt{s}=13$ TeV with the ATLAS detector}},  {\em Phys. Rev.
  Lett.} {\bf 121} (2018), no.~19 191801,
  [\href{https://arxiv.org/abs/1808.00336}{{\tt arXiv:1808.00336}}]. [Erratum:
  Phys. Rev. Lett.122,no.8,089901(2019)].

\bibitem{Sirunyan:2017tqo}
{CMS Collaboration}, {\it {Search for Higgs boson pair production in the
  $bb\tau\tau$ final state in proton-proton collisions at
  $\sqrt{s}=8~\mathrm{TeV}$}},  {\em Phys. Rev.} {\bf D96} (2017), no.~7
  072004, [\href{https://arxiv.org/abs/1707.00350}{{\tt arXiv:1707.00350}}].

\bibitem{Aaboud:2018ewm}
{ATLAS Collaboration}, {\it {Search for Higgs boson pair production in the
  $\gamma\gamma WW^{*}$ channel using $pp$ collision data recorded at $\sqrt{s}
  = 13$ TeV with the ATLAS detector}},  {\em Eur. Phys. J.} {\bf C78} (2018),
  no.~12 1007, [\href{https://arxiv.org/abs/1807.08567}{{\tt
  arXiv:1807.08567}}].

\bibitem{Aaboud:2018ftw}
{ATLAS Collaboration}, {\it {Search for Higgs boson pair production in the
  $\gamma\gamma b\bar{b}$ final state with 13 TeV $pp$ collision data collected
  by the ATLAS experiment}},  {\em JHEP} {\bf 11} (2018) 040,
  [\href{https://arxiv.org/abs/1807.04873}{{\tt arXiv:1807.04873}}].

\bibitem{Khachatryan:2016sey}
{CMS Collaboration}, {\it {Search for two Higgs bosons in final states
  containing two photons and two bottom quarks in proton-proton collisions at 8
  TeV}},  {\em Phys. Rev.} {\bf D94} (2016), no.~5 052012,
  [\href{https://arxiv.org/abs/1603.06896}{{\tt arXiv:1603.06896}}].

\bibitem{Aad:2019uzh}
{ATLAS Collaboration}, {\it {Combination of searches for Higgs boson pairs in
  $pp$ collisions at $\sqrt{s} = $13 TeV with the ATLAS detector}},  {\em Phys.
  Lett.} {\bf B800} (2020) 135103,
  [\href{https://arxiv.org/abs/1906.02025}{{\tt arXiv:1906.02025}}].

\bibitem{Sirunyan:2018two}
{CMS Collaboration}, {\it {Combination of searches for Higgs boson pair
  production in proton-proton collisions at $\sqrt{s} = $ 13 TeV}},  {\em Phys.
  Rev. Lett.} {\bf 122} (2019), no.~12 121803,
  [\href{https://arxiv.org/abs/1811.09689}{{\tt arXiv:1811.09689}}].

\bibitem{ATLAS-CONF-2019-049}
{ATLAS Collaboration}, {\it {Constraints on the Higgs boson self-coupling from
  the combination of single-Higgs and double-Higgs production analyses
  performed with the ATLAS experiment}},  tech. rep., CERN, 2019.
\newblock ATLAS-CONF-2019-049.

\bibitem{ATL-PHYS-PUB-2018-053}
{ATLAS Collaboration}, {\it {Measurement prospects of the pair production and
  self-coupling of the Higgs boson with the ATLAS experiment at the HL-LHC}},
  Tech. Rep. ATL-PHYS-PUB-2018-053, 2018.

\bibitem{CMS-PAS-FTR-18-019}
{CMS Collaboration}, {\it {Prospects for HH measurements at the HL-LHC}},
  Tech. Rep. CMS-PAS-FTR-18-019, 2018.

\bibitem{Butterworth:2008iy}
J.~M. Butterworth, A.~R. Davison, M.~Rubin, and G.~P. Salam, {\it {Jet
  substructure as a new Higgs search channel at the LHC}},  {\em Phys. Rev.
  Lett.} {\bf 100} (2008) 242001, [\href{https://arxiv.org/abs/0802.2470}{{\tt
  arXiv:0802.2470}}].

\bibitem{ATL-PHYS-PUB-2014-013}
{ATLAS Collaboration}, {\it {Flavor Tagging with Track Jets in Boosted
  Topologies with the ATLAS Detector}},  Tech. Rep. ATL-PHYS-PUB-2014-013,
  2014.

\bibitem{Asquith:2018igt}
R.~Kogler et~al., {\it {Jet Substructure at the Large Hadron Collider:
  Experimental Review}},  {\em Rev. Mod. Phys.} {\bf 91} (2019), no.~4 045003,
  [\href{https://arxiv.org/abs/1803.06991}{{\tt arXiv:1803.06991}}].

\bibitem{Aad:2019uoz}
{ATLAS Collaboration}, {\it {Identification of boosted Higgs bosons decaying
  into $b$-quark pairs with the ATLAS detector at 13 TeV}},  {\em Eur. Phys.
  J.} {\bf C79} (2019), no.~10 836,
  [\href{https://arxiv.org/abs/1906.11005}{{\tt arXiv:1906.11005}}].

\bibitem{Aaboud:2018zhk}
{ATLAS Collaboration}, {\it {Observation of $H \rightarrow b\bar{b}$ decays and
  $VH$ production with the ATLAS detector}},  {\em Phys. Lett.} {\bf B786}
  (2018) 59--86, [\href{https://arxiv.org/abs/1808.08238}{{\tt
  arXiv:1808.08238}}].

\bibitem{Sirunyan:2018kst}
{CMS Collaboration}, {\it {Observation of Higgs boson decay to bottom quarks}},
   {\em Phys. Rev. Lett.} {\bf 121} (2018), no.~12 121801,
  [\href{https://arxiv.org/abs/1808.08242}{{\tt arXiv:1808.08242}}].

\bibitem{Sirunyan:2017dgc}
{CMS Collaboration}, {\it {Inclusive search for a highly boosted Higgs boson
  decaying to a bottom quark-antiquark pair}},  {\em Phys. Rev. Lett.} {\bf
  120} (2018), no.~7 071802, [\href{https://arxiv.org/abs/1709.05543}{{\tt
  arXiv:1709.05543}}].

\bibitem{ATLAS-CONF-2018-052}
{ATLAS Collaboration}, {\it {Search for boosted resonances decaying to two
  b-quarks and produced in association with a jet at $\sqrt{s}=13$ TeV with the
  ATLAS detector}},  Tech. Rep. ATLAS-CONF-2018-052, 2018.

\bibitem{Aaboud:2018urx}
{ATLAS Collaboration}, {\it {Observation of Higgs boson production in
  association with a top quark pair at the LHC with the ATLAS detector}},  {\em
  Phys. Lett.} {\bf B784} (2018) 173--191,
  [\href{https://arxiv.org/abs/1806.00425}{{\tt arXiv:1806.00425}}].

\bibitem{Aaboud:2017rss}
{ATLAS Collaboration}, {\it {Search for the standard model Higgs boson produced
  in association with top quarks and decaying into a $b\bar{b}$ pair in $pp$
  collisions at $\sqrt{s}$ = 13 TeV with the ATLAS detector}},  {\em Phys.
  Rev.} {\bf D97} (2018), no.~7 072016,
  [\href{https://arxiv.org/abs/1712.08895}{{\tt arXiv:1712.08895}}].

\bibitem{Sirunyan:2018hoz}
{CMS Collaboration}, {\it {Observation of $\mathrm{t\overline{t}}$H
  production}},  {\em Phys. Rev. Lett.} {\bf 120} (2018), no.~23 231801,
  [\href{https://arxiv.org/abs/1804.02610}{{\tt arXiv:1804.02610}}].

\bibitem{Sirunyan:2018mvw}
{CMS Collaboration}, {\it {Search for $
  \mathrm{t}\overline{\mathrm{t}}\mathrm{H} $ production in the $ \mathrm{H}\to
  \mathrm{b}\overline{\mathrm{b}} $ decay channel with leptonic $
  \mathrm{t}\overline{\mathrm{t}} $ decays in proton-proton collisions at $
  \sqrt{s}=13 $ TeV}},  {\em JHEP} {\bf 03} (2019) 026,
  [\href{https://arxiv.org/abs/1804.03682}{{\tt arXiv:1804.03682}}].

\bibitem{ATL-PHYS-PUB-2015-022}
{ATLAS Collaboration}, {\it {Expected performance of the ATLAS $b$-tagging
  algorithms in Run-2}},  Tech. Rep. ATL-PHYS-PUB-2015-022, 2015.

\bibitem{Sirunyan:2017ezt}
{CMS Collaboration}, {\it {Identification of heavy-flavour jets with the CMS
  detector in pp collisions at 13 TeV}},  {\em JINST} {\bf 13} (2018), no.~05
  P05011, [\href{https://arxiv.org/abs/1712.07158}{{\tt arXiv:1712.07158}}].

\bibitem{Aaboud:2018xwy}
{ATLAS Collaboration}, {\it {Measurements of b-jet tagging efficiency with the
  ATLAS detector using $ t\overline{t} $ events at $ \sqrt{s}=13 $ TeV}},  {\em
  JHEP} {\bf 08} (2018) 089, [\href{https://arxiv.org/abs/1805.01845}{{\tt
  arXiv:1805.01845}}].

\bibitem{Lippmann1987AnIT}
R.~Lippmann, {\it An introduction to computing with neural nets},  {\em IEEE
  ASSP Magazine} {\bf 4} (1987) 4--22.

\bibitem{Hornik:1989}
K.~Hornik, M.~Stinchcombe, and H.~White, {\it Multilayer feedforward networks
  are universal approximators},  {\em Neural Netw.} {\bf 2} (July, 1989)
  359--366.

\bibitem{Hornik1991ApproximationCO}
K.~Hornik, {\it Approximation capabilities of multilayer feedforward networks},
   {\em Neural Networks} {\bf 4} (1991) 251--257.

\bibitem{Baldi:2014kfa}
P.~Baldi, P.~Sadowski, and D.~Whiteson, {\it {Searching for Exotic Particles in
  High-Energy Physics with Deep Learning}},  {\em Nature Commun.} {\bf 5}
  (2014) 4308, [\href{https://arxiv.org/abs/1402.4735}{{\tt arXiv:1402.4735}}].

\bibitem{Baldi:2014pta}
P.~Baldi, P.~Sadowski, and D.~Whiteson, {\it {Enhanced Higgs Boson to
  $\tau^+\tau^-$ Search with Deep Learning}},  {\em Phys. Rev. Lett.} {\bf 114}
  (2015), no.~11 111801, [\href{https://arxiv.org/abs/1410.3469}{{\tt
  arXiv:1410.3469}}].

\bibitem{deOliveira:2015xxd}
L.~de~Oliveira, M.~Kagan, L.~Mackey, B.~Nachman, and A.~Schwartzman, {\it
  {Jet-images — deep learning edition}},  {\em JHEP} {\bf 07} (2016) 069,
  [\href{https://arxiv.org/abs/1511.05190}{{\tt arXiv:1511.05190}}].

\bibitem{Baldi:2016fzo}
P.~Baldi, K.~Cranmer, T.~Faucett, P.~Sadowski, and D.~Whiteson, {\it
  {Parameterized neural networks for high-energy physics}},  {\em Eur. Phys.
  J.} {\bf C76} (2016), no.~5 235,
  [\href{https://arxiv.org/abs/1601.07913}{{\tt arXiv:1601.07913}}].

\bibitem{Caron:2016hib}
S.~Caron, J.~S. Kim, K.~Rolbiecki, R.~Ruiz~de Austri, and B.~Stienen, {\it {The
  BSM-AI project: SUSY-AI–generalizing LHC limits on supersymmetry with
  machine learning}},  {\em Eur. Phys. J.} {\bf C77} (2017), no.~4 257,
  [\href{https://arxiv.org/abs/1605.02797}{{\tt arXiv:1605.02797}}].

\bibitem{Chang:2017kvc}
S.~Chang, T.~Cohen, and B.~Ostdiek, {\it {What is the Machine Learning?}},
  {\em Phys. Rev.} {\bf D97} (2018), no.~5 056009,
  [\href{https://arxiv.org/abs/1709.10106}{{\tt arXiv:1709.10106}}].

\bibitem{Lin:2018cin}
J.~Lin, M.~Freytsis, I.~Moult, and B.~Nachman, {\it {Boosting $H\to b\bar b$
  with Machine Learning}},  {\em JHEP} {\bf 10} (2018) 101,
  [\href{https://arxiv.org/abs/1807.10768}{{\tt arXiv:1807.10768}}].

\bibitem{Albertsson:2018maf}
K.~Albertsson et~al., {\it {Machine Learning in High Energy Physics Community
  White Paper}},  {\em J. Phys. Conf. Ser.} {\bf 1085} (2018), no.~2 022008,
  [\href{https://arxiv.org/abs/1807.02876}{{\tt arXiv:1807.02876}}].

\bibitem{Guest:2018yhq}
D.~Guest, K.~Cranmer, and D.~Whiteson, {\it {Deep Learning and its Application
  to LHC Physics}},  {\em Ann. Rev. Nucl. Part. Sci.} {\bf 68} (2018) 161--181,
  [\href{https://arxiv.org/abs/1806.11484}{{\tt arXiv:1806.11484}}].

\bibitem{Abdughani:2019wuv}
M.~Abdughani, J.~Ren, L.~Wu, J.~M. Yang, and J.~Zhao, {\it {Supervised deep
  learning in high energy phenomenology: a mini review}},  {\em Commun. Theor.
  Phys.} {\bf 71} (2019), no.~8 955,
  [\href{https://arxiv.org/abs/1905.06047}{{\tt arXiv:1905.06047}}].

\bibitem{Windischhofer:2019ltt}
P.~Windischhofer, M.~Zgubi\v{c}, and D.~Bortoletto, {\it {Preserving physically
  important variables in optimal event selections: A case study in Higgs
  physics}},  \href{https://arxiv.org/abs/1907.02098}{{\tt arXiv:1907.02098}}.

\bibitem{NIPS2017_7062}
S.~M. Lundberg and S.-I. Lee, {\it A unified approach to interpreting model
  predictions},  in {\em Advances in Neural Information Processing Systems 30}
  (I.~Guyon, U.~V. Luxburg, S.~Bengio, H.~Wallach, R.~Fergus, S.~Vishwanathan,
  and R.~Garnett, eds.), pp.~4765--4774.
\newblock Curran Associates, Inc., 2017.
\newblock \href{https://arxiv.org/abs/1705.07874}{{\tt arXiv:1705.07874}}.

\bibitem{Dolan:2012rv}
M.~J. Dolan, C.~Englert, and M.~Spannowsky, {\it {Higgs self-coupling
  measurements at the LHC}},  {\em JHEP} {\bf 10} (2012) 112,
  [\href{https://arxiv.org/abs/1206.5001}{{\tt arXiv:1206.5001}}].

\bibitem{Papaefstathiou:2012qe}
A.~Papaefstathiou, L.~L. Yang, and J.~Zurita, {\it {Higgs boson pair production
  at the LHC in the $b \bar{b} W^+ W^-$ channel}},  {\em Phys. Rev.} {\bf D87}
  (2013), no.~1 011301, [\href{https://arxiv.org/abs/1209.1489}{{\tt
  arXiv:1209.1489}}].

\bibitem{Barr:2013tda}
A.~J. Barr, M.~J. Dolan, C.~Englert, and M.~Spannowsky, {\it {Di-Higgs final
  states augMT2ed -- selecting $hh$ events at the high luminosity LHC}},  {\em
  Phys. Lett.} {\bf B728} (2014) 308--313,
  [\href{https://arxiv.org/abs/1309.6318}{{\tt arXiv:1309.6318}}].

\bibitem{Chen:2014xra}
C.-R. Chen and I.~Low, {\it {Double take on new physics in double Higgs boson
  production}},  {\em Phys. Rev.} {\bf D90} (2014), no.~1 013018,
  [\href{https://arxiv.org/abs/1405.7040}{{\tt arXiv:1405.7040}}].

\bibitem{Dawson:2015oha}
S.~Dawson, A.~Ismail, and I.~Low, {\it {What's in the loop? The anatomy of
  double Higgs production}},  {\em Phys. Rev.} {\bf D91} (2015), no.~11 115008,
  [\href{https://arxiv.org/abs/1504.05596}{{\tt arXiv:1504.05596}}].

\bibitem{Lu:2015jza}
C.-T. Lu, J.~Chang, K.~Cheung, and J.~S. Lee, {\it {An exploratory study of
  Higgs-boson pair production}},  {\em JHEP} {\bf 08} (2015) 133,
  [\href{https://arxiv.org/abs/1505.00957}{{\tt arXiv:1505.00957}}].

\bibitem{Kling:2016lay}
F.~Kling, T.~Plehn, and P.~Schichtel, {\it {Maximizing the significance in
  Higgs boson pair analyses}},  {\em Phys. Rev.} {\bf D95} (2017), no.~3
  035026, [\href{https://arxiv.org/abs/1607.07441}{{\tt arXiv:1607.07441}}].

\bibitem{Bizon:2016wgr}
W.~Bizon, M.~Gorbahn, U.~Haisch, and G.~Zanderighi, {\it {Constraints on the
  trilinear Higgs coupling from vector boson fusion and associated Higgs
  production at the LHC}},  {\em JHEP} {\bf 07} (2017) 083,
  [\href{https://arxiv.org/abs/1610.05771}{{\tt arXiv:1610.05771}}].

\bibitem{Bishara:2016kjn}
F.~Bishara, R.~Contino, and J.~Rojo, {\it {Higgs pair production in
  vector-boson fusion at the LHC and beyond}},  {\em Eur. Phys. J.} {\bf C77}
  (2017), no.~7 481, [\href{https://arxiv.org/abs/1611.03860}{{\tt
  arXiv:1611.03860}}].

\bibitem{Adhikary:2017jtu}
A.~Adhikary, S.~Banerjee, R.~K. Barman, B.~Bhattacherjee, and S.~Niyogi, {\it
  {Revisiting the non-resonant Higgs pair production at the HL-LHC}},  {\em
  JHEP} {\bf 07} (2018) 116, [\href{https://arxiv.org/abs/1712.05346}{{\tt
  arXiv:1712.05346}}].

\bibitem{Alves:2017ued}
A.~Alves, T.~Ghosh, and K.~Sinha, {\it {Can We Discover Double Higgs Production
  at the LHC?}},  {\em Phys. Rev.} {\bf D96} (2017), no.~3 035022,
  [\href{https://arxiv.org/abs/1704.07395}{{\tt arXiv:1704.07395}}].

\bibitem{Huang:2017jws}
T.~Huang, J.~M. No, L.~Pernié, M.~Ramsey-Musolf, A.~Safonov, M.~Spannowsky,
  and P.~Winslow, {\it {Resonant di-Higgs boson production in the $b{\bar b}WW$
  channel: Probing the electroweak phase transition at the LHC}},  {\em Phys.
  Rev.} {\bf D96} (2017), no.~3 035007,
  [\href{https://arxiv.org/abs/1701.04442}{{\tt arXiv:1701.04442}}].

\bibitem{Kim:2018uty}
J.~H. Kim, Y.~Sakaki, and M.~Son, {\it {Combined analysis of double Higgs
  production via gluon fusion at the HL-LHC in the effective field theory
  approach}},  {\em Phys. Rev.} {\bf D98} (2018), no.~1 015016,
  [\href{https://arxiv.org/abs/1801.06093}{{\tt arXiv:1801.06093}}].

\bibitem{Chang:2018uwu}
J.~Chang, K.~Cheung, J.~S. Lee, C.-T. Lu, and J.~Park, {\it {Higgs-boson-pair
  production $H(\rightarrow b\overline{b})H(\rightarrow\gamma\gamma)$ from
  gluon fusion at the HL-LHC and HL-100 TeV hadron collider}},  {\em Phys.
  Rev.} {\bf D100} (2019), no.~9 096001,
  [\href{https://arxiv.org/abs/1804.07130}{{\tt arXiv:1804.07130}}].

\bibitem{Kim:2018cxf}
J.~H. Kim, K.~Kong, K.~T. Matchev, and M.~Park, {\it {Probing the Triple Higgs
  Self-Interaction at the Large Hadron Collider}},  {\em Phys. Rev. Lett.} {\bf
  122} (2019), no.~9 091801, [\href{https://arxiv.org/abs/1807.11498}{{\tt
  arXiv:1807.11498}}].

\bibitem{Basler:2018dac}
P.~Basler, S.~Dawson, C.~Englert, and M.~Mühlleitner, {\it {Showcasing HH
  production: Benchmarks for the LHC and HL-LHC}},  {\em Phys. Rev.} {\bf D99}
  (2019), no.~5 055048, [\href{https://arxiv.org/abs/1812.03542}{{\tt
  arXiv:1812.03542}}].

\bibitem{Chang:2019ncg}
J.~Chang, K.~Cheung, J.~S. Lee, and J.~Park, {\it {Probing the trilinear Higgs
  boson self-coupling at the high-luminosity LHC via multivariate analysis}},
  {\em Phys. Rev.} {\bf D101} (2020), no.~1 016004,
  [\href{https://arxiv.org/abs/1908.00753}{{\tt arXiv:1908.00753}}].

\bibitem{Arganda:2018ftn}
E.~Arganda, C.~Garcia-Garcia, and M.~J. Herrero, {\it {Probing the Higgs
  self-coupling through double Higgs production in vector boson scattering at
  the LHC}},  {\em Nucl. Phys.} {\bf B945} (2019) 114687,
  [\href{https://arxiv.org/abs/1807.09736}{{\tt arXiv:1807.09736}}].

\bibitem{Cao:2015oxx}
Q.-H. Cao, Y.~Liu, and B.~Yan, {\it {Measuring trilinear Higgs coupling in WHH
  and ZHH productions at the high-luminosity LHC}},  {\em Phys. Rev.} {\bf D95}
  (2017), no.~7 073006, [\href{https://arxiv.org/abs/1511.03311}{{\tt
  arXiv:1511.03311}}].

\bibitem{Chen:2015gva}
C.-Y. Chen, Q.-S. Yan, X.~Zhao, Y.-M. Zhong, and Z.~Zhao, {\it {Probing
  triple-Higgs productions via $4b2\gamma$ decay channel at a 100 TeV hadron
  collider}},  {\em Phys. Rev.} {\bf D93} (2016), no.~1 013007,
  [\href{https://arxiv.org/abs/1510.04013}{{\tt arXiv:1510.04013}}].

\bibitem{Liu:2018peg}
T.~Liu, K.-F. Lyu, J.~Ren, and H.~X. Zhu, {\it {Probing the quartic Higgs boson
  self-interaction}},  {\em Phys. Rev.} {\bf D98} (2018), no.~9 093004,
  [\href{https://arxiv.org/abs/1803.04359}{{\tt arXiv:1803.04359}}].

\bibitem{Bizon:2018syu}
W.~Bizo\'{n}, U.~Haisch, and L.~Rottoli, {\it {Constraints on the quartic Higgs
  self-coupling from double-Higgs production at future hadron colliders}},
  {\em JHEP} {\bf 10} (2019) 267, [\href{https://arxiv.org/abs/1810.04665}{{\tt
  arXiv:1810.04665}}].

\bibitem{Papaefstathiou:2019ofh}
A.~Papaefstathiou, G.~Tetlalmatzi-Xolocotzi, and M.~Zaro, {\it {Triple Higgs
  boson production to six $b$-jets at a 100 TeV proton collider}},  {\em Eur.
  Phys. J.} {\bf C79} (2019), no.~11 947,
  [\href{https://arxiv.org/abs/1909.09166}{{\tt arXiv:1909.09166}}].

\bibitem{Alwall:2011uj}
J.~Alwall, M.~Herquet, F.~Maltoni, O.~Mattelaer, and T.~Stelzer, {\it {MadGraph
  5 : Going Beyond}},  {\em JHEP} {\bf 06} (2011) 128,
  [\href{https://arxiv.org/abs/1106.0522}{{\tt arXiv:1106.0522}}].

\bibitem{Alwall:2014hca}
J.~Alwall, R.~Frederix, S.~Frixione, V.~Hirschi, F.~Maltoni, O.~Mattelaer,
  H.~S. Shao, T.~Stelzer, P.~Torrielli, and M.~Zaro, {\it {The automated
  computation of tree-level and next-to-leading order differential cross
  sections, and their matching to parton shower simulations}},  {\em JHEP} {\bf
  07} (2014) 079, [\href{https://arxiv.org/abs/1405.0301}{{\tt
  arXiv:1405.0301}}].

\bibitem{ATL-HeavyHiggsTHDM}
{B. Hespel and E. Vryonidou}, ``{Higgs pair production heavy scalar model}.''
  \url{https://cp3.irmp.ucl.ac.be/projects/madgraph/wiki/HiggsPairProduction#BSM:Additionalheavyscalarresonance}.

\bibitem{Frederix:2014hta}
R.~Frederix, S.~Frixione, V.~Hirschi, F.~Maltoni, O.~Mattelaer, P.~Torrielli,
  E.~Vryonidou, and M.~Zaro, {\it {Higgs pair production at the LHC with NLO
  and parton-shower effects}},  {\em Phys. Lett.} {\bf B732} (2014) 142--149,
  [\href{https://arxiv.org/abs/1401.7340}{{\tt arXiv:1401.7340}}].

\bibitem{Grzadkowski:2010es}
B.~Grzadkowski, M.~Iskrzynski, M.~Misiak, and J.~Rosiek, {\it {Dimension-Six
  Terms in the Standard Model Lagrangian}},  {\em JHEP} {\bf 10} (2010) 085,
  [\href{https://arxiv.org/abs/1008.4884}{{\tt arXiv:1008.4884}}].

\bibitem{Brivio:2017vri}
I.~Brivio and M.~Trott, {\it {The Standard Model as an Effective Field
  Theory}},  {\em Phys. Rept.} {\bf 793} (2019) 1--98,
  [\href{https://arxiv.org/abs/1706.08945}{{\tt arXiv:1706.08945}}].

\bibitem{Ellis:2018gqa}
J.~Ellis, C.~W. Murphy, V.~Sanz, and T.~You, {\it {Updated Global SMEFT Fit to
  Higgs, Diboson and Electroweak Data}},  {\em JHEP} {\bf 06} (2018) 146,
  [\href{https://arxiv.org/abs/1803.03252}{{\tt arXiv:1803.03252}}].

\bibitem{Ball:2014uwa}
{\bf NNPDF} Collaboration, R.~D. Ball et~al., {\it {Parton distributions for
  the LHC Run II}},  {\em JHEP} {\bf 04} (2015) 040,
  [\href{https://arxiv.org/abs/1410.8849}{{\tt arXiv:1410.8849}}].

\bibitem{Buckley:2014ana}
A.~Buckley, J.~Ferrando, S.~Lloyd, K.~Nordstr{\"o}m, B.~Page, M.~R{\"u}fenacht,
  M.~Sch{\"o}nherr, and G.~Watt, {\it {LHAPDF6: parton density access in the
  LHC precision era}},  {\em Eur. Phys. J.} {\bf C75} (2015) 132,
  [\href{https://arxiv.org/abs/1412.7420}{{\tt arXiv:1412.7420}}].

\bibitem{Chen:2019lzz}
L.-B. Chen, H.~T. Li, H.-S. Shao, and J.~Wang, {\it {Higgs boson pair
  production via gluon fusion at N$^3$LO in QCD}},  {\em Phys. Lett.} {\bf
  B803} (2020) 135292, [\href{https://arxiv.org/abs/1909.06808}{{\tt
  arXiv:1909.06808}}].

\bibitem{Chen:2019fhs}
L.-B. Chen, H.~T. Li, H.-S. Shao, and J.~Wang, {\it {The gluon-fusion
  production of Higgs boson pair: N$^3$LO QCD corrections and top-quark mass
  effects}},  {\em JHEP} {\bf 03} (2020) 072,
  [\href{https://arxiv.org/abs/1912.13001}{{\tt arXiv:1912.13001}}].

\bibitem{Heinrich:2019bkc}
G.~Heinrich, S.~P. Jones, M.~Kerner, G.~Luisoni, and L.~Scyboz, {\it {Probing
  the trilinear Higgs boson coupling in di-Higgs production at NLO QCD
  including parton shower effects}},  {\em JHEP} {\bf 06} (2019) 066,
  [\href{https://arxiv.org/abs/1903.08137}{{\tt arXiv:1903.08137}}].

\bibitem{deFlorian:2013jea}
D.~de~Florian and J.~Mazzitelli, {\it {Higgs Boson Pair Production at
  Next-to-Next-to-Leading Order in QCD}},  {\em Phys. Rev. Lett.} {\bf 111}
  (2013) 201801, [\href{https://arxiv.org/abs/1309.6594}{{\tt
  arXiv:1309.6594}}].

\bibitem{deFlorian:2015moa}
D.~de~Florian and J.~Mazzitelli, {\it {Higgs pair production at
  next-to-next-to-leading logarithmic accuracy at the LHC}},  {\em JHEP} {\bf
  09} (2015) 053, [\href{https://arxiv.org/abs/1505.07122}{{\tt
  arXiv:1505.07122}}].

\bibitem{Borowka:2016ypz}
S.~Borowka, N.~Greiner, G.~Heinrich, S.~P. Jones, M.~Kerner, J.~Schlenk, and
  T.~Zirke, {\it {Full top quark mass dependence in Higgs boson pair production
  at NLO}},  {\em JHEP} {\bf 10} (2016) 107,
  [\href{https://arxiv.org/abs/1608.04798}{{\tt arXiv:1608.04798}}].

\bibitem{Borowka:2016ehy}
S.~Borowka, N.~Greiner, G.~Heinrich, S.~P. Jones, M.~Kerner, J.~Schlenk,
  U.~Schubert, and T.~Zirke, {\it {Higgs Boson Pair Production in Gluon Fusion
  at Next-to-Leading Order with Full Top-Quark Mass Dependence}},  {\em Phys.
  Rev. Lett.} {\bf 117} (2016), no.~1 012001,
  [\href{https://arxiv.org/abs/1604.06447}{{\tt arXiv:1604.06447}}]. [Erratum:
  Phys. Rev. Lett.117,no.7,079901(2016)].

\bibitem{Davies:2019dfy}
J.~Davies, G.~Heinrich, S.~P. Jones, M.~Kerner, G.~Mishima, M.~Steinhauser, and
  D.~Wellmann, {\it {Double Higgs boson production at NLO: combining the exact
  numerical result and high-energy expansion}},  {\em JHEP} {\bf 11} (2019)
  024, [\href{https://arxiv.org/abs/1907.06408}{{\tt arXiv:1907.06408}}].

\bibitem{Baglio:2018lrj}
J.~Baglio, F.~Campanario, S.~Glaus, M.~Mühlleitner, M.~Spira, and
  J.~Streicher, {\it {Gluon fusion into Higgs pairs at NLO QCD and the top mass
  scheme}},  {\em Eur. Phys. J. C} {\bf 79} (2019), no.~6 459,
  [\href{https://arxiv.org/abs/1811.05692}{{\tt arXiv:1811.05692}}].

\bibitem{Baglio:2020ini}
J.~Baglio, F.~Campanario, S.~Glaus, M.~Mühlleitner, J.~Ronca, M.~Spira, and
  J.~Streicher, {\it {Higgs-Pair Production via Gluon Fusion at Hadron
  Colliders: NLO QCD Corrections}},  {\em JHEP} {\bf 04} (2020) 181,
  [\href{https://arxiv.org/abs/2003.03227}{{\tt arXiv:2003.03227}}].

\bibitem{Sjostrand:2007gs}
T.~Sjostrand, S.~Mrenna, and P.~Z. Skands, {\it {A Brief Introduction to PYTHIA
  8.1}},  {\em Comput. Phys. Commun.} {\bf 178} (2008) 852--867,
  [\href{https://arxiv.org/abs/0710.3820}{{\tt arXiv:0710.3820}}].

\bibitem{Collaboration:2285584}
{ATLAS Collaboration}, {\it {Technical Design Report for the Phase-II Upgrade
  of the ATLAS TDAQ System}},  Tech. Rep. CERN-LHCC-2017-020. ATLAS-TDR-029,
  2017.

\bibitem{Collaboration:2623663}
{ATLAS Collaboration}, {\it {Technical Proposal: A High-Granularity Timing
  Detector for the ATLAS Phase-II Upgrade}},  Tech. Rep. CERN-LHCC-2018-023.
  LHCC-P-012, 2018.

\bibitem{CMS:2667167}
{CMS Collaboration}, {\it {A MIP Timing Detector for the CMS Phase-2 Upgrade}},
   Tech. Rep. CERN-LHCC-2019-003. CMS-TDR-020, 2019.

\bibitem{Tseng:2013dva}
J.~Tseng and H.~Evans, {\it {Sequential recombination algorithm for jet
  clustering and background subtraction}},  {\em Phys. Rev.} {\bf D88} (2013)
  014044, [\href{https://arxiv.org/abs/1304.1025}{{\tt arXiv:1304.1025}}].

\bibitem{Bertolini:2014bba}
D.~Bertolini, P.~Harris, M.~Low, and N.~Tran, {\it {Pileup Per Particle
  Identification}},  {\em JHEP} {\bf 10} (2014) 059,
  [\href{https://arxiv.org/abs/1407.6013}{{\tt arXiv:1407.6013}}].

\bibitem{Cacciari:2014gra}
M.~Cacciari, G.~P. Salam, and G.~Soyez, {\it {SoftKiller, a particle-level
  pileup removal method}},  {\em Eur. Phys. J.} {\bf C75} (2015), no.~2 59,
  [\href{https://arxiv.org/abs/1407.0408}{{\tt arXiv:1407.0408}}].

\bibitem{Komiske:2017ubm}
P.~T. Komiske, E.~M. Metodiev, B.~Nachman, and M.~D. Schwartz, {\it {Pileup
  Mitigation with Machine Learning (PUMML)}},  {\em JHEP} {\bf 12} (2017) 051,
  [\href{https://arxiv.org/abs/1707.08600}{{\tt arXiv:1707.08600}}].

\bibitem{Berta:2019hnj}
P.~Berta, L.~Masetti, D.~W. Miller, and M.~Spousta, {\it {Pileup and Underlying
  Event Mitigation with Iterative Constituent Subtraction}},  {\em JHEP} {\bf
  08} (2019) 175, [\href{https://arxiv.org/abs/1905.03470}{{\tt
  arXiv:1905.03470}}].

\bibitem{deFavereau:2013fsa}
{\bf DELPHES 3} Collaboration, J.~de~Favereau, C.~Delaere, P.~Demin,
  A.~Giammanco, V.~Lemaître, A.~Mertens, and M.~Selvaggi, {\it {DELPHES 3, A
  modular framework for fast simulation of a generic collider experiment}},
  {\em JHEP} {\bf 02} (2014) 057, [\href{https://arxiv.org/abs/1307.6346}{{\tt
  arXiv:1307.6346}}].

\bibitem{Cacciari:2008gp}
M.~Cacciari, G.~P. Salam, and G.~Soyez, {\it {The Anti-k(t) jet clustering
  algorithm}},  {\em JHEP} {\bf 04} (2008) 063,
  [\href{https://arxiv.org/abs/0802.1189}{{\tt arXiv:0802.1189}}].

\bibitem{Cacciari:2011ma}
M.~Cacciari, G.~P. Salam, and G.~Soyez, {\it {FastJet User Manual}},  {\em Eur.
  Phys. J.} {\bf C72} (2012) 1896, [\href{https://arxiv.org/abs/1111.6097}{{\tt
  arXiv:1111.6097}}].

\bibitem{ATLAS:ITk:Pixel:TDR}
{ATLAS Collaboration}, {\it {Technical Design Report for the ATLAS Inner
  Tracker Pixel Detector}},  Tech. Rep. CERN-LHCC-2017-021. ATLAS-TDR-030,
  2017.

\bibitem{Krohn:2009zg}
D.~Krohn, J.~Thaler, and L.-T. Wang, {\it {Jets with Variable R}},  {\em JHEP}
  {\bf 06} (2009) 059, [\href{https://arxiv.org/abs/0903.0392}{{\tt
  arXiv:0903.0392}}].

\bibitem{Sirunyan:2019wwa}
{CMS Collaboration}, {\it {A deep neural network for simultaneous estimation of
  b jet energy and resolution}},  \href{https://arxiv.org/abs/1912.06046}{{\tt
  arXiv:1912.06046}}.

\bibitem{chollet2015keras}
F.~Chollet et~al., ``Keras.'' \url{https://keras.io}, 2015.

\bibitem{Aad:2019yxi}
{ATLAS Collaboration}, {\it {Search for non-resonant Higgs boson pair
  production in the $bb\ell\nu\ell\nu$ final state with the ATLAS detector in
  $pp$ collisions at $\sqrt{s} = 13$ TeV}},  {\em Phys. Lett.} {\bf B801}
  (2020) 135145, [\href{https://arxiv.org/abs/1908.06765}{{\tt
  arXiv:1908.06765}}].

\bibitem{pmlr-v15-glorot11a}
X.~Glorot, A.~Bordes, and Y.~Bengio, {\it Deep sparse rectifier neural
  networks},  in {\em Proceedings of the Fourteenth International Conference on
  Artificial Intelligence and Statistics} (G.~Gordon, D.~Dunson, and M.~Dudík,
  eds.), vol.~15 of {\em Proceedings of Machine Learning Research}, (Fort
  Lauderdale, FL, USA), pp.~315--323, PMLR, 11--13 Apr, 2011.

\bibitem{kingma2014adam}
D.~P. Kingma and J.~Ba, {\it Adam: A method for stochastic optimization},  {\em
  3rd International Conference for Learning Representations} (2014)
  [\href{https://arxiv.org/abs/1412.6980}{{\tt arXiv:1412.6980}}].

\bibitem{JMLR:v15:srivastava14a}
N.~Srivastava, G.~Hinton, A.~Krizhevsky, I.~Sutskever, and R.~Salakhutdinov,
  {\it Dropout: A simple way to prevent neural networks from overfitting},
  {\em Journal of Machine Learning Research} {\bf 15} (2014) 1929--1958.

\bibitem{Huffman:2016wjk}
B.~T. Huffman, C.~Jackson, and J.~Tseng, {\it {Tagging $b$ quarks at extreme
  energies without tracks}},  {\em J. Phys.} {\bf G43} (2016), no.~8 085001,
  [\href{https://arxiv.org/abs/1604.05036}{{\tt arXiv:1604.05036}}].

\end{thebibliography}\endgroup

\newpage
\appendix
%--------------------------------
\section{\label{sec:distributions}Additional kinematic distributions}
\begin{figure}
    \centering    
    \begin{subfigure}[b]{0.5\textwidth}
        \includegraphics[width=\textwidth]{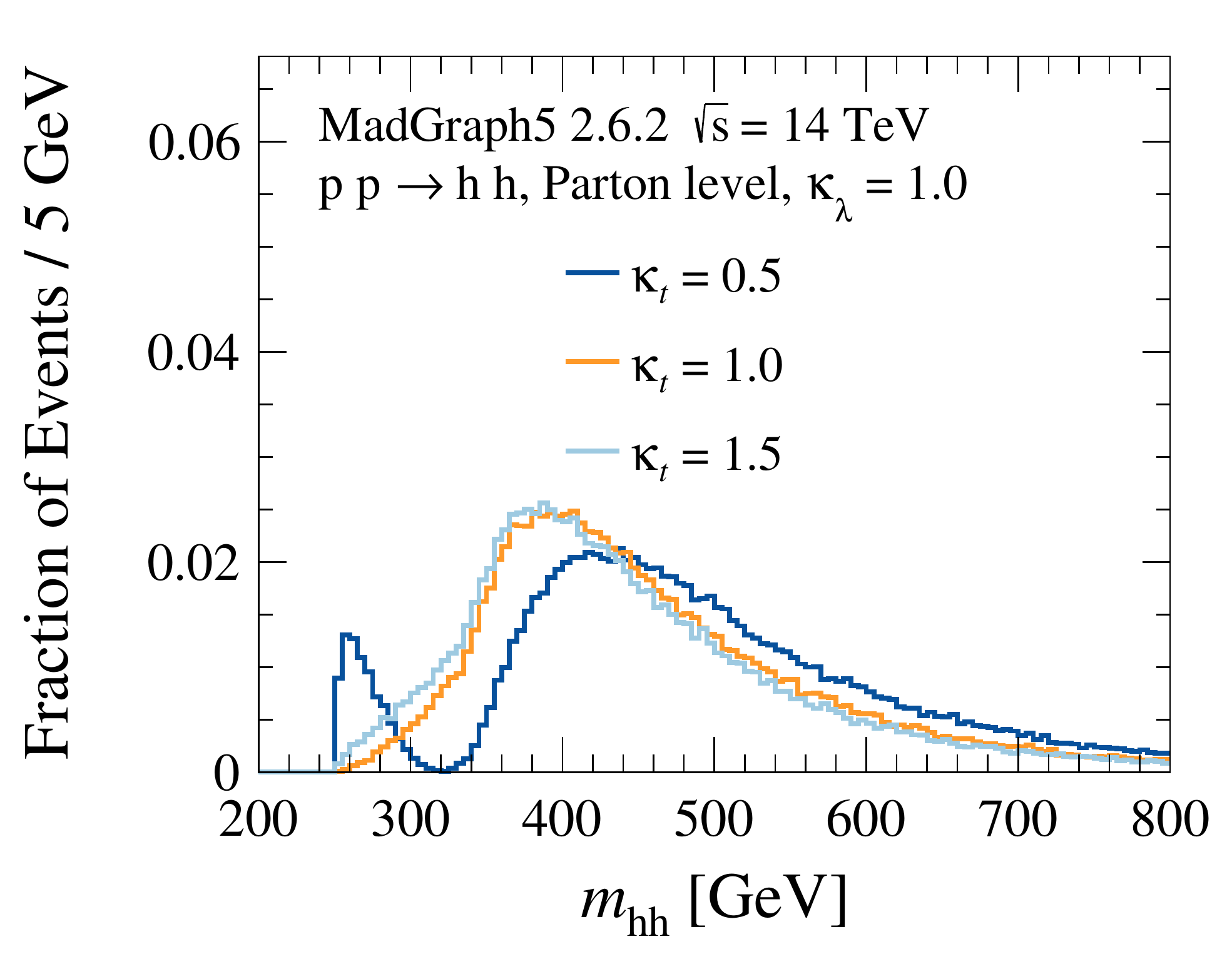}\\
        \includegraphics[width=\textwidth]{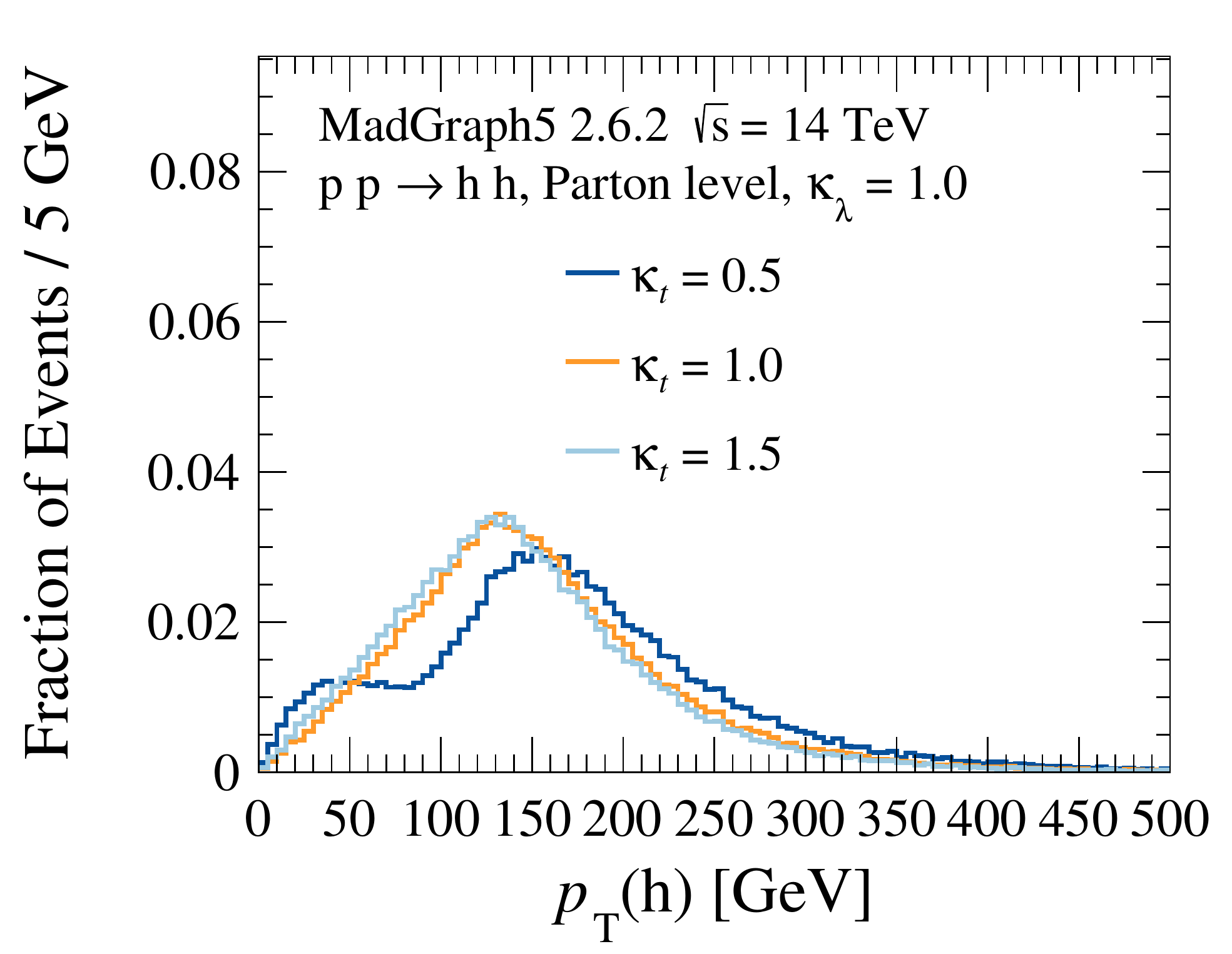}
        \caption{\label{fig:Norm_HiggsPt_DiHiggsDeltaEta_a}Vary top Yukawa \kapt}
    \end{subfigure}%
    \begin{subfigure}[b]{0.5\textwidth}
    \includegraphics[width=\textwidth]{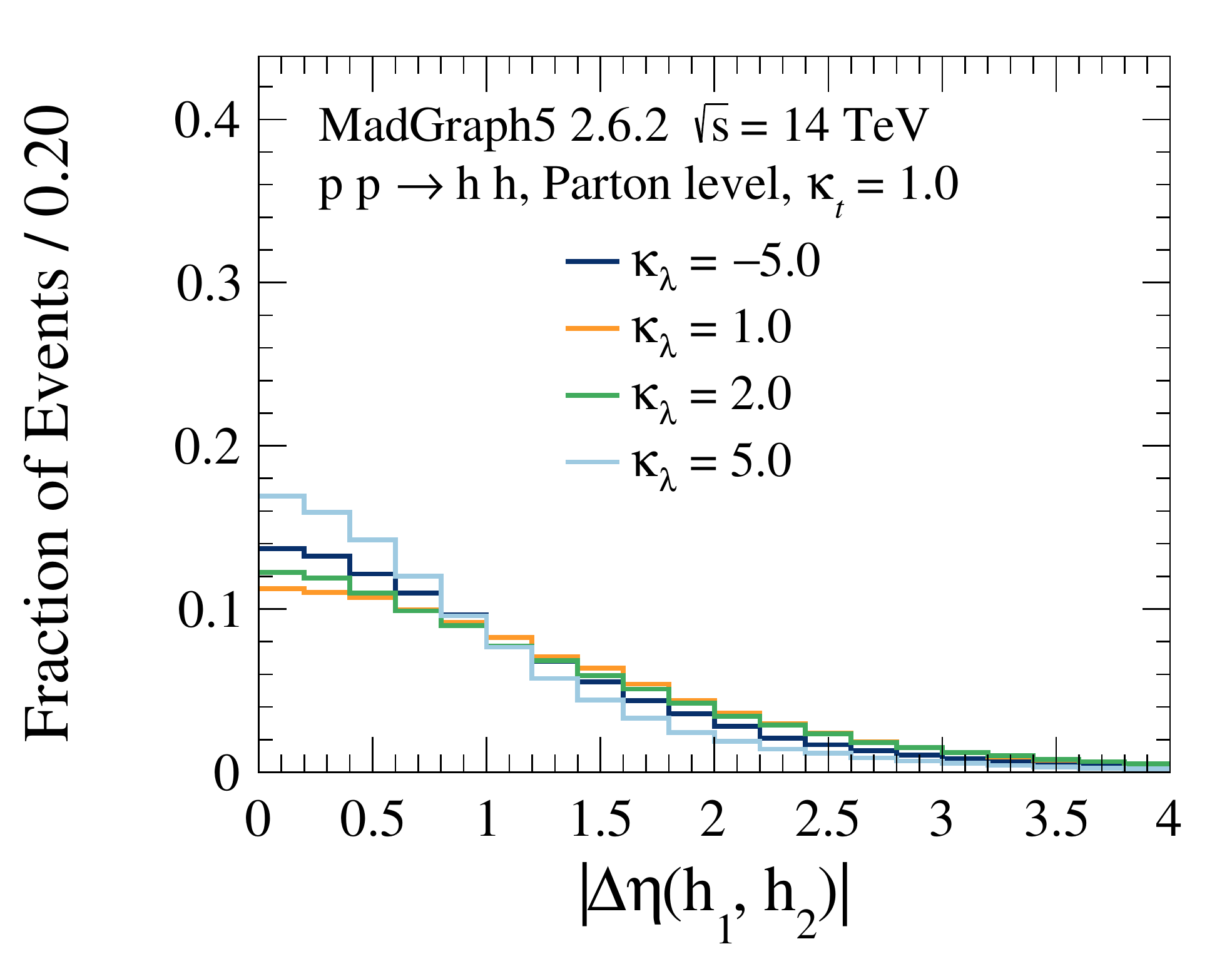}\\
    \includegraphics[width=\textwidth]{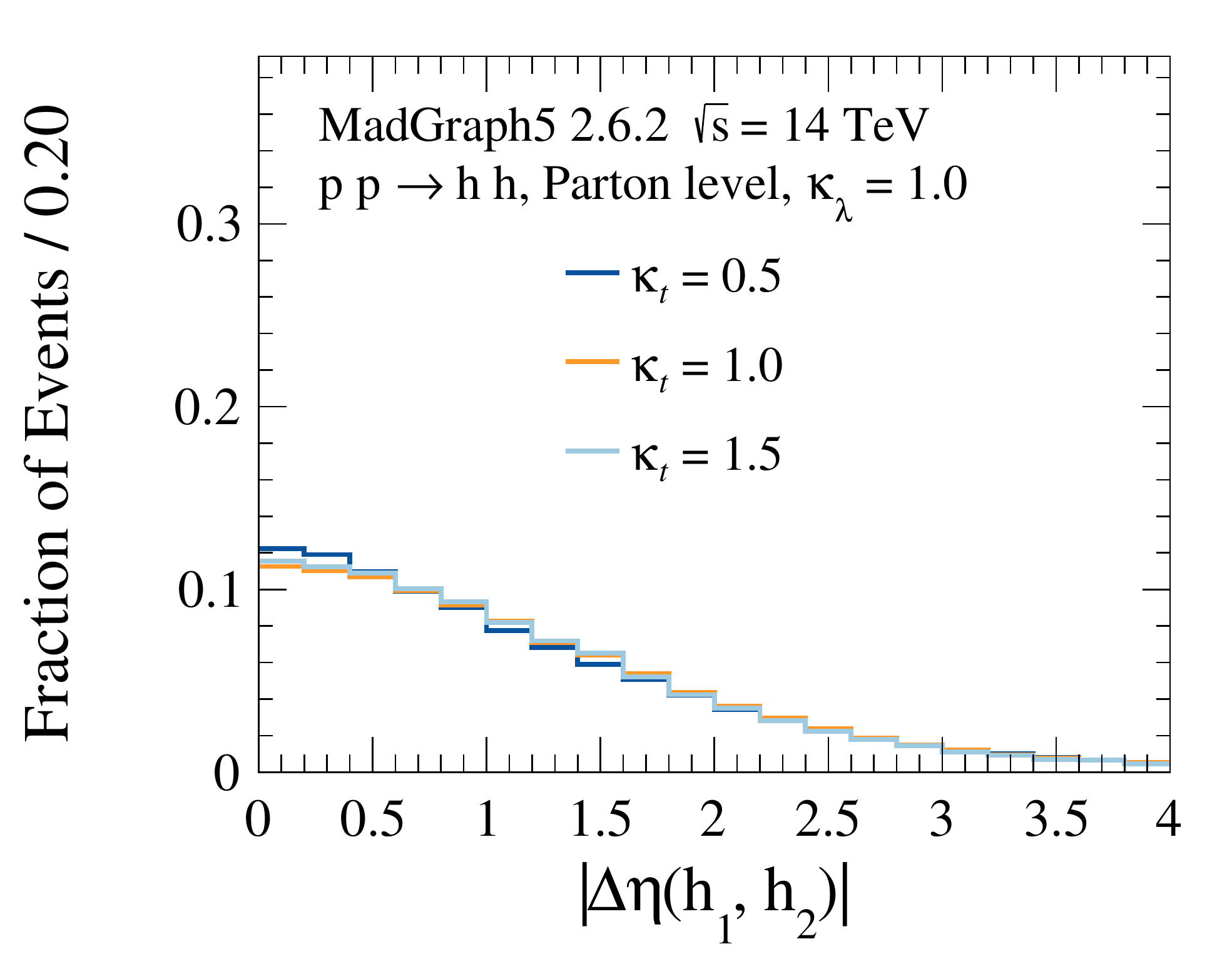}
        \caption{\label{fig:Norm_HiggsPt_DiHiggsDeltaEta_b}Di-Higgs pseudorapidity difference}
    \end{subfigure}%
    \caption{Unit normalised distributions for parton-level signals for different couplings.}
    \label{fig:Norm_HiggsPt_DiHiggsDeltaEta}
\end{figure}

%--------------------------------
\FloatBarrier
This appendix collects additional kinematic distributions. Figure~\ref{fig:Norm_HiggsPt_DiHiggsDeltaEta_a} shows the impact at parton level of top Yukawa variations for fixed $\klam = 1$. In particular, $\kapt = 1.5$ has little impact on the \mhh shape, but lowering \kapt to 0.5 makes the destructive interference induce a similar \mhh shape as $(\klam, \kapt) = (2, 1)$. Figure~\ref{fig:Norm_HiggsPt_DiHiggsDeltaEta_b} displays the rapidity difference between the Higgs bosons $|\eta(h_1, h_2)|$, which has mild discriminating power between couplings.
    
\begin{figure}
    \centering
    \includegraphics[width=0.5\textwidth]{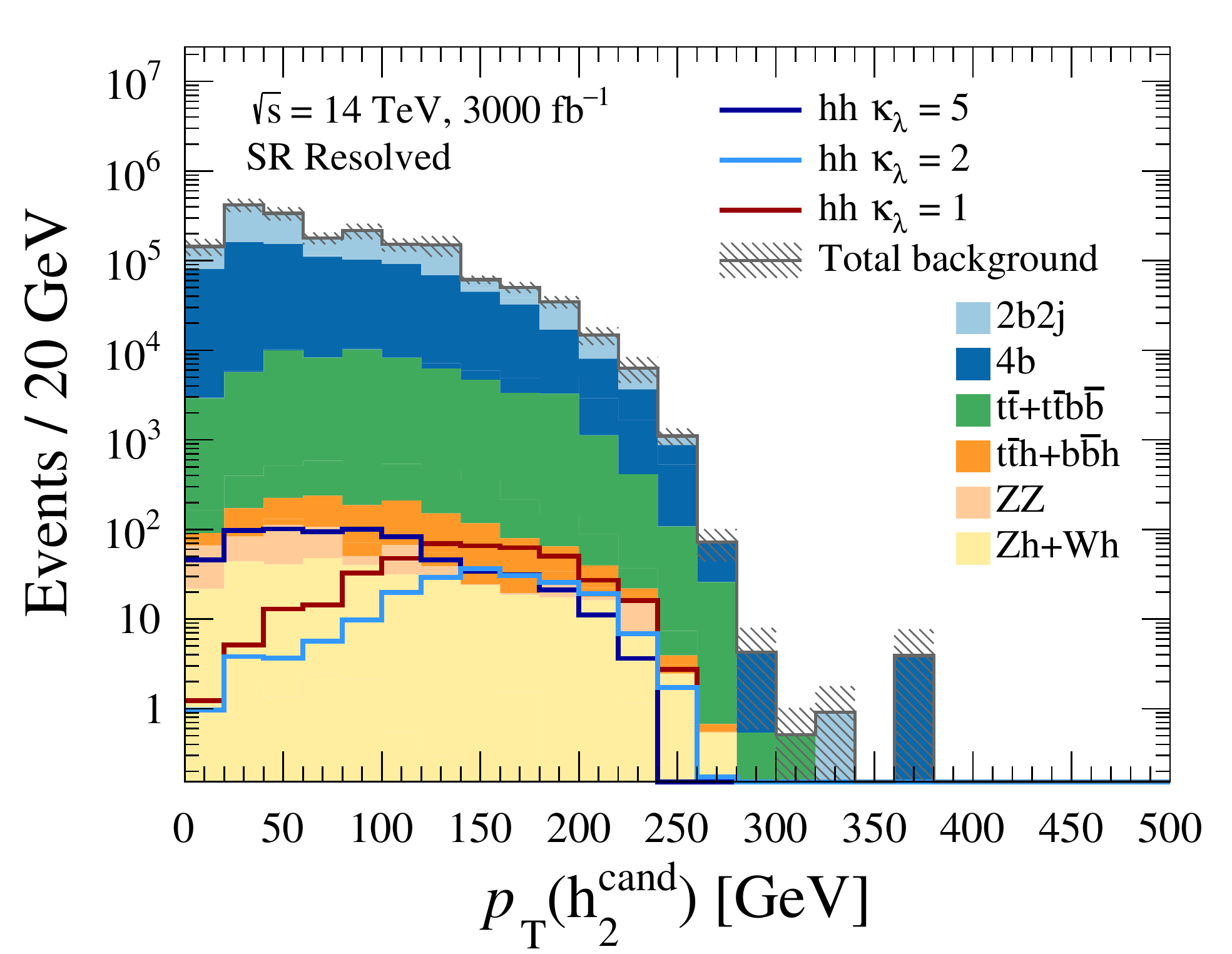}%
    \includegraphics[width=0.5\textwidth]{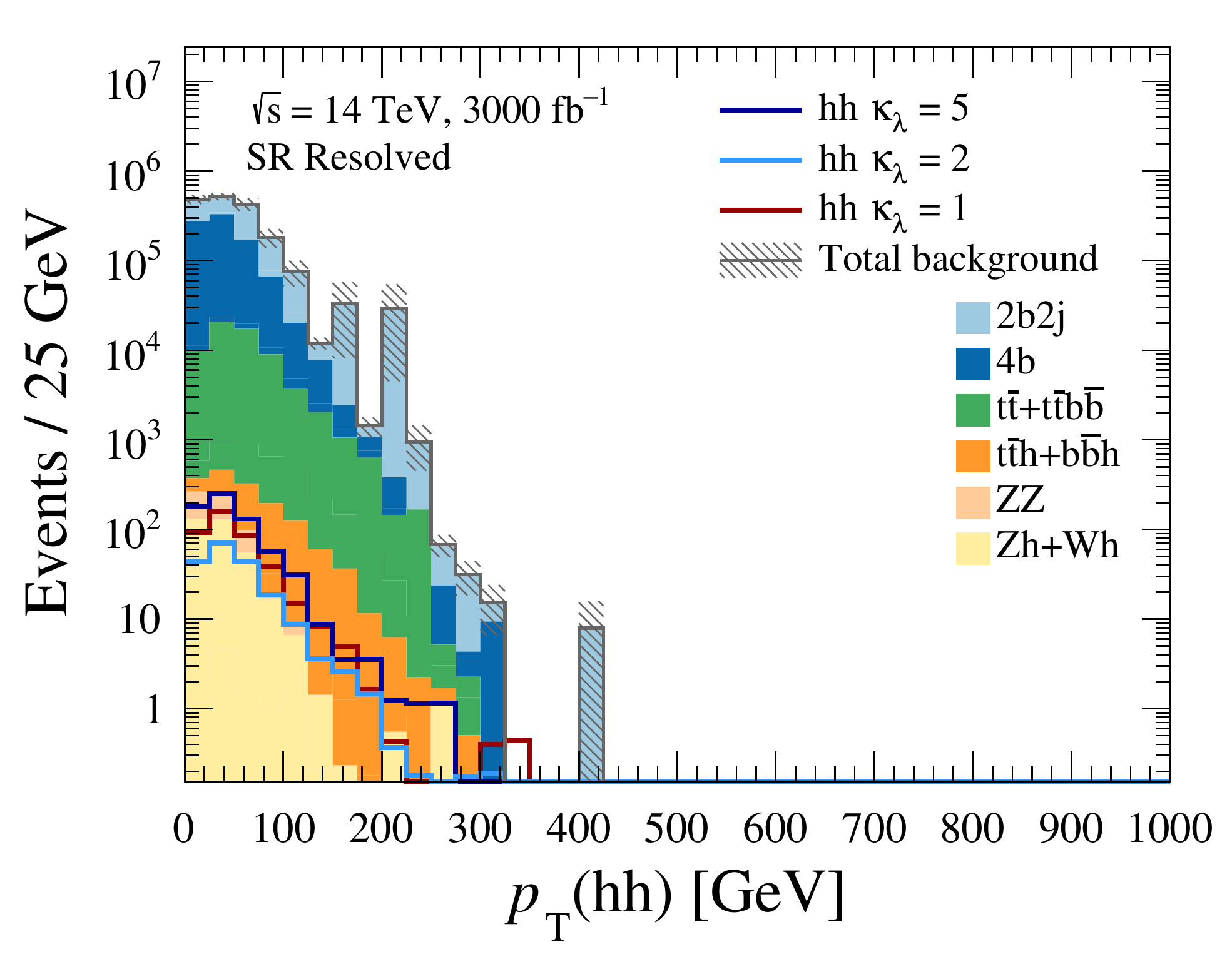}\\
    \includegraphics[width=0.5\textwidth]{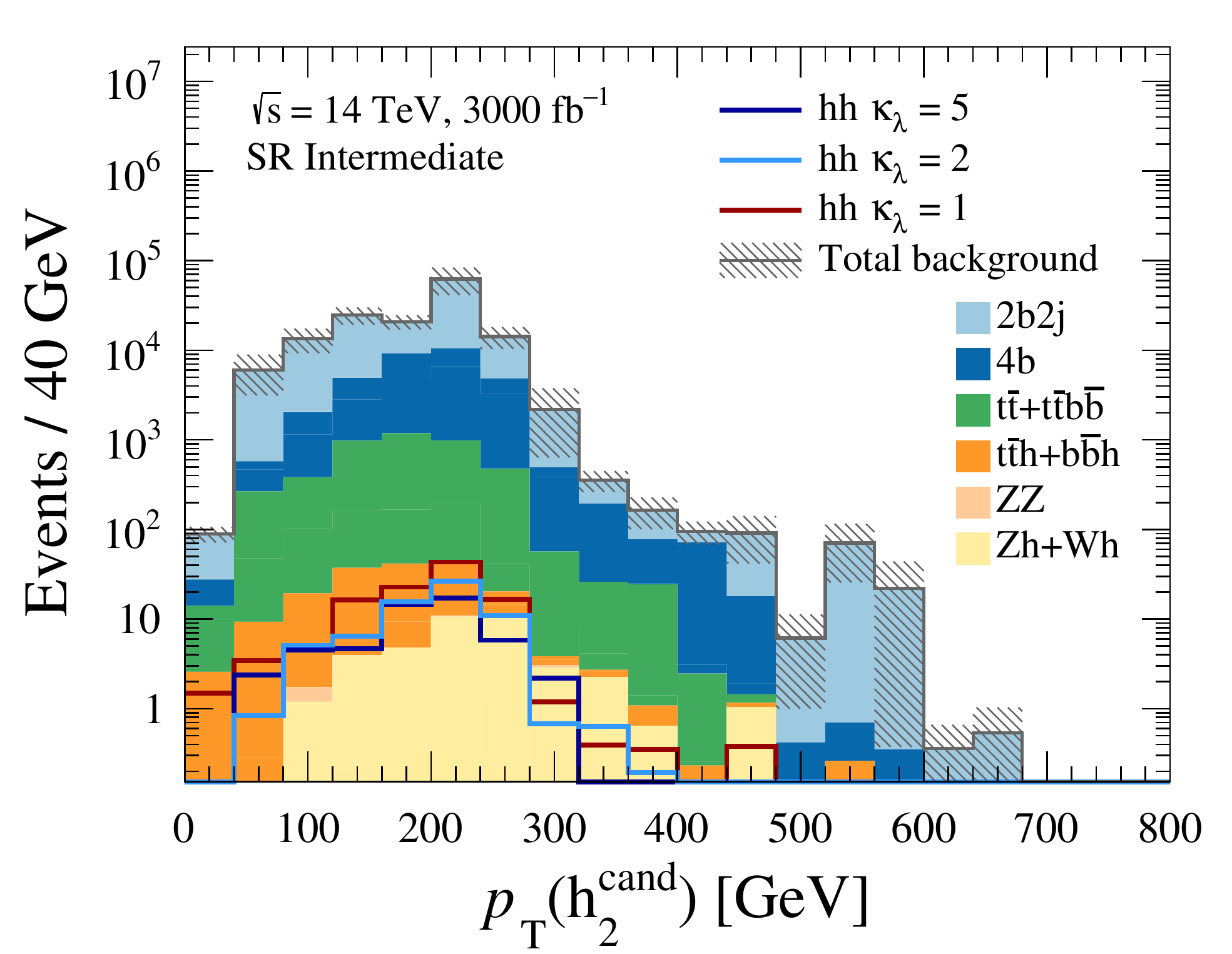}%
    \includegraphics[width=0.5\textwidth]{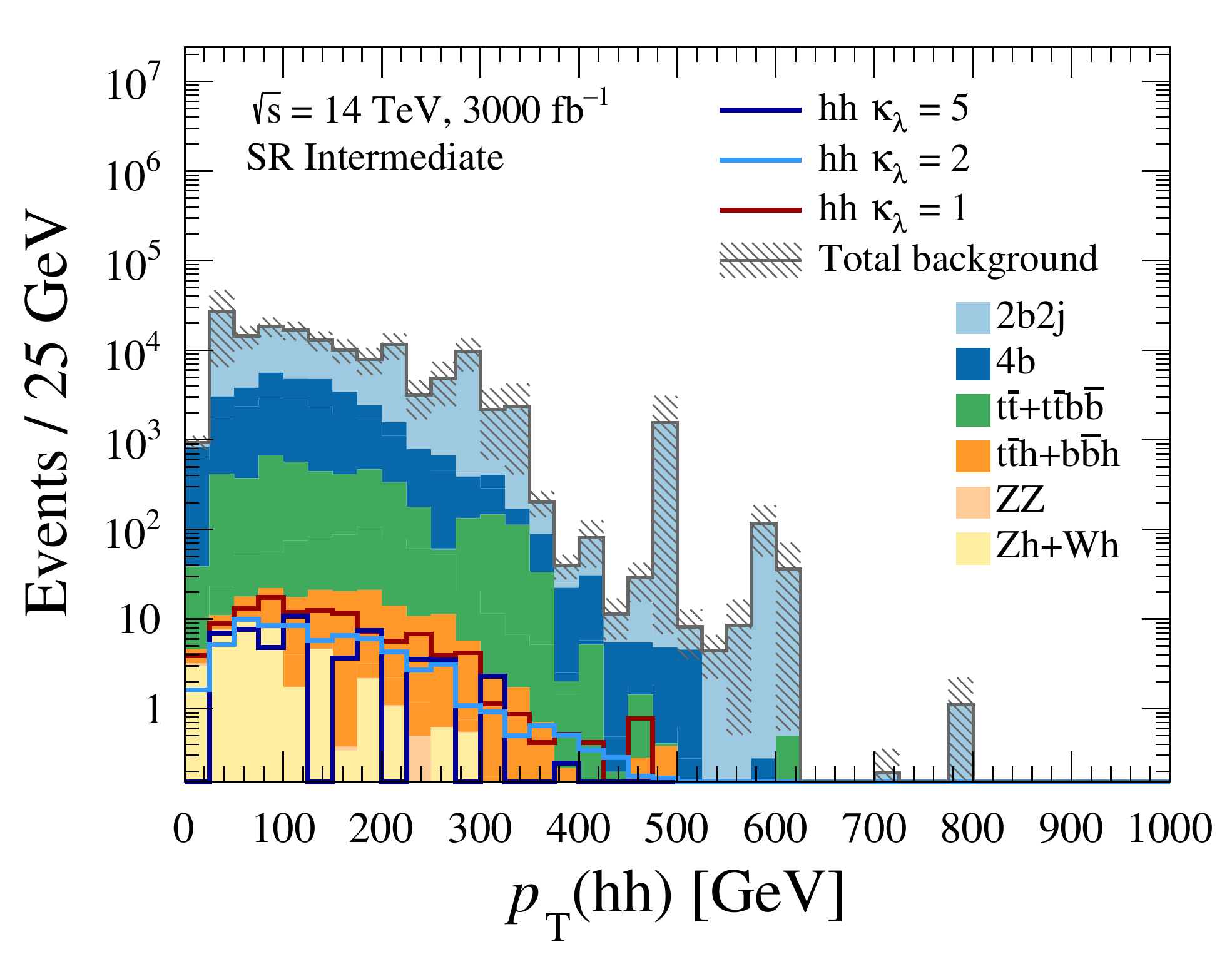}\\
    \includegraphics[width=0.5\textwidth]{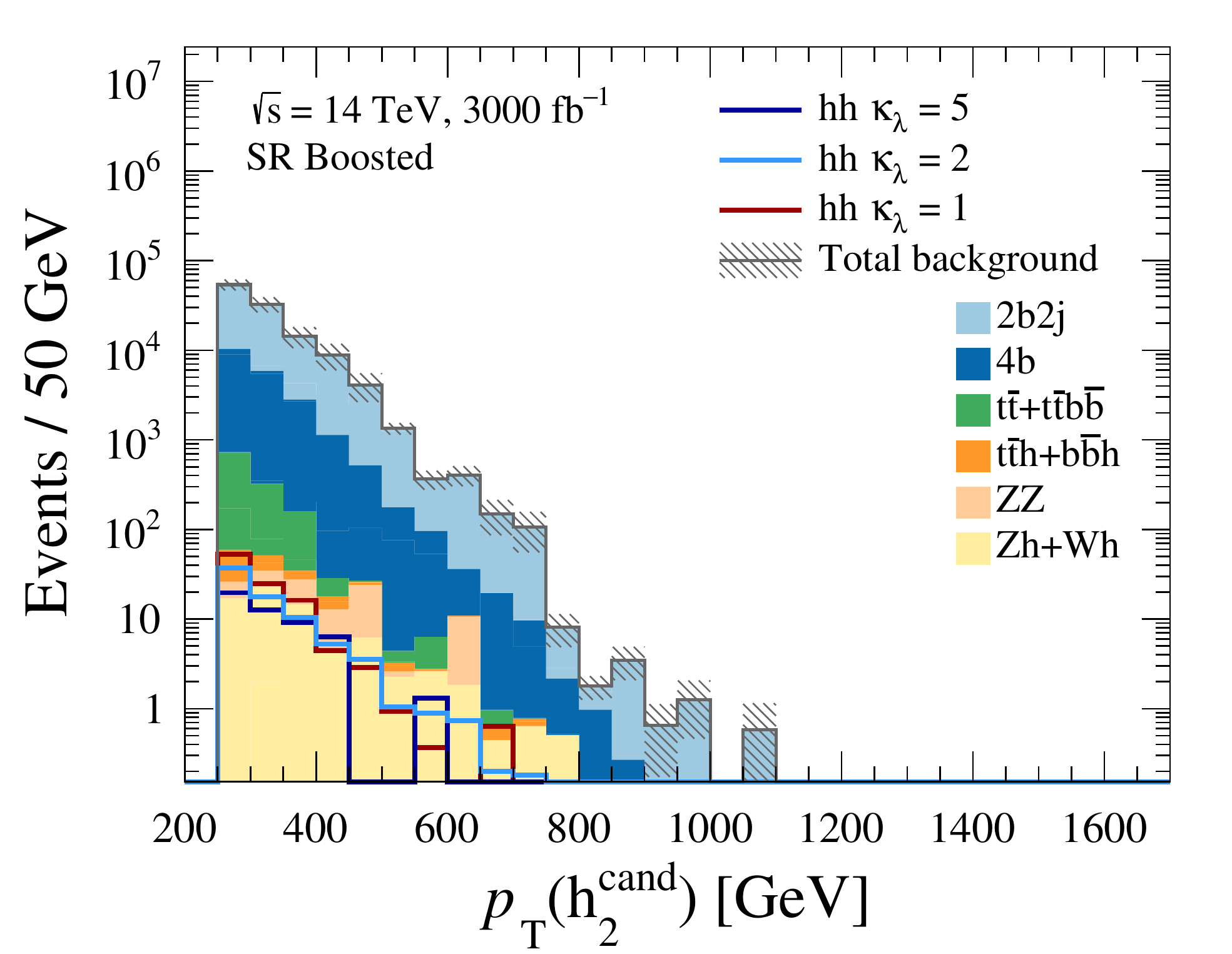}%
    \includegraphics[width=0.5\textwidth]{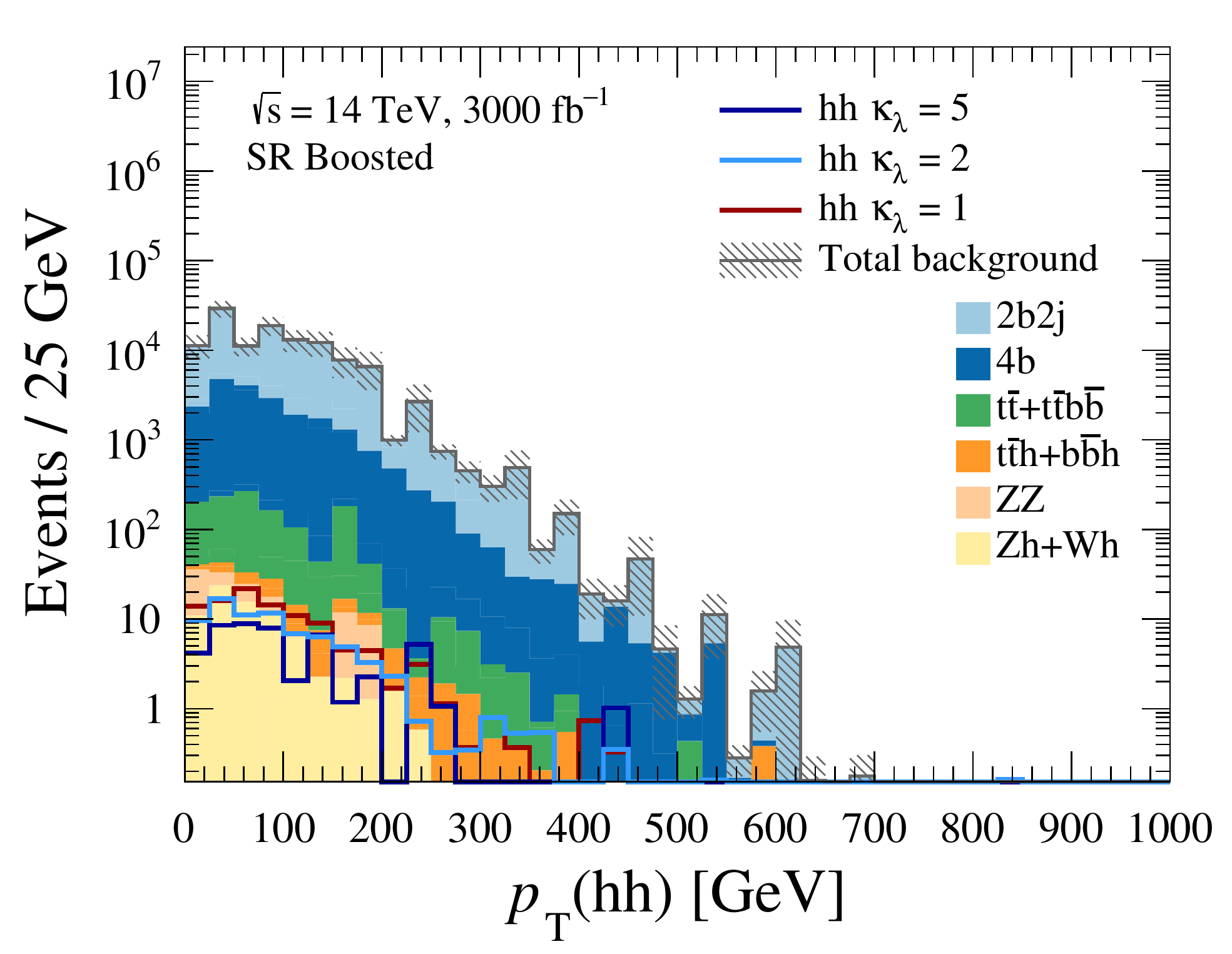}
    \caption{Distributions of signal (lines) and backgrounds (stacked filled) for (left) the transverse momentum of the subleading Higgs candidate $p_\text{T}(h_2^\text{cand})$ and (right) the di-Higgs system $p_\text{T}(hh)$ for the \emph{baseline analysis} in the (upper) resolved, (middle) intermediate and (lower) boosted categories.}
    \label{fig:loose_pTH1_cutana}
\end{figure}

Figure~\ref{fig:loose_pTH1_cutana} displays distributions at reconstructed level for both backgrounds and signals in the \emph{baseline analysis} signal regions normalised to cross-sections. The subleading Higgs candidate $\pt(h_2^\text{cand})$ and transverse momentum of the di-Higgs system $\pt(hh)$ are shown. We note that in the resolved category, the $\pt(h_2^\text{cand})$ retains some discrimination power between different \klam hypotheses. The $\pt(hh)$ is also lower for the resolved than the intermediate and boosted analyses. This suggests that a modest amount of the boost that the Higgs bosons receives arises from the recoil of the di-Higgs system against other activity such as initial state radiation jets.

\begin{figure}
    \centering
    \includegraphics[width=0.5\textwidth]{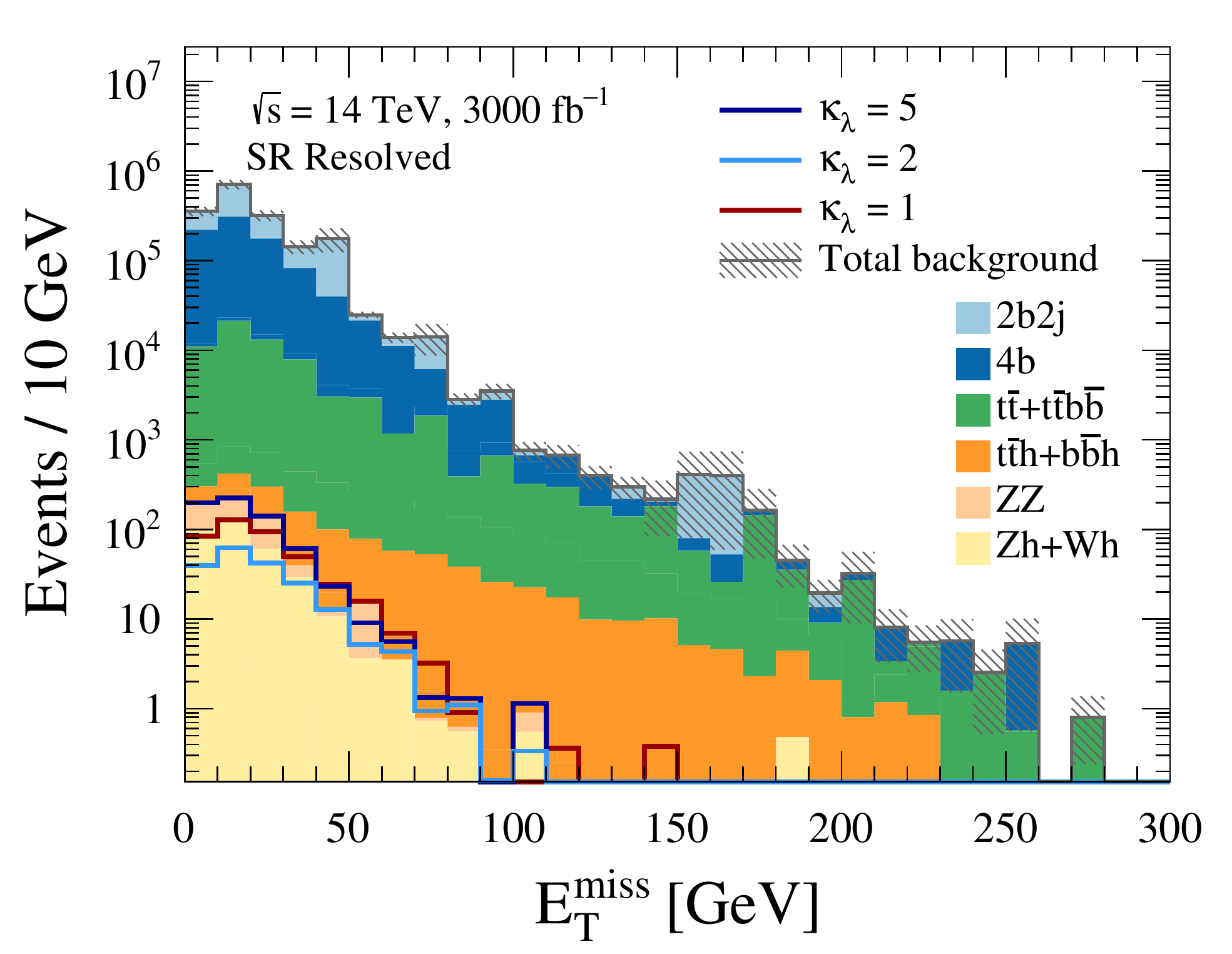}%
    \includegraphics[width=0.5\textwidth]{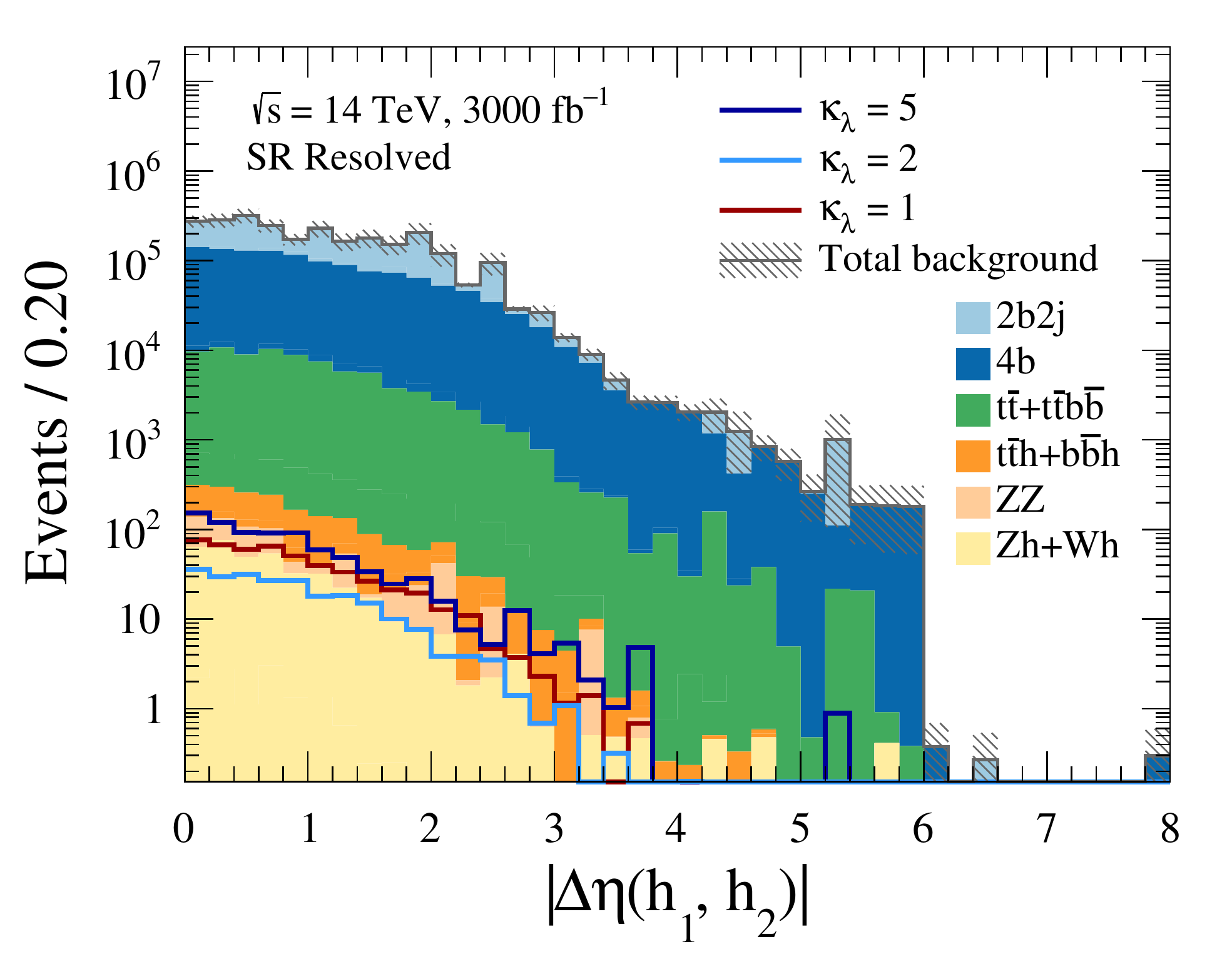}\\
    \includegraphics[width=0.5\textwidth]{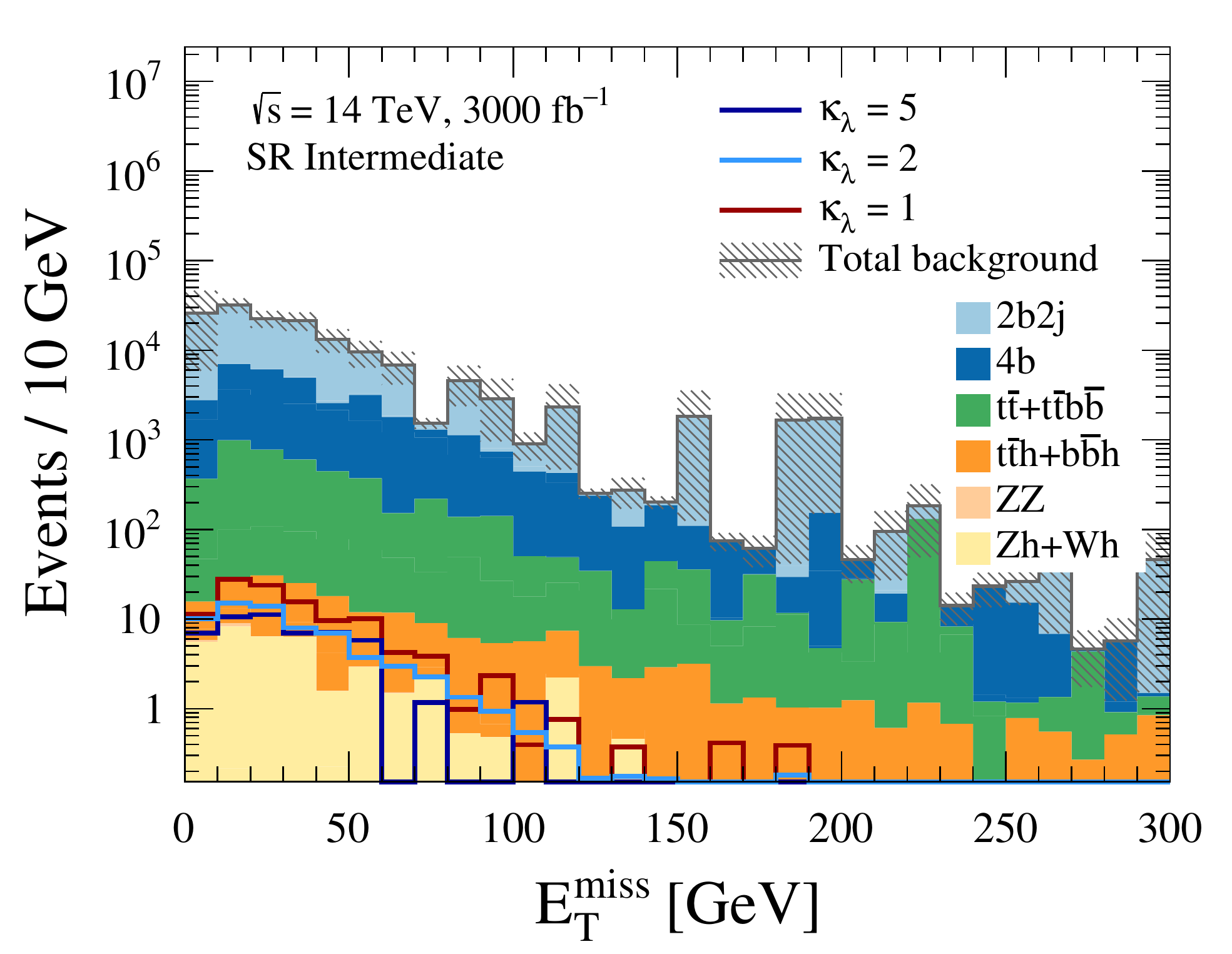}%
    \includegraphics[width=0.5\textwidth]{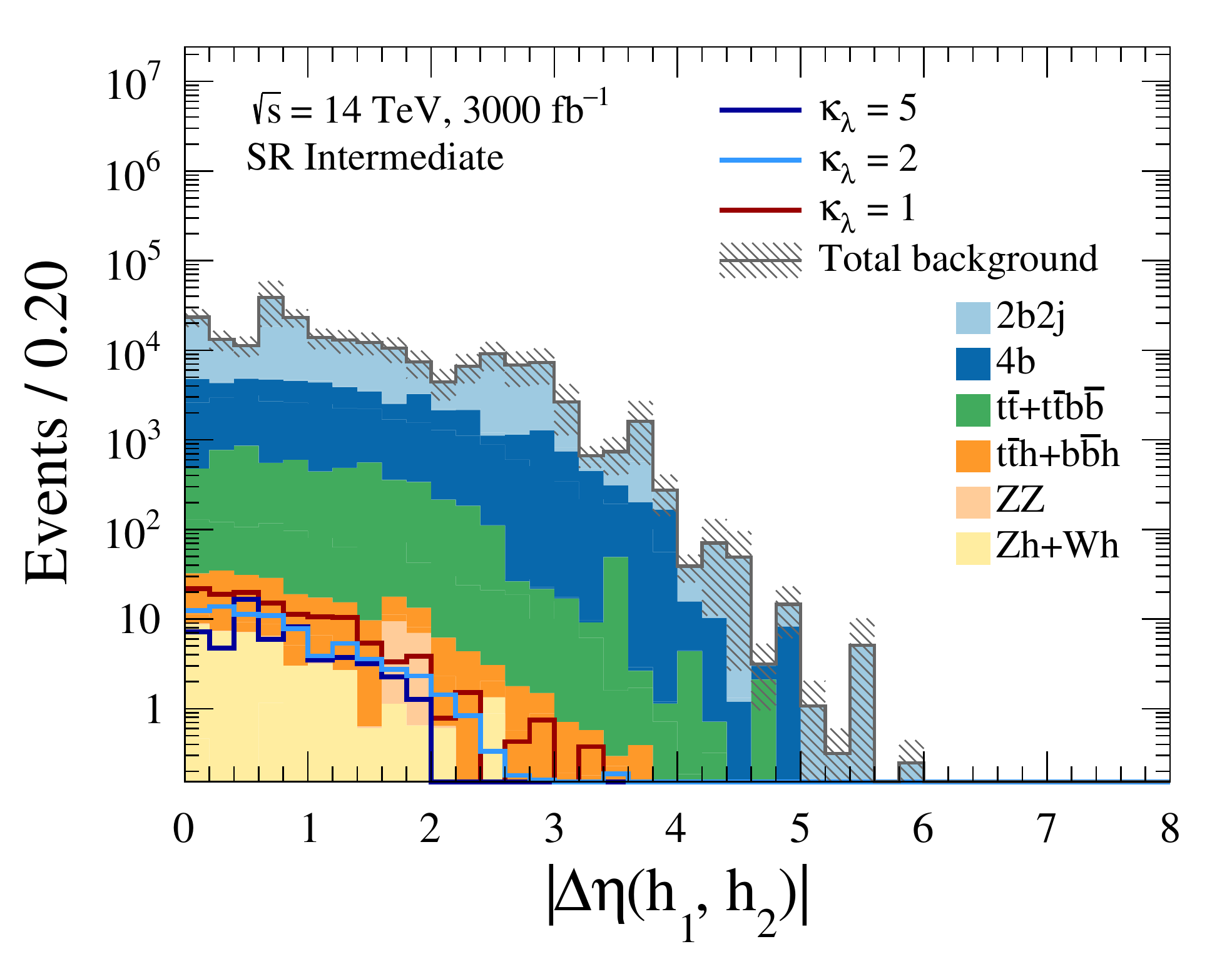}\\
    \includegraphics[width=0.5\textwidth]{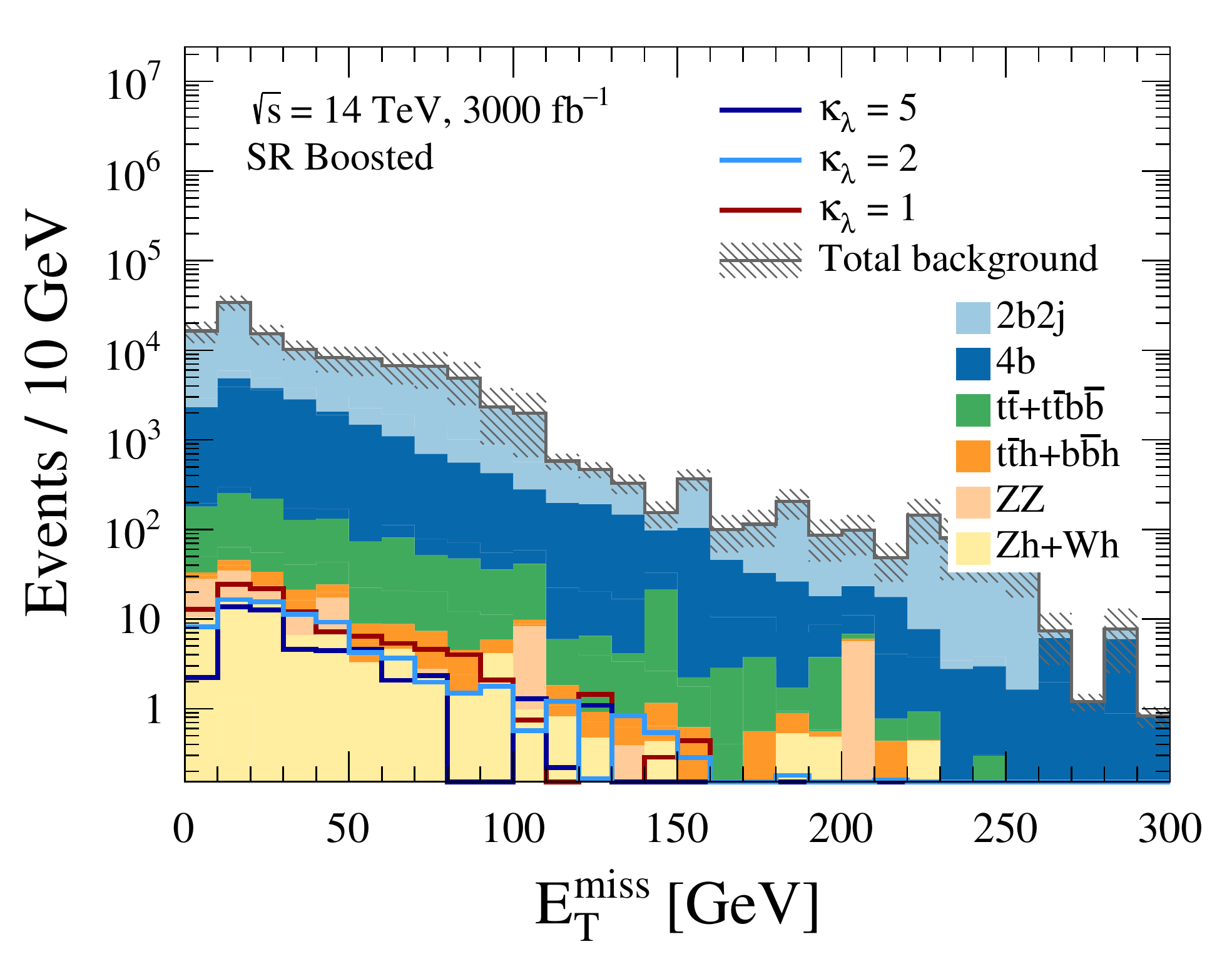}%
    \includegraphics[width=0.5\textwidth]{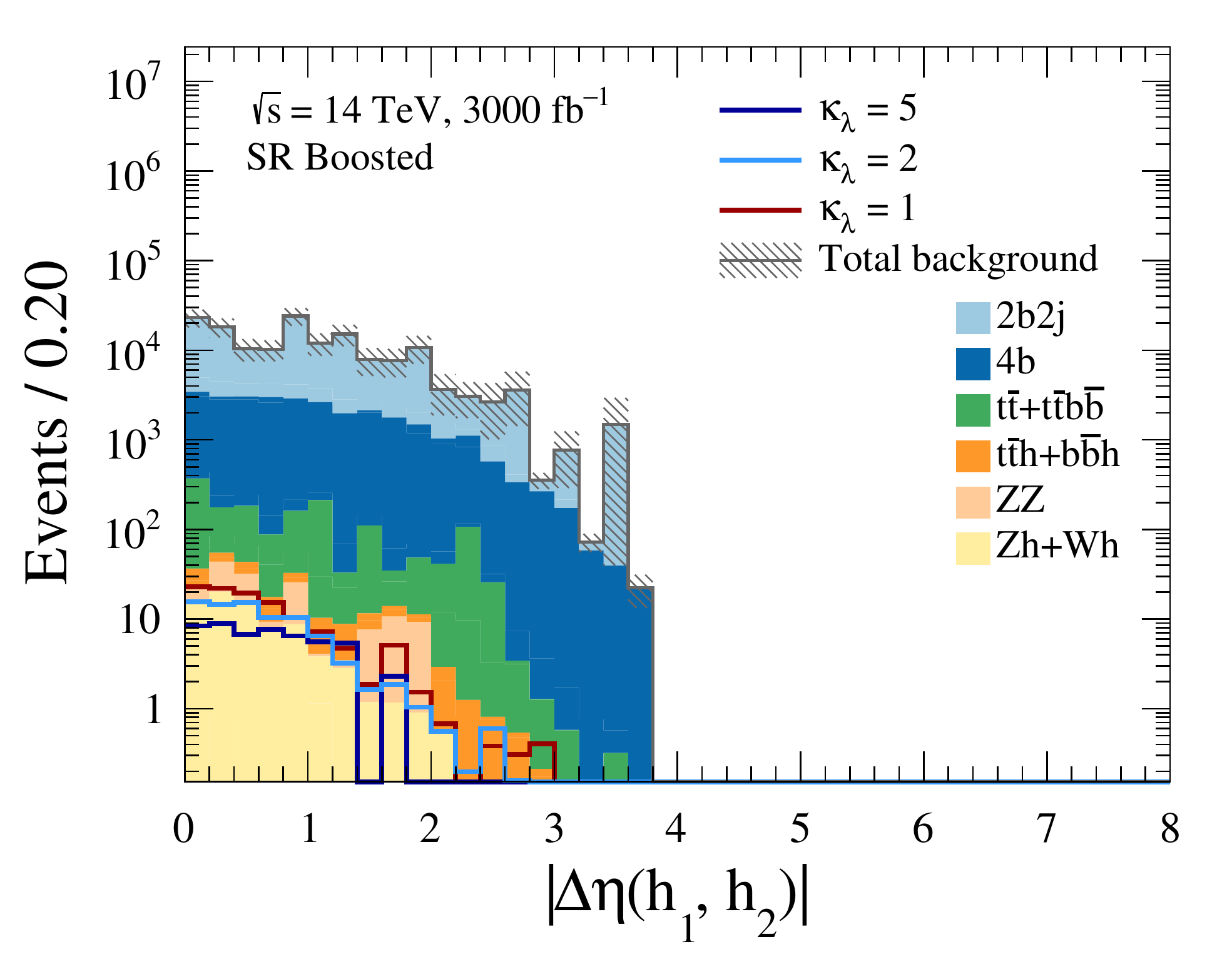}
    \caption{`$N-1$' distributions with all signal region event selection applied except for the variable being plotted. 
    Displayed are various signal (lines) and backgrounds (stacked filled) for (left) the magnitude of the missing transverse momentum $E_\text{T}^\text{miss}$ and (right) the pseudorapidity difference of the two Higgs candidates $\Delta \eta_{hh}$ for the \emph{baseline analysis} in the (upper) resolved, (middle) intermediate and (lower) boosted categories.}
    \label{fig:loose_Met_dEta}
\end{figure}

\FloatBarrier
%--------------------------------
\section{Neural network performance}
\label{sec:NNperform}
%--------------------------------
\FloatBarrier

This appendix provides further performance benchmarks of the \emph{neural network analysis} for the DNN trained on $\klam=1$ and  $\klam=5$ signals. All the plots in this section are made with an independent set of events compared with those used for training. The selection criteria applied for the DNN training, and also for all plots in this section, are looser than the final analysis selection in order to retain sufficient event statistics for reliable training. One significant difference is that this dataset does not include any requirements on the number of $b$-tagged jets. 

\begin{figure}
    \centering
    \begin{subfigure}[b]{0.49\textwidth}
        \includegraphics[width=\textwidth]{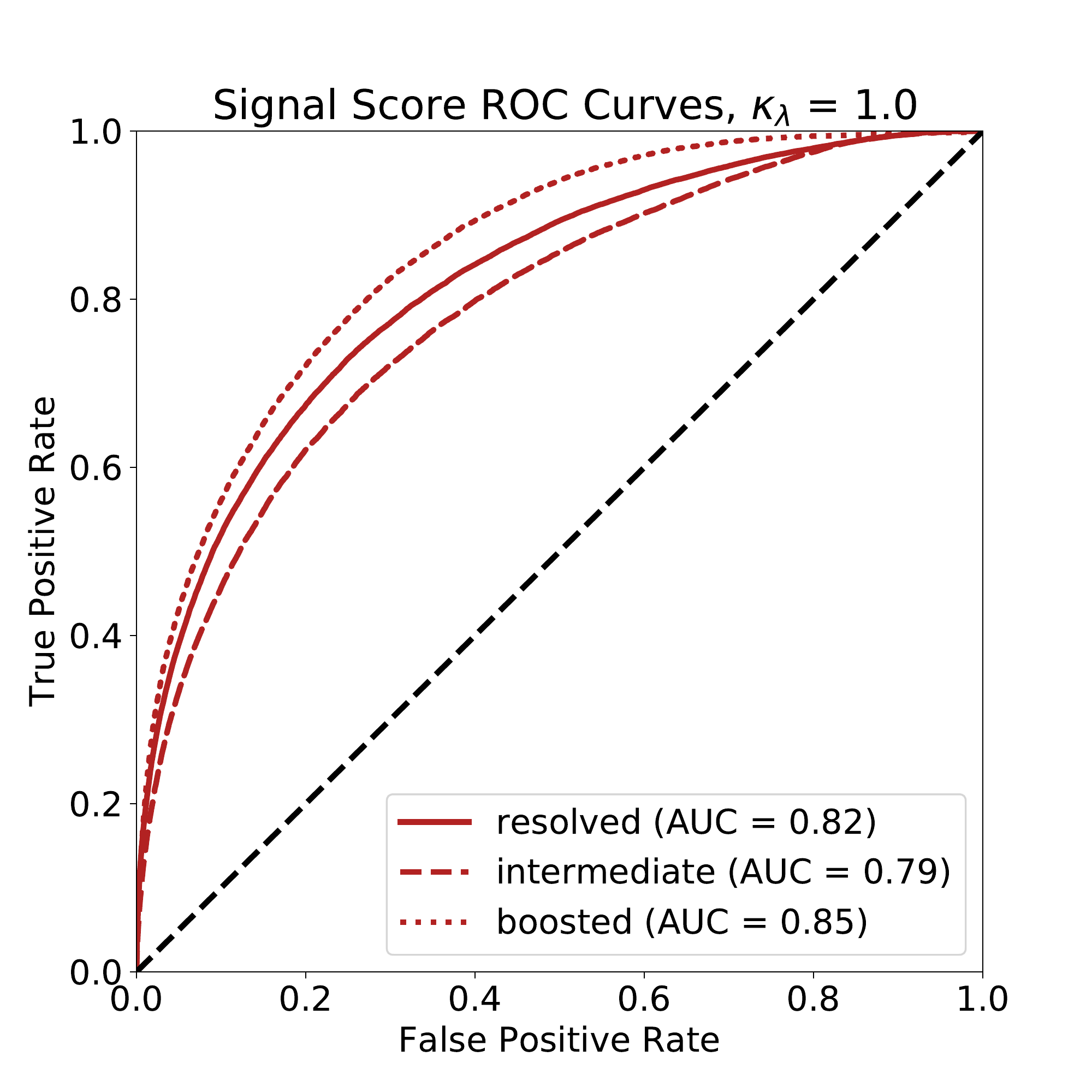}
    \caption{DNN trained on $\klam = 1$}
    \end{subfigure}%
    \begin{subfigure}[b]{0.49\textwidth}
        \includegraphics[width=\textwidth]{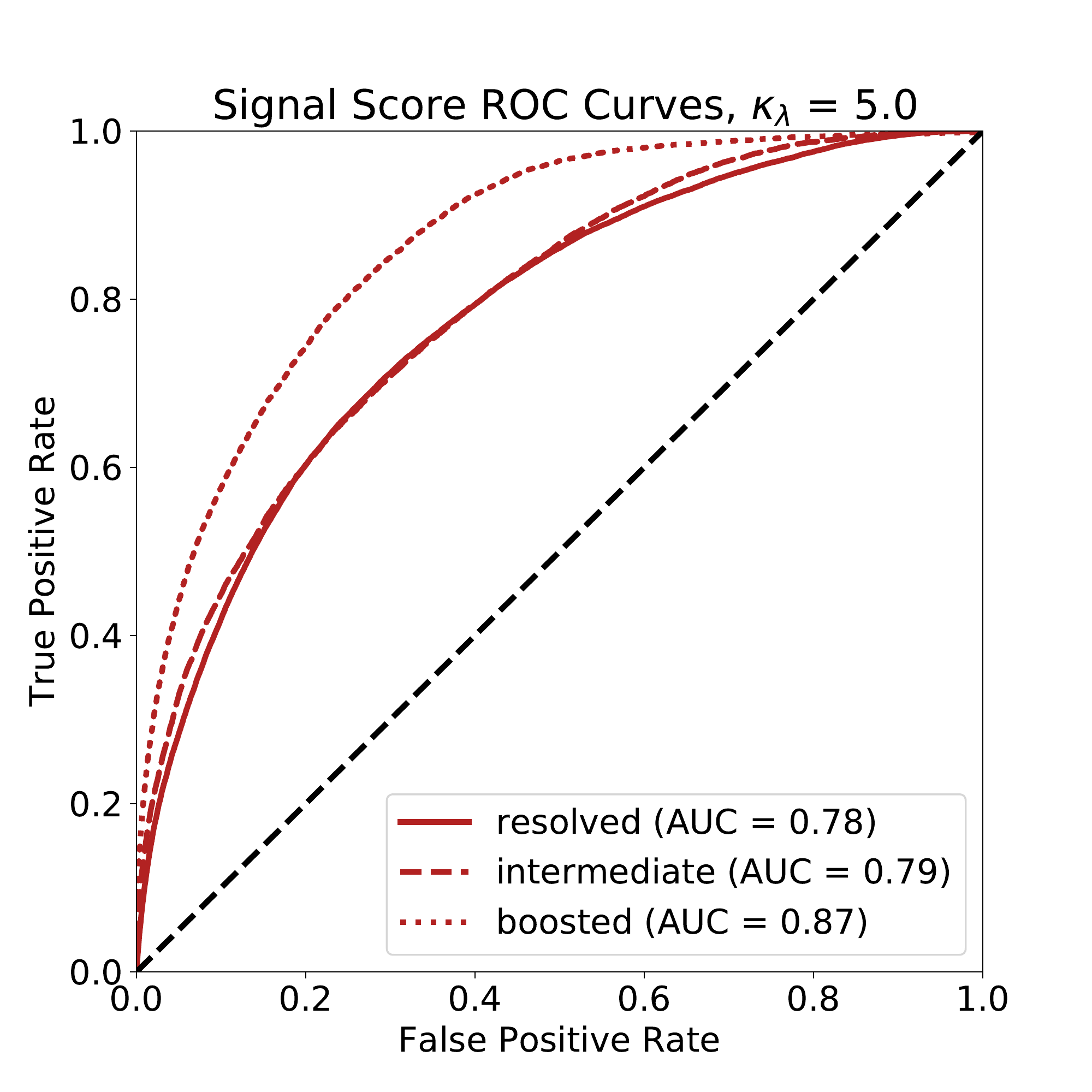}
    \caption{DNN trained on $\klam = 5$}
    \end{subfigure}
    \caption{Receiver operator characteristic (ROC) curves for the DNN trained on (a) $\klam = 1$ and (b) $\klam = 5$. These quantify true positive vs false positive classification rates for the signal events in the resolved (solid), intermediate (dashed), and boosted (dotted) categories. Values for the area under the curve (AUC) are displayed in the legend.
    }
    \label{fig:dnn_roc}
\end{figure}

\begin{figure}
    \centering 
    \begin{subfigure}[b]{0.49\textwidth}
        \includegraphics[width=\textwidth]{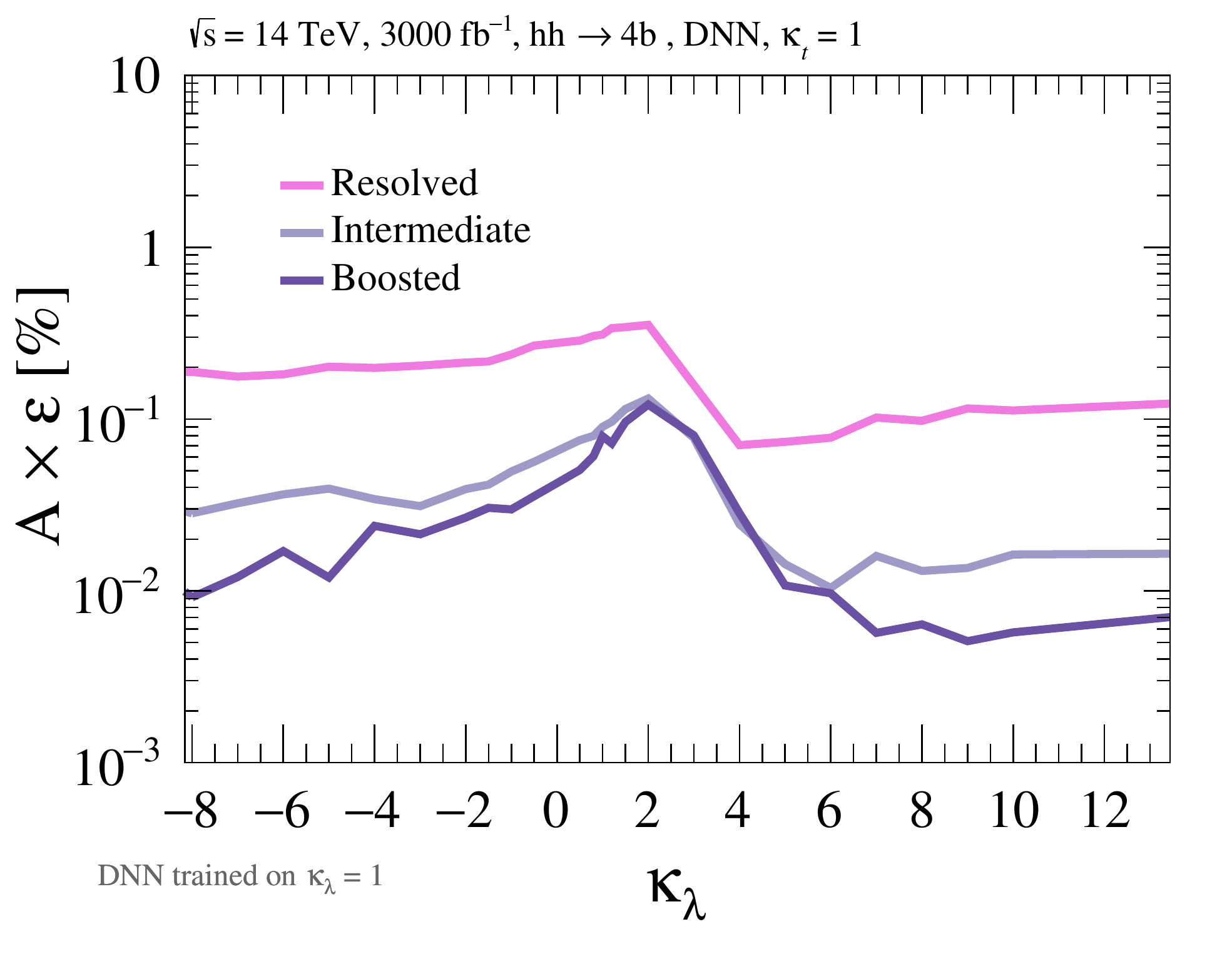}
    \caption{DNN trained on $\klam = 1$}
    \end{subfigure}%
    \begin{subfigure}[b]{0.49\textwidth}
        \includegraphics[width=\textwidth]{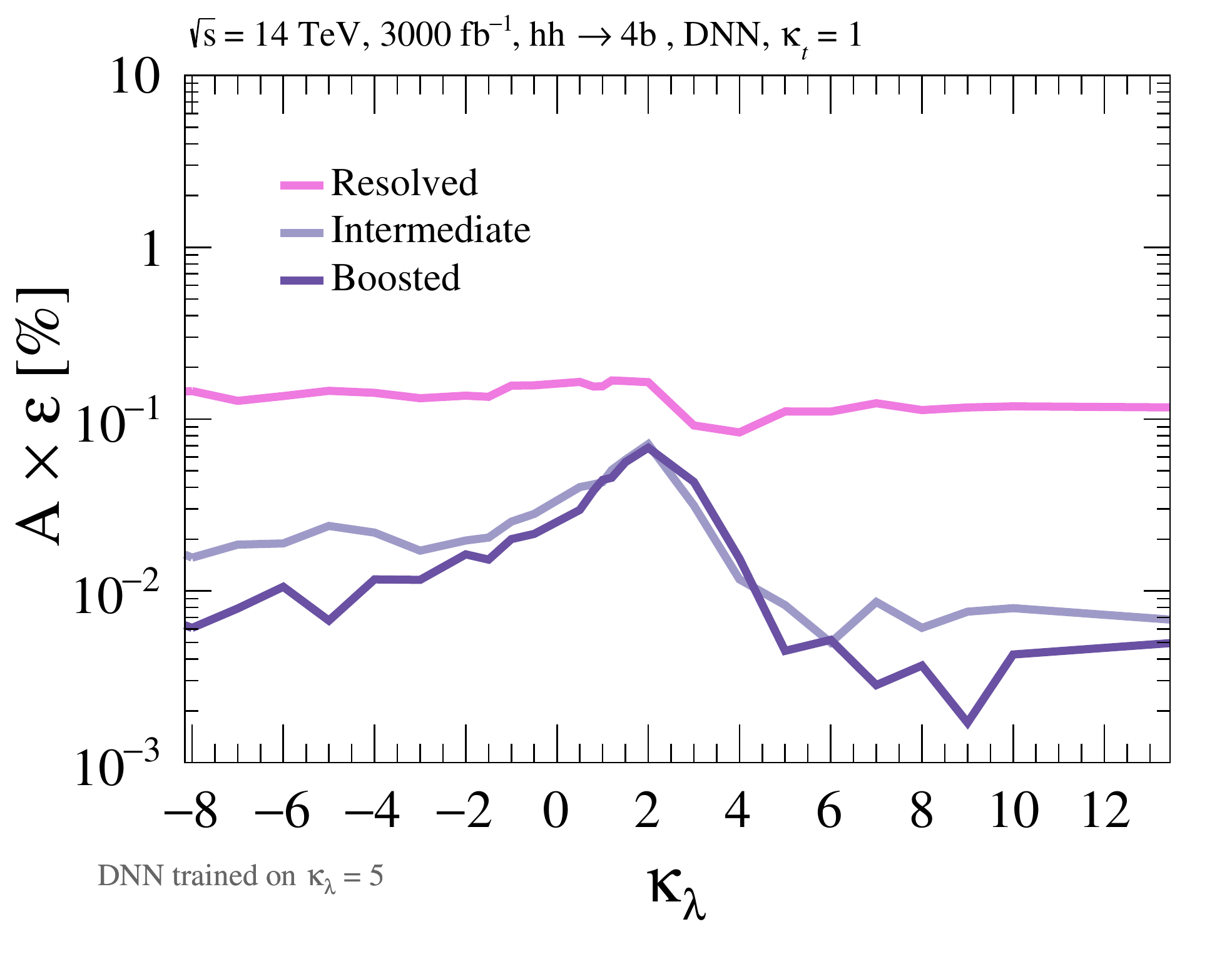}
    \caption{DNN trained on $\klam = 5$}
    \end{subfigure}
    \caption{Signal acceptance times efficiency $A \times \varepsilon$ in percent for the \emph{neural network analysis} trained on (a) $\klam = 1$ and (b) $\klam = 5$ after the signal score requirement $p_\text{signal}^\text{DNN} > 0.75$. This is shown for the resolved (pink), intermediate (lilac) and boosted (purple) channels. The $A \times \varepsilon$ is equivalent to the number of signal events $S$ divided by initial number of events $\sigma \times \mathcal{L}$. }
    \label{fig:dnn_acceptance}
\end{figure}

Figure~\ref{fig:dnn_roc} shows the true positive vs false positive rate known as receiver operator characteristic (ROC) curves. These are computed on the whole test set (i.e.\ the set of events \emph{not} used for training) for these networks in the resolved, intermediate, and boosted categories for signal classification only. In the general case of multi-class classification, the
true positive rate is defined by the fraction of events where an event from sample $X$ is classified correctly as an event from $X$ divided by the total number of events in $X$
\begin{align}
    R_\text{positive}^\text{true} &= \frac{N(\text{$X$ classified as $X$)}}{N(\text{total $X$})}, \quad X \in \{\text{multijet}, t\bar{t}, \text{signal}\}.
\end{align}
The false positive rate is defined by the fraction of events where an event not from sample $X$ is classified as an event from sample $X$ divided by the total number of events not from sample $X$
\begin{align}
    R_\text{positive}^\text{false} &= \frac{N(\text{$Y$ or $Z$ classified as $X$)}}{N(\text{total $Y+Z$})}, \quad X, Y, Z \in \{\text{multijet}, t\bar{t}, \text{signal}\}.
\end{align}
Figures~\ref{fig:dnn_acceptance} also displays the signal acceptance times efficiency $A\times \varepsilon$ as a function of \klam for the different categories in the \emph{neural network analysis}. One can see that the signal acceptance is higher around $\klam\sim 5$ when the DNN is trained on $\klam = 5$.

\begin{figure}
    \centering
    \begin{subfigure}[b]{0.5\textwidth}
       \includegraphics[width=\textwidth]{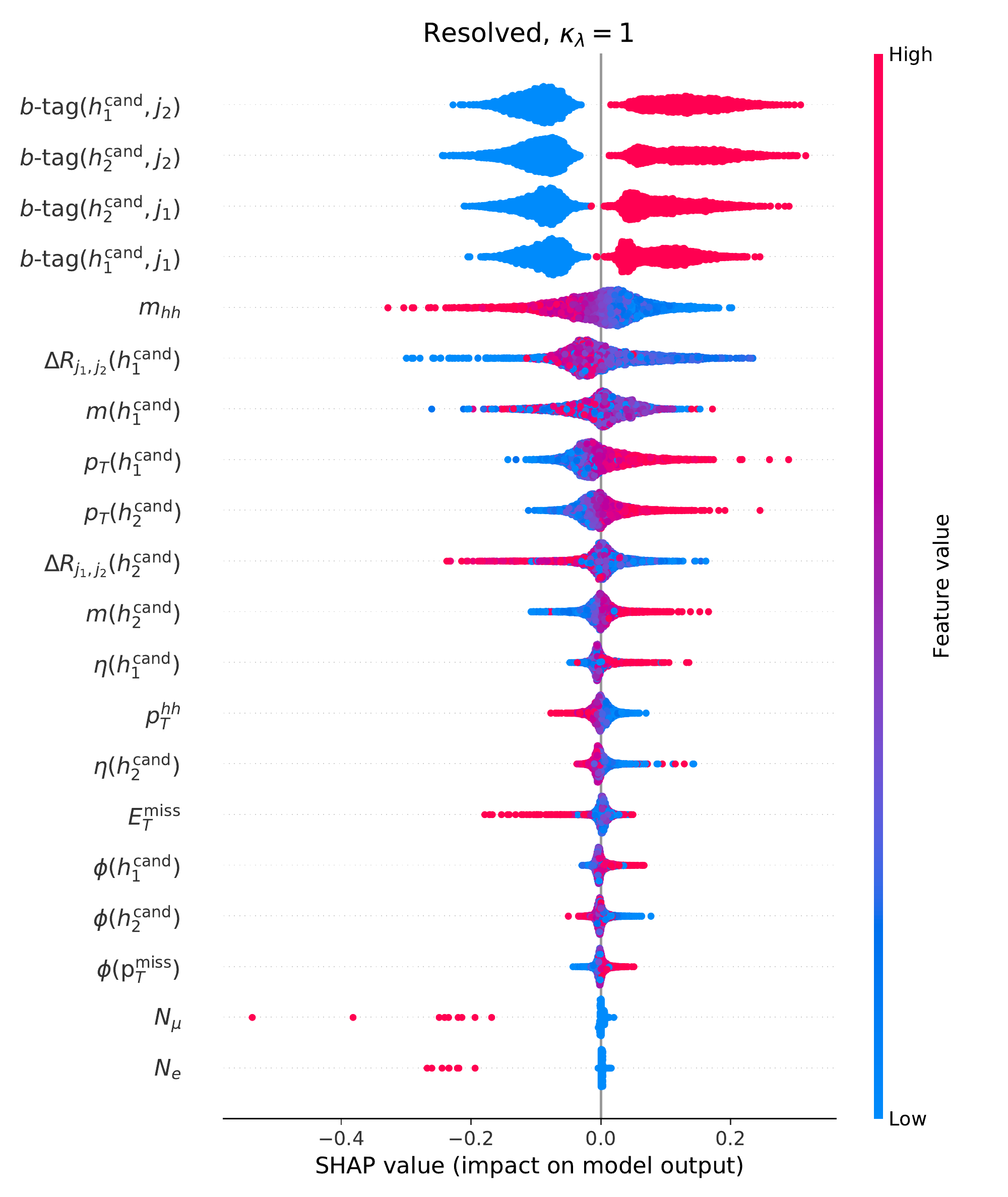}
      \caption{Resolved}
    \end{subfigure}% 
    \begin{subfigure}[b]{0.5\textwidth}
       \includegraphics[width=\textwidth]{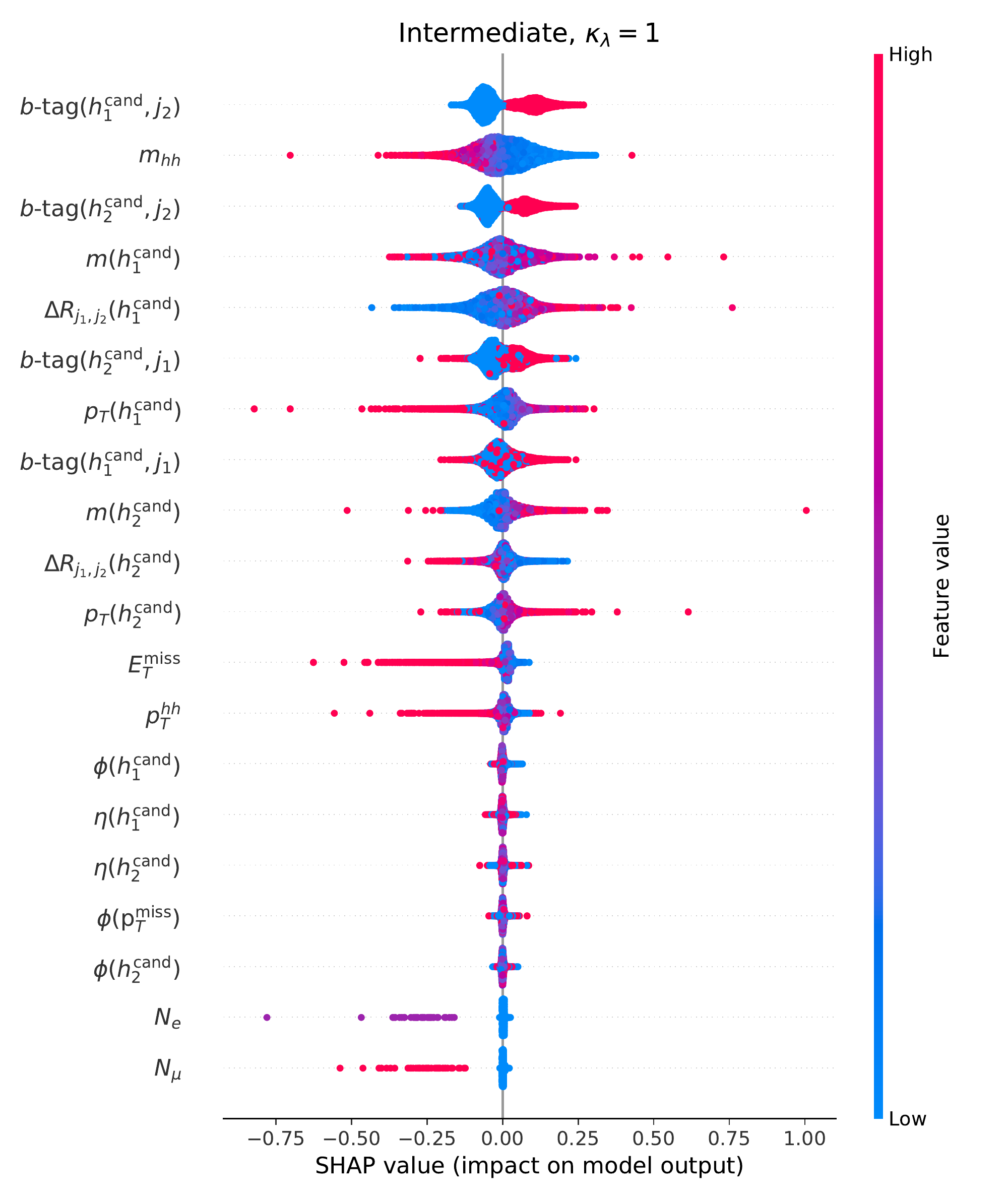}
    \caption{Intermediate}
    \end{subfigure} \\
    \begin{subfigure}[b]{0.5\textwidth}
        \includegraphics[width=\textwidth]{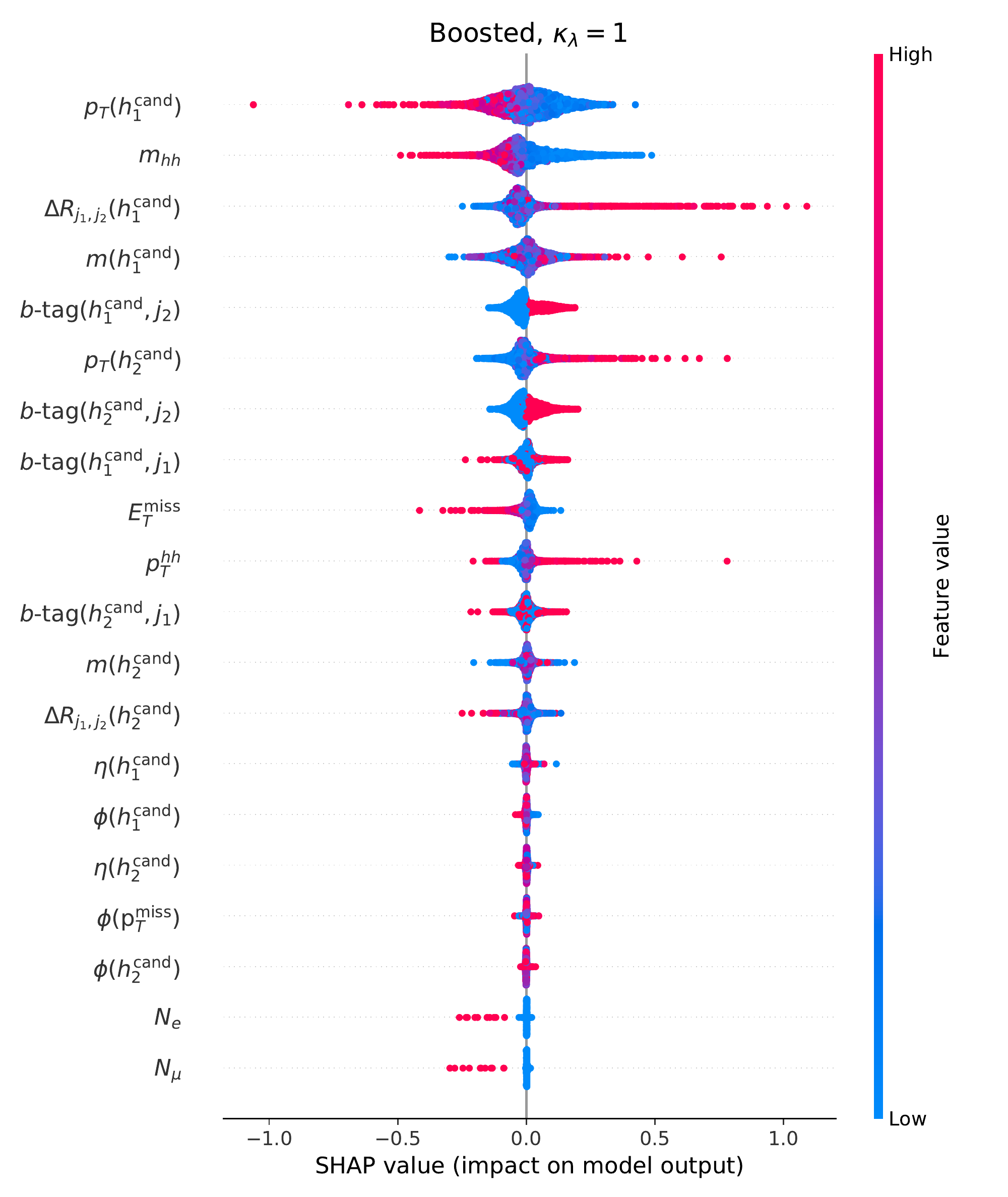}
    \caption{Boosted}
    \end{subfigure}  
    \caption{SHAP value plots representing the ranked variable (feature) importance for the DNN trained on $\klam=1$ signals for the (a) resolved, (b) intermediate and (c) boosted categories. The DNN models are ranked by their average absolute SHAP value, which is plotted on the $x$-axis. The colour scale indicates the value of the feature on the specific event for which the SHAP value is plotted.}
    \label{fig:shap_values_1}
\end{figure}

Finally, Fig.~\ref{fig:shap_values_1} shows the ranked feature importance for the \emph{neural network analysis} trained on $\klam = 1$. Discussion of this figure follows that presented in section~\ref{sec:feature_importance_DNN} of the main text.

%--------------------------
\subsection{\label{sec:shapley_heuristics}Heuristics of Shapley values }

Here we provide a heuristic overview of the SHAP method for model interpretation; for technical details and definitions, see Ref.~\cite{NIPS2017_7062}. The SHAP method is based on Shapley values $\phi$, which were originally introduced in the context of game theory. These calculate the model prediction (neural network score), $f(S \cup x)$, when including a variable $x$ (feature) from a set $S$. The prediction is computed without this variable $f(S)$. From these, the difference, $f(S \cup x) - f(S)$, intuitively characterises the importance of that variable and its impact on the model prediction. Shapley variables satisfy other conditions: a) the total value of $\phi$ is additive in that it equals the sum of the predictions when all the variables are included, b) variables that contribute the same to a prediction have the same to $\phi$, c) if the variable does not contribute any change to the prediction $f$ then it adds nothing to $\phi$. The predictions are computed for each variable in the set to compare their importance systematically. SHAP (SHapley Additive exPlanation) considers Shapley values of a conditional expectation function of the original model. Formal definitions and prescriptions are found in Ref.~\cite{NIPS2017_7062}. 

\FloatBarrier
\subsection{Network correlation plots}

\begin{figure}
    \centering
    \begin{subfigure}[b]{0.5\textwidth}
      \includegraphics[width=\textwidth]{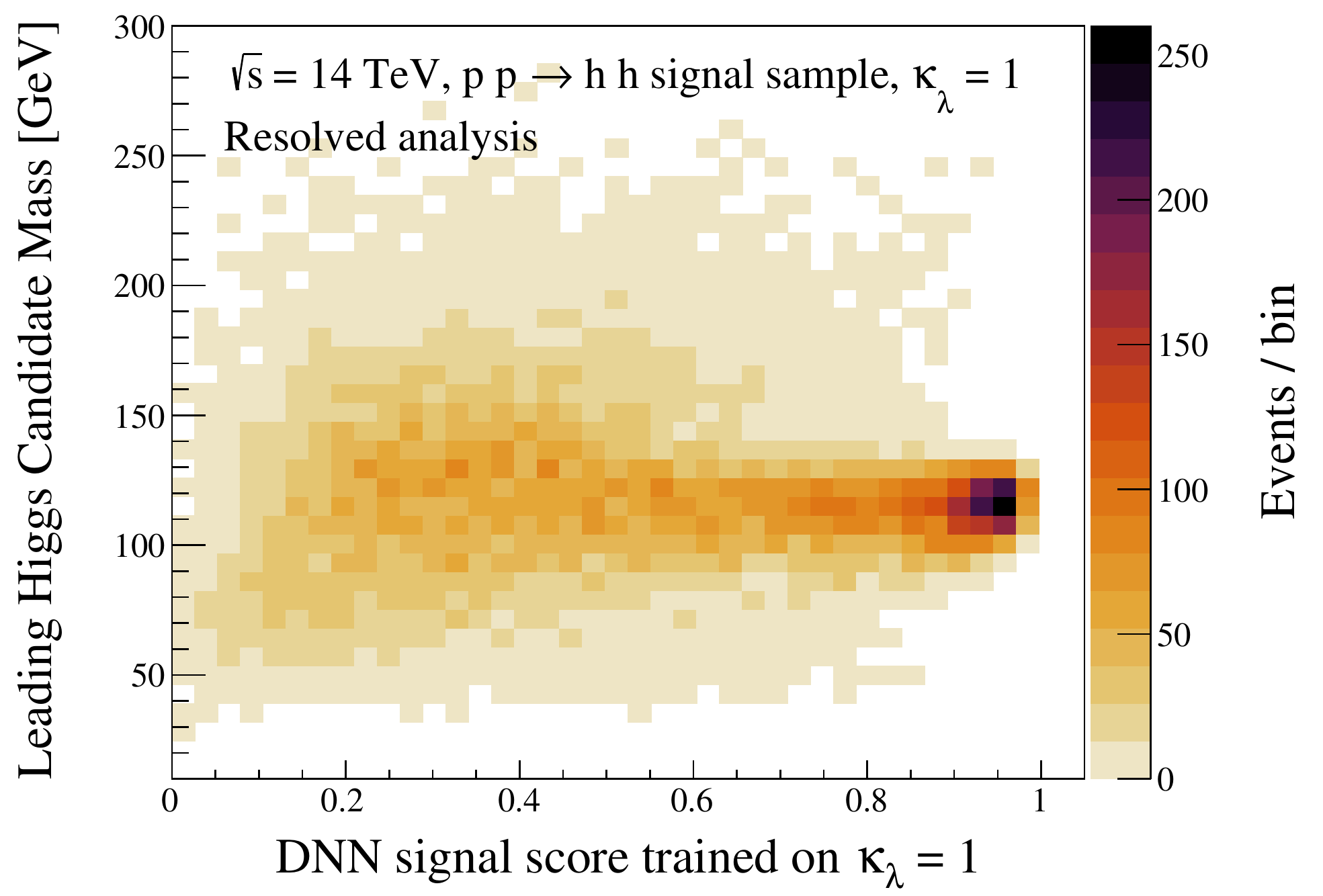}\\
      \includegraphics[width=\textwidth]{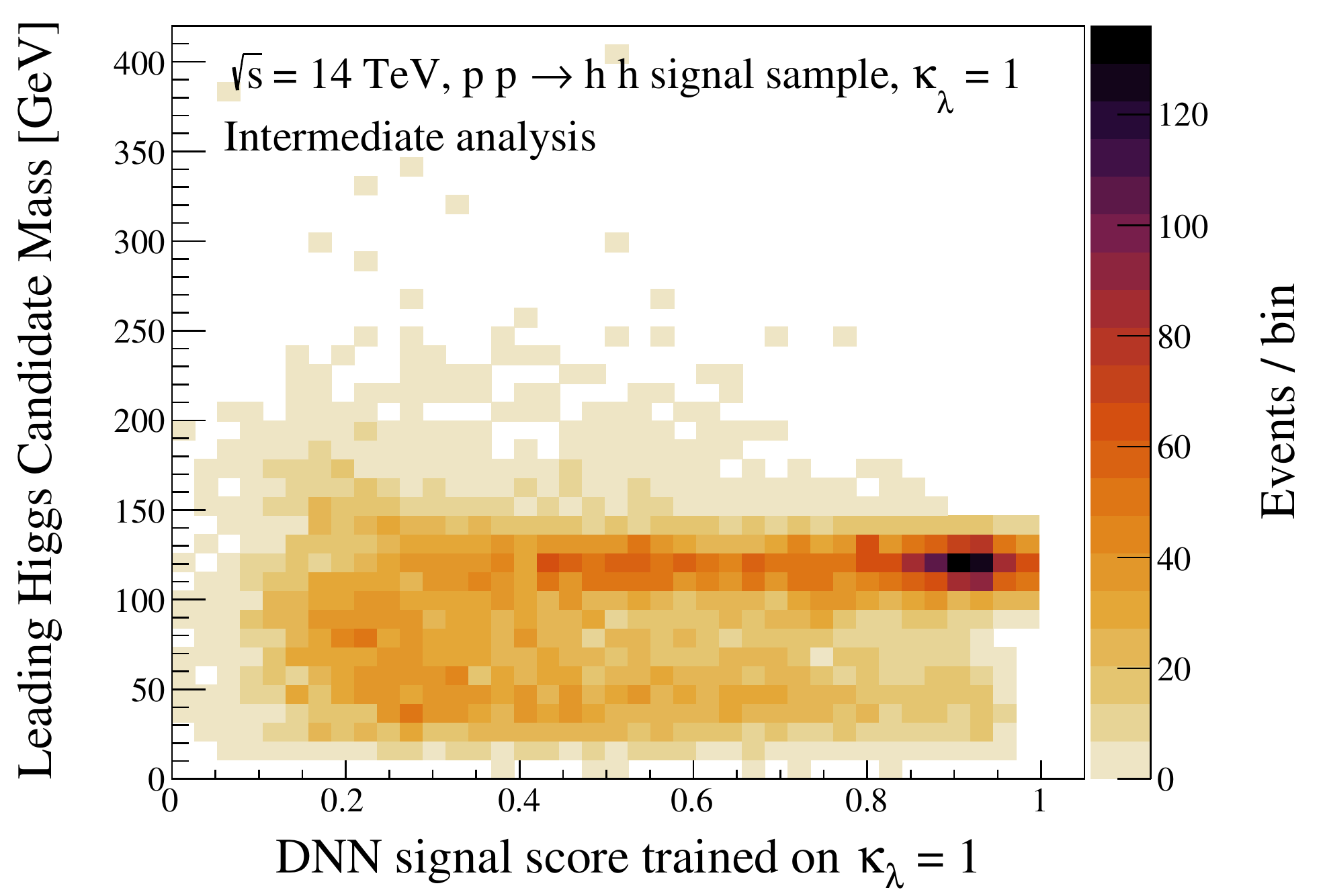}\\
      \includegraphics[width=\textwidth]{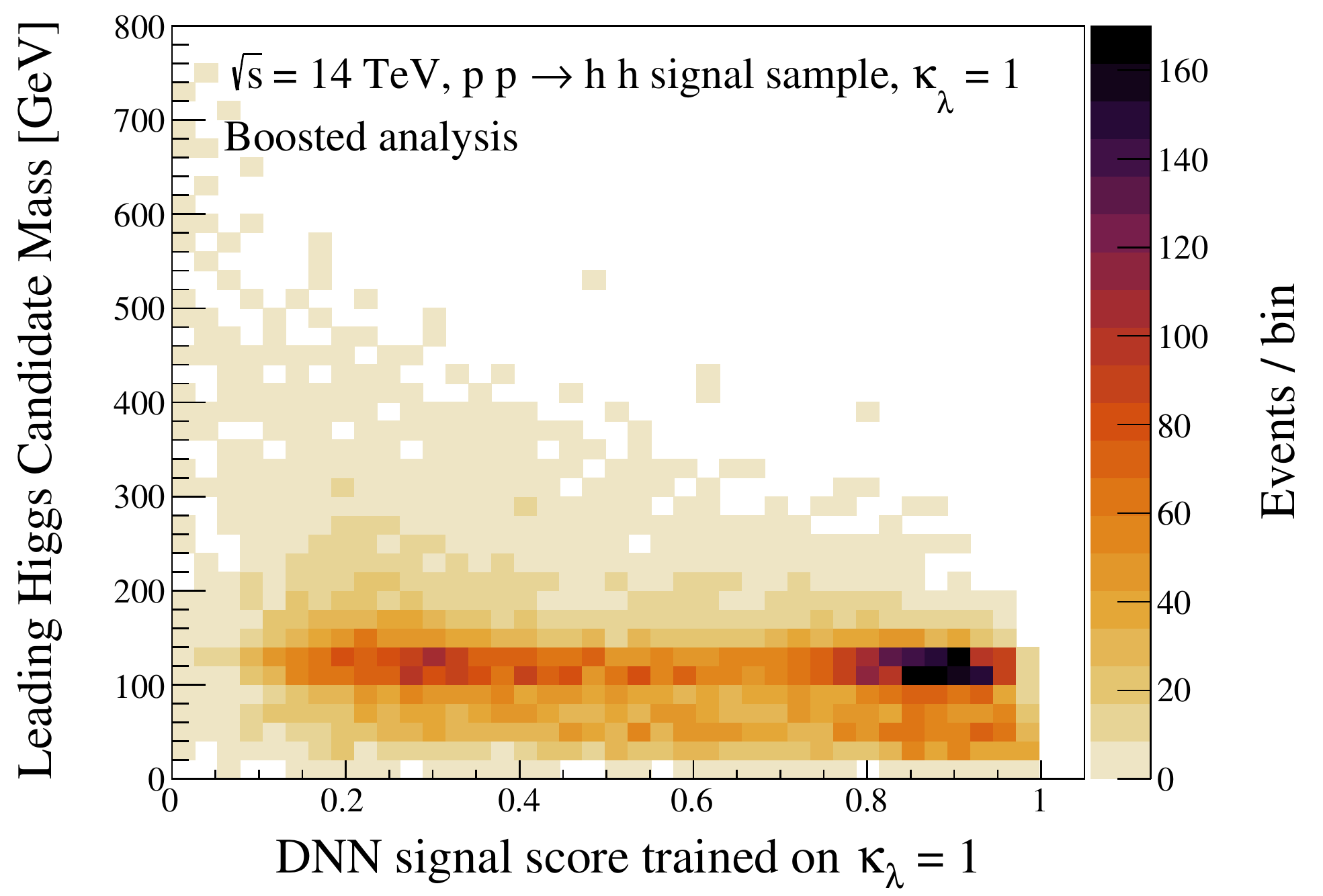} 
    \caption{DNN trained on $\klam = 1$}
    \end{subfigure}%
    \begin{subfigure}[b]{0.5\textwidth}
      \includegraphics[width=\textwidth]{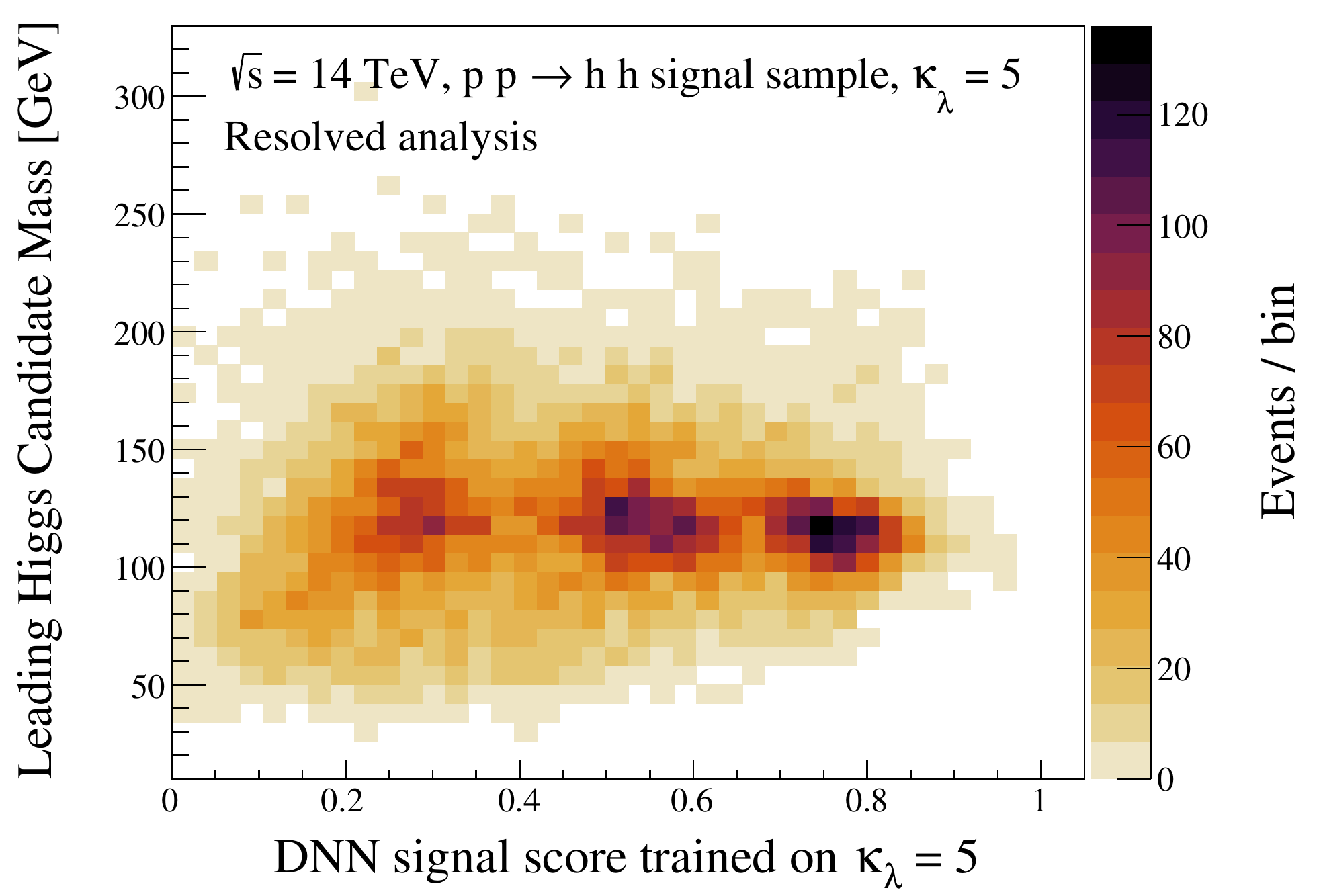}\\
      \includegraphics[width=\textwidth]{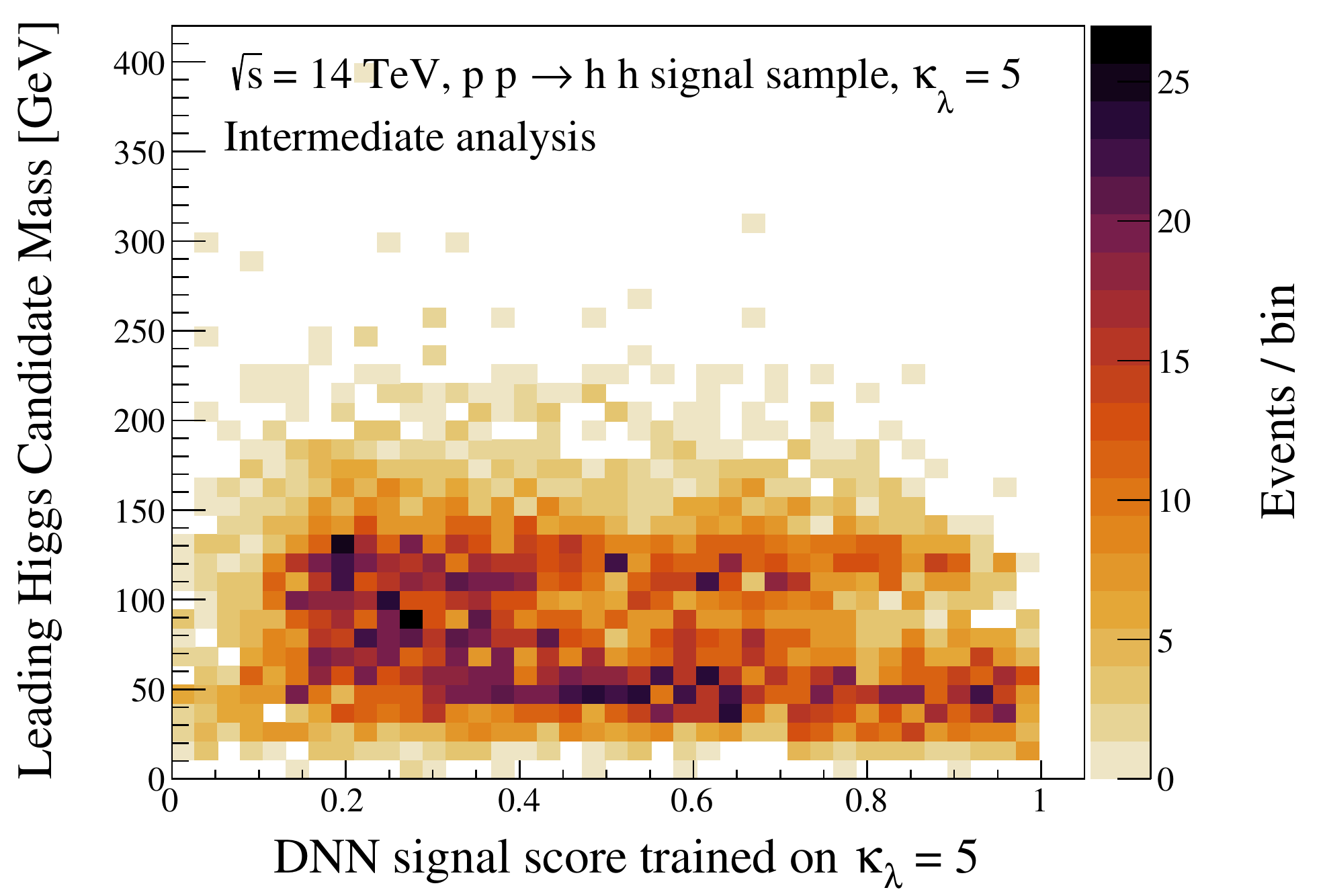}\\
      \includegraphics[width=\textwidth]{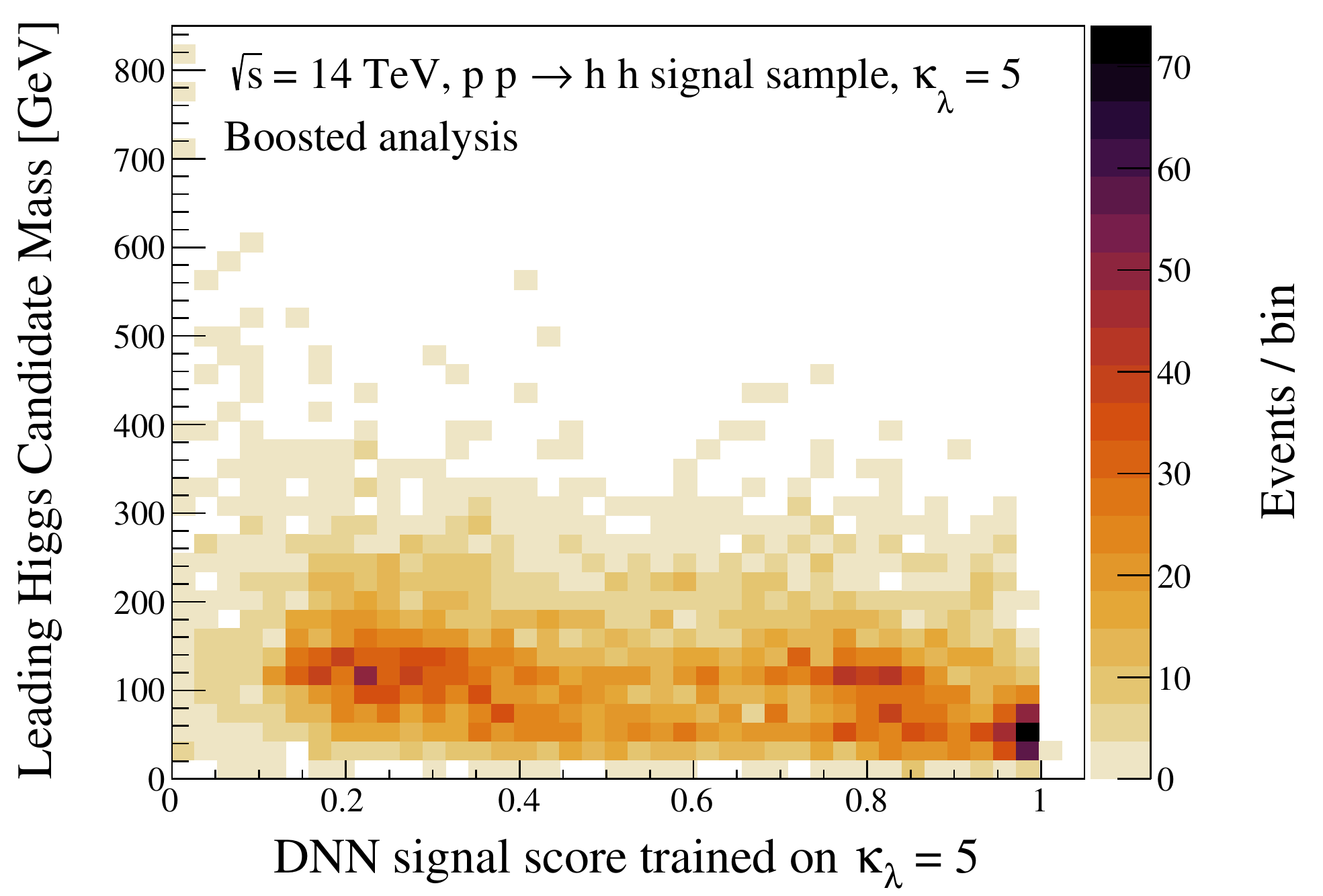}
      \caption{DNN trained on $\klam = 5$}
    \end{subfigure}
    \caption{The leading Higgs candidate mass vs neural network scores trained on (a) $\klam=1$ and (b) $\klam=5$ signal sample for the (upper) resolved, (middle) intermediate, and (lower) boosted analyses. The test samples used to make these distributions are an independent set of (a) $\klam=1$ and (b) $\klam=5$ signal events.
    }
    \label{fig:nnscore_correlation_signal}
\end{figure}

We validate the architecture and training of the \emph{neural network analysis} has the expected behaviour by testing their performance on a statistically independent set of signal and background events. Figure~\ref{fig:nnscore_correlation_signal} show the leading Higgs candidate mass as a function of the signal score trained on $\klam = 1$ and $\klam = 5$ signals when tested on the corresponding $pp\to hh$ signal that exclusively decays to $b$-quarks. For the resolved category, a prominent peak appears at high signal score trained on $\klam =1$ and around $m(h_1^\text{cand}) \sim 125$~GeV. There is a tail of events at lower signal score where it is difficult to kinematically distinguish the signal from background. 

An interesting feature is observed in Fig.~\ref{fig:nnscore_correlation_signal} depicting the $\klam=5$ DNN score computed in the $\klam=5$ signal sample versus the leading Higgs candidate mass. Three distinct peaks can be seen. We find each peak corresponds to events with different numbers of $b$-tags. The most prominent peak with the highest signal score corresponds to events with three or four $b$-tags. Meanwhile, the two peaks with signal scores around 0.3 and 0.5 correspond to events with one and two $b$-tags respectively. 

\begin{figure}
    \centering
    \begin{subfigure}[b]{0.5\textwidth}
        \includegraphics[width=\textwidth]{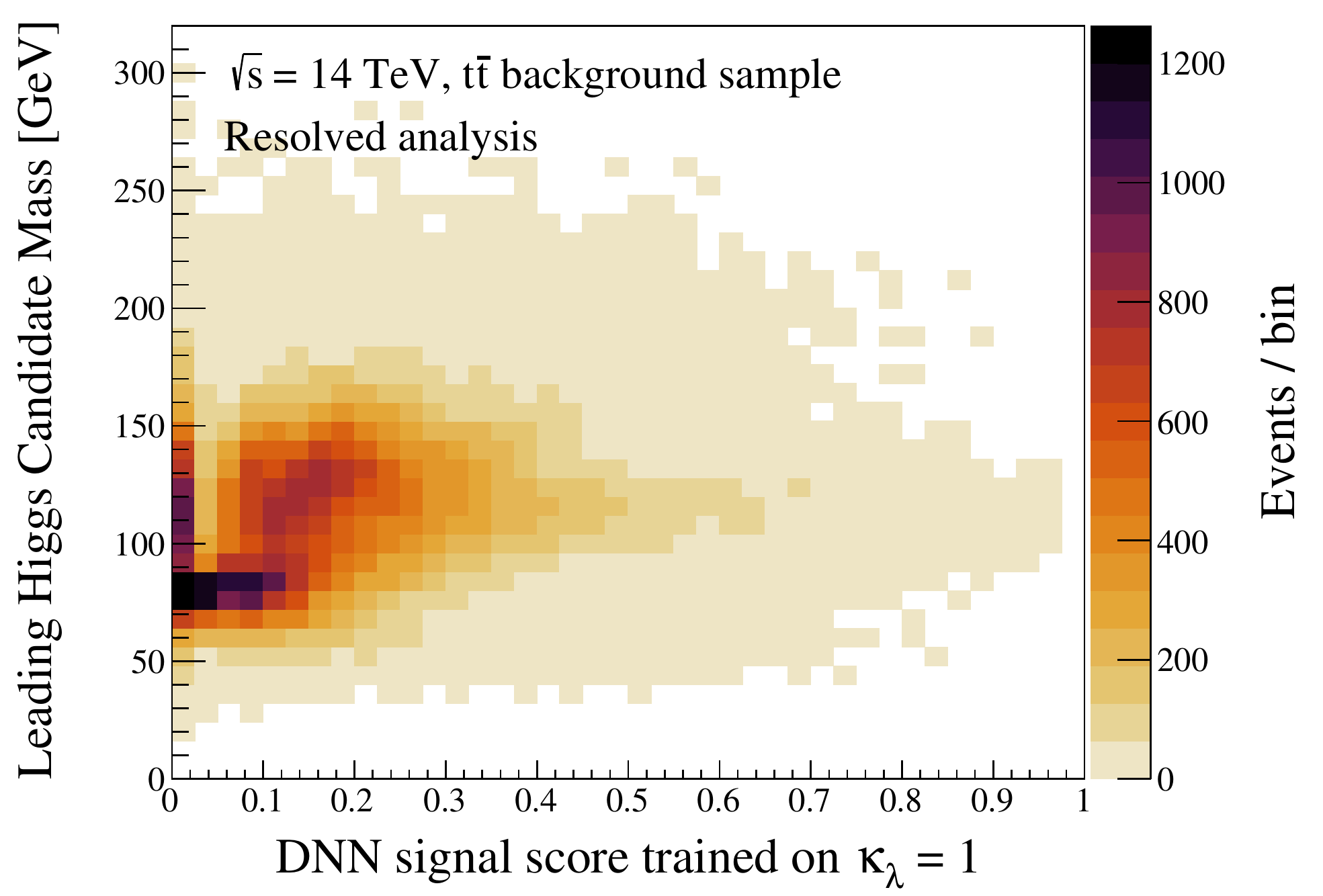}\\
        \includegraphics[width=\textwidth]{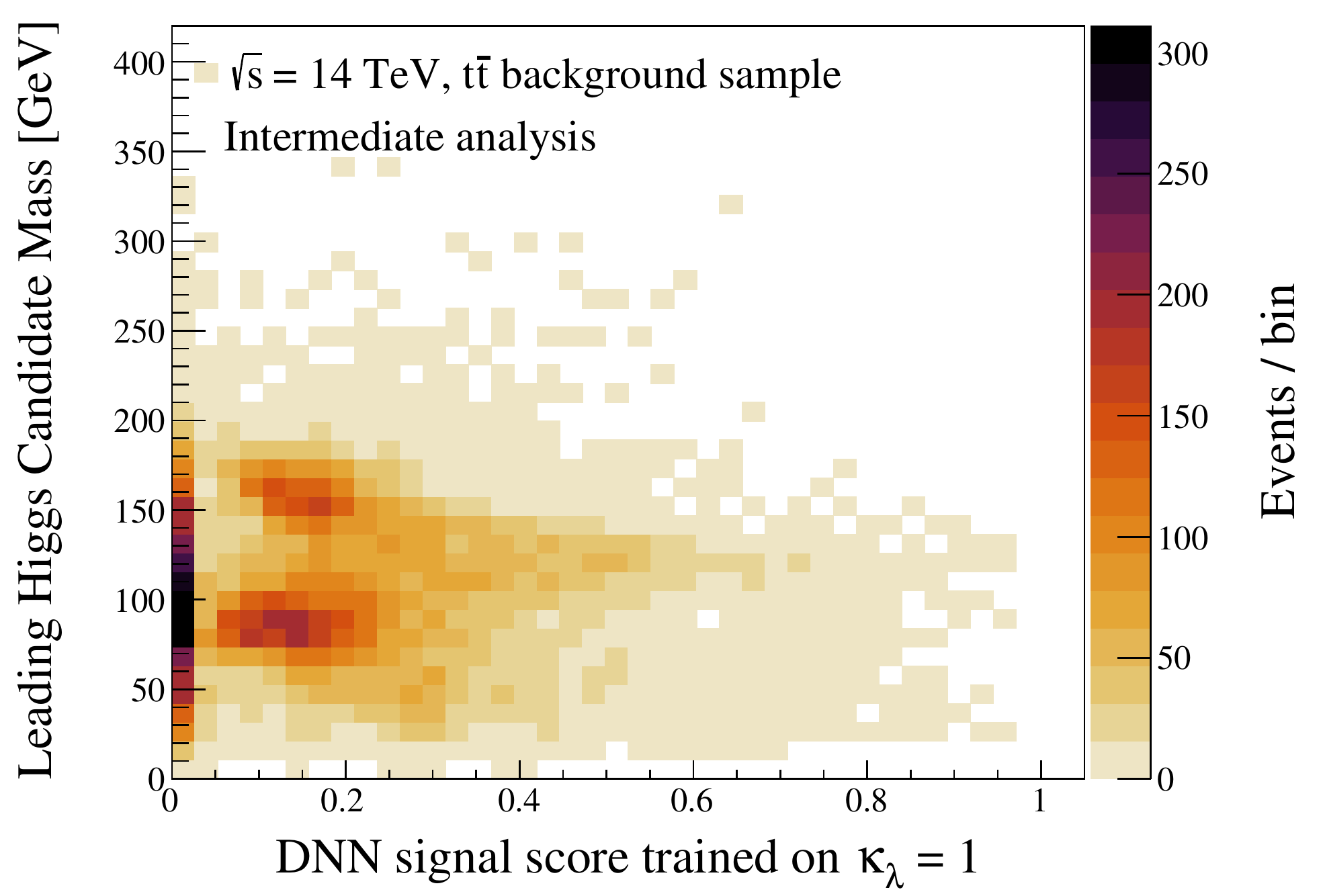}\\
        \includegraphics[width=\textwidth]{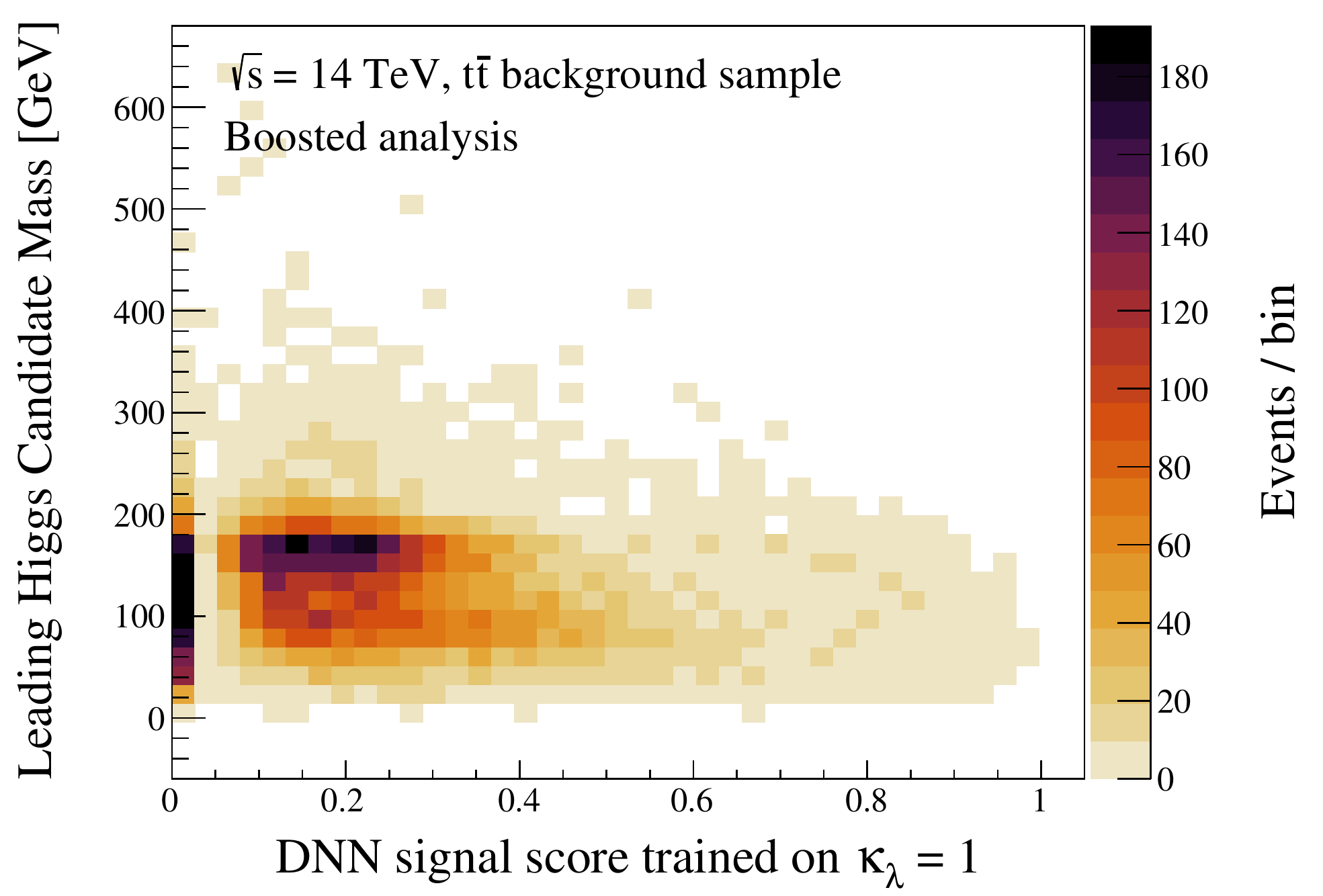}
    \caption{DNN trained on $\klam = 1$}
    \end{subfigure}%
    \begin{subfigure}[b]{0.5\textwidth}
       \includegraphics[width=\textwidth]{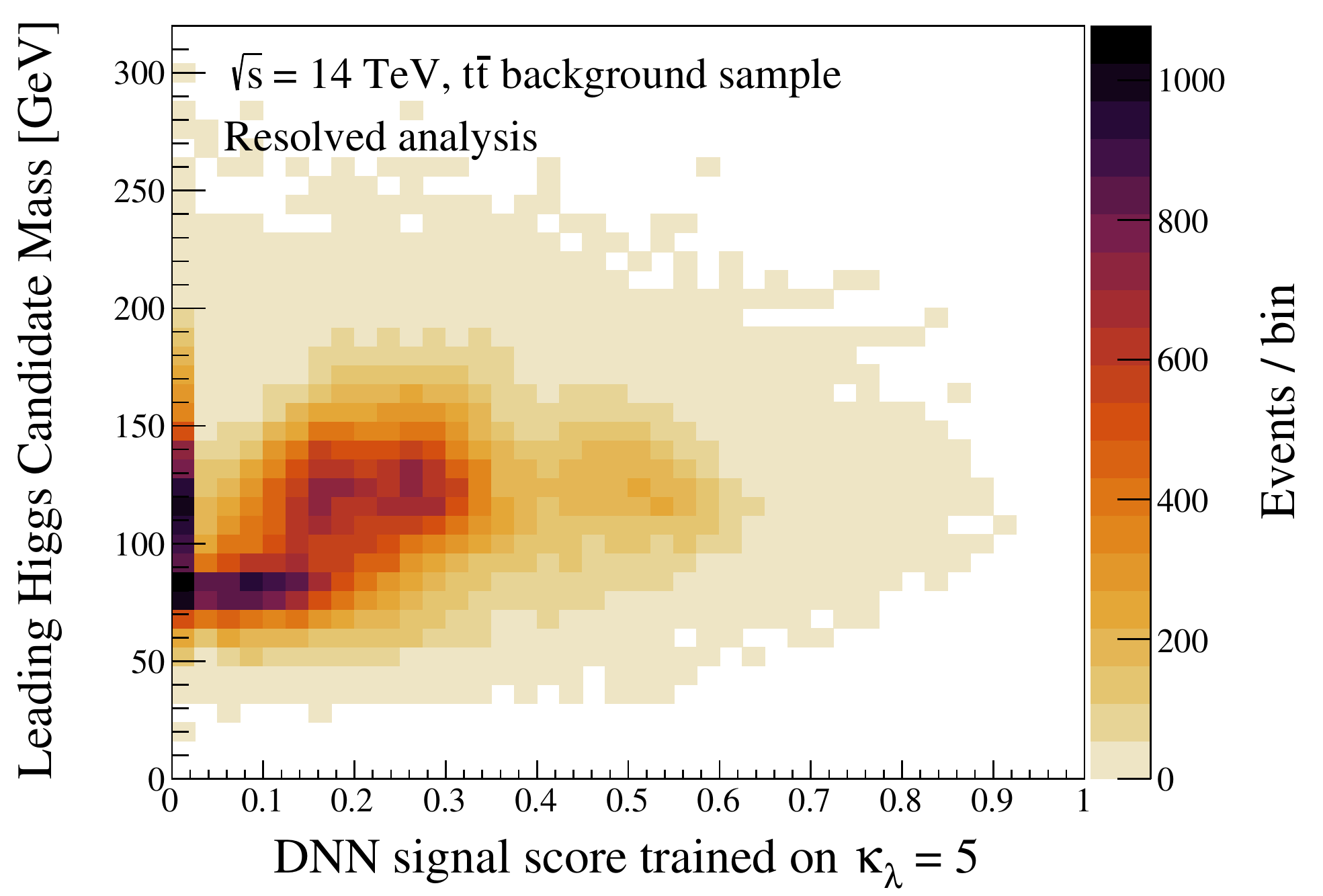}\\
       \includegraphics[width=\textwidth]{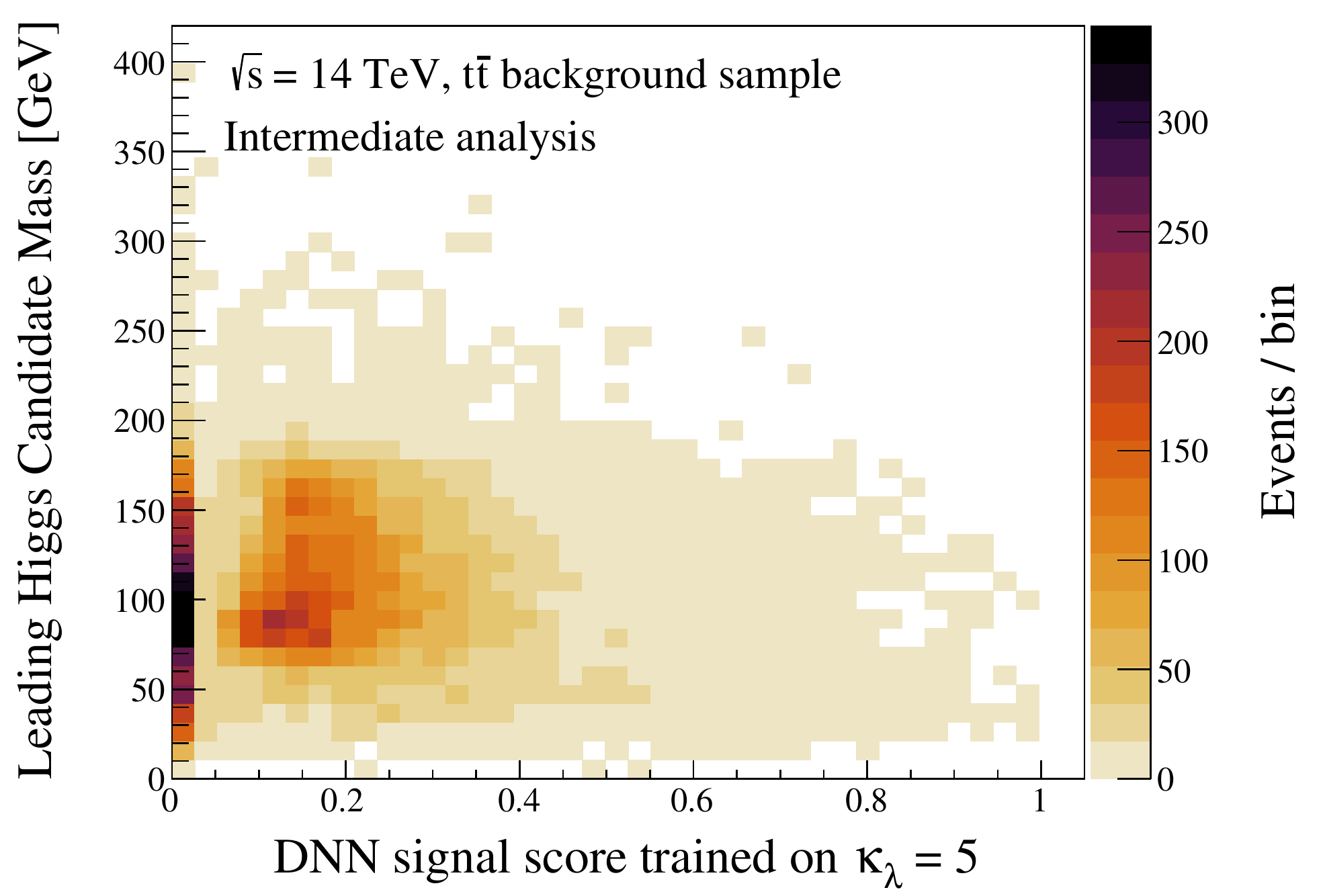}\\
       \includegraphics[width=\textwidth]{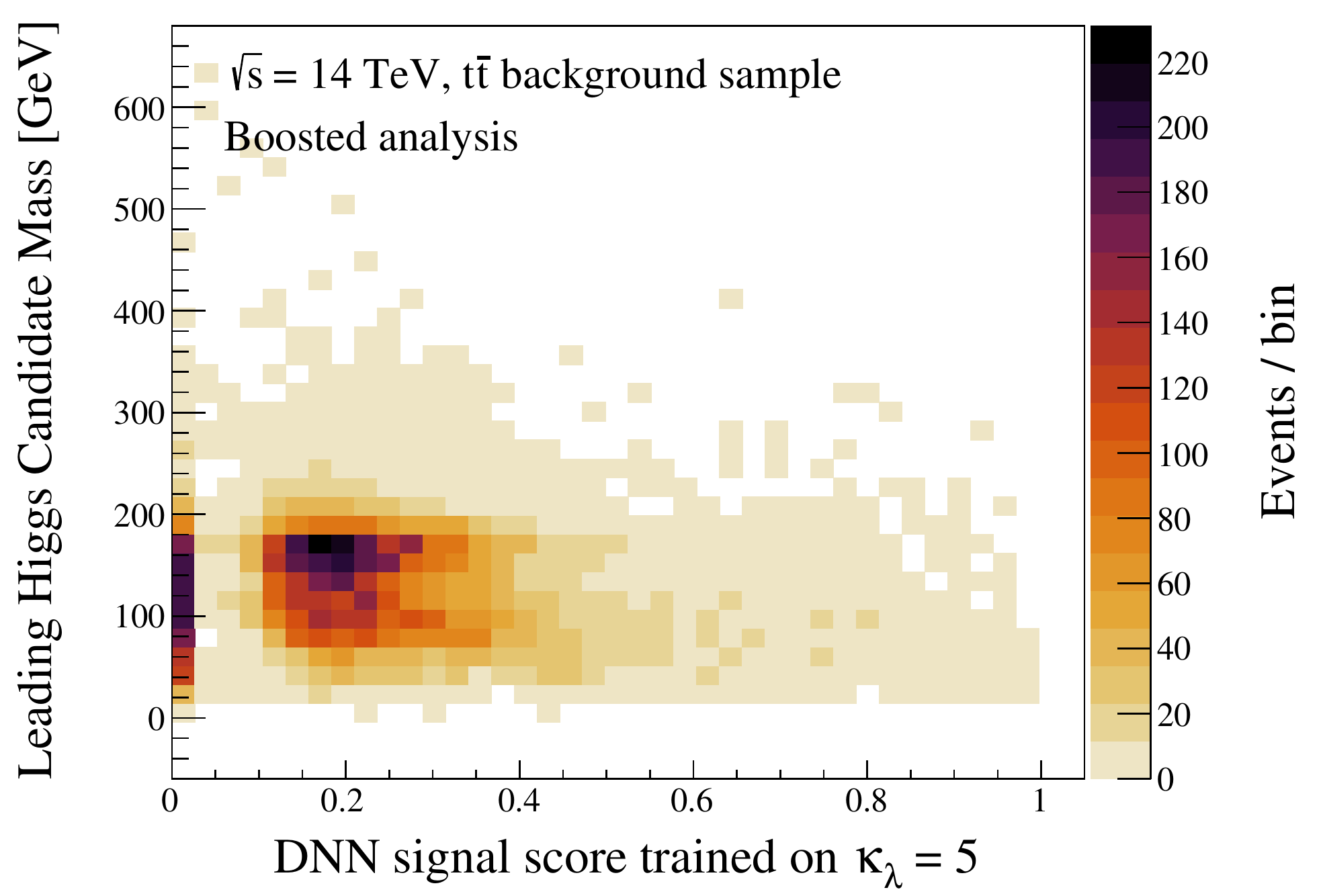}
      \caption{DNN trained on $\klam = 5$}
    \end{subfigure}
    \caption{The leading Higgs candidate mass vs neural network scores trained on (a) $\klam=1$ and (b) $\klam=5$ signal sample for the (upper) resolved, (middle) intermediate, and (lower) boosted analyses. The test samples used to make these distributions are an independent set of $t\bar{t}$ events.
    }
    \label{fig:nnscore_correlation_ttbar}
\end{figure}

Figures~\ref{fig:nnscore_correlation_ttbar} show similar correlation plots, now testing on a $t\bar{t}$ sample. We see low values of signal scores, indicating that the networks are effective at classifying this background. More structure is seen in these two-dimensional plots due to the inclusive decay of the $b$-quarks in this sample. Several of these plots feature two distinct peaks, corresponding to semi-leptonic and hadronic $b$-decays. Hadronic decays correspond to the peak with higher signal score, since these are more similar to the $hh\to 4b$ signal. Furthermore, the plots for the intermediate and boosted categories feature two distinct peaks in the leading Higgs mass around $\sim 80$~GeV and $\sim 170$~GeV. This suggests that the large jet is  capturing the boosted decay products of the $W$ boson and top quarks. 

\begin{figure}
    \centering
    \begin{subfigure}[b]{0.5\textwidth}
      \includegraphics[width=\textwidth]{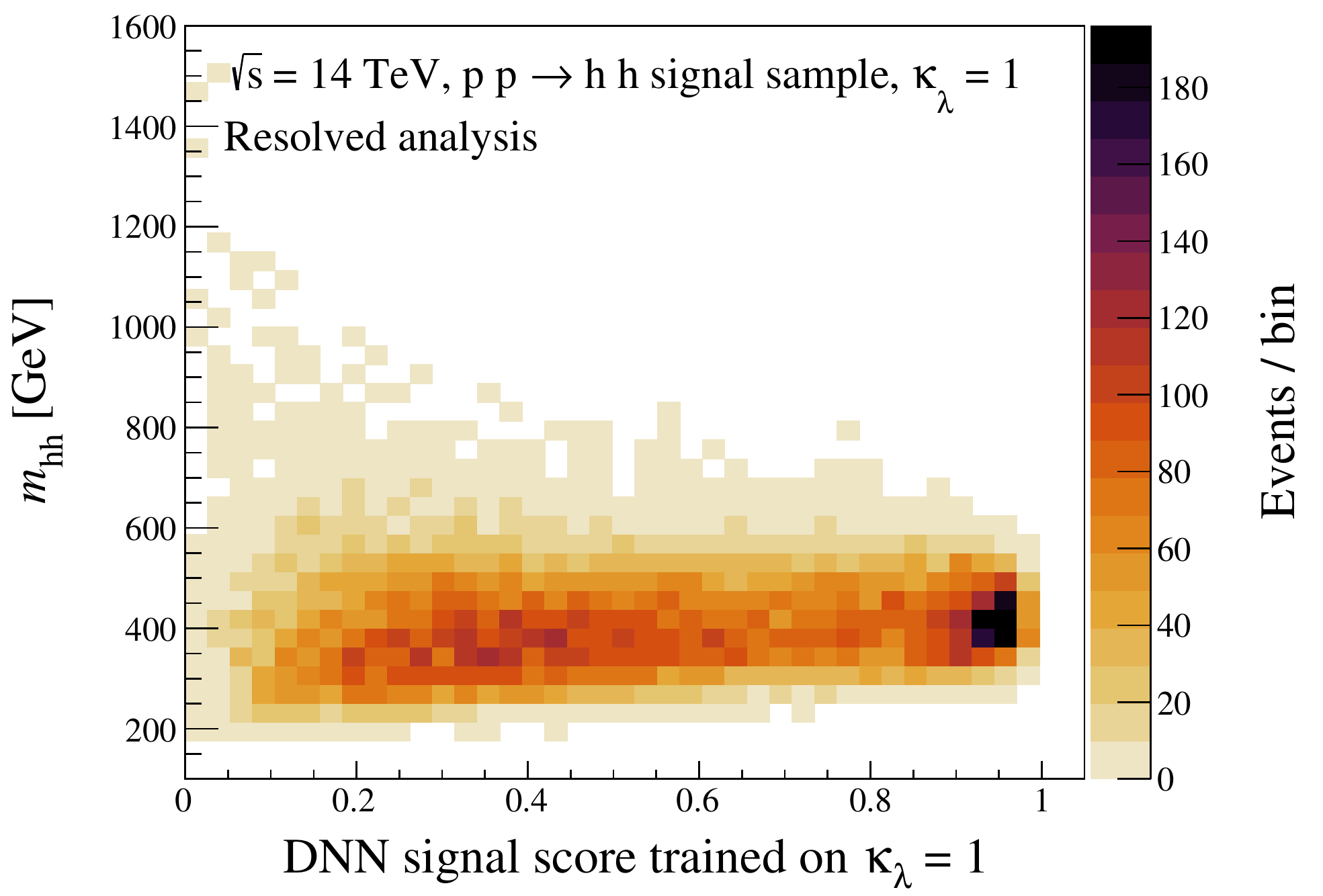}\\
      \includegraphics[width=\textwidth]{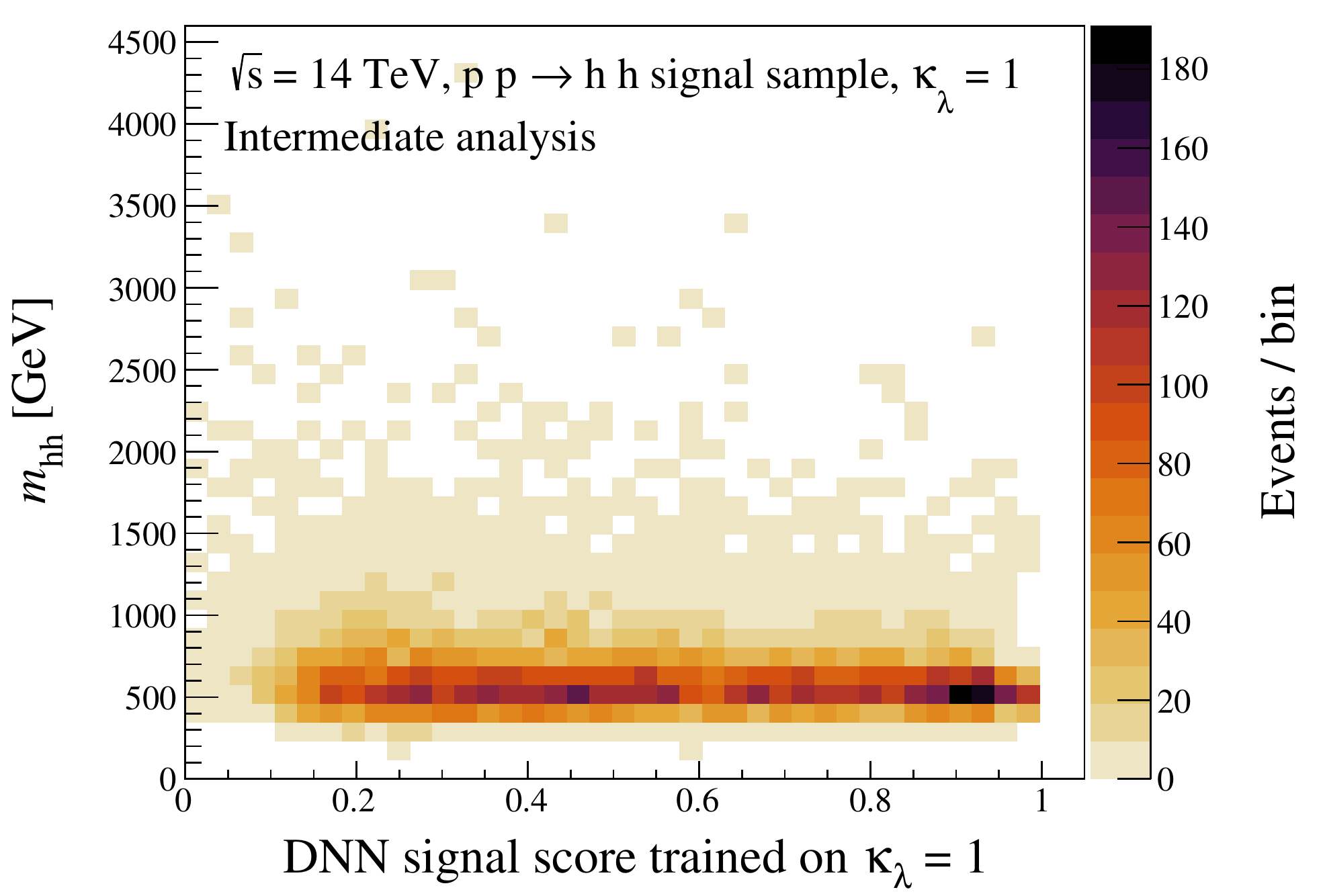}\\
      \includegraphics[width=\textwidth]{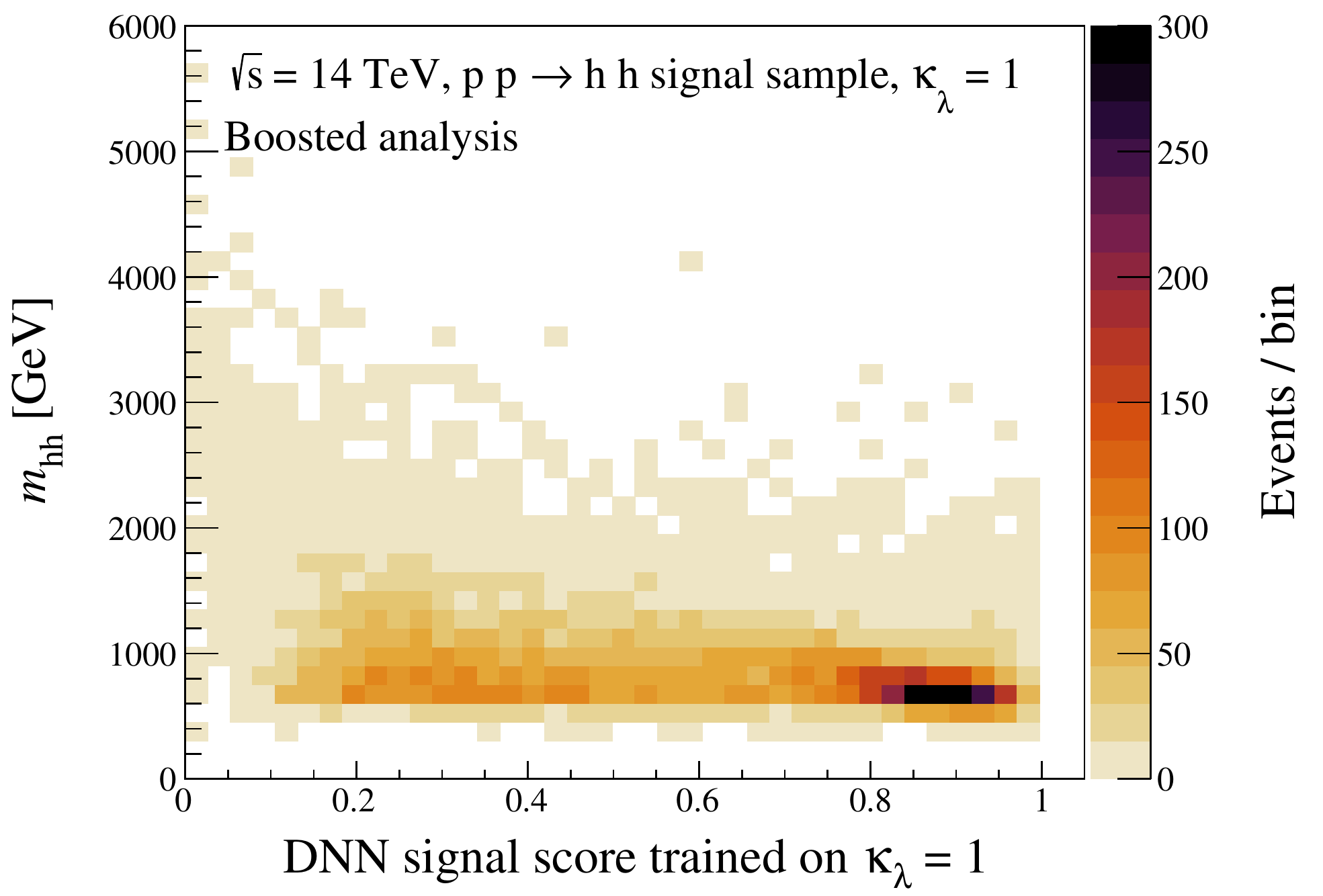} 
    \caption{DNN trained on $\klam = 1$}
    \end{subfigure}%
    \begin{subfigure}[b]{0.5\textwidth}
      \includegraphics[width=\textwidth]{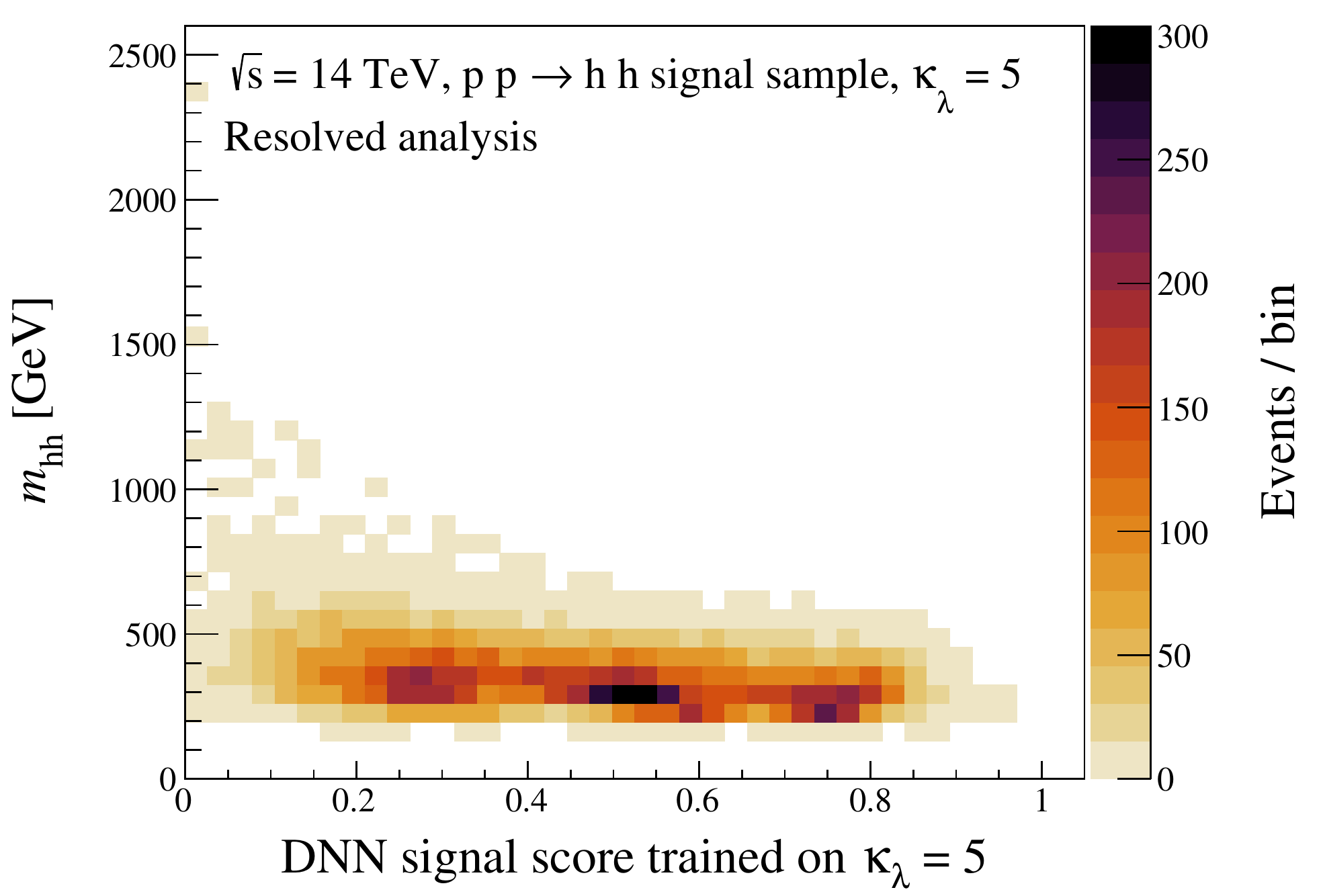}\\
      \includegraphics[width=\textwidth]{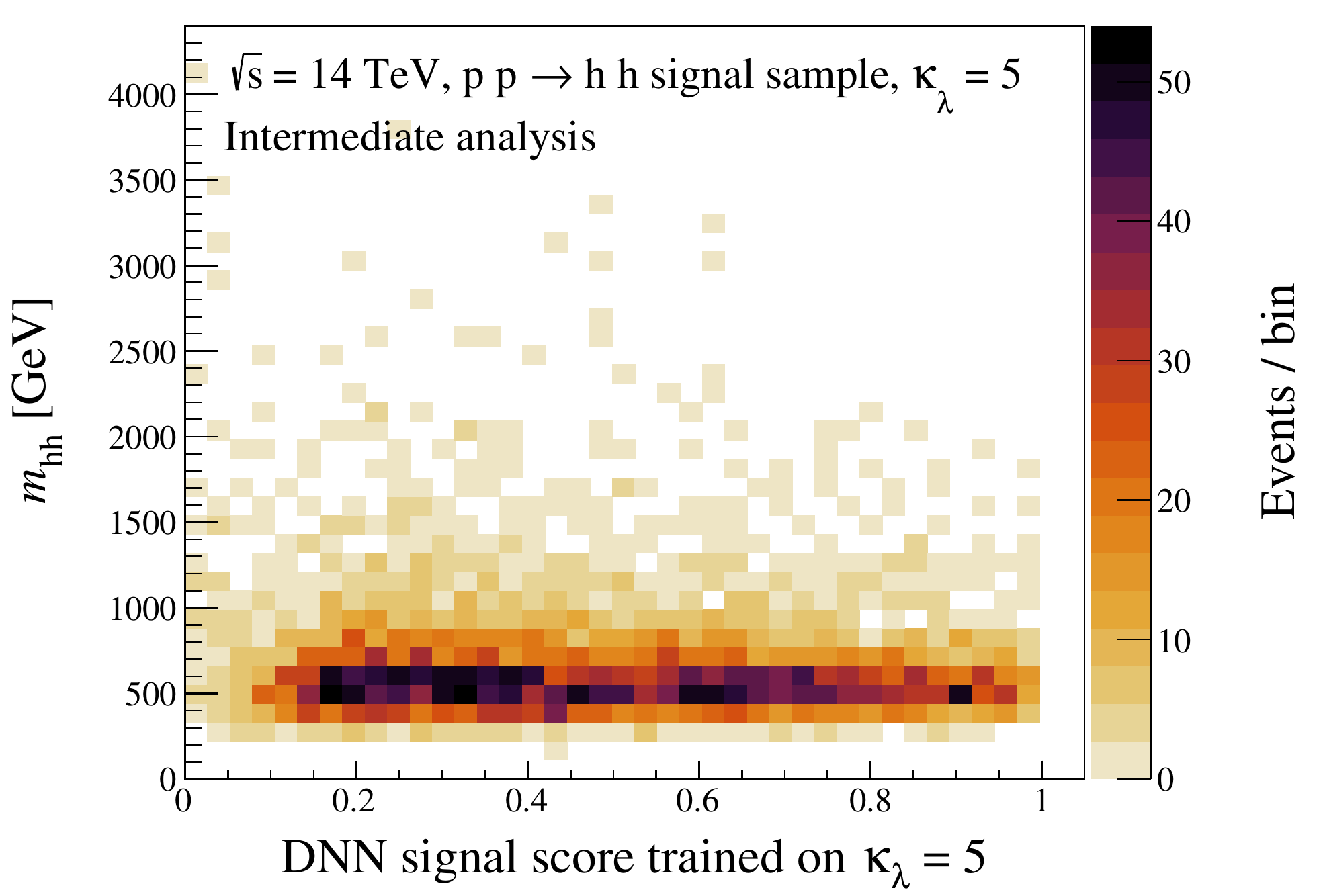}\\
      \includegraphics[width=\textwidth]{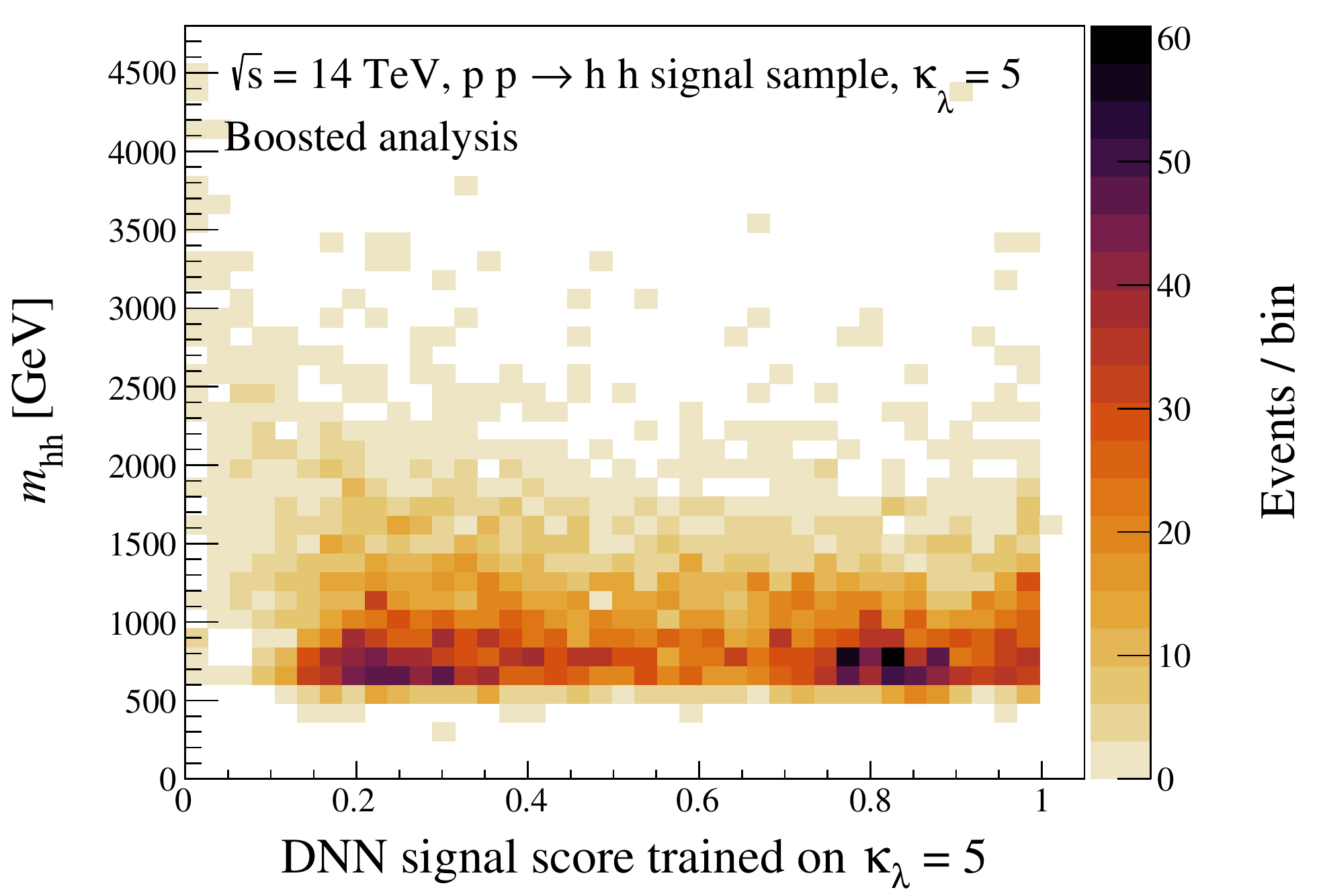}
      \caption{DNN trained on $\klam = 5$}
    \end{subfigure}
    \caption{The di-Higgs invariant mass $\mhh$ vs neural network scores trained on (a) $\klam=1$ and (b) $\klam=5$ signal sample for the (upper) resolved, (middle) intermediate, and (lower) boosted analyses. The test samples used to make these distributions are an independent set of (a) $\klam=1$ and (b) $\klam=5$ signal events.
    }
    \label{fig:nnscore_correlation_signal_mhh}
\end{figure}

\begin{figure}
    \centering
    \begin{subfigure}[b]{0.5\textwidth}
        \includegraphics[width=\textwidth]{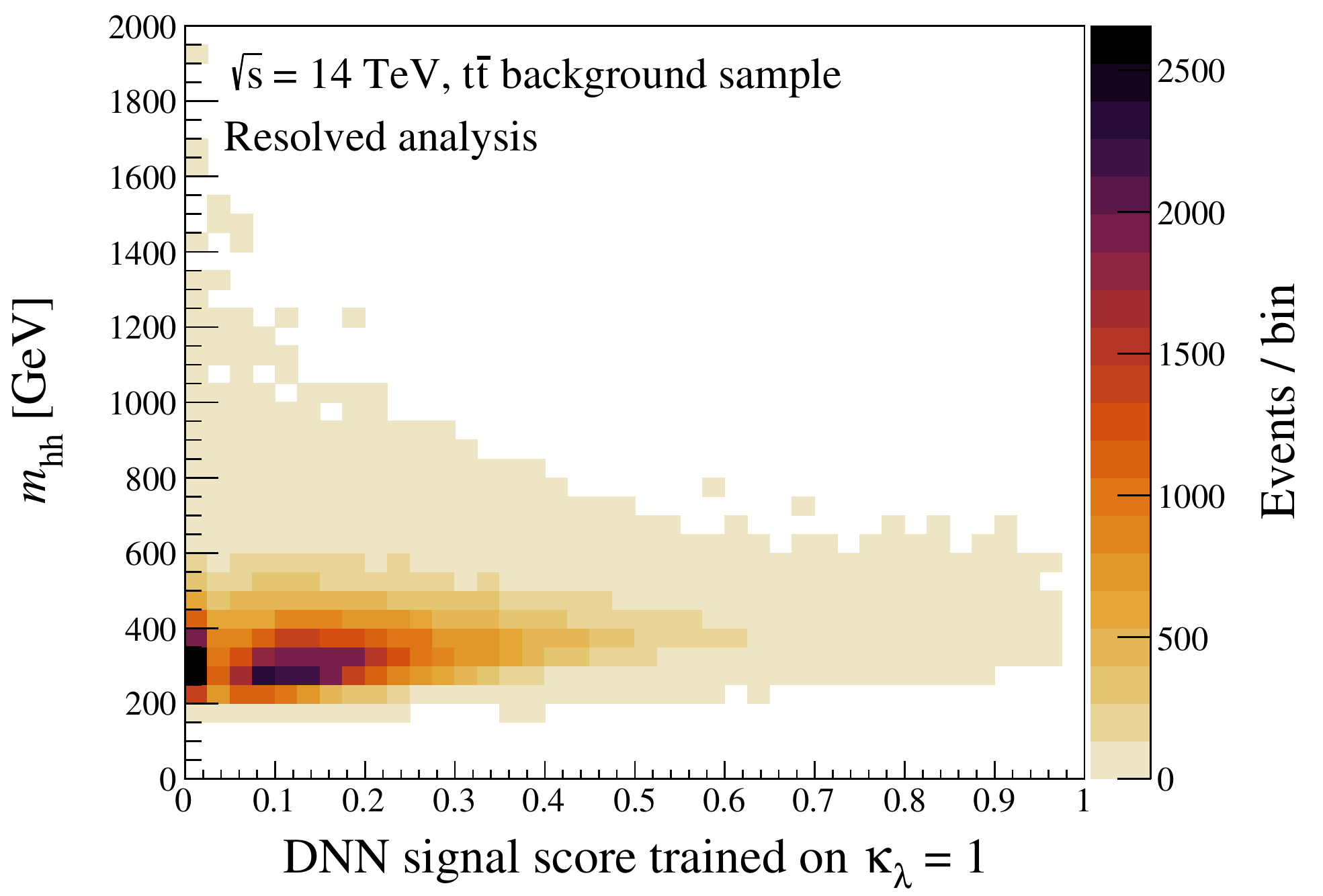}\\
        \includegraphics[width=\textwidth]{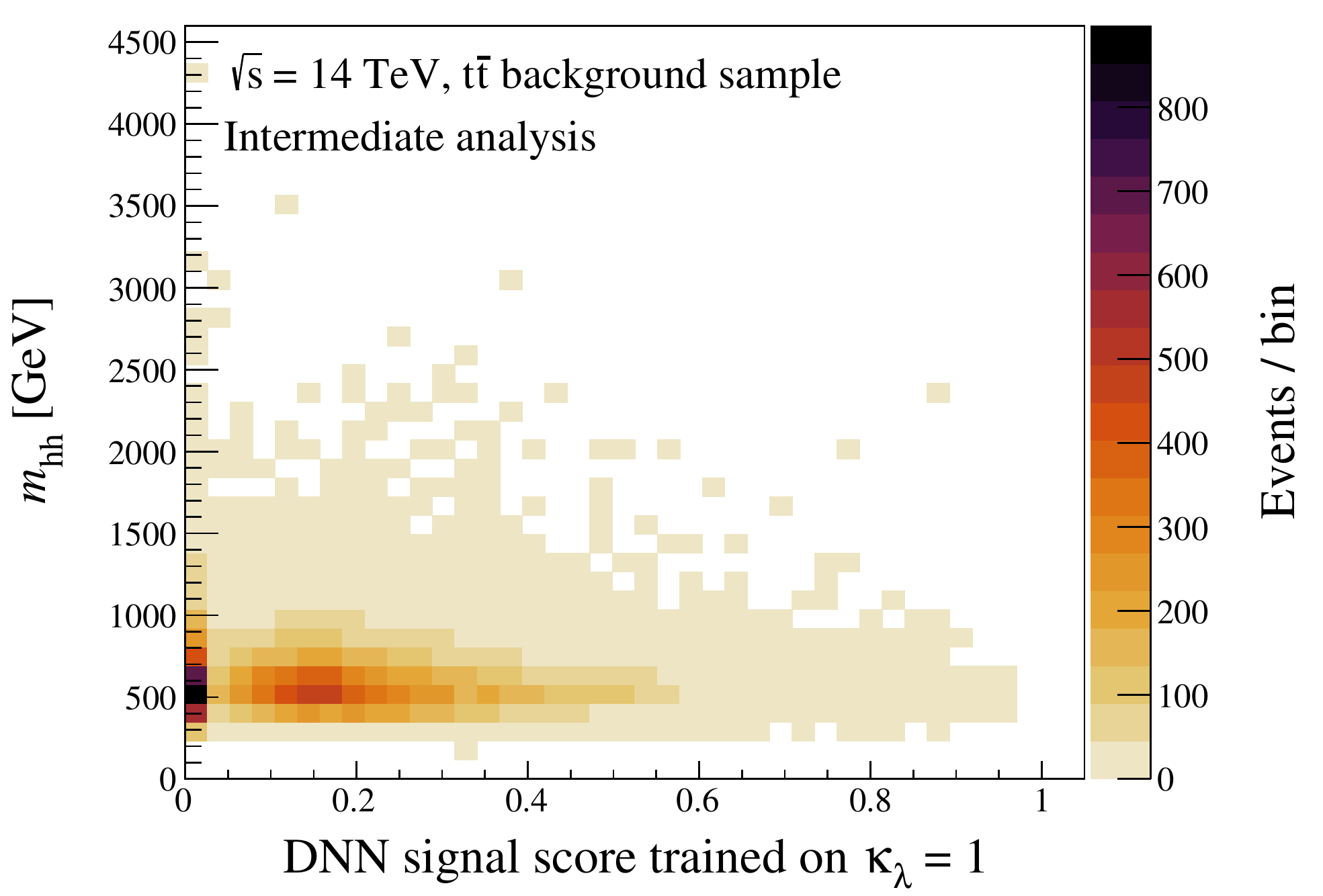}\\
        \includegraphics[width=\textwidth]{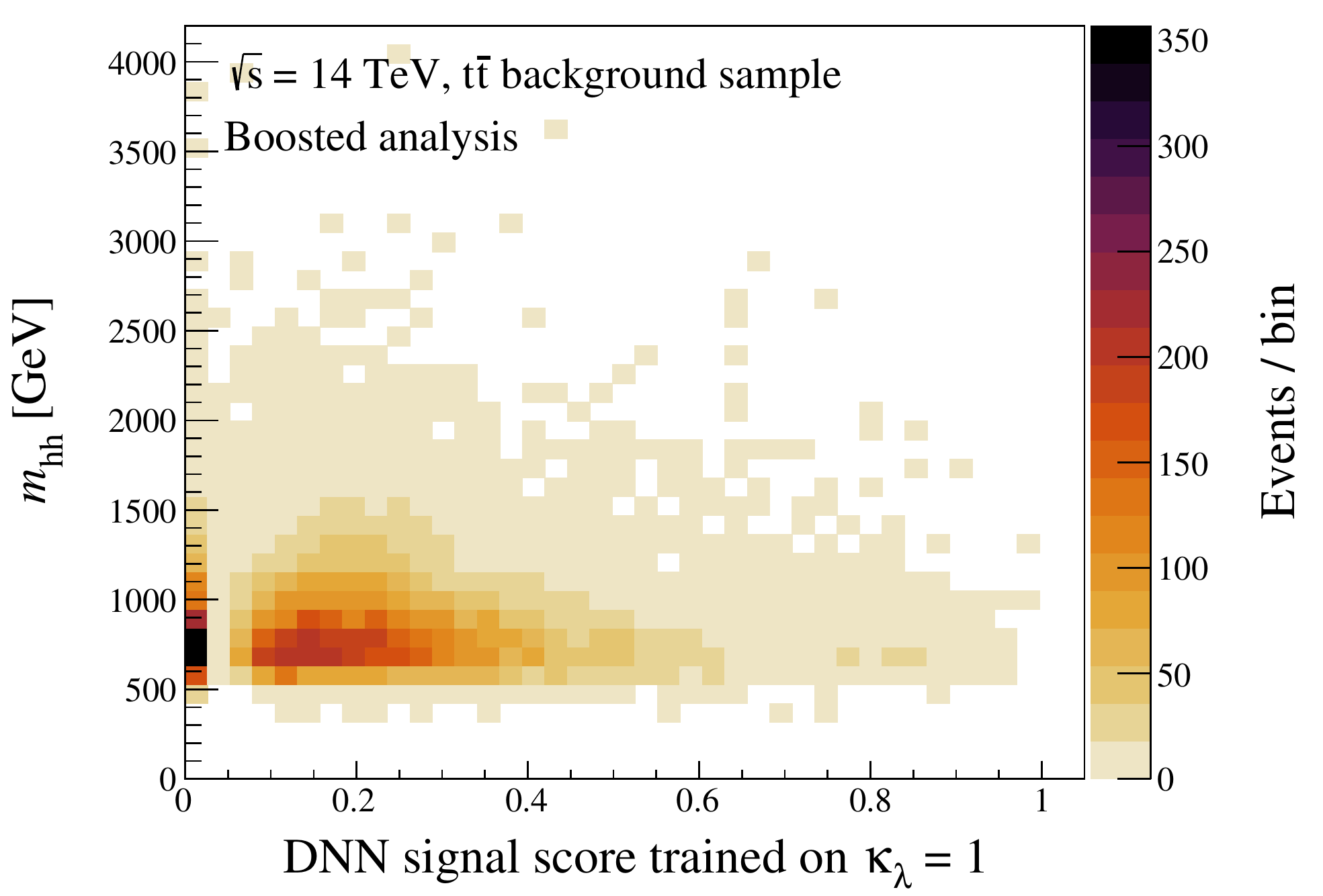}
    \caption{DNN trained on $\klam = 1$}
    \end{subfigure}%
    \begin{subfigure}[b]{0.5\textwidth}
       \includegraphics[width=\textwidth]{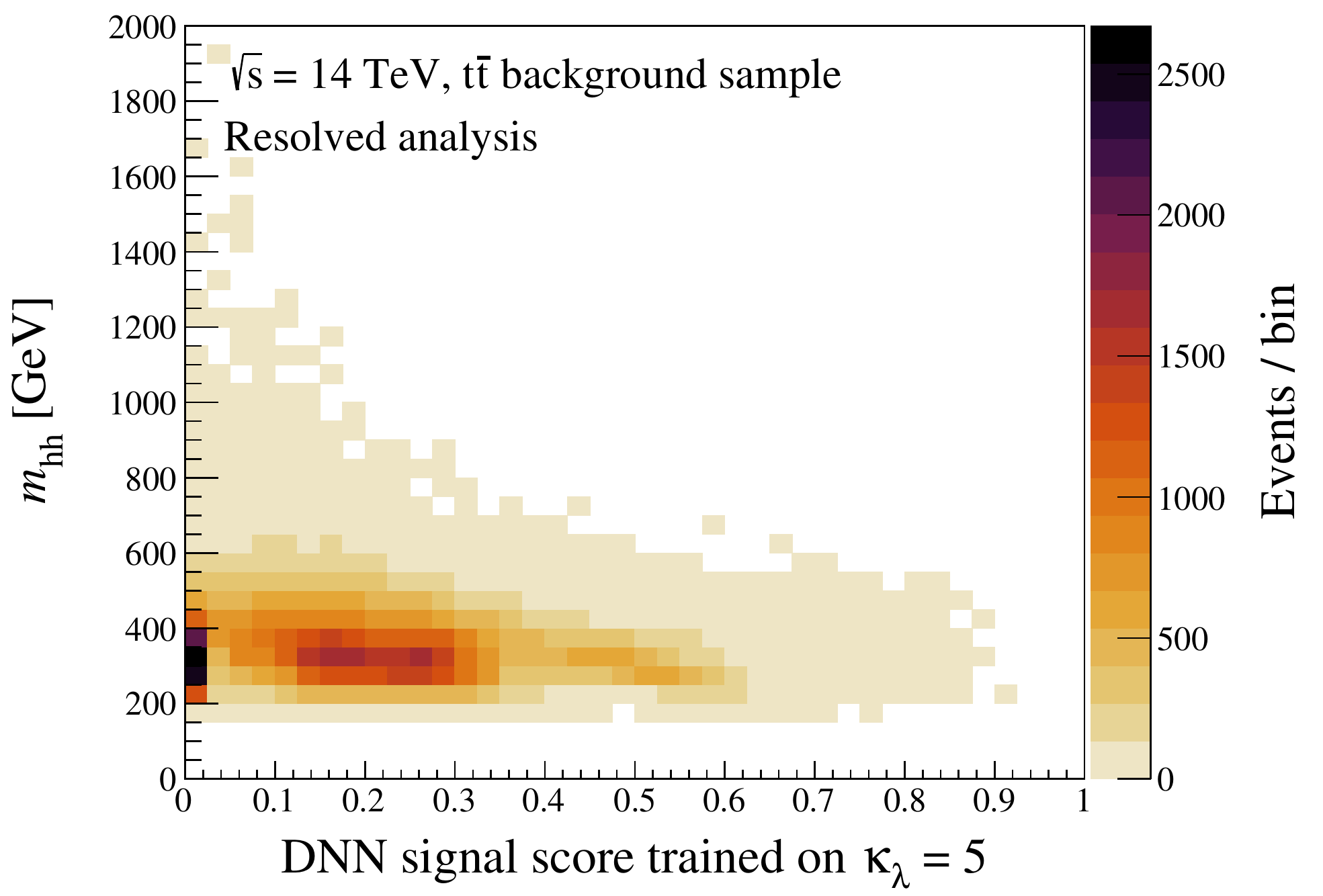}\\
       \includegraphics[width=\textwidth]{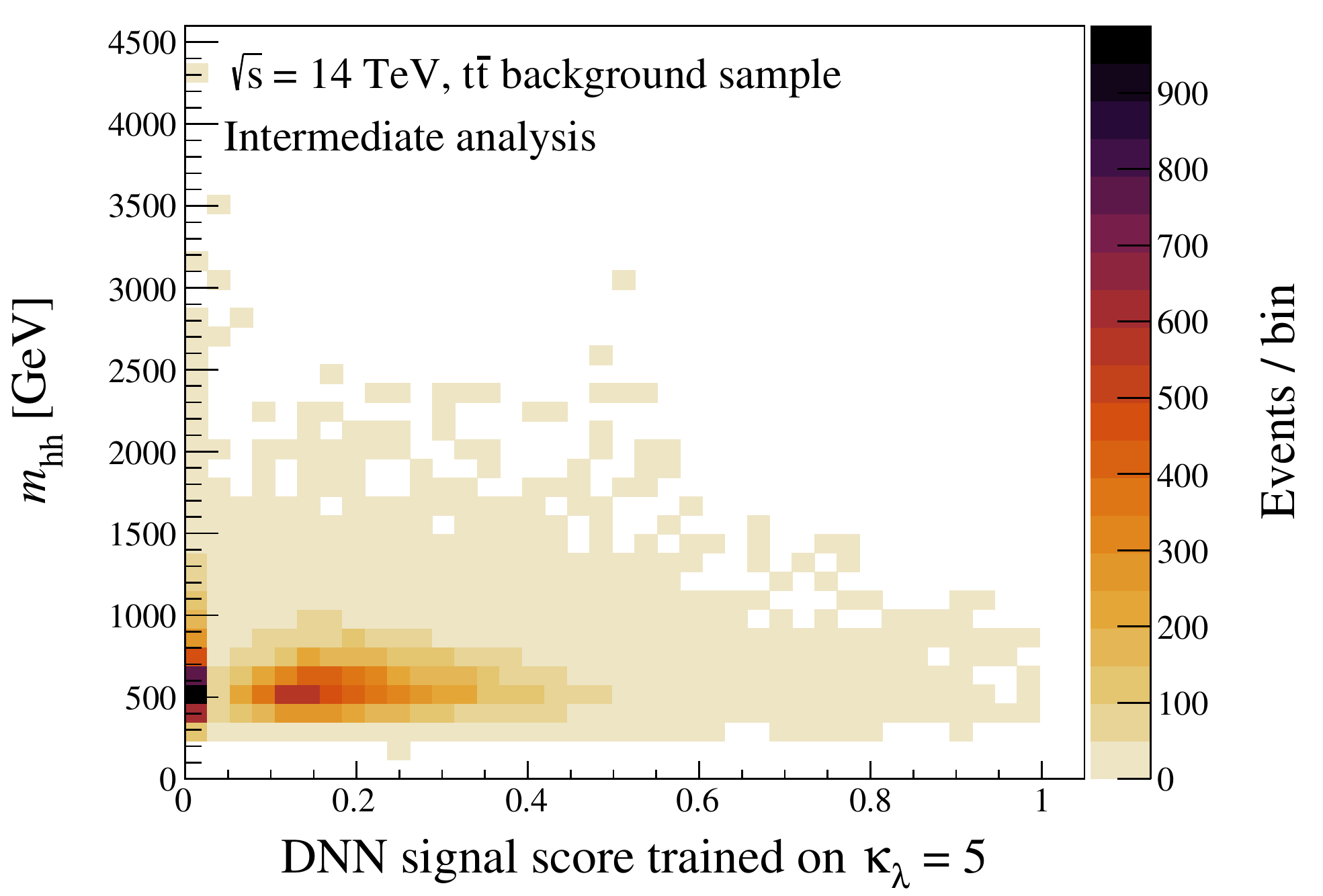}\\
       \includegraphics[width=\textwidth]{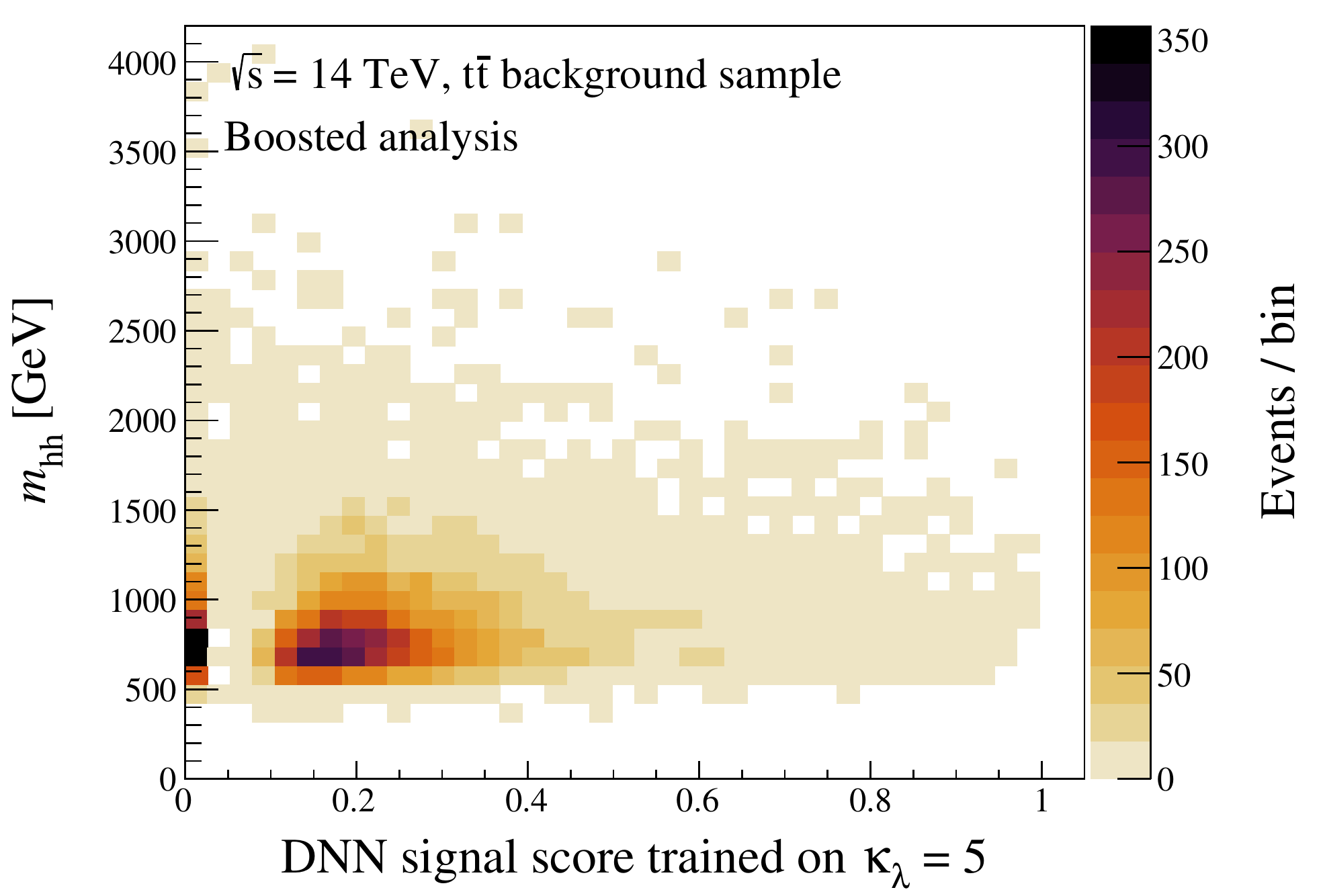}
      \caption{DNN trained on $\klam = 5$}
    \end{subfigure}
    \caption{The di-Higgs invariant mass $\mhh$ vs neural network scores trained on (a) $\klam=1$ and (b) $\klam=5$ signal sample for the (upper) resolved, (middle) intermediate, and (lower) boosted analyses. The test samples used to make these distributions are an independent set of $t\bar{t}$ events.
    }
    \label{fig:nnscore_correlation_ttbar_mhh}
\end{figure}

Figures~\ref{fig:nnscore_correlation_signal_mhh} and \ref{fig:nnscore_correlation_ttbar_mhh} show the corresponding correlation plots for the di-Higgs invariant mass $\mhh$ as a function of DNN signal score.
We see that the signal score is not strongly correlated with \mhh. 

% ttbar sample scores correlations

% ttbar sample scores correlations

\FloatBarrier
%--------------------------------
\section{\label{sec:chiSq}\texorpdfstring{Additional $\chi^2$ distributions}{Additional χ² distributions}}
%--------------------------------

This appendix collects $\chi^2$ distributions supplementing those in the main text.

\begin{figure}
    \centering
    \begin{subfigure}[b]{0.49\textwidth}
        \includegraphics[width=\textwidth]{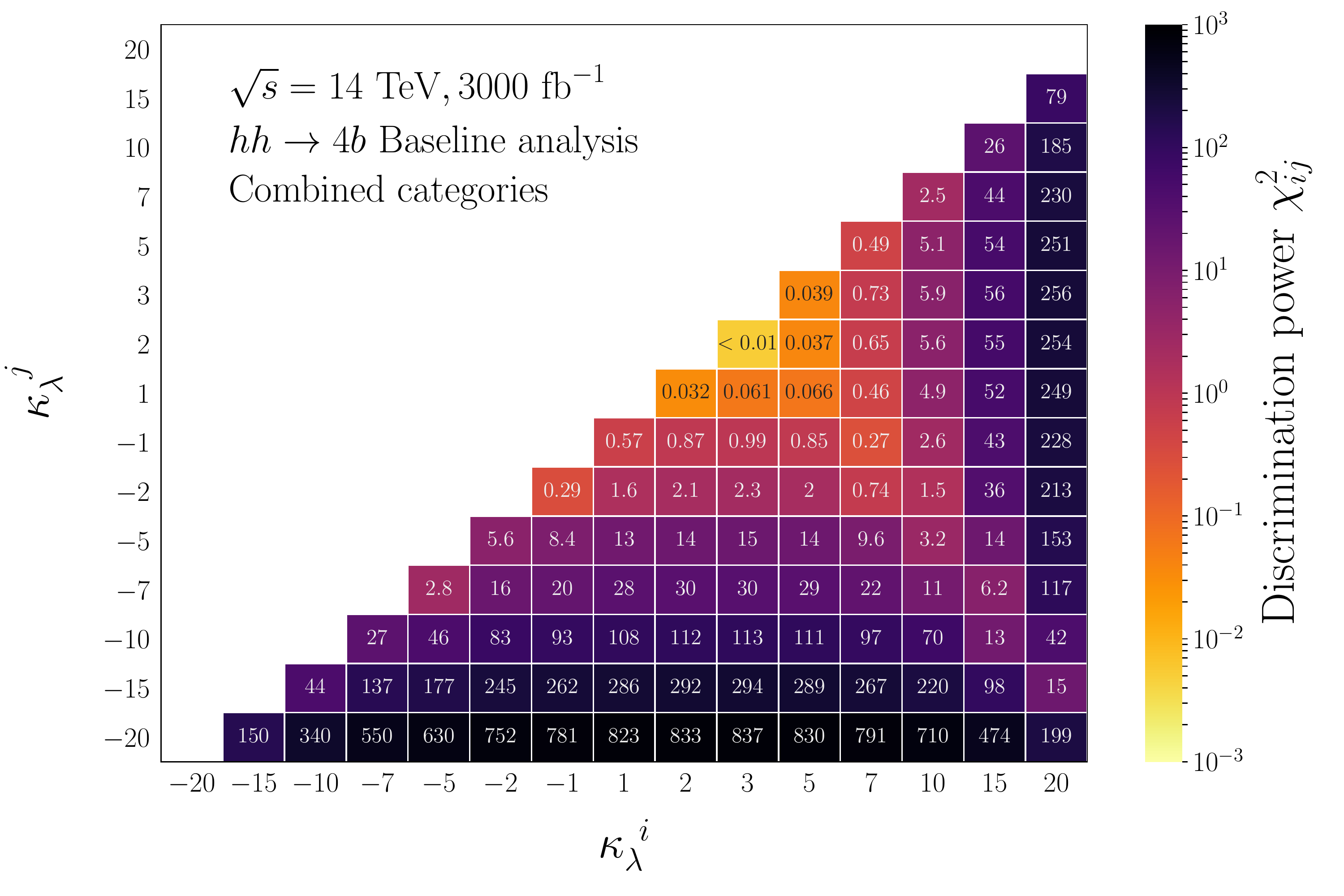}%
    \caption{Baseline analysis}
    \end{subfigure}%
    \begin{subfigure}[b]{0.49\textwidth}
        \includegraphics[width=\textwidth]{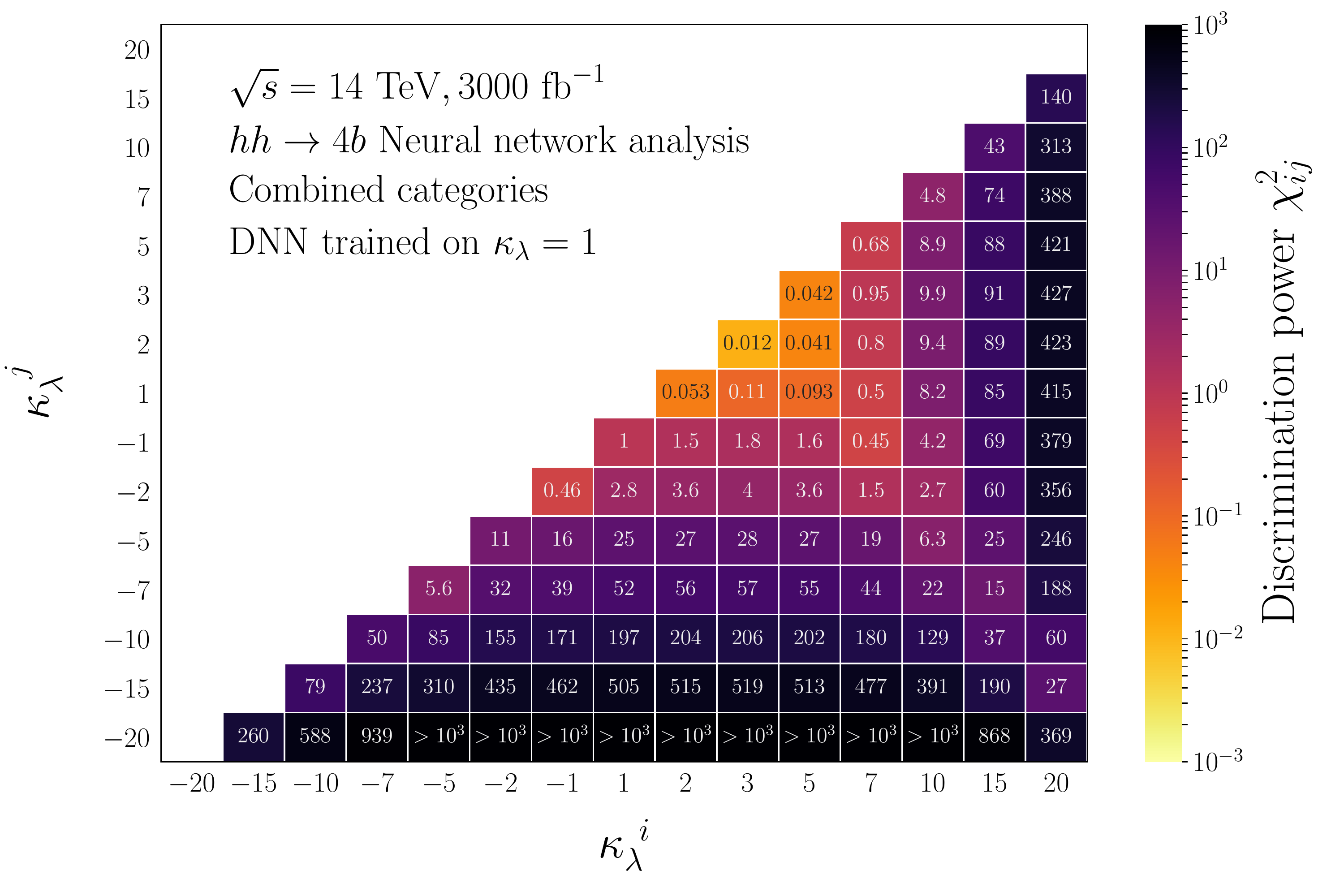}
    \caption{DNN trained on $\klam = 1$}
    \end{subfigure}
    \caption{Discrimination power $\chi^2_{ij}$ (higher is better) between two self-coupling values $\klam^{~i}$ vs $\klam^{~j}$. This is displayed for the (a) \emph{baseline} and (b) \emph{neural network} trained on $\klam = 1$ analyses. The requirement on the signal score $p_\text{signal} > 0.75$ is imposed for the DNN. The $\chi_{ij}^2$ is calculated from Eq.~\eqref{eq:generic_chiSq}. This assumes $\kapt = 1.0$ and background systematics of $\zeta_b=0.3\%$, 1\% and 5\% for the resolved, intermediate and boosted categories, respectively.}
    \label{fig:chiSqij}
\end{figure}

Figure~\ref{fig:chiSqij} shows the discrimination power $\chi_{ij}^2$ matrix between different $\klam$ hypotheses for $\kapt = 1$ for the \emph{baseline analysis} and \emph{neural network analysis} trained on $\klam = 1$.   
We find a similar pattern of regions with lower $\chi^2_{ij}$ values as Fig.~\ref{fig:limit1d_chiSq_2Dlambda_DNN} in the main text, where further discussion can be found.

\begin{figure}
    \centering
    \includegraphics[width=\textwidth]{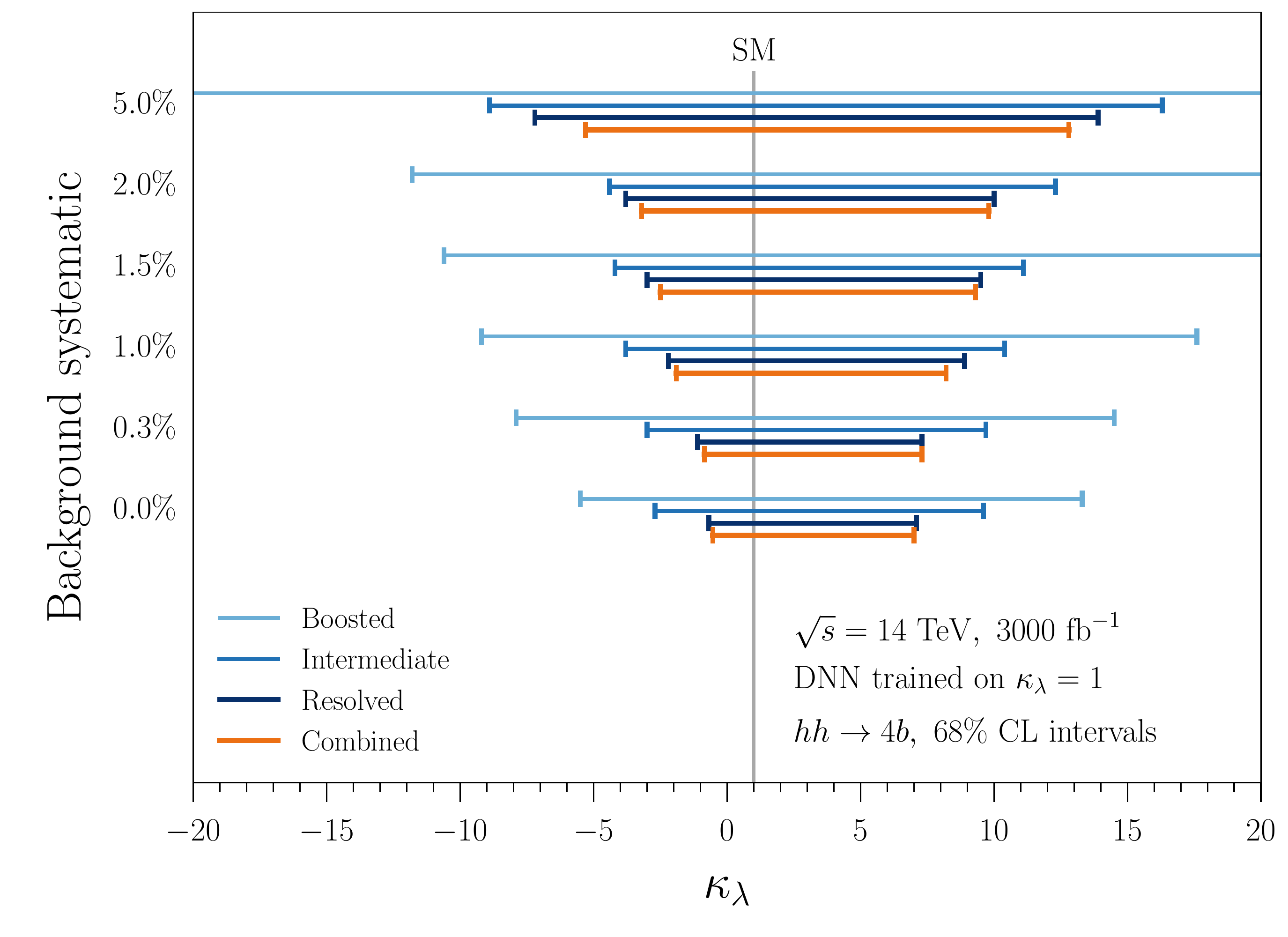}
    \caption{\label{fig:summary_1dlimits_DNNklam1} 
    Summary of 68\% CL intervals ($\chi^2 < 1$) on \klam with fixed $\kapt = 1$ for different assumed background systematics from 0\% to 5\% for the \emph{neural network analysis} trained on the $\klam = 1$ signal. This is shown for the resolved (dark blue), intermediate (medium blue), boosted (light blue) categories along with their combination (orange). The luminosity is assumed to be 3000~fb$^{-1}$. Lines without endcaps mean the 68\% CL limit is outside the considered range of $\klam \in [-20, 20]$. The vertical grey line denotes the SM value of \klam. The nominal systematics assumed are shown in Table~\ref{tab:assumed_syst} of the main text. }
\end{figure}

Figure~\ref{fig:summary_1dlimits_DNNklam1} shows a summary of the 68\% CL limits for different systematics when the DNN is trained on the $\klam = 1$ signal. The qualitative features are similar to the corresponding Fig.~\ref{fig:summary_1dlimits} where the DNN is trained on the $\klam = 5$ signal. As noted in the main text, the DNN trained on $\klam = 1$ for the intermediate and boosted categories are slightly more constraining than that trained on $\klam = 5$.
    
\begin{figure}
    \centering
    \begin{subfigure}[b]{0.49\textwidth}
        \includegraphics[width=\textwidth]{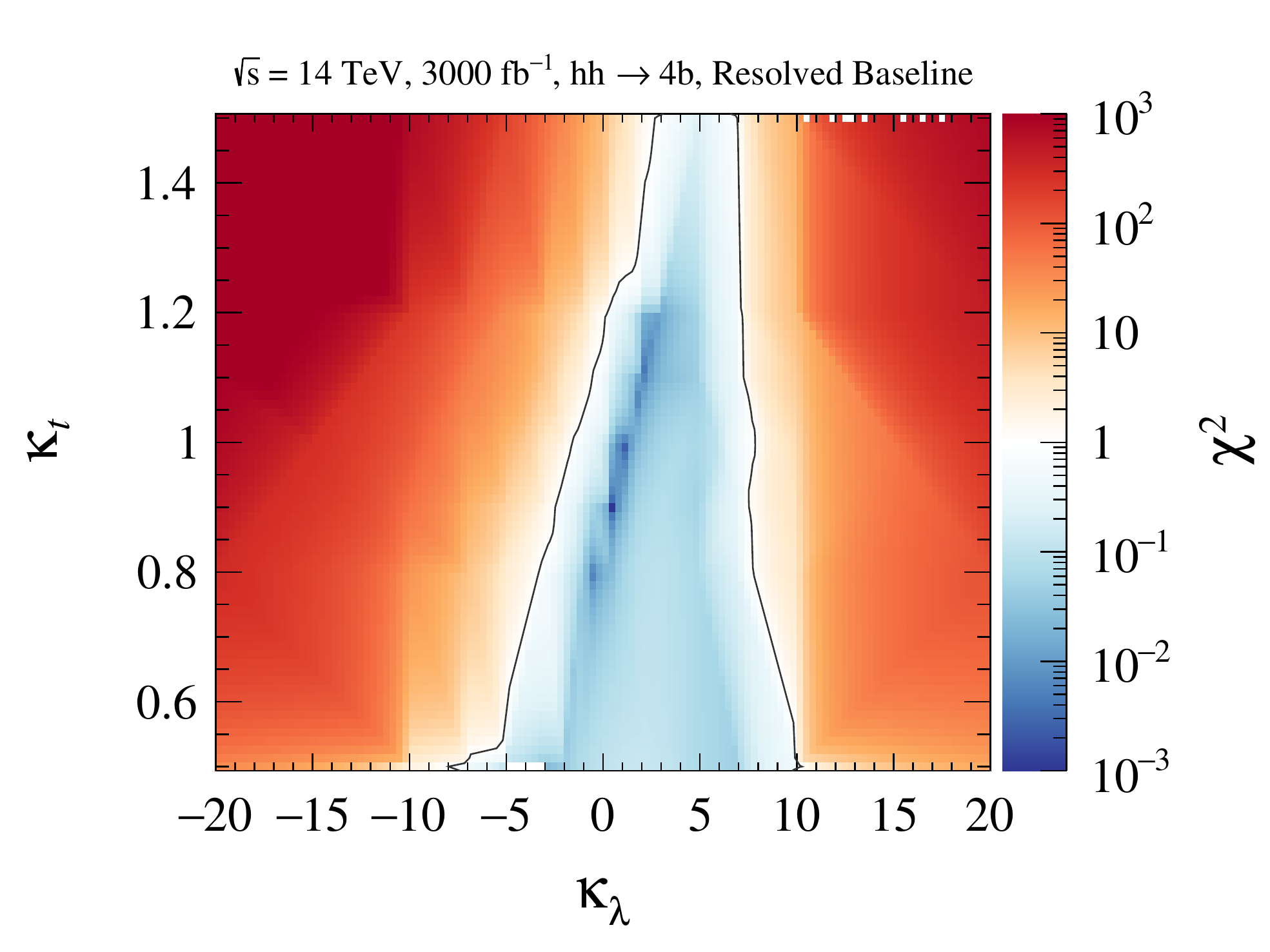}\\
        \includegraphics[width=\textwidth]{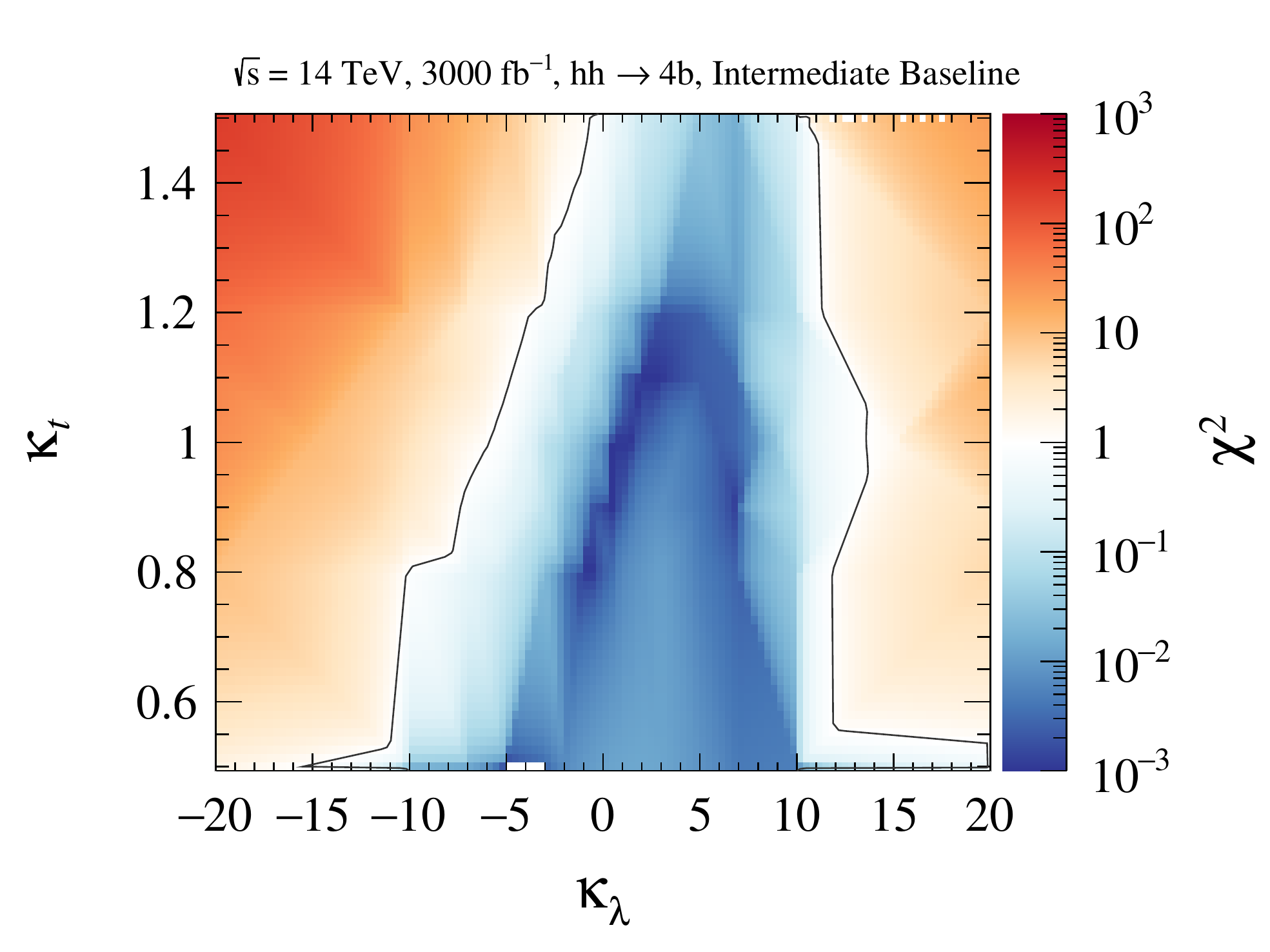}\\
        \includegraphics[width=\textwidth]{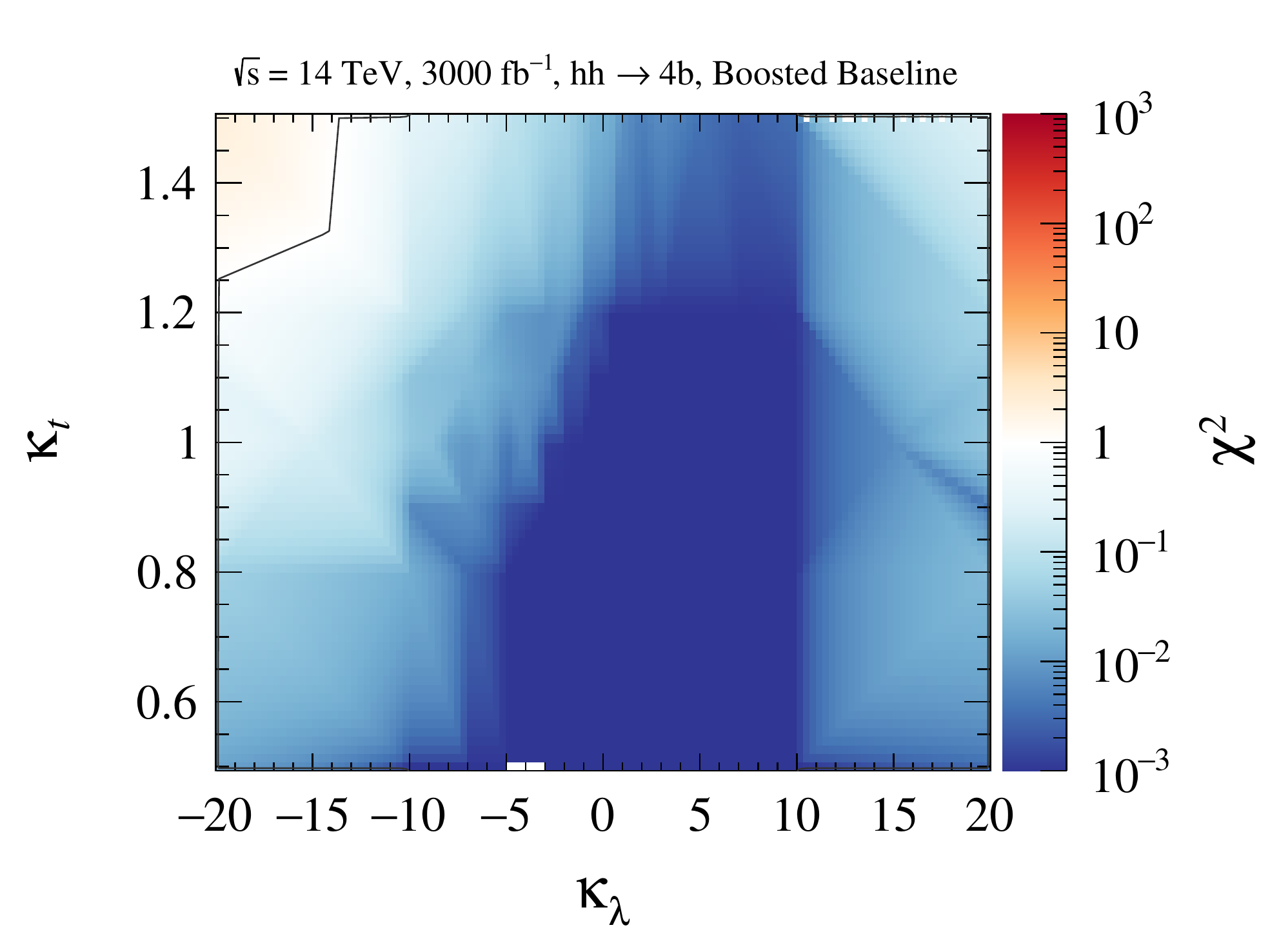}
    \caption{Baseline analysis}
    \end{subfigure}%
    \begin{subfigure}[b]{0.49\textwidth}
        \includegraphics[width=\textwidth]{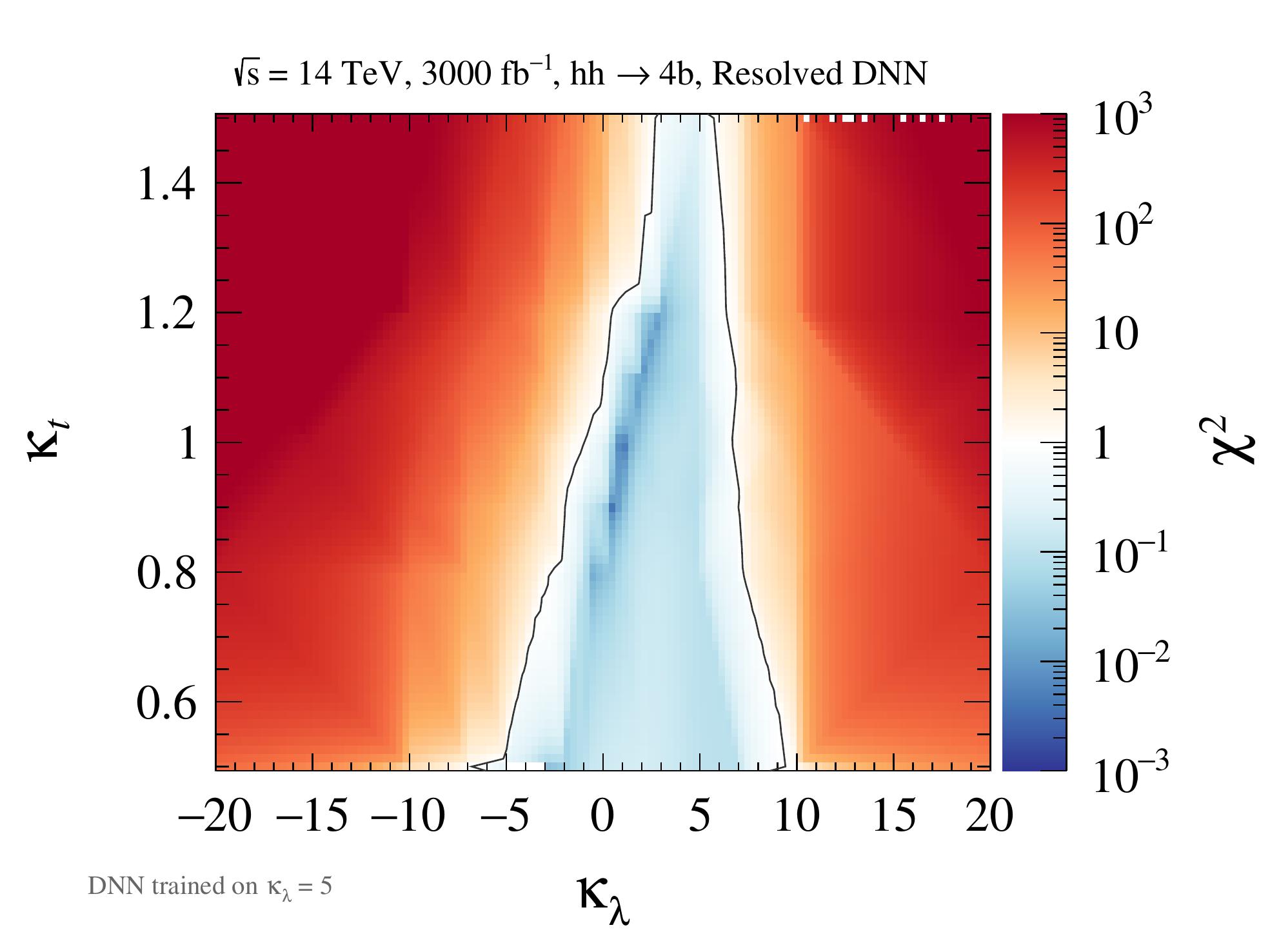}\\
        \includegraphics[width=\textwidth]{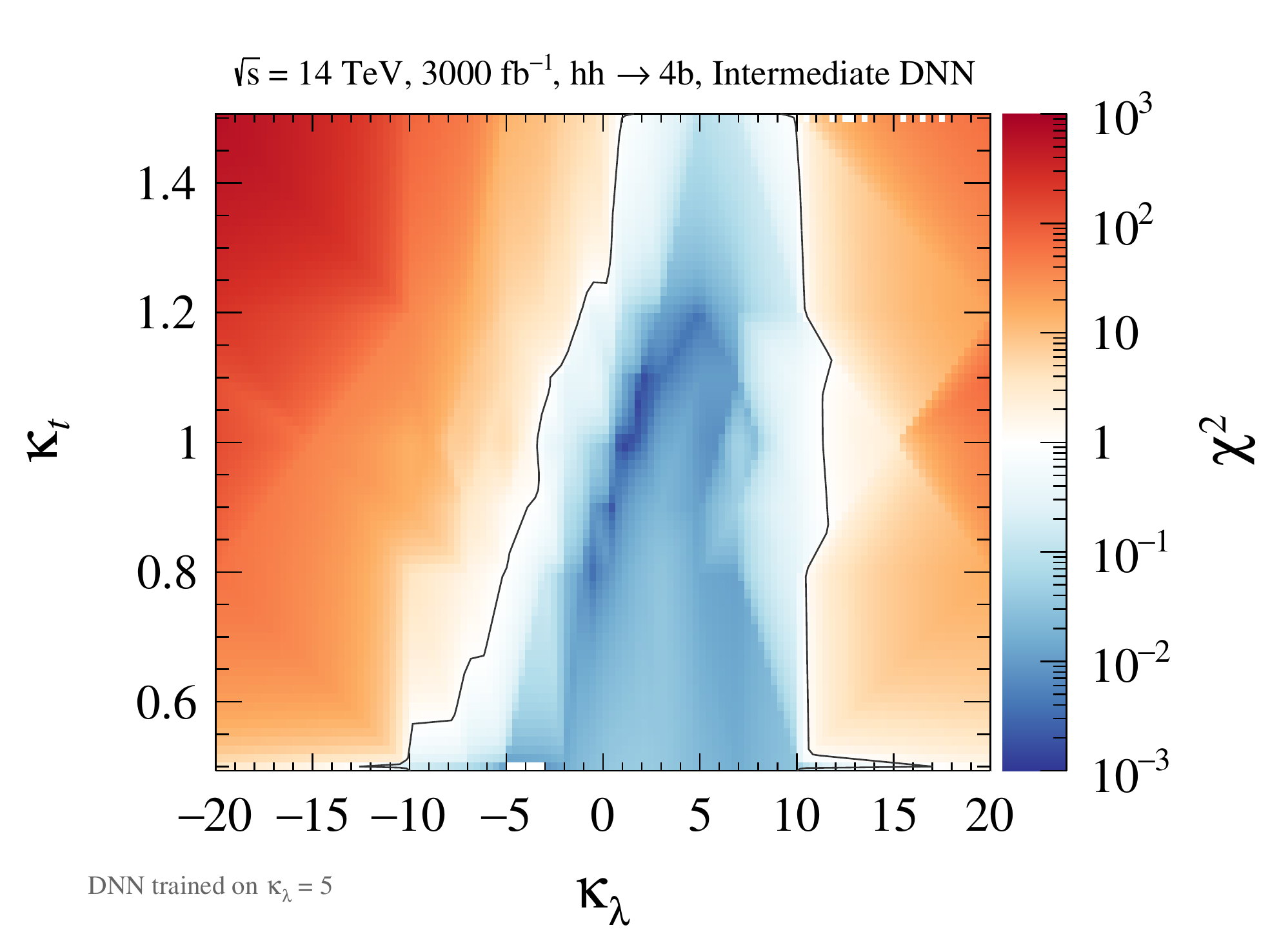}\\
        \includegraphics[width=\textwidth]{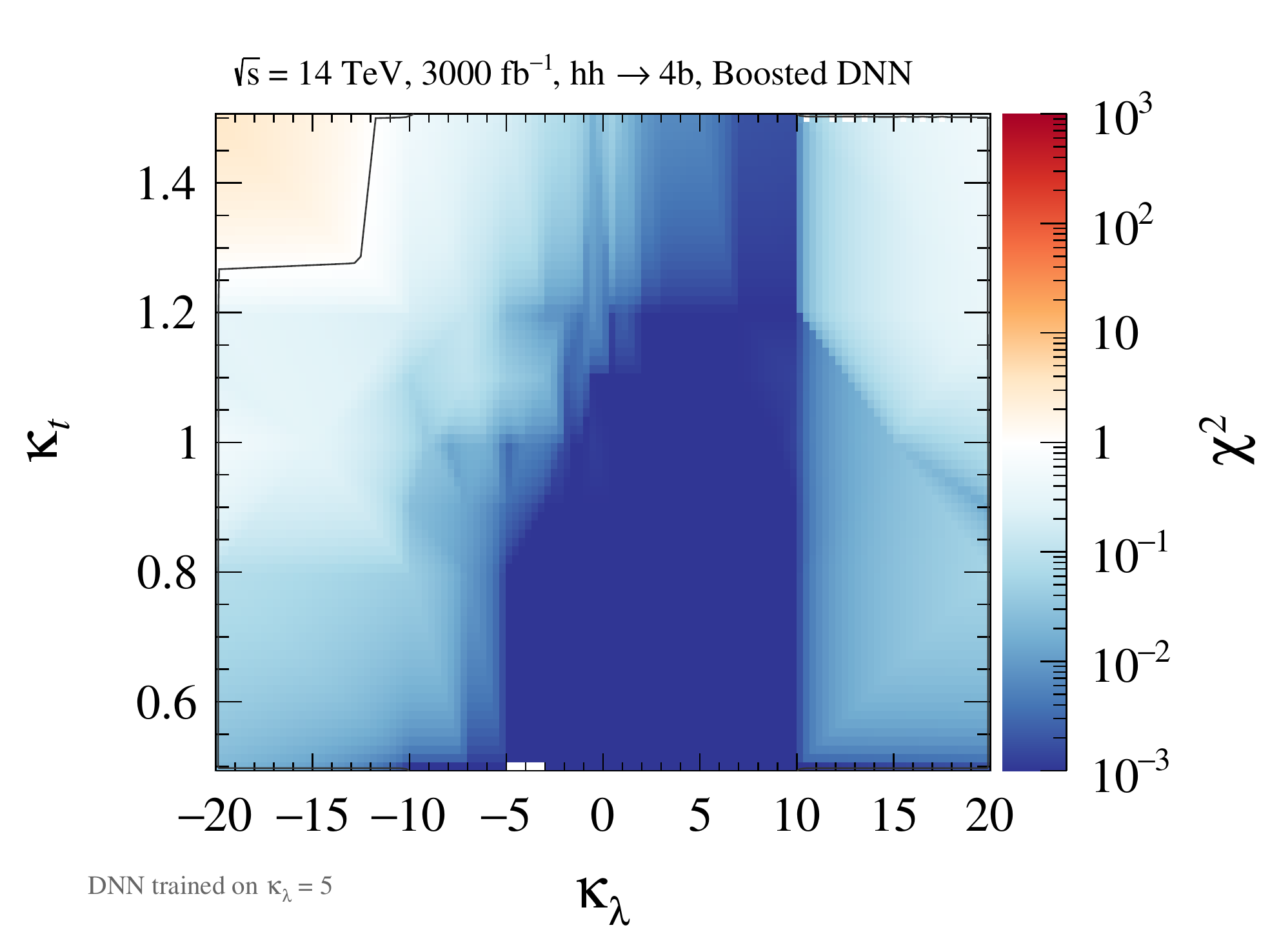}
    \caption{DNN trained on $\klam = 5$}
    \end{subfigure}
    \caption{The $\chi^2$ distributions for the (a) \emph{baseline analysis} and (b) \emph{neural network analysis} trained on the \klam = 5 signal. This is displayed in the $\kapt$ vs $\klam$ plane at $\mathcal{L} = 3000$~fb$^{-1}$. This is shown for the resolved (upper), intermediate (middle),  boosted (lower) categories assuming 0.3\%, 1\% and 5\% systematic uncertainties, respectively. The grey contour indicates $\chi^2=1$ corresponding to 68\% CL. }
    \label{fig:limit2d_chiSq_separate}
\end{figure}

Figure~\ref{fig:limit2d_chiSq_separate} shows the $\chi^2$ distributions for the \emph{baseline} and \emph{neural network} analyses trained on \klam = 5 separated by resolved, intermediate and boosted categories. They assume 3000~fb$^{-1}$ of luminosity and 0.3\%, 1\% and 5\% systematic uncertainties for the resolved, intermediate and boosted categories, respectively. The statistical combination of the three categories results in the $\chi^2$ distributions displayed in Fig.~\ref{fig:limit2d_chiSq} of the main text, where further discussion is found.

\end{document}